\documentclass[a4paper,11pt]{article}
\pdfoutput=1 

\usepackage{jheppub} 

\usepackage[T1]{fontenc} 
\usepackage{mathrsfs}
\usepackage{enumitem}
\usepackage{shuffle}
\usepackage{slashed}
\usepackage{tikz}
\usepackage{adjustbox}
\usepackage{array}
\usepackage{cancel}
\usepackage{bbm}
\usepackage{shuffle}
\usepackage{comment}

\usetikzlibrary{patterns,decorations.pathmorphing}

\def\W #1{\widetilde{#1}}
\def\WH #1{\widehat{#1}}

\def\W #1{\widetilde{#1}}
\def\WH #1{\widehat{#1}}
\def\eps{\epsilon}
\def\Tr{{\rm Tr}}

\newcommand {\Pf}  {\,\text{Pf}\,}
\newcommand {\Pfp}  {\,\text{Pf}\,'}

\def\Label#1{\label{#1}}%

\def\Spaa#1{\left\langle #1 \right\rangle}
\def\a{{\alpha}}

\def\b{{\beta}}

\newcommand{\Sl}{\sum\limits}

\newcommand{\bea}{\begin{eqnarray}}
\newcommand{\eea}{\end{eqnarray}}
\newcommand{\bean}{\begin{eqnarray*}}
\newcommand{\eean}{\end{eqnarray*}}
\newcommand{\nn}{\nonumber \\}

\title{\boldmath Expansion of All Multitrace Tree Level EYM Amplitudes}

\author[a,b]{Yi-Jian Du,}
\emailAdd{yijian.du@whu.edu.cn}
\author[c,d]{Bo Feng,}
\emailAdd{fengbo@zju.edu.cn}
\author[e,f]{and Fei Teng  \footnote{The corresponding author.}}
\emailAdd{Fei.Teng@utah.edu}

\affiliation[a]{Center for Theoretical Physics, School of Physics and Technology,
Wuhan University, \\
No.299 Bayi Road, Wuhan 430072, P.R. China}
\affiliation[b]{Suzhou Institute of Wuhan University, \\
	377 Linquan Street, Suzhou, 215123, P.R. China}
\affiliation[c]{Zhejiang Institute of Modern Physics, Department of Physics,
 Zhejiang University,\\
 No.38 Zheda Road, Hangzhou 310027, P.R. China}
\affiliation[d]{Center of Mathematical Science,
  Zhejiang University,\\
  No.38 Zheda Road, Hangzhou 310027, P.R. China}
\affiliation[e]{Department of Physics and Astronomy, University of Utah,\\ 115 South 1400 East, Salt Lake City, UT 84112, USA}
\affiliation[f]{Department of Physics and Astronomy, Uppsala University, 75108 Uppsala, Sweden}

\abstract{In this paper, we investigate the expansion of tree level multitrace Einstein-Yang-Mills (EYM) amplitudes. First, we propose two types of recursive expansions of tree level EYM amplitudes with an arbitrary number of gluons, gravitons and traces by those amplitudes with fewer traces or/and gravitons. Then we give many support evidence, including proofs using the Cachazo-He-Yuan (CHY) formula and Britto-Cachazo-Feng-Witten (BCFW) recursive relation. As a byproduct, two types of generalized BCJ relations for multitrace EYM are further proposed, which will be useful in the BCFW proof.
After one applies the recursive expansions repeatedly, any multitrace EYM amplitudes can be given in the Kleiss-Kuijf (KK) basis of tree level color ordered Yang-Mills (YM) amplitudes. Thus the Bern-Carrasco-Johansson (BCJ) numerators, as the expansion coefficients, for all multitrace EYM amplitudes are naturally constructed.}

\begin{document}
\maketitle
\flushbottom

\section{Introduction}
The modern study of scattering amplitudes has revealed a lot of hidden structures within the perturbative gauge and gravity theories, many of which are obscure from the point of view of the Lagrangian formalism. One remarkable feature, as first pointed out by Bern, Carrasco and Johansson (BCJ)~\cite{Bern:2008qj} for Yang-Mills (YM) theory, is that it is possible to make the kinematic numerator associated to each trivalent diagram satisfy the same Jacobi identity as the color factor. Then a double copy of such numerators gives directly the Einstein gravity amplitudes~\cite{Bern:2010ue}. This scheme is believed to hold for all loops {(see~\cite{Bern:2012uf} for a four-loop four-point construction)}. At tree level, the existence of such numerators are equivalent to the fact that the color ordered amplitudes satisfy the BCJ relation~\cite{Bern:2008qj}, which have been proved both in string theory and field theory context~\cite{BjerrumBohr:2009rd,Stieberger:2009hq,Feng:2010my,Chen:2011jxa}. Very surprisingly, the duality between kinematic numerator and color factor actually appears in a large variety of theories with matter interaction and supersymmetry. Moreover, the double copy construction can be applied to numerators of two different theories, which gives a family of gravities theories~\cite{Johansson:2014zca,Chiodaroli:2014xia,Chiodaroli:2015rdg,Chiodaroli:2015wal,Chiodaroli:2017ngp}. Recently, the Cachazo-He-Yuan (CHY) formalism~\cite{Cachazo:2013gna,Cachazo:2013hca,Cachazo:2013iea,Cachazo:2014nsa,Cachazo:2014xea} has made the double copy relation very explicit, and even extended it to effective field theories.

On the other hand, explicit construction of the BCJ numerators is far from trivial, even at tree level. The approach we focus on in this paper is to express the gravity amplitude (possibly with matter) in terms of the gauge amplitudes (possibly with matter) in the Kleiss-Kuijf (KK) basis.\footnote{Using the property of color factor, one can always fix two points in the gauge amplitudes through the KK relation~\cite{Kleiss:1988ne}.} Then according to~\cite{DelDuca:1999rs}, the expansion coefficients can be associated to the family of half ladder diagrams, so that we can generate the kinematic numerators associated to all trivalent diagrams using just the Jacobi identity These numerators satisfy the color kinematic duality by construction~\cite{Bern:2010yg,KiermaierTalk,Mafra:2011kj,Cachazo:2013iea}. A very straightforward realization of this idea is to use the Kawai-Lewellen-Tye relation~\cite{Kawai:1985xq}. When summing over one part of amplitudes with the momentum kernel~\cite{Bern:1998ug,BjerrumBohr:2010ta,BjerrumBohr:2010zb,BjerrumBohr:2010yc,BjerrumBohr:2010hn}, we get the desired BCJ numerators corresponding to half-ladder diagrams~\cite{KiermaierTalk,BjerrumBohr:2010hn,Fu:2014pya,Naculich:2014rta}, called  Del Duca-Dixon-Maltoni (DDM) form of BCJ numerators.
However, such numerators are in general not polynomials, possibly containing nonlocal poles. A more promising approach to obtain polynomial numerators is the proper recursive expansion of the corresponding CHY integrand. In~\cite{Nandan:2016pya}, such expansion was performed for Einstein-Yang-Mills (EYM) amplitudes with up to three gravitons and two gluon traces.\footnote{For one graviton case, see also~\cite{Stieberger:2016lng,delaCruz:2016gnm}. String theory approaches can be found in  \cite{Stieberger:2016lng,Schlotterer:2016cxa}.} Shortly after, the expansion was done for an arbitrary number of gravitons in both EYM~\cite{Teng:2017tbo} and Einstein gravity~\cite{Du:2017kpo}. Meanwhile, by choosing a proper ansatz, one can show that such recursive expansion scheme can be uniquely fixed only by gauge invariance~\cite{Berg:2016fui,Arkani-Hamed:2016rak,Fu:2017uzt,Chiodaroli:2017ngp}.

The goal of this paper is to provide a complete recursive expansion scheme for generic EYM amplitudes with an arbitrary number of gluon traces and gravitons. Namely, given an EYM amplitude, we can express it by a linear combination of those amplitudes with fewer gluon traces and gravitons, where the coefficients are polynomials of kinematic variables. As a result, after carrying out the recursive expansion down to the pure YM level (single trace, no graviton), one can simply obtain the polynomial DDM form of BCJ numerators {for YM-scalar with both $\phi^3$ and $\phi^4$ interactions}. By analyzing the multitrace EYM integrand obtained from generalized compactification, we derive two recursive expansions, later called type-I and type-II. The type-I expansion is applicable if there is at least one graviton, while type-II is applicable if there are at least two gluon traces. Therefore, to reach the final form that involves only KK basis of pure YM amplitudes, in general both type-I and II expansions should be used. Very interestingly, through the trick ``turning a graviton into a trace of gluons'', one can ``derive'' both recursive expansions from the single trace expansion as given in~\cite{Fu:2017uzt,Teng:2017tbo}. We also give a proof using the BCFW recursive relation~\cite{Britto:2004ap,Britto:2005fq}. Furthermore, in appendix~\ref{sec:unify}, we show that starting from the Einstein gravity expansion~\cite{Fu:2017uzt,Du:2017kpo}, in principle one can apply the transmutation operator~\cite{Cheung:2017ems} to understand our recursive expansions for multitrace EYM amplitudes by the unifying relations, although the calculation is nontrivial. We demonstrate this point by an example in appendix~\ref{sec:unify}.


Another important result is that there are also two types of multitrace BCJ relations (i.e., the BCJ relations for multitrace EYM amplitudes) associated with the two types of recursive expansions. The type-I BCJ relation is actually the gauge invariance identity of the type-I recursive expansion. The reason is that to write down the expansion, we have to single out a graviton whose gauge invariance is not explicit. On the other hand, for the type-II expansion, all the particles are manifestly gauge invariant, while the type-II BCJ relation can be understood as the consistency condition at a special collinear limit. Alternatively, we can also understand it as a consistency condition under a cyclic permutation of a gluon trace. If we completely expand every multitrace amplitude involved into KK basis of pure YM ones, both the multitrace BCJ relations can be proved using just the generalized BCJ relation \cite{BjerrumBohr:2009rd,Chen:2011jxa} of the pure YM amplitudes.

Besides the generalized compactification, one can also obtain the generic EYM integrand through squeezing gluons into traces. Actually, the first double expansion were derived using just the squeezing-form integrand~\cite{Nandan:2016pya}. In this paper, we also derive a recursive expansion using the squeezing-form integrand for generic pure gluon multitrace EYM amplitudes, later called the squeezing-form recursive expansion. 
If we carry out this expansion into pure YM amplitudes, the result is of a new type, which is related to the KK basis by some algebraic transformations.
Of course, as we will show in appendix~\ref{sec:equiv}, this recursive expansion is equivalent to our aforementioned type-II expansion.

The structure of this paper is the following. In section~\ref{sec:integrand}, we give a brief review of the CHY formalism, including both the generalized compactification and squeezing formulation of EYM integrand. In section~\ref{sec:recursion}, we present the main result: the type-I and II recursive expansions, using the trick of ``turning a graviton into a trace of gluons''. Then in section~\ref{sec:compact}, we provide an understanding of these recursive expansions through a dimensional reduction scheme. Next, we study pure gluon EYM amplitudes under the squeezing formulation. We present examples up to four gluon traces in section~\ref{sec:multi_example} and the general recursive expansion in section~\ref{sec:genexp}. In section~\ref{sec:generalBCJ}, we study the type-I and II multitrace BCJ relations obtained from the corresponding recursive expansions. The correctness of these BCJ relations give a strong support to our recursive expansions. Further consistency checks are given in section~\ref{sec:check}. Then in section~\ref{sec:BCFW}, we give a BCFW proof to our recursive expansions. In section~\ref{sec:YMexpansion}, we present the algorithm to write down directly the coefficients of the final KK basis pure YM expansion, which is nothing but the BCJ numerators {(for YM-scalar with $\phi^3$ and $\phi^4$ interactions)} we are pursuing all the way through. We draw our conclusion and discuss some future directions in section~\ref{sec:conclusion} and leave some technical details used in our calculation into the appendices.

In this paper, we give a lot of details, including many examples, in order to help the readers to understand our recursive expansions. However, we give the following shortcut to those readers who want to quickly grab our main results. We define our notations at the beginning of section~\ref{sec:recursion}, while the type-I and type-II recursive expansions are given in Eq.~\eqref{eq:type1step1} and \eqref{eq:type2step1}. Then the squeezing-form recursive expansion for pure gluon EYM amplitudes is given in Eq.~\eqref{eq:puregluonexp}. The equivalence to the type-II expansion is proved in appendix~\ref{sec:equiv}. Our type-I and type-II expansions lead to two interesting generalized BCJ relations for multitrace EYM amplitudes, which are presented in Eq.~\eqref{Eq:GeneralizedBCJType1} and \eqref{Eq:GeneralizedBCJType2}. Iterating these recursive expansions, we can finally obtain an expansion in terms of pure YM amplitudes in the KK basis. We give a set of graphic rules to directly write down these expansion coefficients in section~\ref{sec:graphicrules}. These rules generalize the ones for single trace EYM amplitudes given in~\cite{Fu:2017uzt,Teng:2017tbo}.


\section{A brief review of CHY integrands for multitrace EYM amplitudes}\label{sec:integrand}

The CHY
formalism~\cite{Cachazo:2013gna,Cachazo:2013hca,Cachazo:2013iea,Cachazo:2014nsa,Cachazo:2014xea}
gives a novel representation to field theory amplitudes. For a large
class of theories, we can write the tree level amplitudes as
\begin{equation}
\label{eq:integral}
    A_{n}=(\text{phase})\times\int{d\Omega_{\text{CHY}}}\,\mathcal{I}_{\text{CHY}}(n)\,,
\end{equation}
where $\mathcal{I}_{\text{CHY}}(n)$ is the CHY integrand and the
integral measure  $d\Omega_{\text{CHY}}$ imposes the tree level
$n$-point scattering equations:
\begin{align}
  \sum_{b=1\,,b\neq a}^{n}\frac{k_a\cdot k_b}{\sigma_{ab}}=0\qquad\qquad\sigma_{ab}\equiv\sigma_a-\sigma_b\,, & & a\in\{1,2\ldots n\}\,.
\end{align}
To compare results from different approaches, we need a phase factor between the physical amplitude and the CHY integral. Such a phase factor has no physical consequence if it depends only on particle numbers and species. For example, we have:\footnote{{These overall phases do not change the cross sections, but they will result in a simpler expression for the coefficients of the recursive expansions to be discussed later in this work.}}
\begin{align}
\label{eq:phase}
&\text{YM:}\quad(\text{phase})=(-1)^{\frac{(n+1)(n+2)}{2}}\nonumber\\
&\text{single trace EYM:}\quad(\text{phase})=(-1)^{\frac{(n+1)(n+2)}{2}}(-1)^{\frac{|\mathsf{H}|(|\mathsf{H}|+1)}{2}}\,,
\end{align}
where $n$ is the total number of particles and $|\mathsf{H}|$ is the number of gravitons. In general, the integrand $\mathcal{I}_{\text{CHY}}(n)$ can
be factorized into two parts:
\begin{equation}
\mathcal{I}_{\text{CHY}}(n)=\mathcal{I}_{L}(n)\mathcal{I}_{R}(n)\,.
\end{equation}
This factorization makes it manifestly the double copy
construction~\cite{Bern:2010ue} for many theories, such as the original KLT relation~\cite{Kawai:1985xq}. It also provides natural frame where one kind of amplitudes is expanded by another kind of
amplitudes as discussed in recent papers~\cite{Fu:2017uzt,Teng:2017tbo,Du:2017kpo}

In this work, we are particularly interested in the YM and
multitrace EYM integrand,  both of which involve the $2n\times 2n$
antisymmetric matrix $\Psi$ defined as:
\begin{equation}
    \Psi=\left(\begin{array}{cc}
        A & -C^{T} \\
        C & B \\
    \end{array}\right)\,.
\end{equation}
Each block of $\Psi$ is an $n\times n$ matrix given below as:
\begin{align}
& A_{ab}=\left\{\begin{array}{>{\displaystyle}l @{\hspace{1em}} >{\displaystyle}l}
\frac{k_a\cdot k_b}{\sigma_{ab}} & a\neq b\\
0 & a=b \\
\end{array}\right.&
& B_{ab}=\left\{\begin{array}{>{\displaystyle}l @{\hspace{1em}} >{\displaystyle}l}
\frac{\epsilon_{a}\cdot\epsilon_{b}}{\sigma_{ab}} & a\neq b\\
0 & a=b \\
\end{array}\right.&
& C_{ab}=\left\{\begin{array}{>{\displaystyle}l @{\hspace{1em}} >{\displaystyle}l}
\frac{\epsilon_{a}\cdot k_{b}}{\sigma_{ab}} & a\neq b\\
-\sum_{c\neq a}\frac{\epsilon_{a}\cdot k_{c}}{\sigma_{ac}} & a=b \\
\end{array}\right.\,,
\label{eq:ABC}
\end{align}
with $1\leqslant a,b,c\leqslant n$. On the support of the CHY
measure $d\Omega_{\text{CHY}}$, $\Psi$ is a co-rank two matrix. We
can easily show that the \emph{reduced Pfaffian}:
\begin{equation}
\label{eq:Pfp}
    \Pfp(\Psi)=\frac{(-1)^{i+j}}{\sigma_{ij}}\Pf(\Psi_{ij}^{ij})\qquad 1\leqslant i<j\leqslant n\,,
\end{equation}
where $\Psi_{ij}^{ij}$ is obtained from $\Psi$ by deleting the $i$-th and $j$-th rows and columns, is nonzero in general, and independent of the rows and columns deleted, when evaluated on the support of the CHY measure.
If all the $n$ particles are gluons, the following CHY integrand
\begin{align}
\text{YM:} & &\mathcal{I}_{L}=\frac{1}{\langle 12\ldots n\rangle}\equiv\frac{1}{\sigma_{12}\sigma_{23}\ldots\sigma_{n-1,n}\sigma_{n1}} & &\mathcal{I}_{R}=\Pfp(\Psi)
\end{align}
leads to the tree level color ordered YM amplitudes upon the integration of
\eqref{eq:integral}.  Interestingly, if we replace the Parke-Taylor
factor $\langle 12\ldots n\rangle^{-1}$ by another copy of $\mathcal{I}_{R}$, we
get the Einstein gravity integrand
\begin{align}
\text{Einstein gravity:} & &\mathcal{I}_{L}=\Pfp(\Psi) & &\mathcal{I}_{R}=\Pfp(\Psi)\,.
\end{align}
In this way, the CHY formalism manifests the double copy relation
between gauge and  gravity theory, providing a way to directly obtain the
BCJ numerators that satisfy the color-kinematic
duality~\cite{Cachazo:2013iea,Bjerrum-Bohr:2016axv,Du:2017kpo}.

Next, we consider the scattering between gluons and gravitons,
described by  the Einstein-Yang-Mills (EYM) interaction:
\begin{equation}
\label{eq:minimal}
  \sqrt{-g}\,\mathcal{L}_{\text{YM}}\;\xrightarrow{g_{\mu\nu}=\eta_{\mu\nu}+h_{\mu\nu}}\;\frac{1}{2}h^{\mu\nu}T_{\mu\nu}+\mathcal{O}(h^2)\,,
\end{equation}
where $h_{\mu\nu}$ (graviton) is the metric perturbation around the
flat one $\eta_{\mu\nu}$ and $T_{\mu\nu}$ is the energy momentum
tensor of the YM. Generic tree level  EYM amplitudes can be
decomposed into color ordered partial EYM amplitudes classified by
the number of gluon color traces and gravitons. The single trace partial
amplitudes are contributed by those Feynman diagrams in which gluons
are all connected (or equivalently, all the gravitons are external).
In general, we are interested in the $n$-point tree level EYM
amplitude with graviton set $\mathsf{H}$ and gluon set $\mathsf{G}$,
where the gluons fall into $m$ color traces:
%
\begin{equation}
  \mathsf{G}=\pmb{1}\cup\pmb{2}\cup\ldots\cup\pmb{m}\,.
\end{equation}
The CHY integrand should still have $\mathcal{I}_{R}=\Pfp(\Psi)$, which encodes all the gluon polarizations and half of the graviton polarizations. The rows and columns of $\Psi$ take values in two copies of $\mathsf{G}\cup\mathsf{H}$. We can thus decompose the blocks of $\Psi$ as
\begin{equation}
\Psi\equiv\Psi_{\mathsf{H},\mathsf{G}|\mathsf{H},\mathsf{G}}=\left(\begin{array}{cccc}
A_{\mathsf{HH}} & A_{\mathsf{HG}} & -(C_{\mathsf{HH}})^{T} & -(C_{\mathsf{GH}})^{T} \\
A_{\mathsf{GH}} & A_{\mathsf{GG}} & -(C_{\mathsf{HG}})^{T} & -(C_{\mathsf{GG}})^{T} \\
C_{\mathsf{HH}} & C_{\mathsf{HG}} & B_{\mathsf{HH}} & B_{\mathsf{HG}} \\
C_{\mathsf{GH}} & C_{\mathsf{GG}} & B_{\mathsf{GH}} & B_{\mathsf{GG}}
\end{array}\right)\,.
\label{eq:Psidecompose}
\end{equation}
In this expression, for example, $A_{\mathsf{HH}}$ is the submatrix
of $A$  with rows and columns taking values in $\mathsf{H}$, and the
rows of $A_{\mathsf{GH}}$ take values in $\mathsf{G}$ while the
columns in $\mathsf{H}$. Starting from this $\Psi$, there exist two
equivalent ways to construct $\mathcal{I}_{L}$ from $\Pfp(\Psi)$: compactification inspired construction and squeezing~\cite{Cachazo:2014xea}.
They will play complementary roles in this work, since they make
manifest different aspects of generic EYM amplitudes. Here, we give
them a brief introduction.\footnote{{When using the CHY formalism of EYM, besides the minimal coupling given in Eq.~\eqref{eq:minimal}, one also has dilaton and $B$-field appearing as internal states.}}


\subsection{Compactification inspired construction}
The first proposal for $\mathcal{I}_{L}$ comes from a generalization
of  the Einstein-Maxwell integrand, which is obtained by
compactifying the $\Pfp(\Psi)$ in higher dimensions. In the current
work, we call this {\emph{compactification inspired construction (CIC)}}.
The resultant $m$-trace color ordered integrand has the following
form~\cite{Cachazo:2014xea}:
\begin{equation}
\label{eq:generalcompact}
\mathcal{I}_L(n)
=\frac{1}{\langle\pmb{1}\rangle}\sum_{a_2<b_2\in\pmb{2}}\ldots\sum_{a_{m}<b_{m}\in\pmb{m}}\left[\,\prod_{i=2}^{m}\frac{\sigma_{a_ib_i}}{\langle\pmb{i}\rangle}\,\right]\Pf\!\left(\Psi_{\mathsf{H},\{a_2,b_2\ldots a_{m},b_{m}\}|\mathsf{H}}\right)\,.
\end{equation}
Here is a digestion of the symbols used in this expression:
\begin{itemize}

    \item We first construct the ordered list
    \begin{equation}
    \{a_2,b_2,a_3,b_3\ldots a_{m-1},b_{m-1},a_m,b_m\}\subset\mathsf{G}\,,
    \end{equation}
    where $a_i<b_i$ are in trace $\pmb{i}$.

    \item 
    The matrix $\Psi_{\mathsf{H},\{a,b\}|\mathsf{H}}$ can be construct as follows. We first keep only the rows (columns) in the lower (right) half of $\Psi_{\mathsf{H},\mathsf{G}|\mathsf{H},\mathsf{G}}$ that are in $\mathsf{H}$, which gives:
    \begin{equation}
    \Psi_{\mathsf{H},\mathsf{G}|\mathsf{H}}=\left(\begin{array}{ccc}
    A_{\mathsf{HH}} & A_{\mathsf{HG}} & -(C_{\mathsf{HH}})^{T} \\
    A_{\mathsf{GH}} & A_{\mathsf{GG}} & -(C_{\mathsf{HG}})^{T} \\
    C_{\mathsf{HH}} & C_{\mathsf{HG}} & B_{\mathsf{HH}}
    \end{array}\right)\,.
    \end{equation}
    To construct $\Psi_{\mathsf{H},\{a_2,b_2\ldots a_m,b_m\}|\mathsf{H}}$ from $\Psi_{\mathsf{H},\mathsf{G}|\mathsf{H}}$, we only keep those rows and columns of $\mathsf{G}$ that take values in $\{a_2,b_2\ldots a_m,b_m\}$, and \emph{order them accordingly}. In~\cite{Cachazo:2014xea}, the integrand carries an alternating sign function $\text{sgn}(\{a,b\})$. It will appear if we always keep $\{a,b\}$ in $\Psi_{\mathsf{H},\{a,b\}|\mathsf{H}}$ in the canonical order.
    %
    \item Finally, although in Eq.~\eqref{eq:generalcompact}, the first trace $\pmb{1}$ is treated differently from the others. Actually, $\mathcal{I}_{L}(n)$ is invariant when we interchange the role of $\pmb{1}$ with any other traces, as long as $\mathcal{I}_{L}(n)$ is evaluated on the solutions to the scattering equations.
\end{itemize}
%
In this expression, we note that $\Pf\left(\Psi_{\mathsf{H},\{a_2,b_2\ldots a_{m},b_{m}\}|\mathsf{H}}\right)$ gives  $SL(2,\mathbb{C})$ weight two to all the gravitons and weight
one to the selected gluons $\{a_2,b_2\ldots a_{m},b_{m}\}$. Then together with the factor $\prod_{i}\sigma_{a_ib_i}$,
we have weight zero for all the gluons and weight two for all the
gravitons in the integrand. We now see that with the
Parke-Taylor string $\prod\langle\pmb{i}\rangle^{-1}$, the integrand
$\mathcal{I}_{L}$ has weight two for all the particles.

\subsection{Squeezing}
The second construction of $\mathcal{I}_{L}$ is to ``squeeze'' all
the  gluon rows (columns) that belong to the trace $\pmb{i}$ into a
single row (column) labeled by the trace. This process reduces the
$2n\times 2n$ matrix $\Psi$ into the following
$2(|\mathsf{H}|+m)\times 2(|\mathsf{H}|+m)$ antisymmetric matrix:
\begin{equation}
\label{eq:Pihm}
  \Pi_m\left(\pmb{1},\pmb{2}\ldots\pmb{m}\,|\,\mathsf{H}\right)=\left(\begin{array}{cccc}
    A_{\mathsf{H}} & A_{\mathsf{H,Tr}} & -(C_{\mathsf{H}})^T & -(C_{\mathsf{Tr,H}})^T \\
    A_{\mathsf{Tr,H}} & {A}_{\mathsf{Tr}} & -(C_{\mathsf{H,Tr}})^T & -({C}_{\mathsf{Tr}})^{T} \\
    C_{\mathsf{H}} & C_{\mathsf{H,Tr}} & B_{\mathsf{H}} & B_{\mathsf{H,Tr}} \\
    C_{\mathsf{Tr,H}} & {C}_{\mathsf{Tr}} & B_{\mathsf{Tr,H}} & {B}_{\mathsf{Tr}} \\
    \end{array}\right)\,.
\end{equation}
The blocks contained in $\Pi$ are defined as follows
\begin{itemize}

\item The $|\mathsf{H}|\times|\mathsf{H}|$ matrix $A_{\mathsf{H}}$,
$B_{\mathsf{H}}$ and $C_{\mathsf{H}}$ are respectively the
submatrix  of $A$, $B$ and $C$ given in Eq.~\eqref{eq:ABC} with
rows and columns ranging in $\mathsf{H}$. Compared to
Eq.~\eqref{eq:Psidecompose}, we now use a simplified notation in
which
\begin{align*}
& A_{\mathsf{H}}\equiv A_{\mathsf{HH}} & & B_{\mathsf{H}}\equiv B_{\mathsf{HH}}
& & C_{\mathsf{H}}\equiv C_{\mathsf{HH}}\,.
\end{align*}

\item The $m\times|\mathsf{H}|$ matrix $A_{\mathsf{Tr,H}}$, $B_{\mathsf{Tr,H}}$
and $C_{\mathsf{Tr,H}}$ are respectively given by
  \begin{align}
    (A_{\mathsf{Tr,H}})_{ib}=\sum_{c\in\pmb{i}}\frac{k_c\cdot k_b}{\sigma_{cb}} & & (B_{\mathsf{Tr,H}})_{ib}=\sum_{c\in\pmb{i}}\frac{\sigma_c k_c\cdot\epsilon_b}{\sigma_{cb}} & & (C_{\mathsf{Tr,H}})_{ib}=\sum_{c\in\pmb{i}}\frac{\sigma_c k_c\cdot k_b}{\sigma_{cb}}\,,
  \end{align}
  where $1\leqslant i\leqslant m$ is the gluon trace index and $b\in\mathsf{H}$ is
   the graviton index. 

\item The $m\times m$ matrix $A_{\mathsf{Tr}}$, ${B}_{\mathsf{Tr}}$, ${C}_{\mathsf{Tr}}$
are respectively given by
  \begin{align}
    (A_{\mathsf{Tr}})_{ij}=\left\{\begin{array}{>{\displaystyle}c @{\hspace{1em}} >{\displaystyle}l}
    \sum_{c\in\pmb{i}}\sum_{d\in\pmb{j}}\frac{k_c\cdot k_d}{\sigma_{cd}} & i\neq j\\
    0 & i=j \\
    \end{array}\right. & & (B_{\mathsf{Tr}})_{ij}=\left\{\begin{array}{>{\displaystyle}c @{\hspace{1em}} >{\displaystyle}l}
    \sum_{c\in\pmb{i}}\sum_{d\in\pmb{j}}\frac{\sigma_c(k_c\cdot k_d)\sigma_d}{\sigma_{cd}} & i\neq j\\
    0 & i=j \\
    \end{array}\right.
  \end{align}
  \begin{equation}
    (C_{\mathsf{Tr}})_{ij}=\sum_{\substack{c\in\pmb{i}~d\in\pmb{j} \\ c\neq d}}\frac{\sigma_c k_c\cdot k_d}{\sigma_{cd}}\,,
  \end{equation}
  where $1\leqslant i,j\leqslant m$. It is important to notice that the diagonal
  entries of $C_{\mathsf{Tr}}$ can be written as
  \begin{align}
    & (C_{\mathsf{Tr}})_{ii}=\sum_{\substack{c\in\pmb{i}~d\in\pmb{i} \\ c\neq d}}\frac{\sigma_c k_c\cdot k_d}{\sigma_{cd}}=-\sum_{\substack{c\in\pmb{i}~d\in\pmb{i} \\ c\neq d}}\frac{\sigma_d k_c\cdot k_d}{\sigma_{cd}}=\frac{1}{2}\sum_{\substack{c\in\pmb{i}~d\in\pmb{i} \\ c\neq d}}\frac{\sigma_{cd} k_c\cdot k_d}{\sigma_{cd}}=\frac{1}{2}\left(\sum_{c\in\pmb{i}}k_c\right)^2\equiv\frac{1}{2}(k_{\pmb{i}})^{2}\,,
  \end{align}
  namely, it does not depends on $\sigma$'s.

\item The $|\mathsf{H}|\times m$ matrix $C_{\mathsf{H,Tr}}$ is given by
  \begin{equation}
    (C_{\mathsf{H,Tr}})_{ai}=\sum_{c\in\pmb{i}}\frac{\epsilon_a\cdot k_c}{\sigma_{ac}}\,,
  \end{equation}
  where $1\leqslant i\leqslant m$ and $a\in\mathsf{H}$.

\end{itemize}

The matrix $\Pi$ has co-rank two, and according
to~\cite{Cachazo:2014xea},  we can define the following
\emph{reduced Pfaffian}:
\begin{equation}
\label{eq:PfpPi}
  \Pfp(\Pi_m)=(-1)^{i+j}\Pf(\Pi_{s+i,2s+m+j}^{s+i,2s+m+j})=\frac{(-1)^{a+b}}{\sigma_{ab}}\Pf(\Pi_{ab}^{ab})\qquad\begin{array}{c}
  a,b\in\mathsf{H}\text{ and }a<b \\
  1\leqslant i,j\leqslant m
  \end{array}\,,
\end{equation}
such that $\Pfp(\Pi)$ is independent of the rows and columns
deleted, when evaluated on the support of $d\Omega_{\text{CHY}}$. In other words,
we can choose the deleted rows (and the corresponding columns) in
the following two ways and obtain the same reduced Pfaffian:
\begin{enumerate}

\item Both of the deleted rows are in the submatrix
  \begin{equation*}
    \left(\begin{array}{cccc}
      A_{\mathsf{H}} & -(A_{\mathsf{Tr,H}})^T & -(C_{\mathsf{H}})^T & -(C_{\mathsf{Tr,H}})^T
      \end{array}\right)\,,
  \end{equation*}
  in this case, we need to supplement the factor $(-1)^{a+b}/\sigma_{ab}$, as in
  the definition of $\Pfp(\Psi)$.

\item One of the deleted rows lies in the submatrix
  \begin{equation*}
    \left(\begin{array}{cccc}
      A_{\mathsf{Tr,H}} & A_{\mathsf{Tr}} & -(C_{\mathsf{H,Tr}})^T & -(C_{\mathsf{Tr}})^T
      \end{array}\right)
  \end{equation*}
  while the other one lies in the submatrix
  \begin{equation*}
    \left(\begin{array}{cccc}
      C_{\mathsf{Tr,H}} & C_{\mathsf{Tr}} & B_{\mathsf{Tr,H}} & B_{\mathsf{Tr}}
      \end{array}\right)\,.
  \end{equation*}
  We only need to supplement the phase $(-1)^{i+j}$ in this case.
\end{enumerate}
In particular, if we only have external gluons, the matrix \eqref{eq:Pihm} reduces to
\begin{equation}
  \Pi_{m}(\pmb{1},\pmb{2}\ldots\pmb{m}\,|\,\mathsf{H}=\varnothing)\equiv\Pi_{m}{(\pmb{1,2\ldots m})}=\begin{pmatrix}
  A_{\mathsf{Tr}} & -(C_{\mathsf{Tr}})^{T} \\
  C_{\mathsf{Tr}} & B_{\mathsf{Tr}}
  \end{pmatrix}\,,
\end{equation}
where the subscript $m$ gives the number of gluon traces involved, and the  argument $(\pmb{1,2\ldots m})$ gives the traces that the rows and columns of $A_{\mathsf{tr}}$, $B_{\mathsf{tr}}$ and $C_{\mathsf{tr}}$ take values in. In the following sections, when the reduced Pfaffian is involved, we always delete the two rows and columns corresponding to the trace $\pmb{m}$ by convention. Namely,
we will consistently use:
\begin{equation}
  \Pfp[\Pi_{m}{(\pmb{1,2\ldots m})}]\equiv\Pf[\Pi_{m}{(\pmb{1,2\ldots m-1})}]\,.
\end{equation}
Finally, the CHY integrand for tree level $m$-trace EYM amplitudes constructed by squeezing is given by:\footnote{The phase factor was not present in the original CHY proposal~\cite{Cachazo:2014xea}. However, this sign is important to establish the equivalence between Eq.~\eqref{eq:mtraceEYM} and \eqref{eq:generalcompact} under our convention.}
\begin{align}
\label{eq:mtraceEYM}
  &\text{$m$-trace EYM:} & &\mathcal{I}_{L}(n)={(-1)^{\frac{(2|\mathsf{H}|+m)(m-1)}{2}}}\left[\prod_{i=1}^{m}\frac{1}{\langle\pmb{i}\rangle}\right]\Pfp[\Pi_m(\pmb{1},\pmb{2}\ldots\pmb{m}\,|\,\mathsf{H})]\,,
  & &\mathcal{I}_{R}(n)=\Pfp(\Psi)\,.
\end{align}
The equivalence of this result and Eq.~\eqref{eq:generalcompact} has been proved  in~\cite{Cachazo:2014xea}. However, it is not manifest in Eq.~\eqref{eq:mtraceEYM} that the CHY integrand has the correct weight:
\begin{equation}
\label{eq:Mobius}
\mathcal{I}_{L}(n)\;\rightarrow\;\mathcal{I}_{L}(n)\prod_{a=1}^{n}(\gamma\sigma_a+\delta)^2
\end{equation}
under the world sheet $SL(2,\mathbb{C})$ transformation
\begin{equation}
\sigma_a\;\rightarrow\;\frac{\alpha\sigma_a+\beta}{\gamma\sigma_a+\delta}\qquad(\alpha\delta-\beta\gamma=1)\,.
\end{equation}
We will give a direct proof in appendix~\ref{sec:PfPi}.

Before moving on, we note that for generic $n$-point EYM integrands with $|\mathsf{H}|$ gravitons, we need to {multiply} the phase factor:
\begin{equation}
\label{eq:genericphase}
\text{Generic EYM:}\qquad(\text{phase})=(-1)^{\frac{(n+1)(n+2)}{2}}(-1)^{\frac{|\mathsf{H}|(|\mathsf{H}|+1)}{2}}\,,
\end{equation}
to obtain amplitudes from integrands. This phase factor is again of no physical importance, but settles our recursive expansion into a very well-organized form.

Finally, in our discussion of expansion, the
key part is the CHY integrand $\mathcal{I}_{L}(n)$. In fact, we can replace
$\mathcal{I}_{R}(n)=\Pfp(\Psi) $ by $\mathcal{I}_{R}(n)=\langle 12\ldots
n\rangle^{-1} $ to find the relation between another two theories, namely, the YM-scalar and bi-adjoint scalar~\cite{Chiodaroli:2014xia,Cachazo:2014xea,Chiodaroli:2015rdg}.
It is crucial that the YM-scalar involves both scalar $\phi^3$ and
$\phi^4$ interactions\footnote{By including only the $\phi^3$
interaction, we can only work out the correct single trace EYM
amplitudes~\cite{Bern:1999bx}.}  and its CHY integrand is:
\begin{align}
\label{eq:YMS}
  \text{YM-scalar:} & &\mathcal{I}_{L}(n)={(-1)^{\frac{(2|\mathsf{H}|+m)(m-1)}{2}}}\left[\prod_{i=1}^{m}\frac{1}{\langle\pmb{i}\rangle}\right]
  \Pfp[\Pi_{m}(\pmb{1},\pmb{2}\ldots\pmb{m}\,|\,\mathsf{H})]\,,
  & &\mathcal{I}_{R}(n)=\frac{1}{\langle 12\ldots n\rangle}\,.
\end{align}
The double copy relation can be easily observed by comparing Eq.~\eqref{eq:mtraceEYM} with \eqref{eq:YMS}.

\section{Recursive expansion of multitrace EYM amplitudes}\label{sec:recursion}
In this section, we introduce two types of general recursive expansion relations  
for multitrace EYM amplitudes with arbitrary number of gravitons,
gluons  and color traces. We will show how to write down them by
starting from the recursive expansion of single trace EYM amplitudes
with appropriately replacing gravitons by color traces. For the sake
of a clear presentation, we first give a summary of the notations to
be used later:
\begin{itemize}

\item we use the boldface $\pmb{A}$ to denote an ordered set in the sense that physical
amplitudes depend on the ordering of its elements. Meanwhile, we
use the serif style $\mathsf{A}$ to denote an unordered set in
the same sense. For example, we use $\mathsf{H}$ to denote the set
of gravitons, and $|\mathsf{H}|$ the number of gravitons in
$\mathsf{H}$. Similarly, we use $\mathsf{Tr}$ to denote the set
of gluon traces. We will focus on the generic $m$-trace case,
namely, $\mathsf{Tr}=\{\pmb{1},\pmb{2}\ldots \pmb{m}\}$.

\item For a given set $\mathsf{A}$ and one of its subset $\mathsf{a}$, the notation
$\mathsf{A}\backslash\mathsf{a}$ stands for the complement of $\mathsf{a}$ in $\mathsf{A}$.
If $\mathsf{a}$ contains only one element: $\mathsf{a}=\{a\}$, then we just simplify the
notation to $\mathsf{A}\backslash a$.

\item The notation $\sum_{\pmb{a}\,|\,\mathsf{b}=\mathsf{A}}$ means that we sum over
all \emph{partial ordered} bi-partitions $\pmb{a}$ and $\mathsf{b}$ of the set $\mathsf{A}$. The partial ordering means
that different orderings in the first subset $\pmb{a}$ should be treated as different partitions, while the ordering in $\mathsf{b}$ does not matter. Both subsets $\pmb{a}$ and $\mathsf{b}$ are allowed to be empty.

\item Given two ordered sets $\pmb{A}$ and $\pmb{B}$, the shuffle product $\pmb{A}\shuffle\pmb{B}$ gives a sum of all permutations of $\pmb{A}\cup\pmb{B}$ that conserves the relative ordering inside each set $\pmb{A}$ and $\pmb{B}$ respectively. Suppose we have a function $f$ written as $ f(\pmb{A}\shuffle\pmb{B}\shuffle\pmb{C}\shuffle\ldots)$, what we mean is:
 %
    \begin{equation}
    \label{eq:sumovershuffle}
    f(\pmb{A}\shuffle\pmb{B}\shuffle\pmb{C}\shuffle\ldots)\equiv\sum_{\pmb{\rho}\in\pmb{A}\shuffle\pmb{B}\shuffle\pmb{C}\ldots}f(\pmb{\rho})\,,
    \end{equation}
    namely, we always keep the sum over the shuffle products implicit. Sometimes, we will meet the shuffle over different levels. For example, the notation $\pmb{A}\shuffle\{\alpha,\pmb{B}\shuffle \pmb{C},\beta\}$ means we should carry out the shuffle $\pmb{G}=\{\pmb{B}\shuffle \pmb{C}\}$ first, put it back to form a new set $\pmb{\W G}=\{\alpha, \pmb{G},\beta\}$, and then do the second shuffle $\pmb{A}\shuffle \pmb{\W G}$.

\item We define the field strength tensor of particle $i$ as
\begin{equation}
\label{F-def}
(F_i)^{\mu\nu}=(k_i)^{\mu}(\epsilon_i)^{\nu}-(k_i)^{\nu}(\epsilon_i)^{\mu}\,,
\end{equation}
where $k_i$ and $\epsilon_i$ are respectively the momentum and polarization vector of particle $i$.


\item Given a gluon trace $\pmb{i}=\{i_1,i_2\ldots i_s\ldots i_t\}$, the notation $Y_{i_s}$
stands for the sum of the \emph{original gluon momenta} at the left hand side of $i_s$ in trace $\pmb{i}$:
    \begin{equation}
    \label{eq:Ysymbol}
    Y_{i_s}=k_{i_1}+k_{i_2}+\ldots +k_{i_s}\,.
    \end{equation}
At some intermediate steps of our recursive expansion, we may meet the situation that some particles considered as gluons in trace $\pmb{i}$ at this step are actually original gravitons.
Thus we define another symbol $X_{i_s}$ to stand for the sum of all the momenta at the left hand side of $i_s$, regardless of their origins.

%
%
%
    %

\item Given a cycle $\langle\pmb{a}\rangle=\langle a_1a_2\ldots a_t\rangle$, we can always
 anchor two particles, say $a_i$ and $a_j$, to the first and
 last position through the KK relation~\cite{Kleiss:1988ne},
 written as:
    \begin{equation}
    \frac{1}{\langle\pmb{a}\rangle}=\frac{1}{\langle a_i,\pmb{\alpha},a_j,\pmb{\beta}\rangle}=(-1)^{|\pmb{a}_{j,i}|}\sum_{\pmb{\rho}\in\mathsf{KK}[\pmb{a},a_i,a_j]}\frac{1}{\langle a_i,\pmb{\rho},a_j\rangle}\,.
    \end{equation}
 In the second equality, we have used the cyclicity of
 $\langle\pmb{a}\rangle$ to put $a_i$ at the first position.
 Then the ordered set $\pmb{\alpha}$ and $\pmb{\beta}$ are
 defined as:
    \begin{align}
    \pmb{\alpha}=\{a_{i+1},a_{i+2}\ldots a_{j-1}\}\equiv\pmb{a}_{i,j}& &\pmb{\beta}=\{a_{j+1},a_{j+2}\ldots a_{i-1}\}\equiv \pmb{a}_{j,i}\,,
    \end{align}
 in which the cyclic continuation $a_{t+1}=a_1$ is understood.
 In the third equality, the symbol
 $\mathsf{KK}[\pmb{a},a_i,a_j]$ stands for the set
    \begin{equation}
    \mathsf{KK}[\pmb{a},a_i,a_j]=\pmb{\alpha}\shuffle\pmb{\beta}^{T}=
    \pmb{a}_{i,j}\shuffle(\pmb{a}_{j,i})^{T}\,.~~~\Label{KK-symbol-1}
    \end{equation}
 Finally, we keep the sum over the above orderings implicit for
 the sake of simplicity, namely, we define:
    \begin{equation}
    \label{eq:KKsymbol}
    (-1)^{|\pmb{a}_{j,i}|}\sum_{\pmb{\rho}\in\mathsf{KK}[\pmb{a},a_i,a_j]}\frac{1}{\langle a_i,\pmb{\rho},a_j\rangle}\equiv\frac{(-1)^{|\pmb{a}_{j,i}|}}{\langle a_i,\mathsf{KK}[\pmb{a},a_i,a_j],a_j\rangle}\,,
    \end{equation}
    where it is very important to notice the sign $(-1)^{|\pmb{a}_{j,i}|}$ in the definition. {In many situations, we can even omit the two anchors in $\mathsf{KK}[\ldots]$, which are usually the two particles that sandwich $\mathsf{KK}[\ldots]$. For convenience, we define the list
    \begin{equation}
    \label{eq:Ksymbol}
    K^{\pmb{i}}_{a_i,b_i}\equiv\{a_i,\mathsf{KK}[\pmb{i},a_i,b_i],b_i\}
    \end{equation}
    as a single object. The final form of the KK relation is  thus:
    \begin{equation}
    \frac{1}{\langle a_i,\pmb{\alpha},a_j,\pmb{\beta}\rangle}=\frac{(-1)^{|\pmb{a}_{j,i}|}}{\langle a_i,\mathsf{KK}[\pmb{a},a_i,a_j],a_j\rangle}=(-1)^{|\pmb{a}_{j,i}|}\frac{1}{\langle K^{\pmb{a}}_{a_i,a_j}\rangle}\,.
    \end{equation}}

\item We denote the generic tree level $m$-trace EYM amplitude with graviton set $\mathsf{H}$
 as
 $A_{m,|\mathsf{H}|}(\pmb{1}|\pmb{2}|\ldots|\pmb{m}\,\Vert\,\mathsf{H})$,
 where we have used a single vertical line to separate gluon
 traces, and a double vertical line to separate gluons and
 gravitons. The number of traces and gravitons are shown in the
 subscript. By convention, we always write the first gluon trace
 as $\pmb{1}=\{1,2\ldots r\}$.
\end{itemize}

In order to get familiar with these notations, we rewrite the
recursive expansion  of the single trace EYM amplitude
$A_{1,|\mathsf{H}|}(1,2\ldots r\,\Vert\,\mathsf{H})$, which has
already been given in~\cite{Fu:2017uzt,Teng:2017tbo}, as:
\begin{align}
\label{eq:singletrace}
A_{1,|\mathsf{H}|}(1,2\ldots r\,\Vert\,\mathsf{H})=\sum_{\pmb{h}\,|\,\mathsf{\W h}
=\mathsf{H}\backslash h_1}\Big[C_{h_1}(\pmb{h})A_{1,|\mathsf{\W h}|}
(1,\{2\ldots r-1\}\shuffle\{\pmb{h},h_1\},r\,\Vert\,\mathsf{\W h})\Big]\,.
\end{align}
Suppose $\pmb{h}=\{i_s,i_{s-1}\ldots i_1\}$, the coefficient $C_{h_1}$ can be written as:
\begin{equation}
C_{h_1}(\pmb{h})=\epsilon_{h_1}\cdot F_{i_1}\cdot F_{i_{2}}\ldots F_{i_{s-1}}\cdot F_{i_s}\cdot Y_{i_s}\,,
\end{equation}
while for $\pmb{h}=\varnothing$, we have $C_{h_1}(\varnothing)=\epsilon_{h_1}\cdot Y_{h_1}$. The expression \eqref{eq:singletrace} is manifestly invariant under the permutations and gauge transformations of the gravitons $\mathsf{H}\backslash h_1$. In the following, we call this $h_1$ \emph{the fiducial graviton} in order to distinguish it from the other gravitons (called \emph{regular gravitons} in the following). The full graviton permutation and gauge invariance, although not explicit, is guaranteed by the generalized BCJ relation~\cite{BjerrumBohr:2009rd,Chen:2011jxa}.

\subsection{Some explicit examples}

As the main story of this section, we are going to present two types of recursive expansions  of generic $m$-trace EYM amplitudes. In this subsection, we show how to construct the expansion from the known single trace results~\cite{Fu:2017uzt,Teng:2017tbo} through the trick \emph{``turning a graviton into a trace of gluons''}, illustrated by several explicit examples.\footnote{If we ``turn a graviton into a trace of gluons'', the particle number must change. In this sense, the construction presented here should only be understood as a trick observed from certain patterns of the amplitudes. We will shown in section~\ref{sec:compact} how this nice pattern emerges.} The generic algorithms will be given in the next subsection, and we defer the proof to the later sections.

As the first example, we show how to obtain the expansion of double
trace  EYM amplitude with one graviton from the known result
\eqref{eq:singletrace} of the single trace amplitude with two
gravitons:
\begin{align}
\label{eq:t1h2}
A_{1,2}(1,2\ldots r\,\Vert\,\{h_1,h_2\})&=\left(\epsilon_{h_1}\cdot Y_{h_1}\right)A_{1,1}(1,\{2\ldots r-1\}\shuffle\{h_1\},r\,\Vert\,\{h_2\})\nonumber\\
&\quad+\left(\epsilon_{h_1}\cdot F_{h_2}\cdot Y_{h_2}\right)A_{1,0}(1,\{2\ldots r-1\}\shuffle\{h_2,h_1\},r)\,.
\end{align}
Here $h_1$ is treated as fiducial, and $h_2$ as regular. Now at the
left hand side  of Eq.~\eqref{eq:t1h2}, we turn the graviton $h_2$ into the
gluon trace $\pmb{2}$, namely, we make the formal replacement:
\begin{equation}
A_{1,2}(1,2\ldots r\,\Vert\,\{h_1,h_2\})\;\rightarrow\;A_{2,1}(1,2\ldots r\,|\,\pmb{2}\,\Vert\,\{h_1\})\,.
\end{equation}
At the right hand side of Eq.~\eqref{eq:t1h2}, the first term
contains $h_2$ as a graviton,  thus we make the same formal
replacement to reach
\begin{equation}
A_{1,1}(1,\{2\ldots r-1\}\shuffle\{h_1\},r\,\Vert\,\{h_2\})\;\rightarrow\;A_{2,0}(1,\{2\ldots r-1\}\shuffle\{h_1\},r\,|\,\pmb{2})\,.
\end{equation}
We can summarize the above manipulation into the following
\emph{replacing rule for a regular graviton $h_i$, which still
remains to be a graviton in the recursive expansion:}
\begin{equation}
\label{eq:gravitonrule}
h_i\;\rightarrow\;\pmb{i}\,.
\end{equation}
For the second term, where the $h_2$ has been treated as a gluon in
the single trace expansion, we have to modify both the coefficient
and the amplitude according to the following \emph{replacing rule
for a regular graviton $h_i$, which has been treated as a gluon:}
\begin{align}
\label{eq:gluonrule}
& (1) \qquad h_i\;\rightarrow\;\{a_i,\mathsf{KK}[\pmb{i},a_i,b_i],b_i\}\nonumber\\
& (2) \qquad (F_{h_i})^{\mu\nu}\;\rightarrow\;-(k_{b_i})^{\mu}(k_{a_i})^{\nu}\nonumber\\
& (3) \qquad Y_{h_i}\;\rightarrow\;Y_{a_i}\quad(\text{if $Y_{h_i}$ appears})\nonumber\\
& (4) \qquad \text{sum over all \emph{ordered pairs} }\{a_i,b_i\}\subset\pmb{i}\text{ together with the sign }(-1)^{|\pmb{i}_{b_i,a_i}|}\,.
\end{align}
{We note that although the replacement $(2)$ breaks the antisymmetry of the $F$ tensor, it will lead to the correct result if all the appearance of $F_{h_i}$ has been taken care.} The whole replacement package effectively ``turns a graviton into a trace of gluons''. Applying these rules to the second term at the right hand side of Eq.~\eqref{eq:t1h2}, we get:
\begin{align}
&\quad\left(\epsilon_{h_1}\cdot F_{h_2}\cdot Y_{h_2}\right)A_{1,0}(1,\{2\ldots r-1\}\shuffle\{h_2,h_1\},r)\nonumber\\
&\rightarrow\underset{\{a_2,b_2\}\subset\pmb{2}}{\widetilde{\sum}}(-)\left(\epsilon_{h_1}\cdot k_{b_2}\right)\left(k_{a_2}\cdot Y_{a_2}\right)A_{1,0}(1,\{2\ldots r-1\}\shuffle\{a_2,\mathsf{KK}[\pmb{2},a_2,b_2],b_2,h_1\},r)\,,
\end{align}
where to simplify the notation, we have defined  $\widetilde{\sum}$
over the \emph{ordered pair} $\{a_2,b_2\}$ to include the sign appearing as a result of Eq.~\eqref{eq:KKsymbol}:\footnote{The reason we define the symbol $\widetilde{\sum}$ is to emphasize the relative sign between the amplitudes that interfere with each other, although one can modify the definition of the $\mathsf{KK}[\ldots]$ symbol in Eq.~\eqref{eq:KKsymbol} by including such a sign. We feel that to avoid mistakes when using our formula, the definition \eqref{tilde-sum} maybe a better choice.}
\begin{equation}
\sum_{\{a_2,b_2\}\subset\pmb{2}}\equiv\sum_{\substack{a_2,b_2\in\pmb{2} \\ a_2\neq b_2}}\,\,,\qquad\qquad\underset{\{a_2,b_2\}\subset\pmb{2}}{\widetilde{\sum}}\equiv\sum_{\substack{a_2,b_2\in\pmb{2} \\  a_2\neq b_2}}(-1)^{|\pmb{2}_{b_2,a_2}|}\,.\Label{tilde-sum}
\end{equation}
Putting all together,  we get the following recursive expansion for
the double trace amplitude with one graviton, which is equivalent to the expression in~\cite{Nandan:2016pya}:
\begin{align}
\label{eq:t2h1}
& A_{2,1}(1,2\ldots r\,|\,\pmb{2}\,\Vert\,\{h_1\})=\left(\epsilon_{h_1}\cdot Y_{h_1}\right)A_{2,0}(1,\{2\ldots r-1\}\shuffle\{h_1\},r\,|\,\pmb{2})\nonumber\\
& \qquad+\underset{\{a_2,b_2\}\subset\pmb{2}}{\widetilde{\sum}}\left(-\epsilon_{h_1}\cdot k_{b_2}\right)\left(k_{a_2}\cdot Y_{a_2}\right)A_{1,0}(1,\{2\ldots r-1\}\shuffle\{a_2,\mathsf{KK}[\pmb{2},a_2,b_2],b_2,h_1\},r)\,.
\end{align}
As a slightly nontrivial example, we construct the expansion of
triple trace amplitude with  one graviton from the single trace
expansion \eqref{eq:singletrace} with three gravitons:
\begin{align}
A_{1,3}(1,2\ldots r\,\Vert\,\{h_1,h_2,h_3\})&=\left(\epsilon_{h_1}\cdot Y_{h_1}\right)A_{1,2}(1,\{2\ldots r-1\}\shuffle\{h_1\},r\,\Vert\,\{h_2,h_3\})\nonumber\\
&\quad+\left(\epsilon_{h_1}\cdot F_{h_2}\cdot Y_{h_1}\right)A_{1,1}(1,\{2\ldots r-1\}\shuffle\{h_2,h_1\},r\,\Vert\,\{h_3\})\nonumber\\
&\quad+\left(\epsilon_{h_1}\cdot F_{h_3}\cdot Y_{h_1}\right)A_{1,1}(1,\{2\ldots r-1\}\shuffle\{h_3,h_1\},r\,\Vert\,\{h_2\})\nonumber\\
&\quad+\left(\epsilon_{h_1}\cdot F_{h_2}\cdot F_{h_3}\cdot Y_{h_1}\right)A_{1,0}(1,\{2\ldots r-1\}\shuffle\{h_3,h_2,h_1\},r)\nonumber\\
&\quad+\left(\epsilon_{h_1}\cdot F_{h_3}\cdot F_{h_2}\cdot Y_{h_1}\right)A_{1,0}(1,\{2\ldots r-1\}\shuffle\{h_2,h_3,h_1\},r)\,.
\end{align}
Now by applying the rule \eqref{eq:gravitonrule} and
\eqref{eq:gluonrule} to both $h_2$ and  $h_3$, we get the desired
expansion:
\begin{align}
& A_{3,1}(1,2\ldots r\,|\,\pmb{2}\,|\,\pmb{3}\,\Vert\,\{h_1\})=\left(\epsilon_{h_1}\cdot Y_{h_1}\right)A_{3,0}(1,\{2\ldots r-1\}\shuffle\{h_1\},r\,|\,\pmb{2}\,|\,\pmb{3})\nonumber\\
&\qquad+\underset{\{a_2,b_2\}\subset\pmb{2}}{\widetilde{\sum}}\left(-\epsilon_{h_1}\cdot k_{b_2}\right)\left(k_{a_2}\cdot Y_{a_2}\right)A_{2,0}(1,\{2\ldots r-1\}\shuffle\{a_2,\mathsf{KK}[\pmb{2},a_2,b_2],b_2,h_1\},r\,|\,\pmb{3})\nonumber\\
&\qquad+\underset{\{a_3,b_3\}\subset\pmb{3}}{\widetilde{\sum}}\left(-\epsilon_{h_1}\cdot k_{b_3}\right)\left(k_{a_3}\cdot Y_{a_3}\right)A_{2,0}(1,\{2\ldots r-1\}\shuffle\{a_3,\mathsf{KK}[\pmb{3},a_3,b_3],b_3,h_1\},r\,|\,\pmb{2})\nonumber\\
&\qquad+\underset{\{a_2,b_2\}\subset\pmb{2}}{\widetilde{\sum}}~\underset{\{a_3,b_3\}\subset\pmb{3}}{\widetilde{\sum}}\left(-\epsilon_{h_1}\cdot k_{b_2}\right)\left(k_{a_2}\cdot k_{b_3}\right)\left(-k_{a_3}\cdot Y_{a_3}\right)\nonumber\\
&\qquad\qquad\qquad\times A_{1,0}(1,\{2\ldots r-1\}\shuffle\{a_3,\mathsf{KK}[\pmb{3},a_3,b_3],b_3,a_2,\mathsf{KK}[\pmb{2},a_2,b_2],b_2,h_1\},r)\nonumber\\
&\qquad+\underset{\{a_3,b_3\}\subset\pmb{3}}{\widetilde{\sum}}~\underset{\{a_2,b_2\}\subset\pmb{2}}{\widetilde{\sum}}\left(-\epsilon_{h_1}\cdot k_{b_3}\right)\left(k_{a_3}\cdot k_{b_2}\right)\left(-k_{a_2}\cdot Y_{a_2}\right)\nonumber\\
&\qquad\qquad\qquad\times A_{1,0}(1,\{2\ldots r-1\}\shuffle\{a_2,\mathsf{KK}[\pmb{2},a_2,b_2],b_2,a_3,\mathsf{KK}[\pmb{3},a_3,b_3],b_3,h_1\},r)\,.
\end{align}
From these two examples, a general pattern starts to emerge. We will
summarize it into the  \emph{type-I recursive expansion} in the next
subsection. However, there is a small caveat in the usage
of this construction: we need at least one fiducial graviton to make
this scheme work. Consequently, it cannot be applied to the
expansion of pure gluon amplitudes.

The above issue can be resolved once we know how to convert the
fiducial $h_1$ into a gluon  trace. Very remarkably, we find that
the following \emph{replacing rule for turning the fiducial graviton to a gluon trace}:
\begin{align}
\label{eq:lasthrule}
&\text{if $h_i$ is fiducial:} & & (1) \quad h_i\;\rightarrow\;\{c_i,\mathsf{KK}[\pmb{i},c_i,d_i],d_i\}\nonumber\\
& & & (2) \quad \epsilon_{h_i}\;\rightarrow\;-k_{c_i}; \quad Y_{h_i}\;\rightarrow\;Y_{c_i}\text{ if appears} \nonumber\\
& & & (3) \quad \text{keep the arbitrarily chosen }d_i\in\pmb{i}\text{ fixed,}\nonumber\\
& & & \quad\quad \text{ and sum over all the other $c_i\in\pmb{i}$}\text{ together with the sign }(-1)^{|\pmb{i}_{d_i,c_i}|}\,.
\end{align}
This rule will result in the \emph{type-II recursive expansion} given in the next subsection.
As applications, we present two examples. The first one is pure double trace case. Starting
from the single trace EYM expansion with just one graviton:
\bea
A_{1,1}(1,2\ldots r\,\Vert\,\{h_1\})=(\epsilon_{h_1}\cdot Y_{h_1})A_{1,0}(1,\{2,\dots,r-1\}\shuffle\{h_1\},r).
\eea
we get immediately the following result:
\begin{equation}
A_{2,0}(1,2\ldots
r\,|\,\pmb{2})=~\underset{\{c_2,\underline{d_2}\}\subset\pmb{2}}{\widetilde{\sum}}\left(-k_{c_2}\cdot
Y_{c_2}\right)A_{1,0}(1,\{2\ldots r-1\}\shuffle\{c_2,\mathsf{KK}[\pmb{2},c_2,d_2],d_2\},r\,),~~~\Label{Double-result}
\end{equation}
after using rule \eqref{eq:lasthrule}, where the tilde-sum is given by:
\begin{equation}
\underset{\{c_2,\underline{d_2}\}\subset\pmb{2}}{\widetilde{\sum}}\equiv\sum_{\substack{c_2\in\pmb{2} \\ c_2\neq d_2}}(-1)^{|\pmb{2}_{d_2,c_2}|}\,,
\end{equation}
namely, the underlined index is fixed. This result is also equivalent to the one given in~\cite{Nandan:2016pya}. Another example is the
expansion of pure triple  trace EYM amplitudes. One can start from
Eq.~\eqref{eq:t1h2} and turn two gravitons to two gluon traces. Or
one can  apply rule \eqref{eq:lasthrule} to the double
trace one graviton expansion given in \eqref{eq:t2h1}. Both methods give the same
answer, and  the expansion of pure  triple trace EYM amplitudes is:
\begin{align}
& A_{3,0}(1,2\ldots r\,|\,\pmb{2}\,|\,\pmb{3})=\underset{\{c_2,\underline{d_2}\}\subset\pmb{2}}{\widetilde{\sum}}\left(-k_{c_2}\cdot Y_{c_2}\right)A_{2,0}(1,\{2\ldots r-1\}\shuffle\{c_2,\mathsf{KK}[\pmb{2},c_2,d_2],d_2\},r\,|\,\pmb{3})\nonumber\\
& +\underset{\{c_2,\underline{d_2}\}\subset\pmb{2}}{\widetilde{\sum}}~\underset{\{a_3,b_3\}\subset\pmb{3}}{\widetilde{\sum}}\left(k_{c_2}\cdot k_{b_3}\right)\left(k_{a_3}\cdot Y_{a_3}\right)A_{1,0}(1,\{2\ldots r-1\}\shuffle\{a_3,\mathsf{KK}[\pmb{3},a_3,b_3]b_3,c_2,\mathsf{KK}[\pmb{2},c_2,d_2],d_2\},r)\,.
\end{align}
 This expansion holds for any choice of $d_2\in\pmb{2}$. In the future, we will call $d_2$ the \emph{fiducial gluon} of the type-II recursive expansion.


\subsection{Generic recursive expansion}

In this subsection, we present two general formulas to recursively expand
the $m$-trace EYM amplitude  with $|\mathsf{h}|$ gravitons from the
corresponding one of the single trace EYM amplitude with
$|\mathsf{H}|=m-1+|\mathsf{h}|$ gravitons. The starting point is again
Eq.~\eqref{eq:singletrace}, with $h_1$ chosen as the fiducial
graviton.

We first separate $\mathsf{H}$ into two parts,
$\mathsf{h}_{\mathsf{g}}$ and $\mathsf{h}$. Then we turn
the gravitons in $\mathsf{h}_{\mathsf{g}}$ into $(m-1)$ gluon traces
through the one-to-one map:
\begin{eqnarray}
\mathsf{h}_{\mathsf{g}}=\{h_{i_1},h_{i_2}\ldots h_{i_{m-1}}\}&\;\longrightarrow\;&\mathsf{Tr}_{m-1}=\{\pmb{t}_1,\pmb{t}_2\ldots \pmb{t}_{m-1}\}\nonumber\\
A(1,2\ldots r\,\Vert\,\mathsf{H})&\;\longrightarrow\;&A(1,2\ldots r\,|\,\pmb{t}_1\,|\,\pmb{t}_2\,|\ldots|\,\pmb{t}_{m-1}\,\Vert\,\mathsf{h})\,,
\end{eqnarray}
while those in $\mathsf{h}$ remain as gravitons. 
Depending on whether the fiducial graviton $h_1$ is contained in
$\mathsf{h}_{\mathsf{g}}$, we have two types of recursive
expansions.

\paragraph{Type-I recursive expansion:}

When $h_1\notin\mathsf{h}_{\mathsf{g}}$, namely, all the gravitons
in $\mathsf{h}_{\mathsf{g}}$ are regular, we can simply invoke the
replacing rule \eqref{eq:gravitonrule} and \eqref{eq:gluonrule} onto
$\mathsf{h}_{\mathsf{g}}$, leading to the \emph{type-I recursive
expansion}. After a suitable relabeling of symbols, we can settle
this expansion into the following closed form:\footnote{According to
our convention, all the boldface sets are ordered.}
\begin{align}
\label{eq:type1step1}
&\text{Type-I recursive expansion:}\nonumber\\
&\quad A(1,2\ldots r\,|\,\pmb{2}\,|\ldots|\,\pmb{m}\,\Vert\,\mathsf{H})=\sum_{\substack{\pmb{h}| \mathsf{h}=\mathsf{H}\backslash h_1 \\ \pmb{\mathsf{Tr}}_s| \mathsf{Tr}_p=\mathsf{Tr}\backslash 1}}\Bigg[\,\underset{\{a_i,b_i\}\subset\pmb{t}_i}{\widetilde{\sum}}\,\Bigg]_{i=1}^s\Big[\mathcal{A}_{1}(\pmb{1}_{\pmb{h},\pmb{K}[
\pmb{\mathsf{Tr}}_s,a,b],h_1}|\,\mathsf{Tr}_p\,\Vert\,\mathsf{h})\Big]\,,
\end{align}
where $\pmb{\mathsf{Tr}}_{s}=\{t_1,t_2\ldots t_s\}$, $\mathsf{Tr}_{p}=\{r_1,r_2\ldots r_p\}$. We emphasize that different orderings of $\pmb{h}$ and $\pmb{\mathsf{Tr}}_s$ are treated as different partitions, and get summed over.
Each $\mathcal{A}_1$ is a linear combination of the EYM amplitudes with $p+1$ gluon traces and $|\mathsf{h}|$ gravitons:\footnote{According to our notation \eqref{eq:sumovershuffle}, here $C_{h_1}(\pmb{h}\,\W{\shuffle}\,\pmb{K})A(\ldots\pmb{h}\,\W{\shuffle}\,\pmb{K}\ldots)$ stands for $\sum_{\pmb{\rho}\in\pmb{h}\W{\shuffle}\pmb{K}}C_{h_1}(\pmb{\rho})A(\ldots\pmb{\rho}\ldots)$.}
%
\begin{align}
\label{eq:type1step2}
\mathcal{A}_{1}(\pmb{1}_{\pmb{h},\pmb{K}[\pmb{\mathsf{Tr}}_s,a,b],h_1}|\,\mathsf{Tr}_p\,\Vert\,\mathsf{h})=C_{h_1}(\pmb{h}\,\W{\shuffle}\,\pmb{K})A_{p+1,|\mathsf{h}|}(1,\{2\ldots r-1\}\shuffle\{\pmb{h}\,\W{\shuffle}\,\pmb{K}(\pmb{\mathsf{Tr}}_s,a,b),h_1\},r|\pmb{r}_{1}|\ldots|\pmb{r}_{p}\Vert\,\mathsf{h})\,.
\end{align}
The notations involved are explained as follows:
\begin{itemize}

\item The $[\,\widetilde{\sum}\,]^s$ notation stands for:
\begin{equation} \Bigg[\,\underset{\{a_i,b_i\}\subset\pmb{t}_i}{\widetilde{\sum}}\,\Bigg]_{i=1}^s\equiv\underset{\{a_1,b_1\}\subset\pmb{t}_1}{\widetilde{\sum}}~\underset{\{a_2,b_2\}\subset\pmb{t}_2}{\widetilde{\sum}}
~....~\underset{\{a_s,b_s\}\subset\pmb{t}_s}{\widetilde{\sum}}~~~\Label{tilde-sum-S}
\end{equation}

\item The ordered set $\pmb{K}(\pmb{\mathsf{Tr}}_{s},a,b)$ is defined as:
\begin{equation}
\pmb{K}(\pmb{\mathsf{Tr}}_{s},a,b)=\{K_{a_{1},b_{1}}^{\pmb{t}_1},K_{a_{2},b_{2}}^{\pmb{t}_2}\ldots K_{a_{s},b_{s}}^{\pmb{t}_s}\}\,,
\end{equation}
where $K_{a_i,b_i}^{\pmb{t}_i}$ is defined in Eq.~\eqref{eq:Ksymbol}.

\item In $\pmb{h}\,\W{\shuffle}\,\pmb{K}$, we treat $K_{a_i,b_i}^{\pmb{t}_i}$ as a single object, which is why we use $\W{\shuffle}$ to emphasize the difference.
For example, if $\pmb{h}=\{h\}$ has only one object, we have:
\begin{align}
\pmb{h}\,\W{\shuffle}\,\pmb{K}(\pmb{\mathsf{Tr}}_{s},a,b)&=\{h,K_{a_{1},b_{1}}^{\pmb{t}_1},K_{a_{2},b_{2}}^{\pmb{t}_2}\ldots K_{a_{s},b_{s}}^{\pmb{t}_s}\}+\{K_{a_{1},b_{1}}^{\pmb{t}_1},h,K_{a_{2},b_{2}}^{\pmb{t}_2}\ldots K_{a_{s},b_{s}}^{\pmb{t}_s}\}+\ldots\nonumber\\
&=\{h,a_1,\mathsf{KK}[\pmb{t}_1],b_1,a_2,\mathsf{KK}[\pmb{t}_2],b_2\ldots\}+\{a_1,\mathsf{KK}[\pmb{t}_1],b_1,h,a_2,\mathsf{KK}[\pmb{t}_2],b_2\ldots\}+\ldots
\end{align}
%
%
%

\item Finally, for a given ordering $\pmb{\rho}=\{\rho_1,\rho_2\ldots\rho_{|\pmb{h}|+s}\}\in\pmb{h}\,\W{\shuffle}\,\pmb{K}(\pmb{\mathsf{Tr}}_{s},a,b)$, the coefficient $C_{h_1}$ is given by:
\begin{align}
\label{eq:Ch_1}
&C_{h_1}(\pmb{\rho})=\epsilon_{h_1}\cdot \mathcal{T}_{\rho_{|\pmb{h}|+s}}\ldots\mathcal{T}_{\rho_2}\cdot \mathcal{T}_{\rho_1}\cdot Y_{\rho_1}\,,\nonumber\\
&\mathcal{T}_{\rho_i}=
\left\{\begin{array}{l@{\hskip 1.5em}l}
(F_{h_a})^{\mu\nu} & \rho_{i}=h_a\in\pmb{h} \\
-(k_{b_j})^{\mu}(k_{a_j})^{\nu} & \rho_{i}=K_{a_j,b_j}^{\pmb{t}_j}\in\pmb{K}(\pmb{\mathsf{Tr}}_{s},a,b)
\end{array}\right.\,.
\end{align}
We also define $\WH{C}_{h_1}$ as:
\begin{equation}
\label{eq:Chat}
\WH{C}_{h_1}(\pmb{\rho})=-k_{h_1}\cdot \mathcal{T}_{\rho_{|\pmb{h}|+s}}\ldots\mathcal{T}_{\rho_2}\cdot \mathcal{T}_{\rho_1}\cdot Y_{\rho_1}\,.
\end{equation}
It will be immediately useful in our type-II recursive expansion, as well as the discussion of gauge invariance.
\end{itemize}
In Eq.~\eqref{eq:type1step2}, the fiducial graviton $h_1$ is treated differently, and its
existence is essential to make this recursive expansion work. Therefore, we require
$|\mathsf{H}|\geqslant 1$ for the type-I recursive expansion.

\paragraph{Type-II recursive expansion:}
When $h_1\in\mathsf{h}_{\mathsf{g}}$, we can still apply the rules
\eqref{eq:gravitonrule}  and \eqref{eq:gluonrule} onto those regular
gravitons, while the rule \eqref{eq:lasthrule} should be used to
$h_1$. This results in the \emph{type-II recursive expansion},
which, after a suitable relabeling of symbols, has the following
closed form:
\begin{align}
\label{eq:type2step1}
\text{Type-II recursive expansion:}&\nonumber\\
A(1,2\ldots r\,|\,\pmb{2}\,|\ldots|\,\pmb{m}\,\Vert\,\mathsf{H})&=
\sum_{\substack{\pmb{h}| \mathsf{h}=\mathsf{H} \\ \pmb{\mathsf{Tr}}_s| \mathsf{Tr}_p=\mathsf{Tr}
\backslash\{1,2\}}}\Bigg[\,\underset{\{a_i,b_i\}\subset\pmb{t}_i}{\widetilde{\sum}}
\,\Bigg]_{i=1}^s~\underset{\{c_2,\underline{d_2}\}\subset\pmb{2}}{\widetilde{\sum}}\Big[\mathcal{A}_{2}(\pmb{1}_{\pmb{h},\pmb{K}[\pmb{\mathsf{Tr}}_s,a,b],\mathsf{KK}[\pmb{2},c_2,d_2]}|\,\mathsf{Tr}_p\,\Vert\,\mathsf{h})\Big]\,,
\end{align}
where $d_2\in\pmb{2}$ is an arbitrary fixed particle (fiducial gluon) in trace
$\pmb{2}$. Each $\mathcal{A}_{2}$ is also a linear combination of the
EYM amplitudes with $p+1$ traces and $|\mathsf{h}|$ gravitons:
\begin{align}
\label{eq:type2step2}
&\quad\mathcal{A}_{2}(\pmb{1}_{\pmb{h},\pmb{K}[\pmb{\mathsf{Tr}}_s,a,b],\mathsf{KK}[\pmb{2},c_2,d_2]}|\,\mathsf{Tr}_p\,\Vert\,\mathsf{h})\nonumber\\
&=\WH{C}_{c_2}(\pmb{h}\,\W{\shuffle}\,\pmb{K})A_{p+1,|\mathsf{h}|}(1,\{2\ldots r-1\}\shuffle\{\pmb{h}\,\W{\shuffle}\,\pmb{K}(\pmb{\mathsf{Tr}}_s,a,b),c_2,\mathsf{KK}[\pmb{2},c_2,d_2],d_2\},r|\pmb{r}_{1}|\ldots|\pmb{r}_{p}\Vert\,\mathsf{h})\,.
\end{align}
According to Eq.~\eqref{eq:Chat}, for $\pmb{\rho}\in\pmb{h}\,\W{\shuffle}\,\pmb{K}$, we have $\WH{C}_{c_2}(\pmb{\rho})=-k_{c_2}\cdot \mathcal{T}_{\rho_{|\pmb{h}|+s}}\ldots\mathcal{T}_{\rho_2}\cdot \mathcal{T}_{\rho_1}\cdot Y_{\rho_1}$. The type-II recursive expansion works for arbitrary
number of gravitons. However, it requires at least two gluon traces.

Having presented the type-I and II recursive expansions, we will devote several sections later in this paper to their understanding
and proof. In particular, we will present supporting evidence from various angles,
such as dimensional reduction, related BCJ identities and consistency checks, etc. Then we will give
a proof using the BCFW recursion relation. Interestingly, knowing the expansion of Einstein gravity~\cite{Fu:2017uzt,Teng:2017tbo}, we can apply the unifying relation~\cite{Cheung:2017ems} to obtain a recursive expansion for EYM amplitudes. We study an example in appendix~\ref{sec:unify} and compare it with our type-I recursive expansion.

\section{Observation from compactification inspired construction}\label{sec:compact}

In this section, we will give our first understanding of type-I and II recursive expansions.
As reviewed in section~\ref{sec:integrand}, the CHY integrand of multitrace EYM amplitudes can be written into two
different but equivalent forms: one from CIC~\eqref{eq:generalcompact}
and the other from squeezing~\eqref{eq:mtraceEYM}. In this section, we will use the form \eqref{eq:generalcompact} to get some physical understanding of the type-I and II recursive expansions.

Originally in~\cite{Cachazo:2014xea}, the expression~\eqref{eq:generalcompact} is conjectured from the CHY integrand of Einstein-Maxwell (EM) theory, while the EM integrand is obtained from Einstein gravity through a dimensional reduction (or compactification). Here, we show that Eq.~\eqref{eq:generalcompact} can also be obtained through a direct dimensional reduction from another integrand with some similarity to the single trace EYM integrand.\footnote{Although this higher dimensional quantity might not be a physical CHY integrand.} The benefit of this new reduction scheme is that one can easily observe the replacing rule~\eqref{eq:gluonrule} and \eqref{eq:gravitonrule} for gravitons. Moreover, the pattern of ``turning a graviton into a trace of gluons'' also emerges naturally.

Our reduction scheme works as follows. We start with the gravitons living in $D=d+m$ dimensions, whose kinematic configuration is given by:
\begin{enumerate}
    \item  All external momenta are in $d$ dimensions, namely:
    $$K_i=(k_i|\underbrace{0\,|\,0\,|\cdots|\,0}_{m \text{ zeros}}\,)\,,$$
    where $k_i$ is the usual $d$ dimensional massless momentum.
    \item In $D$ dimensions, the gravity polarizations are given by $\W{\mathcal{E}}_i\mathcal{E}_i$, where all the $\mathcal{E}$ vectors are in $d$ dimensions, namely:
    $$\mathcal{E}_i=(\epsilon_i|\underbrace{0\,|\,0\,|\cdots|\,0}_{m \text{ zeros}}\,)\,,$$
    where $\epsilon_i$ is the usual polarization vector in $d$ dimensions.
    \item On the other hand, the $\W{\mathcal{E}}$'s have different forms for the would-be $d$ dimensional gravitons and gluons:
    \begin{align}
    \label{E-reduction}
    &\W{\mathcal{E}}_i=(\W{\epsilon}_i|\underbrace{0\,|\,0\,|\cdots|\,0}_{m \text{ zeros}}\,)& &\text{if $i$ is a graviton}\nonumber\\
    &\W{\mathcal{E}_i}=(\,\underbrace{0,0\cdots 0}_{\text{$d$ zeros}}
    |\underbrace{0\,|\cdots|\,0}_{\text{$k-1$ zeros}}|\,1\,|\,0\,|\cdots|\,0\,) & &\text{if $i$ is a gluon and $i\in\pmb{k}$}
    \end{align}
\end{enumerate}
Now we can construct the matrix $\Psi^{D}_{\mathsf{H},\mathsf{\gamma}|\mathsf{H},\mathsf{\gamma}}$ using $K$ and $\W{\mathcal{E}}$ according to Eq.~\eqref{eq:generalcompact} in $D$ dimensions, with $\mathsf{\gamma}=\{a_2,b_2\ldots a_m,b_m\}$.
Under the kinematic configuration specified above, 
one can easily show that:
\begin{equation}
\label{eq:PabDR}
\mathcal{I}_{L}(n)
=\frac{(-1)^{m-1}}{\langle 12\ldots r\rangle}\sum_{a_2<b_2\in\pmb{2}}\ldots\sum_{a_{m}<b_{m}\in\pmb{m}}\left[\,\prod_{i=2}^{m}\frac{\langle a_ib_i\rangle}{\langle\pmb{i}\rangle}\,\right]\Pf\!\left.\left(\Psi^{D}_{\mathsf{H},\mathsf{\gamma}|\mathsf{H},\mathsf{\gamma}}\right)\right|_{\text{dim-red}}\,,
\end{equation}
where $\mathsf{\gamma}=\{a_2,b_2,a_3,b_3\ldots a_{m},b_{m}\}$, and the subscript ``dim-red'' indicates precisely that this equation holds only under the special kinematics for our aforementioned dimensional reduction scheme. We have also used the fact that $\sigma_{a_ib_i}^{2}=-\sigma_{a_ib_i}\sigma_{b_ia_i}=-\langle a_ib_i\rangle$. The advantage of Eq.~\eqref{eq:PabDR} is that we can instead evaluate the $D$ dimensional integrand $\Pf(\Psi^D)$ first with general kinematic configurations, and take the special dimensional reduction scheme at a suitable moment. This enables us to take fully use of the known results.

%
%
%
    %
%


\subsection{The special case: each cycle having only two gluons} 

To demonstrate how this dimensional reduction works, we consider the special case that each gluon trace has only two gluons: $\pmb{i}=\{a_i,b_i\}$, except for the first one. Then the integrand \eqref{eq:generalcompact} has a very simple form:
\begin{equation}
\mathcal{I}_{L}(n)=\frac{(-1)^{m-1}}{\langle 12\ldots r\rangle}\left.\Pf\!\left(\Psi^{D}_{\mathsf{H},\gamma|\mathsf{H},\gamma}\right)\right|_{\text{dim-red}}\,,
\end{equation}
which is nothing but the single trace EYM integrand in $D$ dimensions. We can simply carry out the single trace recursive expansion according to Eq.~\eqref{eq:singletrace}. Then for a given pair $\{a_i,b_i\}$ in the would-be-gluon list $\gamma$, it will have one of the following situations:
%
%
%
%
%
\begin{enumerate}
    \item  For those pairs $\{a_i,b_i\}$ remaining in the graviton Pfaffian, the dimensional reduction leads to:
    \begin{equation}
    \Pf\!\left(\Psi^{D}_{\mathsf{H'},\{\ldots a_i,b_i\ldots\}|\mathsf{H'},\{\ldots a_i,b_i\ldots\}}\right)\xrightarrow{\text{dim-red}} \ldots\frac{-\sigma_{a_ib_i}}{\langle \pmb{i}\rangle}\ldots\Pf(\Psi_{\mathsf{H'},\{\ldots a_i,b_i\ldots\}|\mathsf{H'}})\,,
    \end{equation}
    with $\mathsf{H'}$ a subset of $\mathsf{H}$. Comparing this with the integrand \eqref{eq:generalcompact}, we find that each $\{a_i,b_i\}$ is turned into a trace of gluons. This is nothing but our rule~\eqref{eq:gravitonrule}.

    \item Next, we consider the case that for the trace $\pmb{i}$, both $\{a_i,b_i\}$ have been turned into gluons. First, if both $\{a_i,b_i\}$ are regular gravitons, they must appear in the coefficients of Eq.~\eqref{eq:singletrace} with the form:
    \begin{equation}
    \frac{\W{\mathcal{E}}\cdots F_{b_i}\cdots F_{a_i}\cdots Y}{\langle 1,\{2\ldots r-1\}\shuffle\{\ldots a_i\ldots b_i\ldots\},r\rangle}\qquad (\text{also with $a_i$ and $b_i$ exchanged})\,.
    \end{equation}
    Under our reduction scheme, the combination $F_{x}\cdot F_{a_i}\cdot F_{y}$ is nonzero if and only if either $x$ or $y$ is in the same trace as $a_i$. In our special scenario, we have $\pmb{i}=\{a_i,b_i\}$ for all $i\geqslant 2$, which means that the $F$-chain above is nonzero if and only if $a_i$ and $b_i$ are next to each other:
    \begin{equation}
    \label{Red-special-1}
    \frac{\W{\mathcal{E}}\cdots F_{b_i}\cdot F_{a_i}\cdots Y}{\langle 1,\{2\ldots r-1\}\shuffle\{\ldots a_i, b_i\ldots\},r\rangle}\,\xrightarrow{\text{dim-red}}\,\frac{\epsilon\cdots(-k_{b_i}k_{a_i})\cdots Y}{\langle 1,\{2\ldots r-1\}\shuffle\{\ldots a_i, b_i\ldots\},r\rangle}\,.
    \end{equation}
    This is just the replacing rule for regular gravitons as given in Eq.~\eqref{eq:gluonrule}. In this simple case, however, the list $\mathsf{KK}[\pmb{i},a_i,b_i]$ is empty.

    \item The second scenario is that one of $\{a_i,b_i\}$ is the fiducial graviton. With out losing generality, we assume $b_i$ is fiducial. Similar to the previous case, the only nonzero coefficient looks like
    \begin{equation}
    \label{Red-special-2}
    \frac{\W{\mathcal{E}}_{b_i}\cdot F_{a_i}\cdot F\cdots Y}{\langle 1,\{2,\ldots r-1\}\shuffle\{\ldots a_i,b_i\},r\rangle}\,\xrightarrow{\text{dim-red}}\,\frac{(-k_{a_i})\cdot F\cdots Y}{\langle 1,\{2,\ldots r-1\}\shuffle\{\ldots a_i,b_i\},r\rangle}\,,
    \end{equation}
    which gives us the replacing rule for the fiducial graviton~\eqref{eq:lasthrule}. We note that the dependence of $b_i$ drops out in the numerator.

    \item Finally, for a given trace $\pmb{i}$, if there is only one particle, say $b_i$, has been turned into a gluon, while $a_i$ remains in the graviton Pfaffian, then one can easily see that the expansion coefficient must vanish under the dimensional reduction. The reason is that the only other particle $a_i$ of trace $\pmb{i}$ can never come next to $b_i$ in the chain $\W{\mathcal{E}}\cdots F\cdots F\cdots Y$.


\end{enumerate}
The above example demonstrates how we obtain the replacing rule~\eqref{eq:gravitonrule}, \eqref{eq:gluonrule} and \eqref{eq:lasthrule} from the single trace expansions~\eqref{eq:singletrace}. It also shows for the special case, how the type-I and type-II recursive expansion~\eqref{eq:type1step1} and \eqref{eq:type2step1} emerges from our dimensional reduction scheme~\eqref{E-reduction}.

\subsection{The general case}

Having understood the above special example, we move on to understand the general case. Our purpose is to show how the set $\mathsf{KK}[\pmb{i}]$ emerges from the dimensional reduction, and a further understanding of our replacing rules. For a given list $\gamma=\{a_2,b_2\ldots a_{m},b_{m}\}$, the contribution to the CHY integrand is:
\begin{equation}
\label{Under-red-1}
\frac{(-1)^{m-1}}{\langle 12\ldots r\rangle}\frac{\langle a_2b_2\rangle}{\langle\pmb{2}\rangle}\frac{\langle a_3b_3\rangle}{\langle\pmb{3}\rangle}\ldots\frac{\langle a_mb_m\rangle}{\langle\pmb{m}\rangle}\left.\Pf\!\left(\Psi^{D}_{\mathsf{H},\gamma|\mathsf{H},\gamma}\right)\right|_{\text{dim-red}}\,.
\end{equation}
%
%
%
According to~\cite{Lam:2016tlk}, we can expand $\Pf(\Psi^D)$ in terms of cycles:
\begin{equation}
\Pf\!\left(\Psi^{D}_{\mathsf{H},\gamma|\mathsf{H},\gamma}\right)=(-1)^{\frac{(|\mathsf{H}|+2m-2)(|\mathsf{H}|+2m-1)}{2}}\sum_{p\in\text{perm}(\mathsf{H}\cup\gamma)}(-1)^p\frac{U_{\pmb{I}} U_{\pmb{J}}\ldots U_{\pmb{K}}}{\langle\pmb{I}\rangle\langle\pmb{J}\rangle\ldots\langle\pmb{K}\rangle}\,,
\label{eq:ExpandPf2}
\end{equation}
where the permutation $p$ is composed of the closed cycles $\pmb{I},\pmb{J}\ldots\pmb{K}$. For each closed cycle, $U$ is given by:
\begin{align}
\label{eq:U-cycle}
& U_{\pmb{I}}=\frac{1}{2}\Tr(F_{i_1}\cdot F_{i_2}\cdots F_{i_s})& &\text{for }\pmb{I}=\{i_1,i_2\ldots i_s\}\text{ and } s\geqslant 2 \nonumber\\
& U_{\pmb{I}}=C_{i_1i_1} & &\text{for }\pmb{I}=\{i_1\}\,.
\end{align}
where  $F^{\mu\nu}_{a}$ is defined in Eq.~\eqref{F-def} and $C_{i_1i_1}$ in Eq.~\eqref{eq:ABC}. However, both of them are constructed using $D$ dimensional vectors $K$ and $\W{\mathcal{E}}$.

For simplicity, we again consider a pair $\{a_i,b_i\}\subset\pmb{i}$. Then if $a_i$ and $b_i$ are in different cycles, the contribution vanishes after the dimensional reduction, according to our analysis of the previous subsection. This also applies to the case that $\pmb{I}=\{a_i\}$ since $\W{\mathcal{E}}_{a_i}\cdot K_j=0$ for all $j$ after the dimensional reduction. Next, for $\{a_i,b_i\}\subset\pmb{I}$, they must be next to each other in order to give nonzero contribution. The relevant piece of Eq.~\eqref{eq:ExpandPf2} is
\begin{equation}
\frac{-\langle a_ib_i\rangle}{\langle\pmb{i}\rangle}\frac{U_{\pmb{I}}}{\langle\pmb{I}\rangle}\;\rightarrow\;\frac{-\langle a_ib_i\rangle}{\langle a_i,\pmb{\alpha}_1,b_i,\pmb{\beta}_1\rangle}\frac{1}{2}\frac{\Tr(F_{\pmb{\alpha}_2}\cdot F_{a_i}\cdot F_{b_i}\cdot F_{\pmb{\beta}_2})}{\langle \pmb{\alpha}_2,a_i,b_i,\pmb{\beta}_2\rangle}\,,
\end{equation}
where $F_{\pmb{\alpha}_2}$ (and similarly $F_{\pmb{\beta}_2}$) are defined as the matrix product $\prod_{i\in\pmb{\alpha}_2}F_{i}$. Applying the KK relation to the first Parke-Taylor factor to put $a_i$ and $b_i$ at the two end points, we get
\begin{equation}
\frac{-\langle a_ib_i\rangle}{\langle a_i,\pmb{\alpha}_1,b_i,\pmb{\beta}_1\rangle}\frac{1}{2}\frac{\Tr(F_{\pmb{\alpha}_2}\cdot F_{a_i}\cdot F_{b_i}\cdot F_{\pmb{\beta}_2})}{\langle \pmb{\alpha}_2,a_i,b_i,\pmb{\beta}_2\rangle}=-\frac{1}{2}\frac{(-1)^{|\pmb{\beta}_1|}\Tr(F_{\pmb{\alpha}_2}\cdot F_{a_i}\cdot F_{b_i}\cdot F_{\pmb{\beta}_2})}{\langle\pmb{\alpha}_2,a_i,\pmb{\alpha}_1\shuffle\pmb{\beta}_1^{T},b_i,\pmb{\beta}_2\rangle}\,.
\end{equation}
where $\pmb{\alpha}_1\shuffle\pmb{\beta}_1^{T}=\mathsf{KK}[\pmb{i},a_i,b_i]$ by definition, and the sign $(-1)^{|\pmb{\beta}_1|}$ can be absorbed into the definition of $\W\sum$. Now we can see the pattern: if we treat formally the whole list $\{a_i,\mathsf{KK}[\pmb{i},a_i,b_i],b_i\}$ as a single ``graviton'' $h_i$ and $F_{a_i} \cdot F_{b_i}$ as a single $F_{h_i}$, we can write the above equation as:
\begin{equation}
\label{Du-observe}
\frac{1}{2}\frac{\Tr(F_{\pmb{\alpha}_2}\cdot F_{a_i}\cdot F_{b_i}\cdot F_{\pmb{\beta}_2})}{\langle\pmb{\alpha}_2,a_i,\mathsf{KK}[\pmb{i},a_i,b_i],b_i,\pmb{\beta}_2\rangle}\,\rightarrow\,\frac{1}{2}\frac{\Tr(F_{\pmb{\alpha}_2}\cdot F_{h_i}\cdot F_{\pmb{\beta}_2})}{\langle \pmb{\alpha}_2,h_i,\pmb{\beta}_2\rangle}\equiv\frac{U_{\W{\pmb{I}}}}{\langle\W{\pmb{I}}\rangle}\qquad \left.F_{h_i}\right|_{\text{dim-red}}=-k_{a_i}k_{b_i}\,.
\end{equation}
%
%
Doing it for every pair of $\{a_i,b_i\}$, Eq.~\eqref{Under-red-1} formally becomes
\begin{equation}
\label{Under-red-1-1}
\frac{(-1)^{m-1}(-1)^{\frac{(|\mathsf{H}|+2m-2)(|\mathsf{H}|+2m-1)}{2}}}{\langle 12\ldots r\rangle}\sum_{p}(-1)^{p}\frac{U_{\pmb{\W I}} U_{\pmb{\W J}}\cdots U_{\pmb{\W K}}}{\langle\W{\pmb{I}}\rangle\langle\W{\pmb{J}}\rangle\ldots\langle\W{\pmb{K}}\rangle}
\end{equation}
where $p$ is a permutation of the set $\mathsf{H}\cup\{h_2,h_3\ldots h_{m}\}$ composed of the cycles $\W{\pmb{I}},\W{\pmb{J}}\ldots\W{\pmb{K}}$. Each $h_i$ is formally a trace of gluons in disguise. One can easily see that Eq.~\eqref{Du-observe} and \eqref{Under-red-1-1} are nothing but our familiar single trace EYM integrand, by our replacing rule~\eqref{eq:gluonrule} for regular gravitons.\footnote{Eq.~\eqref{Under-red-1-1} contains a phase factor that depends on the total number of gluon traces. One can show that by following carefully the prescription~\eqref{eq:genericphase} to obtain the amplitudes, the dependence on $m$ will be canceled.}

%
%


In summary, from the point of view given in \eqref{Under-red-1-1}, we have given an explanation of the type-I and type-II recursive expansions \eqref{eq:type1step1} and \eqref{eq:type2step1} as well as the two replacing rules~\eqref{eq:gluonrule} and \eqref{eq:lasthrule}. For the type-I recursion relation and corresponding replacing rule, the understanding is straightforward. For the type-II recursion relation and corresponding replacing rule, thing is a little bit tricky. The reason is that from the form \eqref{Under-red-1-1}, there is no distinction between the regular and fiducial gravitons, while in the single trace recursive expansion, these two types of gravitons are treated very differently. In our paper, although we have derived the \eqref{eq:singletrace} using the CHY-integrand, we did not start from the form \eqref{Under-red-1-1}. How to see the difference between the regular and fiducial gravitons from the form \eqref{Under-red-1-1} is not completely clear to us.

\section{Pure gluon Multitrace EYM amplitudes I: some examples}\label{sec:multi_example}

 In next two
sections, we focus on another special case: pure gluon multitrace  EYM amplitudes (i.e., without any graviton). We will
derive another type of recursive expansion from the squeezing-form integrand~\eqref{eq:mtraceEYM} and show that they are equivalent to the type-II recursive expansion. Thus it provides another support evidence for general type-II recursive expansion. According to Eq.~\eqref{eq:mtraceEYM} and \eqref{eq:genericphase}, we need to add the phase factor
\begin{equation}
\label{eq:phasefactor}
  \text{$m$-trace squeezing-form integrand to amplitude:}\qquad(\text{phase})={(-1)^{\frac{m(m-1)}{2}}}
\end{equation}
to obtain the amplitudes from squeezing-form integrands. The above phase factor is extensively used in both sections.

\subsection{Double-trace}\label{sec:dt}

As first derived from CHY integrands in~\cite{Nandan:2016pya}, pure gluon double-trace
EYM amplitudes can be expanded by pure YM ones. Here, we first repeat
this calculation as our simplest example, in a form more suitable for generalizations.

According to Eq.~\eqref{eq:PfpPi}, if the first trace contains $r$
gluons and  the second contains the rest, the integrand reduces to
\begin{align}
\label{eq:dt1}
\frac{1}{\langle\pmb{1}\rangle\langle\pmb{2}\rangle}\Pfp[\Pi_2(\pmb{1},\pmb{2})]&=-\frac{1}{\langle\pmb{1}\rangle\langle\pmb{2}\rangle}(C_{\mathsf{tr}})_{11}=-\frac{k_{\pmb{1}}^2}{2}\,\frac{1}{\langle\pmb{1}\rangle\langle\pmb{2}\rangle}
\end{align}
Next, we apply the cross-ratio identity~\cite{Cardona:2016gon}:
\begin{equation}
-\frac{k_{\pmb{1}}^2}{2}=\sum_{i\in\pmb{1}}\sum_{j\in\pmb{2}}\left(k_i\cdot k_j\right)\frac{\sigma_{nj}\sigma_{i1}}{\sigma_{ij}\sigma_{n1}}\,,
\end{equation}
where we have assumed implicitly that the two fixed points $1\in\pmb{1}$ and $n\in\pmb{2}$.  To continue, for each term with
$(i,j)$ index, we apply the KK relations to both Parke-Taylor factors:
\begin{align}
\label{eq:KK}
  &\frac{1}{\langle\pmb{1}\rangle}=\frac{1}{\langle 1\ldots i\ldots\rangle}=\frac{(-1)^{|\pmb{1}_{i,1}|}}{\langle 1,\mathsf{KK}[\pmb{1},1,i],i\rangle}\,, \nonumber\\
  &\frac{1}{\langle\pmb{2}\rangle}=\frac{1}{\langle j\ldots n\ldots\rangle}=\frac{(-1)^{|\pmb{2}_{n,j}|}}{\langle j,\mathsf{KK}[\pmb{2},j,n],n\rangle}\,,
\end{align}
where the symbol $\mathsf{KK}[\pmb{1},1,i]$ has been defined in
Eq.~\eqref{KK-symbol-1},
so Eq.~\eqref{eq:dt1} becomes:
\begin{align}
  \frac{1}{\langle\pmb{1}\rangle\langle\pmb{2}\rangle}\Pfp(\Pi)=\underset{\{\underline{1},i\}\in\pmb{1}}{\widetilde{\sum}}~\underset{\{j,\underline{n}\}\subset\pmb{2}}{\widetilde{\sum}}\,\frac{k_i\cdot k_j}{\langle 1,\mathsf{KK}[\pmb{1},1,i],i,j,\mathsf{KK}[\pmb{2},j,n],n\rangle}\,.
\end{align}
Using this result, we get the expansion of pure double-trace EYM
amplitudes by the pure YM ones:
\begin{equation}
\label{eq:dtAn}
  A_{2,0}(\pmb{1}\,|\,\pmb{2})=-\underset{\{\underline{1},i\}\subset\pmb{1}}{\widetilde{\sum}}~\underset{\{j,\underline{n}\}\in\pmb{2}}{\widetilde{\sum}}(k_i\cdot k_j)A_{1,0}(1,\mathsf{KK}[\pmb{1},1,i],i,j,\mathsf{KK}[\pmb{2},j,n],n)\,.
\end{equation}
This expression is different from the one given in Eq.~\eqref{Double-result}, using the type-II recursive expansion.
In appendix~\ref{sec:equiv}, we will show their equivalence through algebraic manipulations. Since our proof using CHY integrand is independent, it gives a strong  support to the proposed type-II recursive expansion \eqref{eq:type2step1}.

\subsection{Triple-trace}\label{sec:tt}
We delete the two rows and columns corresponding to the trace
$\pmb{3}$ in $\Pi_3$, and  expand the reduced Pfaffian of $\Pi_3$ as
\begin{align}
  \Pfp[\Pi_3(\pmb{1},\pmb{2},\pmb{3})]&=\Pf\left(\begin{array}{cccc}
    0 & (A_{\mathsf{tr}})_{12} & -(C_{\mathsf{tr}})_{11} & -(C_{\mathsf{tr}})_{21} \\
    (A_{\mathsf{tr}})_{21} & 0 & -(C_{\mathsf{tr}})_{12} & -(C_{\mathsf{tr}})_{22} \\
    (C_{\mathsf{tr}})_{11} & (C_{\mathsf{tr}})_{12} & 0 & (B_{\mathsf{tr}})_{12} \\
    (C_{\mathsf{tr}})_{21} & (C_{\mathsf{tr}})_{22} & (B_{\mathsf{tr}})_{21} & 0 \\
  \end{array}\right)\nonumber\\
  &=(C_{\mathsf{tr}})_{11}\Pf\left(\begin{array}{cc}
    0 & -(C_{\mathsf{tr}})_{22} \\
    (C_{\mathsf{tr}})_{22} & 0 \\
    \end{array}\right)+(C_{\mathsf{tr}})_{12}(C_{\mathsf{tr}})_{21}-(B_{\mathsf{tr}})_{12}(A_{\mathsf{tr}})_{21}\,.
\end{align}
The triple trace pure gluon EYM integrand thus reduces to
\begin{align}
\label{eq:3tstep1}
\frac{1}{\langle \pmb{1}\rangle\langle \pmb{2}\rangle\langle \pmb{3}\rangle}\Pfp(\Pi_3)&=-\frac{(C_{\mathsf{tr}})_{11}(C_{\mathsf{tr}})_{22}}{\langle \pmb{1}\rangle\langle \pmb{2}\rangle\langle \pmb{3}\rangle}+\frac{(C_{\mathsf{tr}})_{12}(C_{\mathsf{tr}})_{21}-(B_{\mathsf{tr}})_{12}(A_{\mathsf{tr}})_{21}}{\langle \pmb{1}\rangle\langle \pmb{2}\rangle\langle \pmb{3}\rangle}\,.
\end{align}
The second term of the above equation can be further calculated as:
\begin{align}
\label{eq:tt2}
  \mathscr{C}_{12}^{\,\,21}\equiv(C_{\mathsf{tr}})_{12}(C_{\mathsf{tr}})_{21}-(B_{\mathsf{tr}})_{12}(A_{\mathsf{tr}})_{21}&=\sum_{i,b\in\pmb{1}}\sum_{j,a\in\pmb{2}}\frac{\sigma_i\sigma_{aj}(k_i\cdot k_j)(k_a\cdot k_b)}{\sigma_{ij}\sigma_{ab}} \nonumber\\
  &=\frac{1}{2}\sum_{i,b\in\pmb{1}}\sum_{j,a\in\pmb{2}}(k_i\cdot k_j)(k_a\cdot k_b)\frac{\sigma_{aj}\sigma_{ib}}{\sigma_{ij}\sigma_{ab}}\,.
\end{align}
To obtain the last line, we have symmetrized over the pair $(i,b)$ and $(j,a)$.

More importantly, the first term of Eq.~\eqref{eq:3tstep1} can be
processed in the  following way. Assuming the particle $1\in\pmb{1}$ and $n\in\pmb{3}$, we first apply the
cross-ratio identity to $(C_{\mathsf{tr}})_{11}$:
\begin{equation}
\label{eq:C311}
  -(C_{\mathsf{tr}})_{11}=-\frac{k^2_{\pmb{1}}}{2}=\left(\sum_{i\in\pmb{1}}\sum_{j\in\pmb{2}}+\sum_{i\in\pmb{1}}\sum_{j\in\pmb{3}}\right)(k_i\cdot k_j)\frac{\sigma_{nj}\sigma_{i1}}{\sigma_{ij}\sigma_{n1}}\equiv \mathscr{R}(1|2)+\mathscr{R}(1|3)\,.
\end{equation}
As in section~\ref{sec:dt}, acting the second term onto the Parke-Taylor factors, we get
\begin{align}
  \frac{\mathscr{R}(1|3)}{\langle \pmb{1}\rangle\langle \pmb{2}\rangle\langle \pmb{3}\rangle}&=\underset{\{\underline{1},i\}\subset\pmb{1}}{\widetilde{\sum}}~\underset{\{j,\underline{n}\}\subset\pmb{3}}{\widetilde{\sum}}\,\frac{k_i\cdot k_j}{\langle 1,\mathsf{KK}[\pmb{1},1,i],i,j,\mathsf{KK}[\pmb{3},j,n],n\rangle\langle\pmb{2}\rangle}\,.
\end{align}
However, we cannot do this for the $\mathscr{R}(1|2)$ term of \eqref{eq:C311},
since $j$ and $n$ are no longer in the same trace. After the above manipulation, we have
\begin{align}
\label{eq:tt1step1}
  \text{first term of \eqref{eq:3tstep1}}&=-\underset{\{\underline{1},i\}\subset\pmb{1}}{\widetilde{\sum}}~\underset{\{j,\underline{n}\}\subset\pmb{3}}{\widetilde{\sum}}\,\frac{(k_i\cdot k_j)[-(C_{\mathsf{tr}})_{22}]}{\langle 1,\mathsf{KK}[\pmb{1},1,i],i,j,\mathsf{KK}[\pmb{3},j,n],n\rangle\langle\pmb{2}\rangle}
  -\frac{\mathscr{R}(1|2)[-(C_{\mathsf{tr}})_{22}]}{\langle\pmb{1}\rangle\langle\pmb{2}\rangle\langle\pmb{3}\rangle}\,.
\end{align}
Then the first line of the above equation is just a linear combination of double-trace integrands with $\widetilde{\pmb{1}}=\{1,\mathsf{KK}[\pmb{1}],i,j,\mathsf{KK}[\pmb{3}],n\}$ and $\widetilde{\pmb{2}}=\pmb{2}$. Next, we apply the cross-ratio identity to $(C_{\mathsf{tr}})_{22}$ in the second line of \eqref{eq:tt1step1}. However, we choose a different reference point for each term with a given $(i,j)$ pair in the sum of $\mathscr{R}(1|2)$ when using the cross-ratio identity onto $(C_{\mathsf{tr}})_{22}$:
\begin{equation}
\label{eq:R12C22}
  \mathscr{R}(1|2)[-(C_{\mathsf{tr}})_{22}]=\sum_{i\in\pmb{1}}\sum_{j\in\pmb{2}}(k_i\cdot k_j)\frac{\sigma_{nj}\sigma_{i1}}{\sigma_{ij}\sigma_{n1}}\left(\sum_{a\in\pmb{2}}\sum_{b\in\pmb{1}}+\sum_{a\in\pmb{2}}\sum_{b\in\pmb{3}}\right)(k_a\cdot k_b)\frac{\sigma_{nb}\sigma_{aj}}{\sigma_{ab}\sigma_{nj}}\equiv\mathscr{R}[12]+\mathscr{R}(1|2|3)\,,
\end{equation}
where we have defined the following symbols:
\begin{align}
\label{eq:R12}
  \mathscr{R}[12]&=\sum_{i,b\in\pmb{1}}\sum_{j,a\in\pmb{2}}(k_i\cdot k_j)(k_a\cdot k_b)\frac{\sigma_{i1}\sigma_{nb}\sigma_{aj}}{\sigma_{ij}\sigma_{ab}\sigma_{n1}}\nonumber\\
  &=\frac{1}{2}\sum_{i,b\in\pmb{1}}\sum_{j,a\in\pmb{2}}(k_i\cdot k_j)(k_a\cdot k_b)\left(\frac{\sigma_{i1}\sigma_{nb}\sigma_{aj}}{\sigma_{ij}\sigma_{ab}\sigma_{n1}}+\frac{\sigma_{b1}\sigma_{ni}\sigma_{ja}}{\sigma_{ba}\sigma_{ji}\sigma_{n1}}\right)\nonumber\\
  &=\frac{1}{2}\sum_{i,b\in\pmb{1}}\sum_{j,a\in\pmb{2}}(k_i\cdot k_j)(k_a\cdot k_b)\frac{\sigma_{aj}\sigma_{ib}}{\sigma_{ij}\sigma_{ab}}=\mathscr{C}_{12}^{\,\,21}\,,\nonumber\\
  \mathscr{R}(1|2|3)&=\sum_{i\in\pmb{1}}\sum_{j,a\in\pmb{2}}\sum_{b\in\pmb{3}}(k_i\cdot k_j)(k_a\cdot k_b)\frac{\sigma_{i1}\sigma_{nb}\sigma_{aj}}{\sigma_{ij}\sigma_{ab}\sigma_{n1}}\,.
\end{align}
In the argument of $\mathscr{R}$, we use $[\ldots]$ to denote a closed trace cycle and $(a|\ldots|b)$ an open trace cycle with $a$ and $b$ at two ends. 
With the help of the following KK relations:
\begin{align}
  & \frac{1}{\langle\pmb{2}\rangle}=\frac{1}{\langle j\ldots a\ldots\rangle}=\frac{(-1)^{|\pmb{2}_{a,j}|}}{\langle j,\mathsf{KK}[\pmb{2},j,a],a\rangle}\nonumber\\
  & \frac{1}{\langle\pmb{3}\rangle}=\frac{1}{\langle b\ldots n\ldots\rangle}=\frac{(-1)^{|\pmb{3}_{n,b}|}}{\langle b,\mathsf{KK}[\pmb{3},b,n],n\rangle}\,,
\end{align}
we have
\begin{align}
  \frac{\mathscr{R}(1|2|3)}{\langle\pmb{1}\rangle\langle\pmb{2}\rangle\langle\pmb{3}\rangle}&=\underset{\{\underline{1},i\}\subset\pmb{1}}{\widetilde{\sum}}~\underset{\{j,a\}\subset\pmb{2}}{\widetilde{\sum}}~\underset{\{b,\underline{n}\}\subset\pmb{3}}{\widetilde{\sum}}\,\frac{(k_i\cdot k_j)(k_a\cdot k_b)}{\langle1,\mathsf{KK}[\pmb{1},1,i],i,j,\mathsf{KK}[\pmb{2},j,a],a,b,\mathsf{KK}[\pmb{3},b,n],n\rangle}\,.
\end{align}
Consequently, Eq.~\eqref{eq:tt1step1} becomes:
\begin{align}
\label{eq:tt1step2}
\text{first term of \eqref{eq:3tstep1}}&=-\underset{\{\underline{1},i\}\subset\pmb{1}}{\widetilde{\sum}}~\underset{\{j,\underline{n}\}\subset\pmb{3}}{\widetilde{\sum}}\frac{(k_i\cdot k_j)[-(C_{\mathsf{tr}})_{22}]}{\langle 1,\mathsf{KK}[\pmb{1},1,i],i,j,\mathsf{KK}[\pmb{3},j,n],n\rangle\langle\pmb{2}\rangle}
\nonumber\\
  &\quad -\underset{\{\underline{1},i\}\subset\pmb{1}}{\widetilde{\sum}}~\underset{\{j,a\}\subset\pmb{2}}{\widetilde{\sum}}~\underset{\{b,\underline{n}\}\subset\pmb{3}}{\widetilde{\sum}}\,\frac{(k_i\cdot k_j)(k_a\cdot k_b)}{\langle 1,\mathsf{KK}[\pmb{1},1,i],i,j,\mathsf{KK}[\pmb{2},j,a],a,b,\mathsf{KK}[\pmb{3},b,n],n\rangle}\nonumber\\
  &\quad-\frac{\mathscr{R}[12]}{\langle\pmb{1}\rangle\langle\pmb{2}\rangle\langle\pmb{3}\rangle}\,.
\end{align}
Now the second line of the above equation is just a linear combination of pure YM integrands, while the third line cancels
exactly $\mathscr{C}_{12}^{\,\,21}$ given in \eqref{eq:tt2}, since $\mathscr{R}[12]=\mathscr{C}_{12}^{\,\,21}$.
This identity holds at the algebraic level. More generally, we can replace $\pmb{1}$ and  $\pmb{2}$ by two arbitrary traces, and get:
\begin{equation}
\label{eq:Cpq}
  \mathscr{R}[pq]=(C_{\mathsf{tr}})_{pq}(C_{\mathsf{tr}})_{qp}-(B_{\mathsf{tr}})_{pq}(A_{\mathsf{tr}})_{qp}=\mathscr{C}_{pq}^{\,\,qp}\,.
\end{equation}
At this point, we have shown that the triple-trace Pfaffian can be expanded as:
\begin{equation}
  \Pfp(\Pi_3)=-\mathscr{R}(1|3)\Pf[\Pi_2(\pmb{2})]-\mathscr{R}(1|2|3)\,.
\end{equation}
Our final formula follows immediately after we act the above result onto the triple-trace Parke-Taylor factor:
\begin{align}
\label{eq:ttfinal}
  \frac{\Pfp(\Pi_3)}{\langle \pmb{1}\rangle\langle \pmb{2}\rangle\langle \pmb{3}\rangle}&=-\underset{\{\underline{1},i\}\subset\pmb{1}}{\widetilde{\sum}}~\underset{\{j,\underline{n}\}\subset\pmb{3}}{\widetilde{\sum}}\,\frac{(k_i\cdot k_j)[-(C_{\mathsf{tr}})_{22}]}{\langle 1,\mathsf{KK}[\pmb{1},1,i],i,j,\mathsf{KK}[\pmb{3},j,n],n\rangle\langle\pmb{2}\rangle}
  \nonumber\\
  &\quad -\underset{\{\underline{1},i\}\subset\pmb{1}}{\widetilde{\sum}}~\underset{\{j,a\}\subset\pmb{2}}{\widetilde{\sum}}~\underset{\{b,\underline{n}\}\subset\pmb{3}}{\widetilde{\sum}}\frac{(k_i\cdot k_j)(k_a\cdot k_b)}{\langle 1,\mathsf{KK}[\pmb{1},1,i],i,j,\mathsf{KK}[\pmb{2},j,a],a,b,\mathsf{KK}[\pmb{3},b,n],n\rangle}\,.
\end{align}
We remark that when deriving this result, the only place that we require the on-shell condition  is the cross-ratio identity applied to $(C_{\mathsf{tr}})_{11}$ and $(C_{\mathsf{tr}})_{22}$, while all the other transformations are merely algebraic identities. This observation is very important for the generalization to arbitrary number of traces.

To close this subsection, we perform the CHY integration on Eq.~\eqref{eq:ttfinal},  translating it into the amplitude relation between triple-trace, double-trace pure gluon EYM amplitudes, and pure YM ones:
\begin{align}
  A_{3,0}(\pmb{1}\,|\,\pmb{2}\,|\,\pmb{3})&=\underset{\{\underline{1},i\}\subset\pmb{1}}{\widetilde{\sum}}~\underset{\{j,\underline{n}\}\subset\pmb{3}}{\widetilde{\sum}}(-k_j\cdot k_i)A_{2,0}(1,\mathsf{KK}[\pmb{1},1,i],i,j,\mathsf{KK}[\pmb{3},j,n],n\,|\,\pmb{2})\nonumber\\
  &\quad+\underset{\{\underline{1},i\}\subset\pmb{1}}{\widetilde{\sum}}~\underset{\{j,a\}\subset\pmb{2}}{\widetilde{\sum}}~\underset{\{b,\underline{n}\}\subset\pmb{3}}{\widetilde{\sum}}
  (k_b\cdot k_a)(k_j\cdot k_i)
  A_{1,0}(1,\mathsf{KK}[\pmb{1},1,i],i,j,\mathsf{KK}[\pmb{2},j,a],a,b,\mathsf{KK}[\pmb{3},b,n],n)\,.
\end{align}
We note that the sign change is due to our convention
\eqref{eq:phasefactor}. Applying Eq.~\eqref{eq:dtAn} to the first
term, we can expand
$A_{3,0}(\pmb{1}\,|\,\pmb{2}\,|\,\pmb{3})$ in terms of
pure YM amplitudes in KK basis.

\subsection{Four-trace}\label{sec:4trace}
The Laplace expansion of the reduced Pfaffian $\Pfp[\Pi_4(\pmb{1},\pmb{2},\pmb{3},\pmb{4})]$ leads to
\begin{align}
  \label{eq:4t}
  \Pfp(\Pi_4)&=-(C_{\mathsf{tr}})_{11}\Pf[\Pi_3(\pmb{2,3})]\nonumber\\
  &\quad +\left[(C_{\mathsf{tr}})_{12}(C_{\mathsf{tr}})_{21}-(B_{\mathsf{tr}})_{12}(A_{\mathsf{tr}})_{21}\right]\!\Pf[\Pi_2(\pmb{3})]+\left[(C_{\mathsf{tr}})_{13}(C_{\mathsf{tr}})_{31}-(B_{\mathsf{tr}})_{13}(A_{\mathsf{tr}})_{31}\right]\!\Pf[\Pi_2(\pmb{2})]\nonumber\\
  &\quad +\left[(C_{\mathsf{tr}})_{12}(C_{\mathsf{tr}})_{23}-(B_{\mathsf{tr}})_{12}(A_{\mathsf{tr}})_{23}\right](C_{\mathsf{tr}})_{31}-\left[(C_{\mathsf{tr}})_{12}(B_{\mathsf{tr}})_{23}+(B_{\mathsf{tr}})_{12}(C_{\mathsf{tr}})_{32}\right](A_{\mathsf{tr}})_{31}\nonumber\\
  &\quad +\left[(C_{\mathsf{tr}})_{13}(C_{\mathsf{tr}})_{32}-(B_{\mathsf{tr}})_{13}(A_{\mathsf{tr}})_{32}\right](C_{\mathsf{tr}})_{21}-\left[(C_{\mathsf{tr}})_{13}(B_{\mathsf{tr}})_{32}+(B_{\mathsf{tr}})_{13}(C_{\mathsf{tr}})_{23}\right](A_{\mathsf{tr}})_{21}\,,
\end{align}
where the Pfaffians involved are all parts of the EYM integrands with fewer traces:
\begin{align}
  & \Pf[\Pi_3(\pmb{2,3})]=\Pfp[\Pi_3(\pmb{1'},\pmb{2},\pmb{3})]: & &\text{triple-trace integrand with} & & \pmb{1'}=\pmb{1}\cup\pmb{4}\nonumber\\
  & \Pf[\Pi_2(\pmb{3})]=\Pfp[\Pi_2(\pmb{1''},\pmb{2'})]: & &\text{double-trace integrand with} & &\pmb{1''}=\pmb{1}\cup\pmb{2}\cup\pmb{4}\qquad\pmb{2'}=\pmb{3} \nonumber\\
  & \Pf[\Pi_2(\pmb{2})]=\Pfp[\Pi_2(\pmb{1'''},\pmb{2})]: & &\text{double-trace integrand with} & &\pmb{1'''}=\pmb{1}\cup\pmb{3}\cup\pmb{4}\,.
\end{align}
Before we start, it is convenient to list the KK relations we are going to use for the four traces:
\begin{subequations}
\begin{align}
  &\forall i\in\pmb{1}& &\frac{1}{\langle\pmb{1}\rangle}=\frac{1}{\langle 1\ldots i\ldots\rangle}=\frac{(-1)^{|\pmb{1}_{i,1}|}}{\langle 1,\mathsf{KK}[\pmb{1},1,i],i\rangle}\\
  &\forall j,a\in\pmb{2}& &\frac{1}{\langle\pmb{2}\rangle}=\frac{1}{\langle j\ldots a\ldots\rangle}=\frac{(-1)^{|\pmb{2}_{a,j}|}}{\langle j,\mathsf{KK}[\pmb{2},j,a],a\rangle}\\
  &\forall b,c\in\pmb{3}& &\frac{1}{\langle\pmb{3}\rangle}=\frac{1}{\langle b\ldots c\ldots\rangle}=\frac{(-1)^{|\pmb{3}_{c,b}|}}{\langle b,\mathsf{KK}[\pmb{3},b,c],c\rangle}\\
  &\forall d\in\pmb{4}& &\frac{1}{\langle\pmb{4}\rangle}=\frac{1}{\langle d\ldots n\ldots\rangle}=\frac{(-1)^{|\pmb{4}_{n,d}|}}{\langle d,\mathsf{KK}[\pmb{4},d,n],n\rangle}\,.
\end{align}
\end{subequations}
As the first step of this expansion, we apply the cross-ratio identity to $(C_{\mathsf{tr}})_{11}$:
\begin{align}
  -(C_{\mathsf{tr}})_{11}=\sum_{i\in\pmb{1}}\left(\sum_{j\in\pmb{2}}+\sum_{j\in\pmb{3}}+\sum_{j\in\pmb{4}}\right)(k_i\cdot k_j)\frac{\sigma_{nj}\sigma_{i1}}{\sigma_{ij}\sigma_{n1}}\equiv\mathscr{R}(1|2)+\mathscr{R}(1|3)+\mathscr{R}(1|4)\,,
\end{align}
such that the first term of Eq.~\eqref{eq:4t} becomes:
\begin{align}
\label{eq:4tstep1}
-\frac{(C_{\mathsf{tr}})_{11}\Pf[{\Pi}_{3}(\pmb{2,3})]}{\langle\pmb{1}\rangle\langle\pmb{2}\rangle\langle\pmb{3}\rangle\langle\pmb{4}\rangle}&=\frac{\left[\mathscr{R}(1|2)+\mathscr{R}(1|3)+\mathscr{R}(1|4)\right]\Pf[{\Pi}_{3}(\pmb{2,3})]}{\langle\pmb{1}\rangle\langle\pmb{2}\rangle\langle\pmb{3}\rangle\langle\pmb{4}\rangle}\nonumber\\
&=\underset{\{\underline{1},i\}\subset\pmb{1}}{\widetilde{\sum}}~\underset{\{d,\underline{n}\}\subset\pmb{4}}{\widetilde{\sum}}\frac{(k_i\cdot k_d)\Pf[{\Pi}_{3}(\pmb{2,3})]}{\langle 1,\mathsf{KK}[\pmb{1},1,i],i,d,\mathsf{KK}[\pmb{4},d,n],n\rangle\langle\pmb{2}\rangle\langle\pmb{3}\rangle}\nonumber\\
  &\quad+\frac{\left[\mathscr{R}(1|2)+\mathscr{R}(1|3)\right]\Pf[\Pi_{3}(\pmb{2,3})]}{{\langle\pmb{1}\rangle\langle\pmb{2}\rangle\langle\pmb{3}\rangle\langle\pmb{4}\rangle}}\,.
\end{align}
Now the first line of this result is in its final form, giving us a linear combination of triple-trace integrands. We then expand $\Pf[\Pi_3(\pmb{2,3})]$ in the last line of the above equation as follows:
\begingroup
\allowdisplaybreaks
\begin{subequations}
\begin{align}
  \label{eq:t4step2a}
  \mathscr{R}(1|2)\Pf[\Pi_3(\pmb{2,3})]&=\mathscr{R}(1|2)(C_{\mathsf{tr}})_{22}\Pf[\Pi_2(\pmb{3})]+\mathscr{R}(1|2)(C_{\mathsf{tr}})_{23}(C_{\mathsf{tr}})_{32}-\mathscr{R}(1|2)(B_{\mathsf{tr}})_{23}(A_{\mathsf{tr}})_{32}\nonumber\\*
  &=-\mathscr{R}(1|2|4)\Pf[\Pi_2(\pmb{3})]-\mathscr{R}[12]\Pf[\Pi_2(\pmb{3})]\nonumber\\*
  &\quad-\mathscr{R}(1|2|3)\Pf[\Pi_2(\pmb{3})]+\mathscr{R}(1|2)(C_{\mathsf{tr}})_{23}(C_{\mathsf{tr}})_{32}-\mathscr{R}(1|2)(B_{\mathsf{tr}})_{23}(A_{\mathsf{tr}})_{32}\\
  \label{eq:t4step2b}
  \mathscr{R}(1|3)\Pf[\Pi_3(\pmb{2,3})]&=\mathscr{R}(1|3)(C_{\mathsf{tr}})_{33}\Pf[\Pi_2(\pmb{2})]+\mathscr{R}(1|3)(C_{\mathsf{tr}})_{32}(C_{\mathsf{tr}})_{23}-\mathscr{R}(1|3)(B_{\mathsf{tr}})_{32}(A_{\mathsf{tr}})_{23}\nonumber\\*
  &=-\mathscr{R}(1|3|4)\Pf[\Pi_2(\pmb{2})]-\mathscr{R}[13]\Pf[\Pi_2(\pmb{2})]\nonumber\\*
  &\quad-\mathscr{R}(1|3|2)\Pf[\Pi_2(\pmb{3})]+\mathscr{R}(1|3)(C_{\mathsf{tr}})_{32}(C_{\mathsf{tr}})_{23}-\mathscr{R}(1|3)(B_{\mathsf{tr}})_{32}(A_{\mathsf{tr}})_{23}\,,
\end{align}
\end{subequations}
\endgroup
where the cross-ratio identity has been applied to $(C_{\mathsf{tr}})_{22}$ and $(C_{\mathsf{tr}})_{33}$. In the second line of both Eq.~\eqref{eq:t4step2a} and \eqref{eq:t4step2b}, we have
\begin{align}
\label{eq:R124}
  \mathscr{R}(1|2|4)&=\sum_{i\in\pmb{1}}\sum_{a,j\in\pmb{2}}\sum_{d\in\pmb{4}}(k_i\cdot k_j)(k_a\cdot k_d)\frac{\sigma_{i1}\sigma_{nd}\sigma_{aj}}{\sigma_{ij}\sigma_{ad}\sigma_{n1}}\nonumber\\
  \mathscr{R}(1|3|4)&=\sum_{i\in\pmb{1}}\sum_{b,c\in\pmb{3}}\sum_{d\in\pmb{4}}(k_i\cdot k_b)(k_c\cdot k_d)\frac{\sigma_{i1}\sigma_{nd}\sigma_{cb}}{\sigma_{ib}\sigma_{cd}\sigma_{n1}}\,,
\end{align}
each of which leads to a linear combination of double trace integrands:
\begingroup
\allowdisplaybreaks
\begin{subequations}
\begin{align}
  \frac{\mathscr{R}(1|2|4)\Pf[\Pi_2(\pmb{3})]}{\langle\pmb{1}\rangle\langle\pmb{2}\rangle\langle\pmb{3}\rangle\langle\pmb{4}\rangle}&=\underset{\{\underline{1},i\}\subset\pmb{1}}{\widetilde{\sum}}~\underset{\{a,j\}\subset\pmb{2}}{\widetilde{\sum}}~\underset{\{d,\underline{n}\}\subset\pmb{4}}{\widetilde{\sum}}\,\frac{(k_i\cdot k_j)(k_a\cdot k_d)\Pf[\Pi_2(\pmb{3})]}{\langle 1,\mathsf{KK}[\pmb{1},1,i],i,j,\mathsf{KK}[\pmb{2},j,a],a,d,\mathsf{KK}[\pmb{4},d,n],n\rangle\langle\pmb{3}\rangle}\\
  \frac{\mathscr{R}(1|3|4)\Pf[\Pi_3(\pmb{2})]}{\langle\pmb{1}\rangle\langle\pmb{2}\rangle\langle\pmb{3}\rangle\langle\pmb{4}\rangle}&=\underset{\{\underline{1},i\}\subset\pmb{1}}{\widetilde{\sum}}~\underset{\{b,c\}\subset\pmb{3}}{\widetilde{\sum}}~\underset{\{d,\underline{n}\}\in\pmb{4}}{\widetilde{\sum}}\,\frac{(k_i\cdot k_b)(k_c\cdot k_d)\Pf[\Pi_2(\pmb{2})]}{\langle 1,\mathsf{KK}[\pmb{1},1,i],i,b,\mathsf{KK}[\pmb{3},b,c],c,d,\mathsf{KK}[\pmb{4},d,n],n\rangle\langle\pmb{2}\rangle}\,,
\end{align}
\end{subequations}
\endgroup
while $\mathcal{R}[12]$ and $\mathcal{R}[13]$ in the first line of Eq.~\eqref{eq:t4step2a} and \eqref{eq:t4step2b} exactly cancel the two terms in the second line of Eq.~\eqref{eq:4t} because of \eqref{eq:Cpq}.
As the last step, we apply the cross-ratio identity to the last line of both Eq.~\eqref{eq:t4step2a} and \eqref{eq:t4step2b}. Similar to Eq.~\eqref{eq:R12C22}, the result is:
\begin{align}
\label{eq:R123C}
&\mathscr{R}(1|2|3)[-(C_{\mathsf{tr}})_{33}]=\mathscr{R}(1|23|4)+\mathscr{R}(1|23|1)+\mathscr{R}(1|2)\mathscr{R}[23]\nonumber\\
&\mathscr{R}(1|3|2)[-(C_{\mathsf{tr}})_{22}]=\mathscr{R}(1|32|4)+\mathscr{R}(1|32|1)+\mathscr{R}(1|3)\mathscr{R}[32]\,.
\end{align}
The sum of the aforementioned two lines gives:
\begin{align}
\label{eq:t4step3}
  &\text{last line of \eqref{eq:t4step2a}}+\text{last line of \eqref{eq:t4step2b}}=-\mathscr{R}(1|23|4)-\mathscr{R}(1|32|4)-\mathscr{R}[123]\,,
\end{align}
in which we have used the identity \eqref{eq:Cpq}
to reach this result. In Eq.~\eqref{eq:t4step3}, we can use $\mathscr{R}(1|23|4)$ and $\mathscr{R}(1|32|4)$ to connect all the Parke-Taylor factors, resulting in a linear combination of pure YM integrands:
\begingroup
\allowdisplaybreaks
\begin{subequations}
\label{eq:R1234}
\begin{align}
  \mathscr{R}(1|23|4)&=\sum_{i\in\pmb{1}}\sum_{a,j\in\pmb{2}}\sum_{b,c\in\pmb{3}}\sum_{d\in\pmb{4}}(k_i\cdot k_j)(k_a\cdot k_b)(k_c\cdot k_d)\frac{\sigma_{i1}\sigma_{nd}\sigma_{aj}\sigma_{cb}}{\sigma_{ij}\sigma_{ab}\sigma_{cd}\sigma_{n1}}\nonumber\\*
  \frac{\mathscr{R}(1|23|4)}{\langle\pmb{1}\rangle\langle\pmb{2}\rangle\langle\pmb{3}\rangle\langle\pmb{4}\rangle}&=\underset{\{\underline{1},i\}\subset\pmb{1}}{\widetilde{\sum}}~\underset{\{a,j\}\subset\pmb{2}}{\widetilde{\sum}}~\underset{\{b,c\}\subset\pmb{3}}{\widetilde{\sum}}~\underset{\{d,\underline{n}\}\subset\pmb{4}}{\widetilde{\sum}}(k_i\cdot k_j)(k_a\cdot k_b)(k_c\cdot k_d)\nonumber\\*
  &\qquad\quad\times\langle 1,\mathsf{KK}[\pmb{1},1,i],i,j,\mathsf{KK}[\pmb{2},j,a],a,b,\mathsf{KK}[\pmb{3},b,c],c,d,\mathsf{KK}[\pmb{4},d,n],n\rangle^{-1}\\
  \mathscr{R}(1|32|4)&=\sum_{i\in\pmb{1}}\sum_{b,c\in\pmb{3}}\sum_{a,j\in\pmb{2}}\sum_{d\in\pmb{4}}(k_i\cdot k_b)(k_c\cdot k_j)(k_a\cdot k_d)\frac{\sigma_{i1}\sigma_{nd}\sigma_{cb}\sigma_{aj}}{\sigma_{ib}\sigma_{cj}\sigma_{ad}\sigma_{n1}}\nonumber\\*
  \frac{\mathscr{R}(1|32|4)}{\langle\pmb{1}\rangle\langle\pmb{2}\rangle\langle\pmb{3}\rangle\langle\pmb{4}\rangle}&=\underset{\{\underline{1},i\}\subset\pmb{1}}{\widetilde{\sum}}~\underset{\{a,j\}\subset\pmb{2}}{\widetilde{\sum}}~\underset{\{b,c\}\subset\pmb{3}}{\widetilde{\sum}}~\underset{\{d,\underline{n}\}\subset\pmb{4}}{\widetilde{\sum}}(k_i\cdot k_b)(k_c\cdot k_j)(k_a\cdot k_d)\nonumber\\*
&\qquad\quad\times\langle 1,\mathsf{KK}[\pmb{1},1,i],i,b,\mathsf{KK}[\pmb{3},b,c],c,j,\mathsf{KK}[\pmb{2},j,a],a,d,\mathsf{KK}[\pmb{4},d,n],n\rangle^{-1}\,,
\end{align}
\end{subequations}
\endgroup
while $\mathscr{R}[123]\equiv\mathscr{R}(1|23|1)+\mathscr{R}(1|32|1)$ cancel the last two lines of Eq.~\eqref{eq:4t}, because of the identity:
\begin{align}
\label{eq:cancel2}
  \mathscr{R}[pqr]&=\left[(C_{\mathsf{tr}})_{pq}(C_{\mathsf{tr}})_{qr}-(B_{\mathsf{tr}})_{pq}(A_{\mathsf{tr}})_{qr}\right](C_{\mathsf{tr}})_{rp}-\left[(C_{\mathsf{tr}})_{pq}(B_{\mathsf{tr}})_{qr}+(B_{\mathsf{tr}})_{pq}(C_{\mathsf{tr}})_{rq}\right](A_{\mathsf{tr}})_{rp}\nonumber\\
  &\quad +\left[(C_{\mathsf{tr}})_{pr}(C_{\mathsf{tr}})_{rq}-(B_{\mathsf{tr}})_{pr}(A_{\mathsf{tr}})_{rq}\right](C_{\mathsf{tr}})_{qp}-\left[(C_{\mathsf{tr}})_{pr}(B_{\mathsf{tr}})_{rq}+(B_{\mathsf{tr}})_{pr}(C_{\mathsf{tr}})_{qr}\right](A_{\mathsf{tr}})_{qp}\nonumber\\
  &=\mathscr{C}_{pq}^{\,\,qr}(C_{\mathsf{tr}})_{rp}-\mathscr{C}_{pqr}^{\,\,q}(A_{\mathsf{tr}})_{rp}+\mathscr{C}_{pr}^{\,\,rq}(C_{\mathsf{tr}})_{qp}-\mathscr{C}_{prq}^{\,\,r}(A_{\mathsf{tr}})_{qp}\,.
\end{align}
The proof is very similar to that of Eq.~\eqref{eq:Cpq}: the left hand side of Eq.~\eqref{eq:cancel2} gives
\begin{align}
  \mathscr{R}[pqr]&=\sum_{d,i\in\pmb{p}}\sum_{j,a\in\pmb{q}}\sum_{b,c\in\pmb{r}}\left[(k_i\cdot k_j)(k_a\cdot k_b)(k_c\cdot k_d)\frac{\sigma_{i1}\sigma_{aj}\sigma_{nd}\sigma_{cb}}{\sigma_{ij}\sigma_{n1}\sigma_{ab}\sigma_{cd}}+(k_i\cdot k_b)(k_c\cdot k_j)(k_a\cdot k_d)\frac{\sigma_{i1}\sigma_{cb}\sigma_{nd}\sigma_{aj}}{\sigma_{ib}\sigma_{n1}\sigma_{cj}\sigma_{ad}}\right]\nonumber\\
  &=\frac{1}{2}\sum_{d,i\in\pmb{p}}\sum_{j,a\in\pmb{q}}\sum_{b,c\in\pmb{r}}\left[(k_i\cdot k_j)(k_a\cdot k_b)(k_c\cdot k_d)\frac{\sigma_{aj}\sigma_{cb}\sigma_{id}}{\sigma_{ij}\sigma_{ab}\sigma_{cd}}+(k_i\cdot k_b)(k_c\cdot k_j)(k_a\cdot k_d)\frac{\sigma_{aj}\sigma_{cb}\sigma_{id}}{\sigma_{ib}\sigma_{cj}\sigma_{ad}}\right]\,.
\end{align}
On the right hand side of Eq.~\eqref{eq:cancel2}, we have:
\begin{align}
\label{eq:Cpqr}
  &\mathscr{C}_{pq}^{\,\,qr}\equiv(C_{\mathsf{tr}})_{pq}(C_{\mathsf{tr}})_{qr}-(B_{\mathsf{tr}})_{pq}(A_{\mathsf{tr}})_{qr}=\sum_{i\in\pmb{p}}\sum_{j,a\in\pmb{q}}\sum_{b\in\pmb{r}}(k_i\cdot k_j)(k_a\cdot k_b)\frac{\sigma_i\sigma_{ja}}{\sigma_{ij}\sigma_{ab}}\nonumber\\
  &\mathscr{C}_{pqr}^{\,\,q}\equiv(C_{\mathsf{tr}})_{pq}(B_{\mathsf{tr}})_{qr}+(B_{\mathsf{tr}})_{pq}(C_{\mathsf{tr}})_{rq}=\sum_{i\in\pmb{p}}\sum_{j,a\in\pmb{q}}\sum_{b\in\pmb{r}}(k_i\cdot k_j)(k_a\cdot k_b)\frac{\sigma_i\sigma_b\sigma_{ja}}{\sigma_{ij}\sigma_{ab}}\nonumber\\
  &\mathscr{C}_{pr}^{\,\,rq}\equiv(C_{\mathsf{tr}})_{pr}(C_{\mathsf{tr}})_{rq}-(C_{\mathsf{tr}})_{pr}(A_{\mathsf{tr}})_{rq}=\sum_{i\in\pmb{p}}\sum_{j\in\pmb{q}}\sum_{b,c\in\pmb{r}}(k_i\cdot k_b)(k_c\cdot k_j)\frac{\sigma_i\sigma_{bc}}{\sigma_{ib}\sigma_{cj}}\nonumber\\
  &\mathscr{C}_{prq}^{\,\,r}\equiv(C_{\mathsf{tr}})_{pr}(B_{\mathsf{tr}})_{rq}+(B_{\mathsf{tr}})_{pr}(C_{\mathsf{tr}})_{qr}=\sum_{i\in\pmb{p}}\sum_{j\in\pmb{q}}\sum_{b,c\in\pmb{r}}(k_i\cdot k_b)(k_c\cdot k_j)\frac{\sigma_i\sigma_j\sigma_{bc}}{\sigma_{ib}\sigma_{cj}}
\end{align}
Using these new symbols, we can easily see that the right hand side of Eq.~\eqref{eq:cancel2} gives a recursive construction of the $\mathscr{C}$ symbols with more indices:
\begin{align}
\label{eq:Cpqrp}
  \mathscr{C}_{pqr}^{\,\,qrp}&\equiv\mathscr{C}_{pq}^{\,\,qr}(C_{\mathsf{tr}})_{rp}-\mathscr{C}_{pqr}^{\,\,q}(A_{\mathsf{tr}})_{rp}+\mathscr{C}_{pr}^{\,\,rq}(C_{\mathsf{tr}})_{qp}-\mathscr{C}_{prq}^{\,\,r}(A_{\mathsf{tr}})_{qp}\nonumber\\
  &=\left[(C_{\mathsf{tr}})_{pq}(C_{\mathsf{tr}})_{qr}-(C_{\mathsf{tr}})_{pq}(A_{\mathsf{tr}})_{qr}\right](C_{\mathsf{tr}})_{rp}-\left[(C_{\mathsf{tr}})_{pq}(B_{\mathsf{tr}})_{qr}+(B_{\mathsf{tr}})_{pq}(C_{\mathsf{tr}})_{rq}\right](A_{\mathsf{tr}})_{rp}\nonumber\\
  &\quad +\left[(C_{\mathsf{tr}})_{pr}(C_{\mathsf{tr}})_{rq}-(C_{\mathsf{tr}})_{pr}(A_{\mathsf{tr}})_{rq}\right](C_{\mathsf{tr}})_{qp}-\left[(C_{\mathsf{tr}})_{pr}(B_{\mathsf{tr}})_{rq}+(B_{\mathsf{tr}})_{pr}(C_{\mathsf{tr}})_{qr}\right](A_{\mathsf{tr}})_{qp}\nonumber\\
  &=\frac{1}{2}\sum_{i,d\in\pmb{p}}\sum_{a,j\in\pmb{q}}\sum_{b,c\in\pmb{r}}\left[(k_i\cdot k_j)(k_a\cdot k_b)(k_c\cdot k_d)\frac{\sigma_i\sigma_{ja}\sigma_{bc}}{\sigma_{ij}\sigma_{ab}\sigma_{cd}}-(k_i\cdot k_b)(k_c\cdot k_j)(k_a\cdot k_d)\frac{\sigma_d\sigma_{bc}\sigma_{ja}}{\sigma_{ib}\sigma_{cj}\sigma_{ad}}\right]\nonumber\\
  &\quad+\frac{1}{2}\sum_{i,d\in\pmb{p}}\sum_{a,j\in\pmb{q}}\sum_{b,c\in\pmb{r}}\left[(k_i\cdot k_b)(k_c\cdot k_j)(k_a\cdot k_d)\frac{\sigma_i\sigma_{bc}\sigma_{ja}}{\sigma_{ib}\sigma_{cj}\sigma_{ad}}-(k_i\cdot k_j)(k_a\cdot k_b)(k_c\cdot k_d)\frac{\sigma_d\sigma_{ja}\sigma_{bc}}{\sigma_{ij}\sigma_{ab}\sigma_{cd}}\right]\nonumber\\
  &=\frac{1}{2}\sum_{i,d\in\pmb{p}}\sum_{a,j\in\pmb{q}}\sum_{b,c\in\pmb{r}}\left[(k_i\cdot k_j)(k_a\cdot k_b)(k_c\cdot k_d)\frac{\sigma_{aj}\sigma_{cb}\sigma_{id}}{\sigma_{ij}\sigma_{ab}\sigma_{cd}}+(k_i\cdot k_b)(k_c\cdot k_j)(k_a\cdot k_d)\frac{\sigma_{aj}\sigma_{cb}\sigma_{id}}{\sigma_{ib}\sigma_{cj}\sigma_{ad}}\right]\,.
\end{align}
Namely, we have proved another important relation:
\begin{equation}
  \mathscr{R}[pqr]=\mathscr{C}_{pqr}^{\,\,qrp}\,.
\end{equation}
For future convenience, we define the $\mathscr{R}$ symbol symmetrized over the exchange of $p$ and $q$ as:
\begin{equation}
  \mathscr{R}(o\{pq\}r)\equiv\mathscr{R}(o|pq|r)+\mathscr{R}(o|qp|r)\qquad\text{such that}\qquad\mathscr{R}[pqr]=\mathscr{R}(p\{qr\}p)\,.
\end{equation}
Obviously, we have $\mathscr{R}(p\{q\}r)=\mathscr{R}(p|q|r)$ and $\mathscr{R}(p\,\varnothing\,q)=\mathscr{R}(p|q)$. It is important to notice that
\begin{align}
\mathscr{R}(p|qr|p)\neq\mathscr{C}_{pq}^{\,\,qr}(C_{\mathsf{tr}})_{rp}-\mathscr{C}_{pqr}^{\,\,q}(A_{\mathsf{tr}})_{rp}& &\mathscr{R}(p|rq|p)\neq\mathscr{C}_{pr}^{\,\,rq}(C_{\mathsf{tr}})_{qp}-\mathscr{C}_{prq}^{\,\,r}(A_{\mathsf{tr}})_{qp}\,.
\end{align}
Namely, the symmetrization over $(q,r)$ is crucial to make $\mathscr{R}[pqr]=\mathscr{C}_{pqr}^{\,\,qrp}$ hold. Now we can write the final result of the four-trace Pfaffian expansion as:
\begin{align}
\Pfp(\Pi_4)&=\mathscr{R}(1|4)\Pf[\Pi_3(\pmb{2,3})]-\mathscr{R}(1|2|4)\Pf[\Pi_2(\pmb{3})]-\mathscr{R}(1|3|4)\Pf[\Pi_2(\pmb{2})]-\mathscr{R}(1\{23\}4)\,.
\end{align}
If we act these $\mathscr{R}$ symbols onto the Parke-Taylor factors, we get:
\begin{align}
\label{eq:4tresult}
\frac{\Pfp(\Pi_4)}{\langle\pmb{1}\rangle\langle\pmb{2}\rangle\langle\pmb{3}\rangle\langle\pmb{4}\rangle}&=\underset{\{\underline{1},i\}\subset\pmb{1}}{\widetilde{\sum}}~\underset{\{d,\underline{n}\}\subset\pmb{4}}{\widetilde{\sum}}\,\frac{(k_i\cdot k_d)\Pf[{\Pi}_{3}(\pmb{2,3})]}{\langle 1,\mathsf{KK}[\pmb{1},1,i],i,d,\mathsf{KK}[\pmb{4},d,n],n\rangle\langle\pmb{2}\rangle\langle\pmb{3}\rangle}\nonumber\\
&\quad-\underset{\{\underline{1},i\}\subset\pmb{1}}{\widetilde{\sum}}~\underset{\{a,j\}\in\pmb{2}}{\widetilde{\sum}}~\underset{\{d,\underline{n}\}\subset\pmb{4}}{\widetilde{\sum}}\,\frac{(k_i\cdot k_j)(k_a\cdot k_d)\Pf[\Pi_2(\pmb{3})]}{\langle 1,\mathsf{KK}[\pmb{1},1,i],i,j,\mathsf{KK}[\pmb{2},j,a],a,d,\mathsf{KK}[\pmb{4},d,n],n\rangle\langle\pmb{3}\rangle}\nonumber\\
&\quad-\underset{\{\underline{1},i\}\subset\pmb{1}}{\widetilde{\sum}}~\underset{\{a,j\}\subset\pmb{2}}{\widetilde{\sum}}~\underset{\{b,c\}\in\pmb{3}}{\widetilde{\sum}}~\underset{\{d,\underline{n}\}\subset\pmb{4}}{\widetilde{\sum}}(k_i\cdot k_j)(k_a\cdot k_b)(k_c\cdot k_d)\nonumber\\
&\qquad\qquad\quad\times\langle 1,\mathsf{KK}[\pmb{1},1,i],i,j,\mathsf{KK}[\pmb{2},j,a],a,b,\mathsf{KK}[\pmb{3},b,c],c,d,\mathsf{KK}[\pmb{4},d,n],n\rangle^{-1}\nonumber\\
&\quad-\underset{\{\underline{1},i\}\subset\pmb{1}}{\widetilde{\sum}}~\underset{\{b,c\}\subset\pmb{3}}{\widetilde{\sum}}~\underset{\{d,\underline{n}\}\subset\pmb{4}}{\widetilde{\sum}}\,\frac{(k_i\cdot k_b)(k_c\cdot k_d)\Pf[\Pi_2(\pmb{2})]}{\langle 1,\mathsf{KK}[\pmb{1},1,i],i,b,\mathsf{KK}[\pmb{3},b,c],c,d,\mathsf{KK}[\pmb{4},d,n],n\rangle\langle\pmb{2}\rangle}\nonumber\\
&\quad-\underset{\{\underline{1},i\}\subset\pmb{1}}{\widetilde{\sum}}~\underset{\{b,c\}\in\pmb{3}}{\widetilde{\sum}}~\underset{\{a,j\}\subset\pmb{2}}{\widetilde{\sum}}~\underset{\{d,\underline{n}\}\subset\pmb{4}}{\widetilde{\sum}}(k_i\cdot k_b)(k_c\cdot k_j)(k_a\cdot k_d)\nonumber\\
&\qquad\qquad\quad\times\langle 1,\mathsf{KK}[\pmb{1},1,i],i,b,\mathsf{KK}[\pmb{3},b,c],c,j,\mathsf{KK}[\pmb{2},j,a],a,d,\mathsf{KK}[\pmb{4},d,n],n\rangle^{-1}\,.
\end{align}
In terms of physical amplitudes, the above relation reads:
\begin{align}
  &A_{4,0}(\pmb{1}\,|\,\pmb{2}\,|\,\pmb{3}\,|\,\pmb{4})=\underset{\{\underline{1},i\}\subset\pmb{1}}{\widetilde{\sum}}~\underset{\{d,\underline{n}\}\subset\pmb{4}}{\widetilde{\sum}}(-k_d\cdot k_i)A_{3,0}(1,\mathsf{KK}[\pmb{1},1,i],i,d,\mathsf{KK}[\pmb{4},d,n],n\,|\,\pmb{2}\,|\,\pmb{3})\nonumber\\
  &\qquad\quad+\underset{\{\underline{1},i\}\subset\pmb{1}}{\widetilde{\sum}}~\underset{\{a,j\}\in\pmb{2}}{\widetilde{\sum}}~\underset{\{d,\underline{n}\}\subset\pmb{4}}{\widetilde{\sum}}(k_d\cdot k_a)(k_j\cdot k_i)A_{2,0}(1,\mathsf{KK}[\pmb{1},1,i],i,j,\mathsf{KK}[\pmb{2},j,a],a,d,\mathsf{KK}[\pmb{4},d,n],n\,|\,\pmb{3})\nonumber\\
  &\qquad\quad-\underset{\{\underline{1},i\}\subset\pmb{1}}{\widetilde{\sum}}~\underset{\{b,c\}\in\pmb{3}}{\widetilde{\sum}}~\underset{\{a,j\}\subset\pmb{2}}{\widetilde{\sum}}~\underset{\{d,\underline{n}\}\subset\pmb{4}}{\widetilde{\sum}}(k_d\cdot k_c)(k_b\cdot k_a)(k_j\cdot k_i)\nonumber\\
  &\qquad\qquad\qquad\times A_{1,0}(1,\mathsf{KK}[\pmb{1},1,i],i,j,\mathsf{KK}[\pmb{2},j,a],a,b,\mathsf{KK}[\pmb{3},b,c],c,d,\mathsf{KK}[\pmb{4},d,n],n)\nonumber\\
  &\qquad\quad+\underset{\{\underline{1},i\}\subset\pmb{1}}{\widetilde{\sum}}~\underset{\{b,c\}\in\pmb{3}}{\widetilde{\sum}}~\underset{\{d,\underline{n}\}\subset\pmb{4}}{\widetilde{\sum}}(k_d\cdot k_c)(k_b\cdot k_i)A_{2,0}(1,\mathsf{KK}[\pmb{1},1,i],i,b,\mathsf{KK}[\pmb{3},b,c],c,d,\mathsf{KK}[\pmb{4},d,n],n\,|\,\pmb{2})\nonumber\\
  &\qquad\quad-\underset{\{\underline{1},i\}\subset\pmb{1}}{\widetilde{\sum}}~\underset{\{b,c\}\in\pmb{3}}{\widetilde{\sum}}~\underset{\{a,j\}\subset\pmb{2}}{\widetilde{\sum}}~\underset{\{d,\underline{n}\}\subset\pmb{4}}{\widetilde{\sum}}(k_d\cdot k_a)(k_j\cdot k_c)(k_b\cdot k_i)\nonumber\\
&\qquad\qquad\qquad\times A_{1,0}(1,\mathsf{KK}[\pmb{1},1,i],i,b,\mathsf{KK}[\pmb{3},b,c],c,j,\mathsf{KK}[\pmb{2},j,a],a,d,\mathsf{KK}[\pmb{4},d,n],n)\,.
\end{align}
\section{Pure gluon Multitrace EYM amplitudes II: General Expansion}\label{sec:genexp}

In this section, we promote the multitrace pure gluon examples discussed above into a general recursive expansion, which emerges naturally if we expand the pure gluon reduced Pfaffian in a well controlled way:
\begin{equation}
\Pfp[\Pi_{m}(\pmb{1,2\ldots m})]=\Pf[\Pi_{m}(\pmb{1,2\ldots m-1})]\equiv\Pf[\Pi_m(\mathsf{A}_{\cancel{m}})]\,,
\end{equation}
where $\mathsf{A}=\{1,2\ldots m\}$ and $\mathsf{A}_{\cancel{m}}=\mathsf{A}\backslash\{m\}$. First, we define some useful notations. Suppose $\mathsf{a}$ is a subset of $\mathsf{A}_{\cancel{m}}$, we can find a submatrix $\Pi_{|\mathsf{a}|+1}(\mathsf{a})$ of $\Pi_{m}(\mathsf{A}_{\cancel{m}})$ with the form:
\begin{equation}
\Pi_{|\mathsf{a}|+1}(\mathsf{a})=\begin{pmatrix}
A_{\mathsf{a}} & -(C_{\mathsf{a}})^{T} \\ C_{\mathsf{a}} & B_{\mathsf{a}} \\
\end{pmatrix}\,,
\end{equation}
where $A_{\mathsf{a}}$, $B_{\mathsf{a}}$ and $C_{\mathsf{a}}$ are respectively the diagonal submatrices of $A_{\mathsf{tr}}$, $B_{\mathsf{tr}}$ and $C_{\mathsf{tr}}$ whose rows and columns take value only in $\mathsf{a}$. We call $\Pf[\Pi_{|\mathsf{a}|+1}(\mathsf{a})]$ \emph{a blockwise principal minor} of $\Pf[\Pi_{m}(\mathsf{A}_{\cancel{m}})]$. This matrix appears in the pure gluon integrand with $|\mathsf{a}|+1$ traces $\mathsf{a}\cup\{m\}$. Interestingly, we can express $\Pf[\Pi_{m}(\mathsf{A}_{\cancel{m}})]$ in terms of the following linear combination of these blockwise principal minors:
\begin{equation}
\label{eq:Pfexpansion}
  \Pf[\Pi_{m}(\mathsf{A}_{\cancel{m}})]=(-1)^{m+1}(C_{\mathsf{tr}})_{11}\Pf[\Pi_{m-1}(\mathsf{A}_{\cancel{1m}})]+\sum_{\substack{\mathsf{a}|\mathsf{b}=\mathsf{A}_{\cancel{1m}} \\ \mathsf{b}\neq\varnothing}}\mathcal{C}_m[1\,\mathsf{b}]\Pf[\Pi_{|\mathsf{a}|+1}(\mathsf{a})]\,,
\end{equation}
where $\mathsf{A}_{\cancel{1m}}\equiv\mathsf{A}\backslash\{1,m\}$, $m$ is the number of gluon traces, and we define $\Pf[\Pi_1(\mathsf{a}=\varnothing)]=1$. For $\mathsf{b}=\{t_2\ldots t_s\}$, the coefficient $\mathcal{C}_m$ has the following form:
\begin{subequations}
\label{eq:C1s}
\begin{align}
\label{eq:Csign}
  \mathcal{C}_m[t_1t_2\ldots t_s]&=-(-1)^{\frac{s(s+2m-1)}{2}}\mathscr{C}[t_1t_2\ldots t_s]\\
  \mathscr{C}[t_1t_2\ldots t_s]&=\frac{1}{2}\Bigg[\sum_{\{j_\ell,i_\ell\}\subset\pmb{t}_\ell}\sigma_{i_\ell j_\ell}\Bigg]_{\ell=1}^{s}\left[\prod_{c=1}^{s}\frac{k_{i_c}\cdot k_{j_{c+1}}}{\sigma_{i_c}\sigma_{j_{c+1}}}+\text{noncyclic permutations of }(12\ldots s)\right]\,,
\end{align}
\end{subequations}
and we have implicitly $\{i_{s+1},j_{s+1}\}\equiv\{i_1,j_1\}$ in the chain product $\prod_{c=1}^{s}$. By starting the expansion from the row of $(C_{\mathsf{tr}})_{11}$, we essentially fix $t_1=1$. We defer the derivation of this expression into appendix~\ref{sec:Pfaffian1} and~\ref{sec:Pfaffian2}, but only emphasize that Eq.~\eqref{eq:Pfexpansion} is an algebraic identity in the sense that no requirement on kinematics is used to derive it. Thus it can be recursively applied to any blockwise principal minor $\Pf[\Pi_{|\mathsf{a}|+1}(\mathsf{a})]$, with just $\mathcal{C}_{m}$ changed to $\mathcal{C}_{|\mathsf{a}|+1}$. Given the set $\{t_1\ldots t_s\}$, the coefficients satisfy two properties that are useful for later calculations:
\begin{itemize}
\item Both $\mathcal{C}$ and $\mathscr{C}$ are independent of the ordering of $\{t_1\ldots t_s\}$.
\item The total number of gluon traces only appear in the phase factor. We have
  \begin{equation}
    (-1)^{s}\mathcal{C}_{m-1}[t_1\ldots t_s]=\mathcal{C}_m[t_1\ldots t_s]\,.
  \end{equation}
\end{itemize}
Next, we apply the cross-ratio identity to $(C_{\mathsf{tr}})_{11}$. This will generate one term that correctly connects trace $\pmb{1}$ and $\pmb{m}$. For the other terms, we iterate the above operation, just like what we have done in the previous examples. Our final result is:
\begin{align}
\label{eq:finalexpansion}
\Pf[\Pi_{m}(\mathsf{A}_{\cancel{m}})]=\sum_{\mathsf{a}|\mathsf{b}=\mathsf{A}_{\cancel{1m}}}\mathcal{R}(1\,\mathsf{b}\,m)\Pf[\Pi_{|\mathsf{a}|+1}(\mathsf{a})]\,.
\end{align}
For $\mathsf{b}=\{t_2,t_3\ldots t_{s-1}\}$, the coefficient $\mathcal{R}$ is given by:
\begingroup
\allowdisplaybreaks
\begin{subequations}
\label{eq:R1s}
\begin{align}
\label{eq:Rsign}
\mathcal{R}(t_1\{t_2\ldots t_{s}\}t_k)&={(-1)^{\frac{s(s+2m-1)}{2}}}\mathscr{R}(t_1\{t_2\ldots t_{s}\}t_k)\\*
\mathscr{R}(t_1\{t_2\ldots t_{s}\}t_k)&=\sum_{j_k\in\pmb{t}_k}\mathscr{R}(t_1\{t_2\ldots t_{s}\}t_k)_{j_k}\\*
\mathscr{R}(t_1\{t_2\ldots t_{s}\}t_k)_{j_k}&=\mathscr{R}(t_1|t_2\ldots t_{s}|t_k)_{j_k}+\text{permutations of }(2\ldots s)\\*
\mathscr{R}(t_1|t_2\ldots t_{s}|t_k)_{j_k}&=\sum_{i_1\in\pmb{t}_1}\frac{\sigma_{i_11}\sigma_{nj_s}}{\sigma_{n1}}\Bigg[\sum_{\{i_\ell,j_\ell\}\subset\pmb{t}_\ell}\sigma_{i_\ell j_\ell}\Bigg]_{\ell=2}^{s}\frac{k_{i_1}\cdot k_{j_2}}{\sigma_{i_1j_2}}\left(\prod_{c=2}^{s-2}\frac{k_{i_c}\cdot k_{j_{c+1}}}{\sigma_{i_cj_{c+1}}}\right)\frac{k_{i_{s}}\cdot k_{j_k}}{\sigma_{i_{s}j_k}}\,.
\end{align}
\end{subequations}
\endgroup
In particular, we have the identity
\begin{align}
&\mathscr{R}[t_1\ldots t_s]\equiv\mathscr{R}(t_1\{t_2\ldots t_s\}t_1)=\mathscr{C}[t_1\ldots t_s]\nonumber\\
&\mathcal{R}[t_1\ldots t_s]\equiv\mathcal{R}(t_1\{t_2\ldots t_s\}t_1)=-\mathcal{C}_{m}[t_1\ldots t_s]\,.
\end{align}
We need to use both the momentum conservation and scattering equations to derive Eq.~\eqref{eq:finalexpansion}. The proof of Eq.~\eqref{eq:finalexpansion} is a little lengthy, and it will be given in appendix~\ref{sec:CrossRatio}.
We can then obtain \emph{the squeezing-form recursive expansion} of pure gluon multitrace EYM amplitudes by simply acting Eq.~\eqref{eq:finalexpansion} onto the Parke-Taylor factors:
\begin{align}
\label{eq:puregluonexp}
    A_{m,0}(\pmb{1}\,|\,\pmb{2}\,|\ldots|\,\pmb{m})=\sum_{\pmb{\mathsf{Tr}}_s|\mathsf{Tr}_p=\mathsf{Tr}\backslash\{1,m\}}\,\underset{\{\underline{1},i\}\subset\pmb{1}}{\widetilde{\sum}}~\underset{\{j,\underline{n}\}\subset\pmb{m}}{\widetilde{\sum}}\Bigg[\underset{\{a_\ell,b_\ell\}\subset\pmb{t}_\ell}{\widetilde{\sum}}\Bigg]_{\ell=1}^{s}\mathcal{A}_3(\pmb{1}_{\mathsf{KK}[\pmb{1},1,i],\pmb{K}[\pmb{\mathsf{Tr}}_s,a,b],\mathsf{KK}[\pmb{m},j,n]}\,|\,\mathsf{Tr}_p)\,,
\end{align}
where $\pmb{\mathsf{Tr}}_{s}=\{t_1,t_2\ldots t_s\}$ and $\mathsf{Tr}_p=\{r_1,r_2\ldots r_p\}$. Then $\mathcal{A}_3$ is a linear combination of $p+1$ traces pure gluon integrand:
\begin{align}
\mathcal{A}_3(\pmb{1}_{\mathsf{KK}[\pmb{1},1,i],\pmb{K}[\pmb{\mathsf{Tr}}_s,a,b],\mathsf{KK}[\pmb{m},j,n]}\,|\,\mathsf{Tr}_p)&=(-k_j)\cdot(-k_{b_s}k_{a_s})\ldots(-k_{b_2}k_{a_2})\cdot(-k_{b_1}k_{a_1})\cdot k_i\nonumber\\
&\quad\times A_{p+1,0}(1,\mathsf{KK}[\pmb{1},1,i],i,\pmb{K}[\pmb{\mathsf{Tr}}_s,a,b],j,\mathsf{KK}[\pmb{m},j,n],n\,|\,\pmb{r}_1\,|\ldots|\,\pmb{r}_p)\,.
\end{align}
We remark that this $\mathcal{A}_3$ depends on the ordering of $\pmb{\mathsf{Tr}}_s$, since in Eq.~\eqref{eq:puregluonexp} we explicitly write out all the orderings that get summed over in $\mathcal{R}$. This expansion has a different form compared with our type-II one given in Eq.~\eqref{eq:type2step1}. We will show their equivalence in appendix~\ref{sec:equiv}.

At the end, we give a brief discussion on the case with gravitons. Now we have a choice on where to start our Pfaffian expansion: either along a graviton row or a trace row. The former (latter) case will give us a form equivalent to the type-I (type-II) recursive expansion.



\section{General BCJ relations with tree level multitrace EYM amplitudes}\label{sec:generalBCJ}

In this section, we present another consistency check of the type-I and type-II recursive expansions~\eqref{eq:type1step1} and \eqref{eq:type2step1} by considering the related general BCJ relations for tree-level multitrace EYM amplitudes.

\subsection{Type-I generalized BCJ relation}

One obvious consistent condition is the gauge invariance for all gravitons. With the expansion \eqref{eq:type1step1}, one can see the manifest gauge invariance for all gravitons, except the fiducial graviton $h_1$. Thus, if the type-I recursive expansion is correct, we should get zero when we replace $\eps_{h_1}$ by $k_{h_1}$ in Eq.~\eqref{eq:Ch_1}, i.e., we should have the following \emph{type-I generalized BCJ relation:}
\begin{align}
0=\sum_{\substack{\pmb{h}| \mathsf{h}=\mathsf{H}\backslash h_1 \\ \pmb{\mathsf{Tr}}_s| \mathsf{Tr}_p=\mathsf{Tr}\backslash 1}}&\,\Bigg[\,\underset{\{a_i,b_i\}\subset\pmb{t}_i}{\widetilde{\sum}}\,\Bigg]_{i=1}^s \WH C_{h_1}(\pmb{h}\,\W{\shuffle}\,\pmb{K})\nonumber\\
&\times A_{p+1,|\mathsf{h}|}(1,\{2\ldots r-1\}\shuffle\{\pmb{h}\,\W{\shuffle}\,\pmb{K}(\pmb{\mathsf{Tr}}_s,a,b),h_1\},r|\pmb{r}_{1}|\ldots|\pmb{r}_{p}\Vert\,\mathsf{h})\,,
\Label{Eq:GeneralizedBCJType1}
\end{align}
with $\WH C_{h_1}(\pmb{\rho})=-k_{h_1}\cdot \mathcal{T}_{\rho_{|\pmb{h}|+s}}\ldots\mathcal{T}_{\rho_2}\cdot \mathcal{T}_{\rho_1}\cdot Y_{\rho_1}$ defined in~\eqref{eq:Chat} and $\mathcal{T}_{\rho_i}$ defined in~\eqref{eq:Ch_1}.
%
Were there only one trace, this relation would be nothing but the gauge invariant identity (Ward identity) for the single-trace amplitude \eqref{eq:singletrace}. We must note that it is not easy to prove even this very simple case using the generalized BCJ relations of color ordered YM amplitudes~\cite{Chen:2011jxa}. If there are more than one trace, Eq.~\eqref{Eq:GeneralizedBCJType1} provides a set of generalized BCJ relations for multi-trace amplitudes. We now give one example:
\paragraph{A type-I example}
With only one graviton and two gluon traces, the generalized BCJ relation obtained from Eq.~\eqref{Eq:GeneralizedBCJType1} is the same as replacing $\epsilon_{h_1}$ by $-k_{h_1}$ in Eq.~\eqref{eq:t2h1}:
\begin{align}
0 &=-(k_{h_1}\cdot
Y_{h_1})A_{2,0}(1,\{2,\dots,r-1\}\shuffle\{h_1\},r\,|\,\pmb{2})\nn
&\quad+\underset{\{a,b\}\subset\pmb{2}}{\widetilde{\sum}}(k_{h_1}\cdot
k_{b})(k_{a}\cdot Y_{a})A_{1,0}(1,\{2,\dots,r-1\}\shuffle\{a,\mathsf{KK}[\pmb{2},a, b], b, h_1\},r)\,.
\end{align}
An equivalent form of this relation has been proved in~\cite{Nandan:2016pya}.

\subsection{Type-II generalized BCJ relation}

Now we present type-II generalized BCJ relation, which is tightly related to the type-II recursive expansion~\eqref{eq:type2step1}. In this case, all gravitons are manifestly gauge invariant, while there are two different points of view to understand the emerging BCJ relations. We will discuss them one by one.

The first angle to see this relation is from the \emph{collinear limit}. Let us consider, for example, the collinear
limit between $r$ and $d_2$ in \eqref{eq:type2step1}. At the left hand side, $r$ belongs to the first trace while $d_2$ belongs to the second trace. Because they belong to different traces, there is no collinear singularities when $k_r$ is collinear with $k_{d_2}$. However, at the right hand side, the first and the second trace have been merged into a single trace. From the explicit expression in Eq.~\eqref{eq:type2step2}, we see that the expansion of the shuffle product will lead to two different situations: either $(r-1)$ next to $r$ or $d_2$ next to $r$:
\begin{equation*}
\Big\{\{\pmb{x}, r-1\}\shuffle\{\pmb{y}, d_2\},r\Big\}=\Big\{\pmb{x}\shuffle\{\pmb{y},d_2\}, r-1,r\Big\}+\Big\{\{\pmb{x},r-1\}\shuffle\pmb{y}, d_2,r\Big\}\,.
\end{equation*}
Only the latter case might lead to a potential collinear singularity since now $d_2$ and $r$ are next to each other. More explicitly, the collinear limit at the right hand side of~\eqref{eq:type2step1} has the following form:
\begin{align}
&\quad\sum_{\substack{\pmb{h}| \mathsf{h}=\mathsf{H} \\ \pmb{\mathsf{Tr}}_s| \mathsf{Tr}_p=\mathsf{Tr}
\backslash\{1,2\}}}\Bigg[\,\underset{\{a_i,b_i\}\subset\pmb{t}_i}{\widetilde{\sum}}
\,\Bigg]_{i=1}^s~\underset{\{c_2,\underline{d_2}\}\subset\pmb{2}}{\widetilde{\sum}}
\WH{C}_{c_2}(\pmb{h}\,\W{\shuffle}\,\pmb{K})\nn &\qquad\times A_{p+1,|\mathsf{h}|}(1,\{2\ldots r-1\}\shuffle\{\pmb{h}\,\W{\shuffle}\,\pmb{K}(\pmb{\mathsf{Tr}}_s,a,b),c_2,\mathsf{KK}[\pmb{2},c_2,d_2],d_2\},r|\pmb{r}_{1}|
\ldots|\pmb{r}_{p}\Vert\,\mathsf{h})\nn
&\sim \sum_{\substack{\pmb{h}| \mathsf{h}=\mathsf{H} \\ \pmb{\mathsf{Tr}}_s| \mathsf{Tr}_p=\mathsf{Tr}
\backslash\{1,2\}}}\Bigg[\,\underset{\{a_i,b_i\}\subset\pmb{t}_i}{\widetilde{\sum}}
\,\Bigg]_{i=1}^s~\underset{\{c_2,\underline{d_2}\}\subset\pmb{2}}{\widetilde{\sum}}
\WH{C}_{c_2}(\pmb{h}\,\W{\shuffle}\,\pmb{K})\nn &\qquad\times A_{p+1,|\mathsf{h}|}(1,\{2\ldots r-1\}\shuffle\{\pmb{h}\,\W{\shuffle}\,\pmb{K}(\pmb{\mathsf{Tr}}_s,a,b),c_2,\mathsf{KK}[\pmb{2},c_2,d_2]\},d_2,r|\pmb{r}_{1}|
\ldots|\pmb{r}_{p}\Vert\,\mathsf{h})\nn
&\sim\,{\cal S}_{d_2,r;\W{d}_2}\sum_{\substack{\pmb{h}| \mathsf{h}=\mathsf{H} \\ \pmb{\mathsf{Tr}}_s| \mathsf{Tr}_p=\mathsf{Tr}
\backslash\{1,2\}}}\Bigg[\,\underset{\{a_i,b_i\}\subset\pmb{t}_i}{\widetilde{\sum}}
\,\Bigg]_{i=1}^s~\underset{\{c_2,\underline{d_2}\}\subset\pmb{2}}{\widetilde{\sum}}
\WH{C}_{c_2}(\pmb{h}\,\W{\shuffle}\,\pmb{K})\nn &\qquad\times A_{p+1,|\mathsf{h}|}(1,\{2\ldots r-1\}\shuffle\{\pmb{h}\,\W{\shuffle}\,\pmb{K}(\pmb{\mathsf{Tr}}_s,a,b),c_2,\mathsf{KK}[\pmb{2},c_2,\W{d}_2]\},\W{d}_2|\pmb{r}_{1}|
\ldots|\pmb{r}_{p}\Vert\,\mathsf{h})\,.
\end{align}
where the factor ${\cal S}_{d_2,r;\W{d}_2}$ is the universal splitting function at the collinear limit while $\W{d}_2$ represent the combination of $r$ and $d_2$: $k_{\W{d}_2}=k_{d_2}+k_r$. Since the left hand side of~\eqref{eq:type2step1}
is regular and the ${\cal S}_{d_2,r;\W{d}_2}$ is the universal singular piece,
to be consistent, we must have the following relation:
\begin{align} 0 & = \sum_{\substack{\pmb{h}| \mathsf{h}=\mathsf{H} \\ \pmb{\mathsf{Tr}}_s| \mathsf{Tr}_p=\mathsf{Tr}
\backslash\{1,2\}}}\Bigg[\,\underset{\{a_i,b_i\}\subset\pmb{t}_i}{\widetilde{\sum}}
\,\Bigg]_{i=1}^s~\underset{\{c_2,\underline{d_2}\}\subset\pmb{2}}{\widetilde{\sum}}
\WH{C}_{c_2}(\pmb{h}\,\W{\shuffle}\,\pmb{K})\nn &\quad \times A_{p+1,|\mathsf{h}|}(1,\{2\ldots r-1\}\shuffle\{\pmb{h}\,\W{\shuffle}\,\pmb{K}(\pmb{\mathsf{Tr}}_s,a,b),c_2,\mathsf{KK}[\pmb{2},c_2,d_2]\}, r|\pmb{r}_{1}|
\ldots|\pmb{r}_{p}\Vert\,\mathsf{h})\,,\Label{Eq:GeneralizedBCJType2}
\end{align}
namely, the coefficient of $\mathcal{S}_{d_2,r;\W{d}_2}$ has to be zero. It is worth to emphasize that starting from $n$-point recursive expansion, the derived type-II generalized BCJ relation is, in fact, among $(n-1)$-points. Especially, the point $d_2$ in trace $\pmb{2}$ is fictitious since it has been identified with $r$ in trace $\pmb{1}$. However, when we try to write down the type-II BCJ relation~\eqref{Eq:GeneralizedBCJType2}, we do need to insert such a fictitious leg to trace $\pmb{2}$ since it defines the KK basis.


The type-II BCJ relation can also be understood from the fact that the type-II recursive expansion is independent of the fiducial gluon $d_2$. Assuming the second trace $\pmb{2}$ has $(l-1)$ elements $\pmb{2}=\{c_1,c_2\dots c_{l-1}\}$, if we set the spurious point as $d_2=c_l$, Eq.~\eqref{Eq:GeneralizedBCJType2}
can be more explicitly written as:
\begin{align}
0 & = \sum_{\substack{\pmb{h}| \mathsf{h}=\mathsf{H} \\ \pmb{\mathsf{Tr}}_s| \mathsf{Tr}_p=\mathsf{Tr}
\backslash\{1,2\}}}\Bigg[\,\underset{\{a_i,b_i\}\subset\pmb{t}_i}{\widetilde{\sum}}
\,\Bigg]_{i=1}^s~ \Sl_{\substack{j=1\\}}^{l-1}(-1)^{j-1}
\WH{C}_{c_j}(\pmb{h}\,\W{\shuffle}\,\pmb{K})\nn &\quad\times A_{p+1,|\mathsf{h}|}(1,\{2\ldots r-1\}\shuffle\{\pmb{h}\,\W{\shuffle}\,\pmb{K}(\pmb{\mathsf{Tr}}_s,a,b),c_j,\{c_{j-1} \dots c_1\}\shuffle\{c_{j+1}
\dots c_{l-1}\}\}, r|\pmb{r}_{1}|
\ldots|\pmb{r}_{p}\Vert\,\mathsf{h})\,.\Label{Eq:GeneralizedBCJType2-new-1}
\end{align}
For the last shuffle sum, there are two cases: either $c_1$ or $c_{l-1}$ to be the last element. Separating these two
cases, we get two terms:
\begin{align} T_1 & = -\sum_{\substack{\pmb{h}| \mathsf{h}=\mathsf{H} \\ \pmb{\mathsf{Tr}}_s| \mathsf{Tr}_p=\mathsf{Tr}
\backslash\{1,2\}}}\Bigg[\,\underset{\{a_i,b_i\}\subset\pmb{t}_i}{\widetilde{\sum}}
\,\Bigg]_{i=1}^s~ \Sl_{\substack{j=2\\}}^{l-1}(-1)^{j-2}
\WH{C}_{c_j}(\pmb{h}\,\W{\shuffle}\,\pmb{K})\nn &\quad\times A_{p+1,|\mathsf{h}|}(1,\{2\ldots r-1\}\shuffle\{\pmb{h}\,\W{\shuffle}\,\pmb{K}(\pmb{\mathsf{Tr}}_s,a,b),c_j,\{c_{j-1}\dots c_2\}\shuffle\{c_{j+1}\dots c_{l-1}\},c_1\}, r|\pmb{r}_{1}|
\ldots|\pmb{r}_{p}\Vert\,\mathsf{h})\label{Eq:GeneralizedBCJType2-new-1-T1} \\
T_2 & =\sum_{\substack{\pmb{h}| \mathsf{h}=\mathsf{H} \\ \pmb{\mathsf{Tr}}_s| \mathsf{Tr}_p=\mathsf{Tr}
\backslash\{1,2\}}}\Bigg[\,\underset{\{a_i,b_i\}\subset\pmb{t}_i}{\widetilde{\sum}}
\,\Bigg]_{i=1}^s~ \Sl_{\substack{j=1\\}}^{l-2}(-1)^{j-1}
\WH{C}_{c_j}(\pmb{h}\,\W{\shuffle}\,\pmb{K})\nn &\quad\times A_{p+1,|\mathsf{h}|}(1,\{2\ldots r-1\}\shuffle\{\pmb{h}\,\W{\shuffle}\,\pmb{K}(\pmb{\mathsf{Tr}}_s,a,b),c_j,\{c_{j-1}\dots c_1\}\shuffle\{c_{j+1}
\dots c_{l-2}\},c_{l-1}\}, r|\pmb{r}_{1}|
\ldots|\pmb{r}_{p}\Vert\,\mathsf{h}),\label{Eq:GeneralizedBCJType2-new-1-T2}
\end{align}
while it is important to notice the difference in the sum over $j$ for $T_1$ and $T_2$.
Let us compare Eq.~\eqref{Eq:GeneralizedBCJType2-new-1-T1} and~\eqref{Eq:GeneralizedBCJType2-new-1-T2} with
the type-II recursive expansion~\eqref{eq:type2step1} and~\eqref{eq:type2step2}. It is easy to see that
$T_2$ is exact the recursive expansion with $c_{l-1}$ being taken as the fiducial gluon, while $T_1$
is the \emph{negative} of the recursive expansion with $c_{1}$ being taken as the fiducial gluon.
Since the recursive expansion is independent of such a choice, we should have $T_1+T_2=0$ which
is exactly the type-II BCJ relation given in~\eqref{Eq:GeneralizedBCJType2-new-1}.

\subsubsection{Some explicit examples}

Having presented the two different understandings, we give three explicit examples, the first two of which involve only double trace pure gluon amplitudes, while the last one involves one graviton. From now on, until the end of the current section, we suppress the subscript of $A_{m,|\mathsf{H}|}$ that label the number of gluon traces and gravitons.

The first example is the one with the second trace having only two elements. The right hand side of the type-II BCJ relation~\eqref{Eq:GeneralizedBCJType2} is the following combination:
\begin{align}
\label{eq:B2}
{\cal B}_2&\equiv\Sl_{\shuffle}(-1)\left(k_{a_1}\cdot
Y_{a_1}\right)
A\left(1,\{2\dots,r-1\}\shuffle\{a_1,a_2\},r\right)\nn
&\quad +\Sl_{\shuffle}\left(k_{a_2}\cdot Y_{a_2}\right)
A\left(1,\{2\dots,r-1\}\shuffle\{a_2,a_1\},r\right)\,.
\end{align}
To make things clear, we have put back the sum over shuffle products.
To prove $\mathcal{B}_2=0$, we need to use the generalized fundamental BCJ relation for single trace YM amplitudes~\cite{Chen:2011jxa}:
\begin{align}
0 &=\Sl_{\shuffle}\left(\sum_{i=1}^k k_{a_i}\cdot X_{a_i}\right) A\left(1,\{2\dots r-1\}\shuffle\{a_1\ldots a_k\},r\right)
\Label{Gen-BCJ}\\
0 &=\Sl_{\pmb{\rho}\in\{2\ldots r-1\}\shuffle\{a_2\ldots a_k\}}\Sl_{\shuffle}\left(k_{a_1}\cdot X_{a_1}\right)A\left(1,\pmb{\rho}\shuffle\{a_1\},r\right)\nonumber\\
&=\Sl_{\shuffle}\left(k_{a_1}\cdot X_{a_1}\right)A\left(1,\{2\ldots r-1\}\shuffle\{a_2\ldots a_k\}\shuffle\{a_1\},r\right)\,.\Label{Gen-BCJ-b}
\end{align}
With the $\{a_1\ldots a_k\}$ identified as  $\{a_2, a_1\}$, Eq.~\eqref{Gen-BCJ} gives:
\begin{equation}\Sl_{\shuffle}\left(k_{a_2}\cdot Y_{a_2}\right)
A\left(1,\{2\dots,r-1\}\shuffle\{a_2,a_1\},r\right)=
\Sl_{\shuffle}(-1)\left(k_{a_1}\cdot X_{a_1}\right)
A\left(1,\{2\dots,r-1\}\shuffle\{a_2,a_1\},r\right)\,,
\end{equation}
where the $X$ symbol has been defined below Eq.~\eqref{eq:Ysymbol}, and $X_{a_2}= Y_{a_2}$ for this case. Putting it back to Eq.~\eqref{eq:B2}, we get:
\bea -{\cal B}_2&=&\Sl_{\shuffle}\left(k_{a_1}\cdot
Y_{a_1}\right)
A\left(1,\{2\dots r-1\}\shuffle\{a_1,a_2\},r\right) +\Sl_{\shuffle}\left(k_{a_1}\cdot X_{a_1}\right)
A\left(1,\{2\dots r-1\}\shuffle\{a_2,a_1\},r\right)\nn
& = & \Sl_{\shuffle}\left(k_{a_1}\cdot X_{a_1}\right)
A\left(1,\{2\ldots r-1\}\shuffle\{a_2\}\shuffle\{a_1\},r\right)=0\,,\eea
according to the fundamental BCJ relation~\eqref{Gen-BCJ-b} for $\{a_1\}$.

In the second example, we put three elements in trace $\pmb{2}$, still with no graviton. The right hand side of the type-II BCJ relation \eqref{Eq:GeneralizedBCJType2} gives:
\bea {\cal B}_3& =&\Sl_{\shuffle}(-1)\left(k_{a_1}\cdot Y_{a_1}\right)
A\left(1,\{2\dots,r-1\}\shuffle\{a_1,a_2,a_3\},r\right)\nn
&&+\Sl_{\shuffle}\left(k_{a_2}\cdot Y_{a_2}\right)
A\left(1,\{2\dots,r-1\}\shuffle\{a_2,\{a_1\}\shuffle\{a_3\}\},r\right)\nn
&&+\Sl_{\shuffle}(-1)\left(k_{a_3}\cdot Y_{a_3}\right)
A\left(1,\{2\dots,r-1\}\shuffle\{a_3,a_2,a_1\},r\right).\Label{Eq:Nonlocal2}
\eea
To simplify our notation, we define:
\bea {\cal B}_{\{12\ldots r\}}[\{a_1\ldots a_k\}; i]\equiv \Sl_{\shuffle} (k_{a_i}\cdot X_{a_i}) A\left(1,\{2\dots r-1\}\shuffle\{a_1\ldots a_k\},r\right)\Label{BCJ-base}\eea
as our building blocks, where $i$ stands for the element at the $i$-th position in the list $\{a_2\ldots a_k\}$. The generalized fundamental BCJ relation~\eqref{Gen-BCJ} and~\eqref{Gen-BCJ-b} can be simply written as:
\begin{align}
0&=\sum_{i=1}^k {\cal B}_{\{12\ldots r\}}[\{a_1\ldots a_k\}; i]
\Label{Gen-BCJ-1} \\
0&=\sum_{\shuffle}\mathcal{B}_{\{1\{2\ldots r-1\}\shuffle\{a_2\ldots a_k\}r\}}[\{a_1\};1]=\sum_{i=1}^{k-1}\mathcal{B}_{\{12\ldots r\}}[\{a_2\ldots a_i,a_1,a_{i+1}\ldots a_k\},i]\,.\Label{Gen-BCJ-1b}
\end{align}
In the following, we will always use the second form in Eq.~\eqref{Gen-BCJ-1b} for the fundamental BCJ relation, and we will also omit the common subscript $\{12\ldots r\}$. With this new notation, we can do the following manipulations:
\bea {\cal B}_3 & = & -{\cal B}[\{a_1, a_2, a_3\};1]+ {\cal B}[\{a_2, a_1,
a_3\};1]+{\cal B}[\{a_2, a_3, a_1\};1]- {\cal B}[\{a_3, a_2, a_1\};1]\nn
& = & -{\cal B}[\{a_1, a_2, a_3\};1]+ {\cal B}[\{a_2, a_1, a_3\};1]+{\cal B}[\{a_2, a_3,
a_1\};1]+ {\cal B}[\{ a_2,a_3, a_1\};2]+{\cal B}[\{ a_2, a_1,a_3\};3]\nn
& = & -{\cal B}[\{a_1, a_2, a_3\};1]- {\cal B}[\{a_2, a_1, a_3\};2]-{\cal B}[\{a_2, a_3,
a_1\};3]=0\eea
to show the vanishing of $\mathcal{B}_3$. To go from the first line to the second, we have used~\eqref{Gen-BCJ-1b} for $\{a_3\}$ to the last term:
\begin{equation*}
\mathcal{B}[\{a_3,a_2,a_1\};1]+\mathcal{B}[\{a_2,a_3,a_1\};2]+\mathcal{B}[\{a_2,a_1,a_3\};3]=0\,.
\end{equation*}
To go from the second line to the third line, we have used the \eqref{Gen-BCJ-1} for the set $\{a_2, a_1, a_3\}$ to
combine the second and fifth terms, and for the set $\{a_2, a_3, a_1\}$ to
combine the third and fourth terms:
\begin{align*}
&\mathcal{B}[\{a_2,a_1,a_3\};1]+\mathcal{B}[\{a_2,a_1,a_3\};2]+\mathcal{B}[\{a_2,a_1,a_3\};3]=0\\
&\mathcal{B}[\{a_2,a_3,a_1\};1]+\mathcal{B}[\{a_2,a_3,a_1\};2]+\mathcal{B}[\{a_2,a_3,a_1\};3]=0\,.
\end{align*}
Finally, the last line is zero by \eqref{Gen-BCJ-1b} for the set $\{a_1\}$.

The third example is a little bit trivial. If the second trace has only one element $a_1$ and there is another graviton $h$,
the right hand side of the type-II BCJ relation \eqref{Eq:GeneralizedBCJType2} gives:\footnote{A gluon trace in principle should have at least two elements. Here, the other one should be viewed as fictitious, as in~\eqref{Eq:GeneralizedBCJType2}.}
\begin{equation} {\cal B}_{1;1}=\Sl_{\shuffle}\left(k_{a_1}\cdot
Y_{a_1}\right)
 A\left(1,\{2\dots,r-1\}\shuffle\{a_1\},r\,\Vert\,h\right)+\Sl_{\shuffle}\left(k_{a_1}\cdot F_{h}\cdot Y_{h}\right)
A\left(1,\{2\dots,r-1\}\shuffle\{h,a_1\},r\right).
\end{equation}
This is nothing, but the $a_1$ gauge invariance identity of single
trace EYM amplitude with two gravitons $\{a_1, h\}$, and $a_1$ is treated as fiducial [see \eqref{eq:singletrace}].

\subsection{Mixed form relations}

Although the form \eqref{Eq:GeneralizedBCJType2} [or more explicitly, Eq.~\eqref{Eq:GeneralizedBCJType2-new-1}] is very convenient, we do have the freedom to change it into a different form. In particular, this new form is obtained by using both type-I and type-II BCJ relations, so that we call it \emph{mixed form relations}.
To see it, let us start from the case $l=3$ in
Eq.~\eqref{Eq:GeneralizedBCJType2-new-1}. In this case, we have the expansion:\footnote{For simplicity, we will not write all the summation $$\sum_{\substack{\pmb{h}| \mathsf{h}=\mathsf{H} \\ \pmb{\mathsf{Tr}}_s| \mathsf{Tr}_p=\mathsf{Tr}\backslash\{1,2\}}}\Bigg[\,\underset{\{a_i,b_i\}\subset\pmb{t}_i}{\widetilde{\sum}}\,\Bigg]_{i=1}^s~\underset{\{c_2,\underline{d_2}\}\subset\pmb{2}}{\widetilde{\sum}}$$ until the end of this section.}
%
\begin{align}
0&= \WH{C}_{c_1}(\pmb{h}\shuffle\pmb{K})  A(1,\{2\ldots r-1\}\shuffle\{\pmb{h}\,\W{\shuffle}\,\pmb{K}(\pmb{\mathsf{Tr}}_s,a,b),c_1, c_2\}, r|\pmb{r}_{1}|
\ldots|\pmb{r}_{p}\Vert\,\mathsf{h})\nn
&\quad -\WH{C}_{c_2}(\pmb{h}\,\W{\shuffle}\,\pmb{K}) A(1,\{2\ldots r-1\}\shuffle\{\pmb{h}\,\W{\shuffle}\,\pmb{K}(\pmb{\mathsf{Tr}}_s,a,b),c_2, c_1\}, r|\pmb{r}_{1}|
\ldots|\pmb{r}_{p}\Vert\,\mathsf{h})\,,\Label{Newform-1}
\end{align}
where $\pmb\rho=\{\rho_1,..., \rho_{|\pmb{h}|+s}\}\in \pmb{h}\,\W{\shuffle}\,\pmb{K}$. Now we consider the type-I BCJ relation
\eqref{Eq:GeneralizedBCJType1}, where the $c_2$ has been treated as the fiducial graviton, and $c_1$ merged into the first trace:
\begin{align}
&0=\WH{C}_{c_2}(\pmb{h}\,\W{\shuffle}\,\pmb{K}) A(1,\{2\ldots r-1\}\shuffle\{c_1\}\shuffle\{\pmb{h}\,\W{\shuffle}\,\pmb{K}(\pmb{\mathsf{Tr}}_s,a,b),c_2\}, r|\pmb{r}_{1}|
\ldots|\pmb{r}_{p}\Vert\,\mathsf{h})\,.\Label{Newform-2}
\end{align}
%
Using the associativity of the shuffle algebra, we can transform the second line of Eq.~\eqref{Newform-1} by using Eq.~\eqref{Newform-2}, such that Eq.~\eqref{Newform-1} becomes:
%
%
%
\begin{align}
0&= \WH{C}_{c_1}(\pmb{h}\,\W{\shuffle}\,\pmb{K})  A(1,\{2\ldots r-1\}\shuffle\{\pmb{h}\,\W{\shuffle}\,\pmb{K}(\pmb{\mathsf{Tr}}_s,a,b),c_1, c_2\}, r|\pmb{r}_{1}|
\ldots|\pmb{r}_{p}\Vert\,\mathsf{h})\nn
&\quad +\WH{C}_{c_2}(\pmb{h}\,\W{\shuffle}\,\pmb{K}) A(1,\{2\ldots r-1\}\shuffle\{\{c_1\}\shuffle\pmb{h}\,\W{\shuffle}\,\pmb{K}(\pmb{\mathsf{Tr}}_s,a,b),c_2\}, r|\pmb{r}_{1}|
\ldots|\pmb{r}_{p}\Vert\,\mathsf{h})\,.\Label{Newform-1-1}
\end{align}
One advantage of above new form \eqref{Newform-1-1} is that for the special double trace pure gluon case, it reduces to the familiar generalized BCJ relation:
\bea
0=\Sl_{\shuffle}\left(k_{c_1}\cdot Y_{c_1}+k_{c_2}\cdot X_{c_2}\right)A(1,\{2,\dots,r-1\}\shuffle\{c_1,c_2\},r)\,.
\eea
Next, we consider the case with $l=4$. The type-II BCJ relation~\eqref{Eq:GeneralizedBCJType2-new-1} gives:
\begin{align}
0&= \WH{C}_{c_1}(\pmb{h}\,\W{\shuffle}\,\pmb{K})  A(1,\{2\ldots r-1\}\shuffle\{\pmb{h}\,\W{\shuffle}\,\pmb{K}(\pmb{\mathsf{Tr}}_s,a,b),c_1, c_2,c_3\}, r|\pmb{r}_{1}|
\ldots|\pmb{r}_{p}\Vert\,\mathsf{h})\nn
&\quad -\WH{C}_{c_2}(\pmb{h}\,\W{\shuffle}\,\pmb{K})A(1,\{2\ldots r-1\}\shuffle\{\pmb{h}\,\W{\shuffle}\,\pmb{K}(\pmb{\mathsf{Tr}}_s,a,b),c_2, c_1, c_3\}, r|\pmb{r}_{1}|
\ldots|\pmb{r}_{p}\Vert\,\mathsf{h})\nn
&\quad -\WH{C}_{c_2}(\pmb{h}\,\W{\shuffle}\,\pmb{K}) A(1,\{2\ldots r-1\}\shuffle\{\pmb{h}\,\W{\shuffle}\,\pmb{K}(\pmb{\mathsf{Tr}}_s,a,b),c_2, c_3, c_1\}, r|\pmb{r}_{1}|
\ldots|\pmb{r}_{p}\Vert\,\mathsf{h})\nn
&\quad +\WH{C}_{c_3}(\pmb{h}\,\W{\shuffle}\,\pmb{K})  A(1,\{2\ldots r-1\}\shuffle\{\pmb{h}\,\W{\shuffle}\,\pmb{K}(\pmb{\mathsf{Tr}}_s,a,b),c_3, c_2,c_1\}, r|\pmb{r}_{1}|
\ldots|\pmb{r}_{p}\Vert\,\mathsf{h})\,.\Label{Newform-4}
\end{align}
Applying the type-I BCJ relation~\eqref{Eq:GeneralizedBCJType1} to the last term (treating $c_3$ as the fiducial graviton), and using the relation:
\begin{align}
&\quad\{2\ldots r-1\}\shuffle\{c_2,c_1\}\shuffle\{\pmb{h}\,\W{\shuffle}\,\pmb{K},c_3\}\nonumber\\
&=\{2\ldots r-1\}\shuffle\Big(\{\pmb{h}\,\W{\shuffle}\,\pmb{K}\shuffle\{c_2,c_1\},c_3\}+\{\pmb{h}\,\W{\shuffle}\,\pmb{K},c_3,c_2,c_1\}+\{\pmb{h}\,\W{\shuffle}\,\pmb{K}\shuffle\{c_2\},c_3,c_1\}\Big)\,,
\end{align}
we can change it to
\begin{align}
\text{last term of \eqref{Newform-4}}&=-\WH{C}_{c_3}(\pmb{h}\,\W{\shuffle}\,\pmb{K})  A(1,\{2\ldots r-1\}\shuffle\{\pmb{h}\,\W{\shuffle}\,\pmb{K}(\pmb{\mathsf{Tr}}_s,a,b)\shuffle \{c_2\},c_3,c_1\}, r|\pmb{r}_{1}|
\ldots|\pmb{r}_{p}\Vert\,\mathsf{h})\nn
&\quad -\WH{C}_{c_3}(\pmb{h}\,\W{\shuffle}\,\pmb{K})  A(1,\{2\ldots r-1\}\shuffle\{\pmb{h}\,\W{\shuffle}\,\pmb{K}(\pmb{\mathsf{Tr}}_s,a,b)\shuffle \{c_2,c_1\},c_3\}, r|\pmb{r}_{1}|
\ldots|\pmb{r}_{p}\Vert\,\mathsf{h})\,.\Label{Newform-5}
\end{align}
When combining \eqref{Newform-5} with the second term of \eqref{Newform-4},  they provide a  part of $l=3$ identity \eqref{Newform-1-1}, with $\{2\ldots r-1\}\shuffle\{c_1\}$ considered as fixed in the first trace:
\begin{align}
0&= \WH{C}_{c_2}(\pmb{h}\,\W{\shuffle}\,\pmb{K})  A(1,\{2\ldots r-1\}\shuffle\{c_1\}\shuffle\{\pmb{h}\,\W{\shuffle}\,\pmb{K}(\pmb{\mathsf{Tr}}_s,a,b),c_2, c_3\}, r|\pmb{r}_{1}|
\ldots|\pmb{r}_{p}\Vert\,\mathsf{h})\nn
&\quad +\WH{C}_{c_3}(\pmb{h}\,\W{\shuffle}\,\pmb{K}) A(1,\{2\ldots r-1\}\shuffle\{c_1\}\shuffle\{\{c_2\}\shuffle\pmb{h}\,\W{\shuffle}\,\pmb{K}(\pmb{\mathsf{Tr}}_s,a,b),c_3\}, r|\pmb{r}_{1}|
\ldots|\pmb{r}_{p}\Vert\,\mathsf{h})\,,
\end{align}
where the shuffle products can be expanded as:
\begin{align*}
&\{c_1\}\shuffle\{\pmb{h}\,\W{\shuffle}\,\pmb{K},c_2,c_3\}=\{\{c_1\}\shuffle\pmb{h}\,\W{\shuffle}\,\pmb{K},c_2,c_3\}+\{\pmb{h}\,\W{\shuffle}\,\pmb{K},c_2,c_1,c_3\}+\{\pmb{h}\,\W{\shuffle}\,\pmb{K},c_2,c_3,c_1\}\\
&\{c_1\}\shuffle\{\{c_2\}\shuffle\pmb{h}\,\W{\shuffle}\,\pmb{K},c_3\}=\{\{c_2\}\shuffle\pmb{h}\,\W{\shuffle}\,\pmb{K},c_3,c_1\}+\{\{c_2,c_1\}\shuffle\pmb{h}\,\W{\shuffle}\,\pmb{K},c_3\}+\{\{c_1,c_2\}\shuffle\pmb{h}\,\W{\shuffle}\,\pmb{K},c_3\}\,.
\end{align*}
After using the above two identities to rewrite Eq.~\eqref{Newform-4}, we obtain:
\begin{align}
0 & =  \WH{C}_{c_1}(\pmb{h}\,\W{\shuffle}\,\pmb{K})  A(1,\{2\ldots r-1\}\shuffle\{\pmb{h}\,\W{\shuffle}\,\pmb{K}(\pmb{\mathsf{Tr}}_s,a,b),c_1, c_2,c_3\}, r|\pmb{r}_{1}|
\ldots|\pmb{r}_{p}\Vert\,\mathsf{h})\nn
&\quad +   \WH{C}_{c_2}(\pmb{h}\,\W{\shuffle}\,\pmb{K})  A(1,\{2\ldots r-1\}\shuffle\{\{c_1\}\shuffle\pmb{h}\,\W{\shuffle}\,\pmb{K}(\pmb{\mathsf{Tr}}_s,a,b), c_2,c_3\}, r|\pmb{r}_{1}|
\ldots|\pmb{r}_{p}\Vert\,\mathsf{h})\nn
&\quad +  \WH{C}_{c_3}(\pmb{h}\,\W{\shuffle}\,\pmb{K})  A(1,\{2\ldots r-1\}\shuffle\{\{c_1,c_2\}\shuffle\{\pmb{h}\,\W{\shuffle}\,\pmb{K}(\pmb{\mathsf{Tr}}_s,a,b)\}, c_3\}, r|\pmb{r}_{1}|
\ldots|\pmb{r}_{p}\Vert\,\mathsf{h})\,.\Label{Newform-6}
\end{align}
For the special double trace pure gluon case, it again reduces to the familiar generalized BCJ relation:
\bea
0=\Sl_{\shuffle}\left(k_{c_1}\cdot X_{c_1}+k_{c_2}\cdot X_{c_2}+k_{c_3}\cdot X_{c_3}\right)A(1,\{2\dots r-1\}\shuffle\{c_1,c_2,c_3\},r).
\eea
Now we can see the pattern. For an arbitrary second cycle $\pmb{2}=\{c_1\ldots c_j\}$, we can write the mixed form BCJ relation as:
\bea 0 & = & \sum_{\substack{\pmb{h}| \mathsf{h}=\mathsf{H} \\ \pmb{\mathsf{Tr}}_s| \mathsf{Tr}_p=\mathsf{Tr}
\backslash\{1,2\}}}\Bigg[\,\underset{\{a_i,b_i\}\subset\pmb{t}_i}{\widetilde{\sum}}
\,\Bigg]_{i=1}^s~~\sum_{i=1}^j \WH{C}_{c_i}(\pmb{h}\,\W{\shuffle}\,\pmb{K})\nn
%
& & ~~\times A(1,\{2\ldots r-1\}\shuffle\{\{c_1\ldots c_{i-1}\}\shuffle\{\pmb{h}\,\W{\shuffle}\,\pmb{K}(\pmb{\mathsf{Tr}}_s,a,b)\},c_i\ldots c_j \}, r|\pmb{r}_{1}|
\ldots|\pmb{r}_{p}\Vert\,\mathsf{h})\,.\Label{Newform-7} \eea
For the special double trace pure gluon case, it reduces to:
\bea
0=\Sl_{\shuffle}\left(\sum_{i=1}^j k_{c_i}\cdot X_{c_i}\right)A(1,\{2\dots r-1\}\shuffle\{c_1\ldots c_j\},r)\,,
\eea
which is nothing but the generalized fundamental BCJ relation of single trace YM amplitudes. This is the main advantage of the mixed form BCJ relations.


\section{Generalized gauge independence of new recursive relations}\label{sec:check}

It is very clear that both type-I and type-II recursive expansions [see Eq.~\eqref{eq:type1step1} and \eqref{eq:type2step1}] can have different forms since there are some gauge freedom on our choices:
\begin{itemize}
    \item In type-I, we have the freedom to choose one trace as fixed (namely, as trace $\pmb{1}$), and the fiducial graviton as $h_1$.
    \item In type-II, we have the freedom to choose two traces as fixed (namely, as trace $\pmb{1}$ and $\pmb{2}$). In trace $\pmb{2}$, we have the freedom to choose the fiducial gluon $d_2$.
\end{itemize}
If our formulas are right, all these choices must lead to equivalent expansions, including the equivalence between type-I and type-II recursive expansions. {The general consistency check is complicated and we will not do it
in this paper.}
Instead we will present several examples to demonstrate the salient idea.

Our first consideration is exchanging the role of $\pmb{1}$ and $\pmb{2}$ in the type-II recursive expansion.
To have the clear picture, let us start with a concrete example, i.e., the pure double trace amplitude $A(1,2,3\,|\,4,5,6)$. If $\{1,2,3\}$ is taken as the trace $\pmb{1}$, and $6$ is taken as the fixed $d_2$, the type-II recursive expansion~\eqref{eq:type2step1} gives:
\bea
\label{eq:123}
A(1,2,3\,|\,4,5,6)&=&(-k_4\cdot Y_4)A(1,2\shuffle\{4,5,6\},3)+(-k_5\cdot Y_5)(-1)A(1,2\shuffle\{5,4,6\},3),
\eea
where the sum over shuffle is implicit.
On the other hand, if $\{4,5,6\}$ is taken as trace $\pmb{1}$, and $3$ is taken as the fixed $d_2$, the expansion becomes:
\bea
\label{eq:456}
A(4,5,6\,|\,1,2,3)&=&(-k_1\cdot Y_1)A(4,5\shuffle\{1,2,3\},6)+(-k_2\cdot Y_2)(-1)A(4,5\shuffle\{2,1,3\},6),
\eea
To show the equivalence, we carry out the shuffle product of the above expansion as:
\begin{subequations}
\begin{align}
(-k_1\cdot Y_1)A(4,5\shuffle\{1,2,3\},6)&=(-k_1\cdot k_4-k_1\cdot k_5)A(4,5,1,2,3,6)+(-k_1\cdot k_4)A(4,1,2,3,5,6)\nonumber\\
&\quad+(-k_1\cdot k_4)A(4,1,\{5\}\shuffle\{2\},3,6) \\
(-k_2\cdot Y_2)A(4,5\shuffle\{2,1,3\},6)&=(-k_2\cdot k_4-k_2\cdot k_5)A(4,5,2,1,3,6)+(-k_2\cdot k_4)A(4,2,1,3,5,6)\nonumber\\
&\quad+(-k_2\cdot k_4)A(4,2,\{5\}\shuffle\{1\},3,6)\,.
\end{align}
\end{subequations}
%
%
%
%
%
Eq.~\eqref{eq:123} advises us to collect terms according to the coefficients of $(-k_5\cdot k_1)$, $(-k_5\cdot k_2)$, $(-k_4\cdot k_1)$ and $(-k_4\cdot k_2)$ respectively. Meanwhile, we should write all the amplitudes in the KK basis with point $1$ and $3$ fixed at the ends. The $(-k_5\cdot k_1)$ and $(-k_5\cdot k_2)$ terms are straightforward:
\begin{subequations}
\begin{align}
&(-k_1\cdot k_5)A(4,5,1,2,3,6)=(-k_1\cdot k_5)(-1)A(1,2\shuffle\{5,4,6\},3)\\
&(-k_2\cdot k_5)(-1)A(4,5,2,1,3,6)=(-k_2\cdot k_5)(-1)A(1,2,5,4,6,3)\,,
\end{align}
\end{subequations}
while the $(-k_4\cdot k_1)$ and $(k_4\cdot k_2)$ terms are a little involved:
\begin{subequations}
\begin{align}
&\quad (-k_1\cdot k_4)\Bigl[A(4,5,1,2,3,6)+A(4,1,\{5\}\shuffle\{2\},3,6)+A(4,1,2,3,5,6)\Bigr]\nn
&=(-k_1\cdot k_4)\Bigl[-A(1,\{2\}\shuffle\{5,4,6\},3)+A(1,\{2\}\shuffle\{5\}\shuffle\{4,6\},3)-A(1,2\shuffle\{4,6,5\},3)\Bigl]\nn
&=(-k_1\cdot k_4)A(1,\{2\}\shuffle\{4,5,6\},3)\\
&\quad (-k_2\cdot k_4)(-1)\Bigl[A(4,5,2,1,3,6)+A(4,2,\{5\}\shuffle\{1\},3,6)+A(4,2,1,3,5,6)\Bigr]\nn
&=(-k_2\cdot k_4)(-1)\Bigl[A(1,2,5,4,6,3)+A(1,5,2,4,6,3)-A(1,5\shuffle\{2,4,6\},3)+A(1,2,4,6,5,3)\Bigr]\nn
&=(-k_2\cdot k_4)A(1,2,4,5,6,3)\,.
\end{align}
\end{subequations}
Collecting all the above terms, we arrive at
\bea
(-k_4\cdot Y_4)A(1,\{2\}\shuffle\{4,5,6\},3)+(-k_5\cdot Y_5)(-1)A(1,\{2\}\shuffle\{5,4,6\},3)\,,
\eea
which is exactly the first expansion~\eqref{eq:123}. Thus we have proved the equivalence of the two expansions~\eqref{eq:123} and \eqref{eq:456}.

Although the above example is very simple, it does provide the strategy to show the equivalence of the choice of trace $\pmb{1}$ and $\pmb{2}$ for more  generic type-II expansions, i.e., with more gravitons or more traces.
For pure double trace with arbitrary number of points, by picking terms corresponding to $-k_{a\in \pmb{1}}\cdot k_{b\in \pmb{2}}$ and applying KK relation in one type of expansion, one can recover another type of expansion.
For those cases with at least one graviton or/and more than two traces, things become a little more complicated. Now
in the expansion, we have gravitons/traces inserted between $a\in \pmb{1}$ and $b\in \pmb{2}$ in the amplitudes, as well as tensor $\mathcal{T}$ between $-k_{a\in \pmb{1}}\cdot k_{b\in \pmb{2}}$. Nevertheless, one can again show the equivalence  by collecting terms having same kinematic coefficients and then using the KK relation.

The second consideration involves changing the fiducial graviton in the type-I recursive expansion.
For simplicity, we consider the single trace EYM amplitude $A(1,2,3\,\Vert\,h_1,h_2)$ as the example.
Taking $h_1$ as the fiducial graviton, we have the expansion:
\bea
\label{eq:123h1}
A(1,2,3\,\Vert\,h_1,h_2)=(\epsilon_{h_1}\cdot Y_{h_1})A(1,\{2\}\shuffle\{h_1\},3\,\Vert\,h_2)+(\epsilon_{h_1}\cdot F_{h_2}\cdot Y_{h_2})A(1,\{2\}\shuffle\{h_2,h_1\},3)\,.
\eea
If we take $h_2$ as the fiducial graviton, the expansion looks like:
\bea
\label{eq:123h2}
A(1,2,3\,\Vert\,h_1,h_2)=(\epsilon_{h_2}\cdot Y_{h_2})A(1,\{2\}\shuffle\{h_2\},3\}\,\Vert\,h_1)+(\epsilon_{h_2}\cdot F_{h_1}\cdot Y_{h_1})A(1,\{2\}\shuffle\{h_1,h_2\},3)\,.
\eea
To see the equivalence, we need to expand them further to pure YM amplitudes.
Then the first expansion~\eqref{eq:123h1} leads to:
\bea
A(1,2,3\,\Vert\,h_1,h_2)&=&(\epsilon_{h_1}\cdot X_{h_1})(\epsilon_{h_2}\cdot X_{h_2})A(1,\{2\}\shuffle\{h_1,h_2\},3)\nn
&&+(\epsilon_{h_1}\cdot X_{h_1})(\epsilon_{h_2}\cdot X_{h_2})A(1,\{2\}\shuffle\{h_2,h_1\},3)\nn
&&-(\epsilon_{h_1}\cdot \epsilon_{h_2})(k_{h_2}\cdot Y_{h_2})A(1,\{2\}\shuffle\{h_2,h_1\},3)\,,
\eea
while the second expansion gives:
\bea
A(1,2,3\,\Vert\,h_1,h_2)&=&(\epsilon_{h_1}\cdot X_{h_1})(\epsilon_{h_2}\cdot X_{h_2})A(1,\{2\}\shuffle\{h_1,h_2\},3)\nn
&&+(\epsilon_{h_1}\cdot X_{h_1})(\epsilon_{h_2}\cdot X_{h_2})A(1,\{2\}\shuffle\{h_2,h_1\},3)\nn
&&-(\epsilon_{h_2}\cdot \epsilon_{h_1})(k_{h_1}\cdot Y_{h_1})A(1,2\shuffle\{h_1,h_2\},3).
\eea
We remind that $X_h$ sums over all the momenta at the left hand side of $h$, including other gravitons. The difference of these two expansions is\footnote{Using the notation defined in the previous section and appendix~\ref{sec:BCJblock}, we can see  that the part inside the bracket is $-{\cal B}[\{h_2, h_1\}; 1]+{\cal B}[\{h_1, h_2\}; 1]$, so it is zero by the explicit result given in Eq.~\eqref{B-block-n=2}.}
\bea
(\epsilon_{h_1}\cdot \epsilon_{h_2})\left[-(k_{h_2}\cdot Y_{h_2})A(1,2\shuffle\{h_2,h_1\},3)+(k_{h_1}\cdot Y_{h_1})A(1,2\shuffle\{h_1,h_2\},3)\right]\,.
\eea
After we apply the generalized fundamental BCJ relation~\eqref{Gen-BCJ} to the first term in the bracket, the above expression becomes:
\bea
(\epsilon_{h_1}\cdot \epsilon_{h_2})(k_{h_1}\cdot X_{h_1}+k_{h_2}\cdot X_{h_2})A(1,2\shuffle\{h_1,h_2\},3)=0,
\eea
which is zero according to~\eqref{Gen-BCJ}. Thus the equivalence has been proved. Above discussion
can be trivially  generalized to $A(1,2,\dots,n\,\Vert\,h_1,h_2)$, namely, with arbitrary number of gluons. The generalization to arbitrary numbers of gluon traces and gravitons are more complicated, and we may need to use the type-I and type-II BCJ relations.

The third consideration involves the equivalence between the type-I and type-II recursive expansions.  For simplicity, we consider the amplitude $A(1,2\,|\,3,4,5\,\Vert\,h)$ as the example. The type-I recursive expansion gives the following result, which is equivalent to the one given in~\cite{Nandan:2016pya}:
\begin{align}
A(1,2\,|\,3,4,5\,\Vert\,h)&=\left(\epsilon_h\cdot k_1\right)A(1,h,2\,|\,3,4,5)\nn
&\quad +(\epsilon_h\cdot k_5)(-k_3\cdot k_1)A(1,3,4,5,h,2)+(\epsilon_h\cdot k_5)(-k_4\cdot k_1)(-1)A(1,4,3,5,h,2)\nn
&\quad +(\epsilon_h\cdot k_4)(-k_5\cdot k_1)A(1,5,3,4,h,2)+(\epsilon_h\cdot k_4)(-k_3\cdot k_1)(-1)A(1,3,5,4,h,2)\nn
&\quad +(\epsilon_h\cdot k_3)(-k_4\cdot k_1)A(1,4,5,3,h,2)+(\epsilon_h\cdot k_3)(-k_5\cdot k_1)(-1)A(1,5,4,3,h,2).
\end{align}
The first term can be further expanded using the type-II recursive expansion. Finally we reach the pure YM expansion:
\begin{align}
A(1,2\,|\,3,4,5\,\Vert\,h)&=\epsilon_h\cdot(k_5+k_1)(-k_3\cdot k_1)A(1,3,4,5,h,2)+\epsilon_h\cdot(k_5+k_1)(-k_4\cdot k_1)(-1)A(1,4,3,5,h,2)\nn
&\quad +(\epsilon_h\cdot k_1)[(-k_3\cdot k_1)A(1,3,\{h\}\shuffle\{4\},5,2)+(-k_3\cdot k_1-k_3\cdot k_h)A(1,h,3,4,5,2)]\nn
&\quad +(\epsilon_h\cdot k_1)[(-k_4\cdot k_1)A(1,4,\{h\}\shuffle\{3\},5,2)+(-k_4\cdot k_1-k_4\cdot k_h)A(1,h,4,3,5,2)]\nn
&\quad +(\epsilon_h\cdot k_4)(-k_5\cdot k_1)A(1,5,3,4,h,2)+(\epsilon_h\cdot k_4)(-k_3\cdot k_1)(-1)A(1,3,5,4,h,2)\nn
&\quad +(\epsilon_h\cdot k_3)(-k_4\cdot k_1)A(1,4,5,3,h,2)+(\epsilon_h\cdot k_3)(-k_5\cdot k_1)(-1)A(1,5,4,3,h,2)\,.\Label{Eq:GravitonTrace1}
\end{align}
We then repeat the process by using the type-II recursive expansion first:
\bea
A(1,2\,|\,3,4,5\,\Vert\,h)&=&(-k_3\cdot k_1)A(1,3,4,5,2\,\Vert\,h)+(-k_4\cdot k_1)(-1)A(1,4,3,5,2\,\Vert\,h)\nn
&&+(-k_3\cdot F_h\cdot k_1)A(1,h,3,4,5,2)+(-k_4\cdot F_h\cdot k_1)(-1)A(1,h,4,3,5,2)\,.
\eea
Next, we further expand the single trace EYM amplitudes, using the special case of type-I expansion. The final result is:
\begin{align}
&A(1,2\,|\,3,4,5\,\Vert\,h)=(-k_3\cdot k_1-k_3\cdot k_h)(\epsilon_h\cdot k_1)A(1,h,3,4,5,2)+(-k_3\cdot \epsilon_h)(-k_h\cdot k_1)A(1,h,3,4,5,2)\nn
&\qquad+(-k_3\cdot k_1)\epsilon_h\cdot (k_1+k_3)A(1,3,h,4,5,2)+(-k_3\cdot k_1)\epsilon_h\cdot (k_1+k_3+k_4)A(1,3,4,h,5,2)\nn
&\qquad+(-k_3\cdot k_1)\epsilon_h\cdot(k_1+k_3+k_4+k_5)A(1,3,4,5,h,2)\nn
&\qquad+(-k_4\cdot k_1-k_4\cdot k_h)(\epsilon_h\cdot k_1)(-1)A(1,h,4,3,5,2)+(-k_4\cdot \epsilon_h)(-k_h\cdot k_1)(-1)A(1,h,4,3,5,2)\nn
&\qquad+(-k_4\cdot k_1)\epsilon_h\cdot (k_1+k_4)(-1)A(1,4,h,3,5,2)+(-k_4\cdot k_1)\epsilon_h\cdot (k_1+k_4+k_3)(-1)A(1,4,3,h,5,2)\nn
&\qquad+(-k_4\cdot k_1)\epsilon_h\cdot(k_1+k_4+k_3+k_5)(-1)A(1,4,3,5,h,2)\,.\Label{Eq:GravitonTrace2}
\end{align}
The difference of Eq.~\eqref{Eq:GravitonTrace1} and \eqref{Eq:GravitonTrace2} can be group into two terms, one proportional to $(\epsilon_h\cdot k_3)$ and the other to $(-\epsilon_h\cdot k_4)$:
\begin{subequations}
\begin{align}
(\epsilon_h\cdot k_3)&\Bigl\{-(k_h\cdot k_1)A(1,h,3,4,5,2)+(k_3\cdot k_1)
A(1,3,\{h\}\shuffle\{4,5\},2)\nn
&\quad-(k_4\cdot k_1)
A(1,4,\{5\}\shuffle\{3,h\},2)+(k_5\cdot k_1)A(1,5,4,3,h,2)\Bigr\}\label{eq:type2finala}\\
(-\epsilon_h\cdot k_4)&\Bigl\{-(k_h\cdot k_1)A(1,h,4,3,5,2)+(k_4\cdot k_1)
A(1,4,\{h\}\shuffle\{3,5\},2)\nn
&\quad-(k_3\cdot k_1)
A(1,3,\{5\}\shuffle\{4,h\},2)+(k_5\cdot k_1)A(1,5,3,4,h,2)\Bigr\}\label{eq:type2finalb}
\end{align}
\end{subequations}
The curly bracket in \eqref{eq:type2finala} and \eqref{eq:type2finalb} are related by exchanging the label $3$ and $4$, so we only need to deal with one of them. For Eq.~\eqref{eq:type2finala}, we use the fundamental BCJ relation to write the last term as:
\begin{equation}
(k_5\cdot k_1)A(1,5,4,3,h,2)=\left(-k_5\cdot X_5\right)A(1,4,\{5\}\shuffle\{3,h\},2)\,.
\end{equation}
%
%
Putting it back, we write the sum of $(k_5\cdot k_1)$ term and $(k_4\cdot k_1)$ term as:
\bea
-(k_4\cdot k_1+k_5\cdot X_5)A(1,4,\{5\}\shuffle\{3,h\},2)=\left(k_4\cdot X_4+k_5\cdot X_5\right)A(1,3,\{4,5\}\shuffle\{h\},2)\,,
\eea
where a generalized fundamental BCJ relation has been applied.  Now combining the above expression with the $(k_3\cdot k_1)$ term, we get:
\begin{equation}
\left(k_3\cdot k_1+k_4\cdot X_4+k_5\cdot X_5\right)A(1,3,\{4,5\}\shuffle\{h\},2)=-\left(k_3\cdot X_3+k_4\cdot X_4+k_5\cdot X_5\right)A(1,h,3,4,5,2)\,,
\end{equation}
with the help of another generalized fundamental BCJ relation. Finally, the sum of the above expression and the $(k_h\cdot k_1)$ term gives:
\bea
-\left(k_h\cdot X_h+k_3\cdot X_3+k_4\cdot X_4+k_5\cdot X_5\right)A(1,\{h,3,4,5\}\shuffle\varnothing,2)=0\,.
\eea
By exchanging the label $3$ and $4$, we can easily show that the curly bracket of Eq.~\eqref{eq:type2finalb} also vanishes.
%
%
In this way, we have proved that the difference of Eq.~\eqref{Eq:GravitonTrace1} and \eqref{Eq:GravitonTrace2} vanishes, so the two expansions are equivalent.
\section{Proof using BCFW recursion relation}\label{sec:BCFW}

In this section, we will present the proof of type-I recursive expansion \eqref{eq:type1step1} by BCFW recursion relation~\cite{Britto:2004ap,Britto:2005fq}.
The proof for type-II recursive expansion can follow the same line, or alternatively, follow the equivalence studied in the previous section.
We shift the momentum of gluon $1$ and $r$ in the first trace as
\bea
\WH k_1=k_1+zq\qquad\qquad\WH k_r=k_r-zq\qquad\qquad q^2=q\cdot k_1=q\cdot k_2=0\,.\Label{Eq:MomentaDeformation}
\eea
Under such a shift, an amplitude $A$ becomes $A(z)$, a rational function of $z$. The original amplitude $A$ can be recovered by:
\bea
A=\text{Res}_{z= 0}\left[\frac{A(z)}{z}\right]=-\sum_{z_i\in\text{finite poles}}\text{Res}_{z=z_i}\left[\frac{A(z)}{z}\right]+\text{Boundary term}\,.
\eea
In the coming subsections, we will demonstrate that both sides of~\eqref{eq:type1step1} have zero boundary terms, the residues at all the finite physical poles on both sides match with each other, and the spurious poles on the right hand side cancel out. We can prove the recursive expansion~\eqref{eq:type1step1} by showing that all the above statements are true.

\subsection{The cancellation of boundary terms}

We first investigate the cancellation of boundary terms, which can be easily worked out for the most generic cases. It is well known that under the deformation \eqref{Eq:MomentaDeformation} where $1$ and $r$ are adjacent to each other, there is at least one good deformation such that the boundary behavior
of the multitrace tree amplitude $A(\WH 1(z)\dots\WH r(z)\,|\,\pmb{2}\,|\dots|\,\pmb{m}\,\Vert\,{\mathsf{H}})$ can be $z^{-1}$ (see, for example,~\cite{ArkaniHamed:2008yf}), i.e., the boundary contribution is zero. At the right hand side of the type-I recursive expansion~\eqref{eq:type1step1}, there are two origins of the $z$ contribution. The first place is the expansion coefficient. From the expressions, the leading behavior is of order $z^1$ when $z\to \infty$ as seen as
\bea
z^1(\epsilon_{h_1}\cdot \mathcal{T}_{\rho_{|\pmb{h}|+s}}\ldots\mathcal{T}_{\rho_2}\cdot \mathcal{T}_{\rho_1}\cdot q).
\eea
The second is from individual amplitudes, which has the leading large $z$ behavior $z^{-1}$. When combining these two together, we get the leading behavior as $z^0$:
%
\bea
&&\sum_{\substack{\pmb{h}| \mathsf{h}=\mathsf{H}\backslash h_1 \\ \pmb{\mathsf{Tr}}_s| \mathsf{Tr}_p=\mathsf{Tr}\backslash 1}}\Bigg[\,\underset{\{a_i,b_i\}\subset\pmb{t}_i}{\widetilde{\sum}}\,\Bigg]_{i=1}^sz(\epsilon_{h_1}\cdot \mathcal{T}_{\rho_{|\pmb{h}|+s}}\ldots\mathcal{T}_{\rho_2}\cdot \mathcal{T}_{\rho_1}\cdot q)\nn
&&~~~~~~~~~~~~~~~~~~~~~~\times A(\WH 1(z),\{2\ldots r-1\}\shuffle\{\pmb{h}\,\W{\shuffle}\,\pmb{K}(\pmb{\mathsf{Tr}}_s,a,b),h_1\},\WH r(z)\,|\,\pmb{r}_{1}\,|\ldots|\,\pmb{r}_{p}\,\Vert\,\mathsf{h})\,.\Label{Boundary-form}
\eea
Naively, we would expect nonzero boundary contribution. But as we will show, in fact,
after some careful analysis, the leading large $z$ behavior of~\eqref{Boundary-form} is $z^{-1}$, so there is actually no boundary contribution.

To see that, let us consider the part with fixed partitions $\pmb{h}| \mathsf{h}=\mathsf{H}\backslash h_1$,  $\pmb{\mathsf{Tr}}_s| \mathsf{Tr}_p=\mathsf{Tr}\backslash 1$, and the choice of $(a_i, b_i)$. Furthermore, in the shuffle
sum of $\pmb{h}\,\W{\shuffle}\,\pmb{K}(\pmb{\mathsf{Tr}}_s,a,b)$, there are many terms, and we focus to one particular
ordering denoted by $\pmb{\rho}$. Now the key observation is that, with the fixed ordering $\pmb\rho$, all terms in the shuffle sum
$$\{2\ldots r-1\}\shuffle \{\pmb{\rho}, h_1\}$$
will have the same coefficient $(\epsilon_{h_1}\cdot \mathcal{T}_{\rho_{|\pmb{h}|+s}}\ldots\mathcal{T}_{\rho_2}\cdot \mathcal{T}_{\rho_1}\cdot q)$.\footnote{This claim will not be true if $q$ is replaced by $Y_{\rho_1}$.} With this observation, we can see that
\bea & & \sum_{\shuffle}z(\epsilon_{h_1}\cdot \mathcal{T}_{\rho_{|\pmb{h}|+s}}\ldots\mathcal{T}_{\rho_2}\cdot \mathcal{T}_{\rho_1}\cdot q)A(\WH 1(z),\{2\ldots r-1\}\shuffle\{\pmb{\rho},h_1\},\WH r(z)\,|\,\pmb{r}_{1}\,|\ldots|\,\pmb{r}_{p}\,\Vert\,\mathsf{h})\nn
& = & z(\epsilon_{h_1}\cdot \mathcal{T}_{\rho_{|\pmb{h}|+s}}\ldots\mathcal{T}_{\rho_2}\cdot \mathcal{T}_{\rho_1}\cdot q)(-1)^{s+|\pmb{h}|+1}A(\WH 1(z),\{2\ldots r-1\},\WH r(z),h_1, \pmb{\rho}^T\,|\,\pmb{r}_{1}\,|\ldots|\,\pmb{r}_{p}\,\Vert\,\mathsf{h})\,, \eea
where the KK relation has been used. Now it is clear that since the two shifted gluons $1$ and $r$ are not adjacent, the leading large $z$ behavior has been reduced to $z^{-2}$. Thus together with the coefficient $z^1$, the leading behavior is, in fact, the $z^{-1}$. Therefore, the boundary contribution is indeed zero. For the special case with $r=2$, the shifted gluons $1$ and $r$ are always adjacent. However, if there exists another gluon trace that has more than two gluons, we can fixed that one as our trace $\pmb{1}$ instead. The consistency has been discussed in section~\ref{sec:check}. If all traces have only two gluon, we effectively reduce to the EM cases discussed in section~\ref{sec:compact}, which can be understood by compactification of Einstein gravity.


\subsection{Matching finite physical poles}
\begin{figure}[t]
	\centering
	\begin{tikzpicture}
	\draw [very thick] (0,0) node [left=1pt]{$\WH{1}$} -- (8,0) node [right=1pt]{$\WH{r}$};
	\draw [line width=2pt] (4.3,-2) -- ++(-0.3,0) -- ++(0,4) -- ++(-0.3,0);
	\draw [very thick] (0.5,0) -- ++(0,-1) node [below=0pt]{$2$} (1,0) -- ++(0,-1) node [below=0pt]{$3$} (3,0) -- ++(0,-1) node [below=0pt]{$j$} (5,0) -- ++(0,-1) node[below=0pt]{$j+1$} (7.5,0) -- ++(0,-1) node[below=0pt]{$r-1$};
	\node at (2,-0.5) {$\cdots\cdots$};
	\node at (6.5,-0.5) {$\cdots\cdots$};
	\draw [dashed,thick] (0.5,-0.5) -- ++(-1,-1) (3.5,0) -- ++(0,-1) (6,0) -- ++(0,1) (2,0) -- ++(0,1) (5,-0.5) -- ++(1,-1);
	\draw [very thick] (2,1.5) ++(0:0.5) -- ++(0:0.5);
	\draw [very thick] (2,1.5) ++(180:0.5) -- ++(180:0.5);
	\draw [very thick] (2,1.5) ++(90:0.5) -- ++(90:0.5);
	\draw [very thick] (2,1.5) ++(45:0.5) -- ++(45:0.5);
	\draw [very thick] (2,1.5) ++(135:0.5) -- ++(135:0.5);
	\draw [thick,pattern=north east lines] (2,1.5) circle (0.5cm);
	\draw [very thick] (6,1.5) ++(0:0.5) -- ++(0:0.5);
	\draw [very thick] (6,1.5) ++(180:0.5) -- ++(180:0.5);
	\draw [very thick] (6,1.5) ++(90:0.5) -- ++(90:0.5);
	\draw [very thick] (6,1.5) ++(45:0.5) -- ++(45:0.5);
	\draw [very thick] (6,1.5) ++(135:0.5) -- ++(135:0.5);
	\draw [thick,pattern=north east lines] (6,1.5) circle (0.5cm);		
	\node at (2,-2.5) [align=center]{$\mathsf{H}_{L}$~$\mathsf{Tr}_L$ \\ $\pmb{1}_L=\{1,2\ldots j\}$};
	\node at (6,-2.5) [align=center]{$\mathsf{H}_{R}$~$\mathsf{Tr}_R$ \\ $\pmb{1}_R=\{j+1\ldots r\}$};
	\draw [very thick] (9,0) -- ++(1,0) node[right=2pt]{gluon};
	\draw [dashed,thick] (9,1) -- ++(1,0) node[right=2pt]{graviton};
	\draw [thick,pattern=north east lines] (9.5,-1) circle (0.3cm);
	\node at (10,-1) [right=2pt]{gluon trace};
	\end{tikzpicture}
	\caption{A typical BCFW factorization channel of $A(1,2\ldots r\,|\,\pmb{2}\,|\ldots|\,\pmb{m}\,\Vert\,\mathsf{H})$.}
	\label{fig:factorization}
\end{figure}
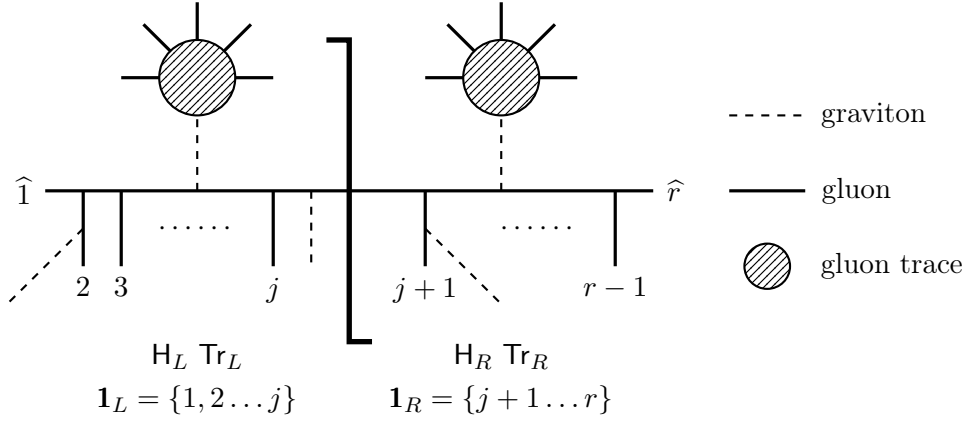

The purpose of this subsection is to show that the right hand side of the type-I expansion~\eqref{eq:type1step1}:
\begin{align}
\label{eq:rightexpansion}
{\tt Expansion}=\sum_{\substack{\pmb{h}| \mathsf{h}=\mathsf{H}\backslash h_1 \\ \pmb{\mathsf{Tr}}_s| \mathsf{Tr}_p=\mathsf{Tr}\backslash 1}}&\Bigg[\,\underset{\{a_i,b_i\}\subset\pmb{t}_i}{\widetilde{\sum}}\,\Bigg]_{\pmb{t}_i\in\pmb{\mathsf{Tr}}_s}\sum_{\pmb\rho\in\pmb{h}\,\W{\shuffle}\,\pmb{K}(\pmb{\mathsf{Tr}}_s,a,b)}C_{h_1}(\pmb\rho) \nonumber\\
&\qquad\times A_{p+1,|\mathsf{h}|}(1,\{2\ldots r-1\}\shuffle\{\pmb\rho,h_1\},r\,|\,\mathsf{Tr}_p\,\Vert\,\mathsf{h})
\end{align}
has the same residue as the left hand side at all the physical poles. In this section, we will sometimes strip off the $Y$ symbol from $C_{h_1}$, so given $\pmb\rho\in\pmb{h}\,\W{\shuffle}\,\pmb{K}(\pmb{\mathsf{Tr}}_s,a,b)$, we define:
\begin{equation}
C_{h_1}(\pmb\rho)=C_{h_1}^{\mu}(\pmb\rho)(Y_{\rho_1})_{\mu}\,.
\end{equation}
With the BCFW shift~\eqref{Eq:MomentaDeformation}, since $1$ and $r$ always belong to same trace, the BCFW recursive expansion does not have any graviton channel (i.e., internal graviton propagator), and only the gluon channels in the first trace survive. With this in mind, let us look at a typical channel that divides the graviton set $\mathsf{H}$ into $\mathsf{H}_L$, $\mathsf{H}_R$ , and the traces $\mathsf{Tr}\backslash 1$ into $\mathsf{Tr}_L$, $\mathsf{Tr}_R$. In particular, the first trace is divided into $\pmb{1}_{L}=\{1,2\ldots j\}$ and $\pmb{1}_{R}=\{j+1\ldots r\}$. On the left hand side of \eqref{eq:type1step1}, the residue of this channel is:
\begin{equation}
\label{eq:resleft}
\text{Res}\left[\frac{1}{z}\,A(\WH 1,2\ldots \WH r\,|\,\pmb{2}\,|\ldots|\,\pmb{m}\,\Vert\,\mathsf{H})\right]=A_{L}\times\frac{i}{P^2_L}\times A_{R}\,,
\end{equation}
where
\begin{align}
A_L=A(\WH{1},2\ldots j,-\WH{P}_{L}\,|\,\mathsf{Tr}_{L}\,\Vert\,\mathsf{H}_{L})& &A_R=(\WH{P}_{L},j+1\ldots \WH{r}\,|\,\mathsf{Tr}_{R}\,\Vert\,\mathsf{H}_{R})\,.
\end{align}
The sum of internal states is kept implicit. This channel is depicted in figure~\ref{fig:factorization}. If we use $K(\mathsf{A})$ to denote the sum of all momenta in the set $\mathsf{A}$, then we have:
\begin{equation}
P_L=k_1+\ldots +k_j+K(\mathsf{H}_{L})+K(\mathsf{Tr}_L)\,.
\end{equation}
We need to check whether the right hand side of the expansion, namely, Eq.~\eqref{eq:rightexpansion}, has the same residue at this channel. We can first formally write
\begin{equation}
\text{Res}\left[\frac{{\tt Expansion}}{z}\right]=\W{A}_L\times\frac{i}{P_L^2}\times\W{A}_R
\end{equation}
at this channel, and then check whether $\W{A}_L$ and $\W{A}_R$ match with $A_L$ and $A_R$ in~\eqref{eq:resleft}. A prominent feature of our expansion~\eqref{eq:rightexpansion} is that some gravitons and gluon traces have been \emph{transmuted} into gluons in the first trace. Depending on how these transmuted particles get divided into the left and right sub-amplitude, we need to investigate two scenarios: the fiducial graviton $h_1\in\mathsf{H}_L$ and $h_1\in\mathsf{H}_R$.

\subsubsection{Case one: \texorpdfstring{$h_1\in\mathsf{H}_L$}{h1 in HL}}
\begin{figure}[t]
	\centering
	\begin{tikzpicture}
	\draw [line width=2pt] (4.3,-2) -- ++(-0.3,0) -- ++(0,4) -- ++(-0.3,0);
	\draw [very thick] (0.5,0) -- ++(0,-1) node [below=0pt]{$2$} (3,0) -- ++(0,-1) node [below=0pt]{$j$} (5,0) -- ++(0,-1) node[below=0pt]{$j+1$} (7.5,0) -- ++(0,-1) node[below=0pt]{$r-1$};
	\draw [blue,very thick,decorate,decoration={coil,aspect=0}] (1,0) -- ++(0,-2);
	\draw [blue,very thick,decorate,decoration={coil,aspect=0}] (1.5,0)  -- ++(0,-2);
	\draw [red,very thick] (2.5,0) -- ++(0,-1.8) node[below=0pt]{$h_1$};
	\node at (6.5,-0.5) {$\cdots\cdots$};
	\draw [dashed,thick] (0.5,-0.5) -- ++(-1,-1) (3.5,0) -- ++(0,-1) (6,0) -- ++(0,1) (2,0) -- ++(0,1) (5,-0.5) -- ++(1,-1);
	\draw [very thick] (2,1.5) ++(0:0.5) -- ++(0:0.5);
	\draw [very thick] (2,1.5) ++(180:0.5) -- ++(180:0.5);
	\draw [very thick] (2,1.5) ++(90:0.5) -- ++(90:0.5);
	\draw [very thick] (2,1.5) ++(45:0.5) -- ++(45:0.5);
	\draw [very thick] (2,1.5) ++(135:0.5) -- ++(135:0.5);
	\draw [thick,pattern=north east lines] (2,1.5) circle (0.5cm);
	\draw [very thick] (6,1.5) ++(0:0.5) -- ++(0:0.5);
	\draw [very thick] (6,1.5) ++(180:0.5) -- ++(180:0.5);
	\draw [very thick] (6,1.5) ++(90:0.5) -- ++(90:0.5);
	\draw [very thick] (6,1.5) ++(45:0.5) -- ++(45:0.5);
	\draw [very thick] (6,1.5) ++(135:0.5) -- ++(135:0.5);
	\draw [thick,pattern=north east lines] (6,1.5) circle (0.5cm);		
	\draw [very thick] (0,0) node [left=1pt]{$\WH{1}$} -- (8,0) node [right=1pt]{$\WH{r}$};
	\node at (2,-3) [align=center]{$\mathsf{H}_{L}$~$\mathsf{Tr}_L$ \\ $\pmb{1}_L=\{1,2\ldots j\}$};
	\node at (6,-3) [align=center]{$\mathsf{H}_{R}$~$\mathsf{Tr}_R$ \\ $\pmb{1}_R=\{j+1\ldots r\}$};
	\draw [very thick] (9,1) -- ++(1,0) node[right=2pt]{gluon};
	\draw [dashed,thick] (9,2) -- ++(1,0) node[right=2pt]{graviton};
	\draw [blue,very thick,decorate,decoration={coil,aspect=0}] (9,-1) -- ++(1,0) node[right=2pt]{transmuted gluon};
	\draw [red,very thick] (9,-2) -- ++(1,0) node[right=2pt]{transmuted graviton};
	\draw [thick,pattern=north east lines] (9.5,0) circle (0.3cm);
	\node at (10,0) [right=2pt]{gluon trace};
	\end{tikzpicture}
	\caption{A typical BCFW factorization channel of~\eqref{eq:rightexpansion} with $h_1\in\mathsf{H}_{L}$.}
	\label{fig:caseone}
\end{figure}
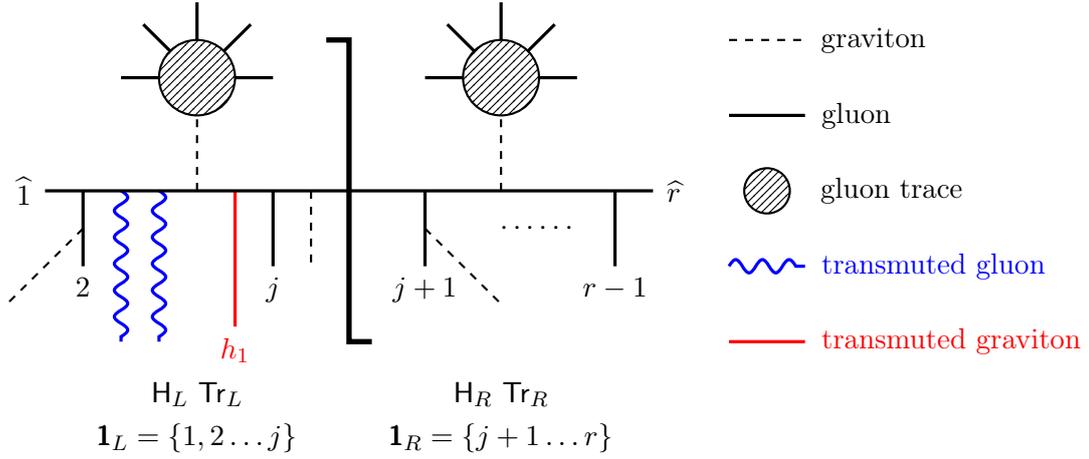

If the fiducial graviton $h_1\in\mathsf{H}_L$, then there is no transmuted particles in the right sub-amplitude, since according to~\eqref{eq:rightexpansion}, $h_1$ is the last transmuted particle. Therefore, we have:
\begin{equation}
\W{A}_{R}=A_{R}(\WH{P}_{L},j+1\ldots \WH{r}\,|\,\mathsf{Tr}_{R}\,\Vert\,\mathsf{H}_{R})\,.
\end{equation}
This factorization channel is depicted in figure~\ref{fig:caseone}. For $\W{A}_L$, we have:
\begin{align}
\W{A}_{L}&=\sum_{\substack{\pmb{h}| \mathsf{h}=\mathsf{H}_L\backslash h_1 \\ \pmb{\mathsf{Tr}}_s| \mathsf{Tr}_p=\mathsf{Tr}_L}}\Bigg[\,\underset{\{a_i,b_i\}\subset\pmb{t}_i}{\widetilde{\sum}}\,\Bigg]_{\pmb{t}_i\in\pmb{\mathsf{Tr}}_s} \sum_{\pmb\rho\in\pmb{h}\,\W{\shuffle}\,\pmb{K}(\pmb{\mathsf{Tr}}_s,a,b)}C_{h_1}(\pmb\rho) A(\WH{1},\{2\ldots j\}\shuffle\{\pmb\rho,h_1\},-\WH{P}_L\,|\,\mathsf{Tr}_p\,\Vert\,\mathsf{h})\nonumber\\
&=A_{L}(\WH{1},2\ldots j,-\WH{P}_L\,|\,\mathsf{Tr}_L\,\Vert\,\mathsf{H}_L)\,,
\end{align}
according to the type-I recursive expansion with fewer points. Therefore, we have shown that:
\begin{equation}
\label{eq:resultcase1}
\text{Res}\left[\frac{{\tt Expansion}}{z}\right]=\text{Res}\left[\frac{1}{z}\,A(\WH 1,2\ldots \WH r\,|\,\pmb{2}\,|\ldots|\,\pmb{m}\,\Vert\,\mathsf{H})\right]
\end{equation}
holds for this case.

\subsubsection{Case two: \texorpdfstring{$h_1\in\mathsf{H}_R$}{h1 in HR}}
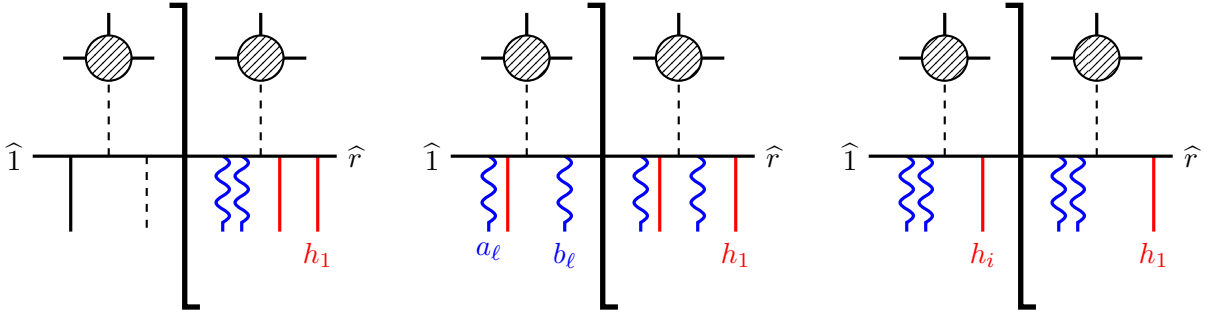
\begin{figure}[t]
	\centering
	\begin{tikzpicture}
	\begin{scope}[xshift=0]
	\draw [line width=2pt] (2.2,-2) -- ++(-0.2,0) -- ++(0,4) -- ++(-0.2,0);
	\foreach \x in {1,3}{
		\draw [dashed,thick] (\x,0) -- ++(0,1);
		\draw [thick,pattern=north east lines] (\x,1.3) circle (0.3cm);
		\draw [very thick] (\x,1.3) ++(0:0.3) --++(0:0.3) (\x,1.3) ++(90:0.3) -- ++(90:0.3) (\x,1.3) ++(180:0.3) -- ++(180:0.3);
	}
	\draw [very thick] (0.5,0) -- ++(0,-1);
	\draw [dashed,thick] (1.5,0) -- ++(0,-1);
	\draw [blue,very thick,decorate,decoration={coil,aspect=0}] (2.5,0) -- ++(0,-1);
	\draw [blue,very thick,decorate,decoration={coil,aspect=0}] (2.75,0) -- ++(0,-1);
	\draw [very thick,red] (3.25,0) -- ++(0,-1) (3.75,0) -- ++(0,-1) node[below=0pt]{$h_1$};
	\draw [very thick] (0,0) node[left=0pt]{$\WH{1}$} -- (4,0) node[right=0pt]{$\WH{r}$};
	\end{scope}
	\begin{scope}[xshift=5.5cm]
	\draw [line width=2pt] (2.2,-2) -- ++(-0.2,0) -- ++(0,4) -- ++(-0.2,0);
	\foreach \x in {1,3}{
		\draw [dashed,thick] (\x,0) -- ++(0,1);
		\draw [thick,pattern=north east lines] (\x,1.3) circle (0.3cm);
		\draw [very thick] (\x,1.3) ++(0:0.3) --++(0:0.3) (\x,1.3) ++(90:0.3) -- ++(90:0.3) (\x,1.3) ++(180:0.3) -- ++(180:0.3);
	}
	\draw [blue,very thick,decorate,decoration={coil,aspect=0}] (1.5,0) -- ++(0,-1) node[below=0pt]{$b_\ell$};
	\draw [blue,very thick,decorate,decoration={coil,aspect=0}] (0.5,0) -- ++(0,-1) node[below=0pt]{$a_\ell$};
	\draw [red,very thick] (0.75,0) -- ++(0,-1) (2.75,0) -- ++(0,-1) (3.75,0) -- ++(0,-1) node[below=0pt]{$h_1$};
	\draw [blue,very thick,decorate,decoration={coil,aspect=0}] (2.5,0) -- ++(0,-1);
	\draw [blue,very thick,decorate,decoration={coil,aspect=0}] (3.25,0) -- ++(0,-1);
	\draw [very thick] (0,0) node[left=0pt]{$\WH{1}$} -- (4,0) node[right=0pt]{$\WH{r}$};
	\end{scope}
	\begin{scope}[xshift=11cm]
	\draw [line width=2pt] (2.2,-2) -- ++(-0.2,0) -- ++(0,4) -- ++(-0.2,0);
	\foreach \x in {1,3}{
		\draw [dashed,thick] (\x,0) -- ++(0,1);
		\draw [thick,pattern=north east lines] (\x,1.3) circle (0.3cm);
		\draw [very thick] (\x,1.3) ++(0:0.3) --++(0:0.3) (\x,1.3) ++(90:0.3) -- ++(90:0.3) (\x,1.3) ++(180:0.3) -- ++(180:0.3);
	}
	\draw [red,very thick] (1.5,0) -- ++(0,-1) node[below=0pt]{$h_i$} (3.75,0) -- ++(0,-1) node[below=0pt]{$h_1$};
	\draw [blue,very thick,decorate,decoration={coil,aspect=0}] (0.5,0) -- ++(0,-1);
	\draw [blue,very thick,decorate,decoration={coil,aspect=0}] (0.75,0) -- ++(0,-1);
	\draw [blue,very thick,decorate,decoration={coil,aspect=0}] (2.5,0) -- ++(0,-1);
	\draw [blue,very thick,decorate,decoration={coil,aspect=0}] (2.75,0) -- ++(0,-1);
	\draw [very thick] (0,0) node[left=0pt]{$\WH{1}$} -- (4,0) node[right=0pt]{$\WH{r}$};
	\end{scope}
	\end{tikzpicture}
	\caption{Three possible factorization channels of~\eqref{eq:rightexpansion} with $h_1\in\mathsf{H}_R$.}
	\label{fig:casetwo}
\end{figure}

If the fiducial graviton $h_1\in\mathsf{H}_R$, the analysis is more complicated. There are three possibilities for the last transmuted particle in the left sub-amplitude: there is no transmuted particle at all, or it can be an original gluon $b_\ell\in\pmb{\ell}$, or an original graviton $h_i$. These three cases are depicted in order in figure~\ref{fig:casetwo}. To find the residue of~\eqref{eq:rightexpansion}, we need to sum up these three contributions.

\paragraph{(a) There is no transmuted particle in the left sub-amplitude}~\\
This is the simplest case among the three, shown in the left panel of figure~\ref{fig:casetwo}. The contribution has the following form:
\begin{equation}
\text{Res}_{a}=\W{A}_{aL}\times\frac{i}{P_L^2}\times \W{A}_{aR}=A_L\times\frac{i}{P_L^2}\times\W{A}_{aR}\,,
\end{equation}
where $\W{A}_{aL}=A(\WH{1},2\ldots j,-\WH{P}_L\,|\mathsf{Tr}_L\,\Vert\,\mathsf{H}_{L})=A_L$, and
\begin{align}
\W{A}_{aR}=\sum_{\substack{\pmb{h}_R| \mathsf{h}_R=\mathsf{H}_R\backslash h_1 \\ \pmb{\mathsf{Tr}}_{Rs}| \mathsf{Tr}_{Rp}=\mathsf{Tr}_R}}&\Bigg[\,\underset{\{a_i,b_i\}\subset\pmb{t}_i}{\widetilde{\sum}}\,\Bigg]_{\pmb{t}_i\in\pmb{\mathsf{Tr}}_{Rs}} ~\sum_{\pmb\rho\in\pmb{h}_R\,\W{\shuffle}\,\pmb{K}(\pmb{\mathsf{Tr}}_{Rs},a,b)}C_{h_1}^{\mu}(\pmb\rho) (\W{Y}_{\rho_1})_{\mu}\nonumber\\
&\quad\times A(\WH{P}_L,\{j+1\ldots r-1\}\shuffle\{\pmb\rho,h_1\},\WH{r}\,|\,\mathsf{Tr}_{Rp}\,\Vert\,\mathsf{h}_R)\,.
\label{eq:ARa}
\end{align}
Here $\W{Y}_{\rho_1}$ is defined as follows. If $\rho_1$ is right after $i\in\{j+1\ldots r-1\}$, then
\begin{equation}
\W{Y}_{\rho_1}=\WH{k}_1+k_2+\ldots +k_j+k_{j+1}+\ldots +k_i\,.
\end{equation}

\paragraph{(b) The last transmuted particle in the left sub-amplitude is $b_\ell\in\pmb{\ell}$}~\\
This case is depicted in the middle panel of figure~\ref{fig:casetwo}. Now we factorize the summations in~\eqref{eq:rightexpansion} as
\begin{align}
\sum_{\pmb{h}|\mathsf{h}=\mathsf{H}\backslash h_1}=\sum_{\pmb{h}_L|\mathsf{h}_L=\mathsf{H}_L}~\sum_{\pmb{h}_R|\mathsf{h}_R=\mathsf{H}_R\backslash h_1}& &\sum_{\pmb{\mathsf{Tr}}_s|\mathsf{Tr}_p=\mathsf{Tr}\backslash 1}=\sum_{\pmb{\mathsf{Tr}}_{Ls}|\mathsf{Tr}_{Lp}=\mathsf{Tr}_L\backslash \ell}~\sum_{\pmb{\mathsf{Tr}}_{Rs}|\mathsf{Tr}_{Rp}=\mathsf{Tr}_R}\,.
\end{align}
In particular, the $\W{\sum}$ part becomes:
\begin{equation}
\Bigg[\,\underset{\{a_i,b_i\}\subset\pmb{t}_i}{\widetilde{\sum}}\,\Bigg]_{\pmb{t}_i\in\pmb{\mathsf{Tr}}_s}=\Bigg[\,\underset{\{a_i,b_i\}\subset\pmb{t}_i}{\widetilde{\sum}}\,\Bigg]_{\pmb{t}_i\in\pmb{\mathsf{Tr}}_{Ls}}\Bigg[\,\underset{\{a_i,b_i\}\subset\pmb{t}_i}{\widetilde{\sum}}\,\Bigg]_{\pmb{t}_i\in\pmb{\mathsf{Tr}}_{Rs}}~\sum_{b_\ell\in\pmb{\ell}}\underset{\{a_\ell,\underline{b_\ell}\}\subset\pmb{\ell}}{\widetilde{\sum}}\,,
\end{equation}
namely, we single out the summation of $a_\ell$ and $b_\ell$ in the trace $\pmb{\ell}$, and write it in a form that we sum over $a_\ell\neq b_\ell$ first for a fixed $b_\ell$, before summing over all the $b_\ell$ in the trace $\pmb{\ell}$. It is crucial to realize that the coefficient $C_{h_1}$ factorizes at $(-k_{b_\ell}k_{a_\ell})$:
\begin{equation}
C_{h_1}(\pmb{\rho})=(\epsilon_{h_1}\ldots k_{b_\ell})(-k_{a_\ell}\ldots Y)=C_{h_1}^{\mu}(\pmb\rho_R)(k_{b_\ell})_{\mu}\WH{C}_{a_\ell}(\pmb{\rho}_{L})\,,
\end{equation}
where $\pmb\rho_{R}\in\pmb{h}_{R}\,\W{\shuffle}\,\pmb{K}(\pmb{\mathsf{Tr}}_{Rs},a,b)$ and $\pmb\rho_{L}\in\pmb{h}_{L}\,\W{\shuffle}\,\pmb{K}(\pmb{\mathsf{Tr}}_{Ls},a,b)$ are the restriction of $\pmb\rho$ to the right and left sub-amplitude respectively.
The contribution to the residue has the following form:
\begin{align}
&\text{Res}_{b}=\sum_{\ell\in\mathsf{Tr}_L}\W{A}_{L\ell}\times\frac{i}{P_L^2}\times \W{A}_{R\ell}\,.
\end{align}
By properly distributing the summations and coefficients, we get
\begin{align}
\W{A}_{L\ell}&=\sum_{\substack{\pmb{h}_L| \mathsf{h}_L=\mathsf{H}_L \\ \pmb{\mathsf{Tr}}_{Ls}| \mathsf{Tr}_{Lp}=\mathsf{Tr}_L}}\Bigg[\,\underset{\{a_i,b_i\}\subset\pmb{t}_i}{\widetilde{\sum}}\,\Bigg]_{\pmb{t}_i\in\pmb{\mathsf{Tr}}_{Ls}}~\underset{\{a_\ell,\underline{b_\ell}\}\subset\pmb{\ell}}{\widetilde{\sum}}~\sum_{\pmb\rho_L\in\pmb{h}_L\,\W{\shuffle}\,\pmb{K}(\pmb{\mathsf{Tr}}_{Ls},a,b)}\WH{C}_{a_\ell}(\pmb\rho_L)\nonumber\\
&\qquad\qquad\times A(\WH{1},\{2\ldots j\}\shuffle\{\pmb\rho_L,a_\ell,\mathsf{KK}[\pmb\ell,a_\ell,b_\ell],b_\ell\},-\WH{P}_L\,|\,\mathsf{Tr}_{Lp}\,\Vert\,\mathsf{h}_L)\nonumber\\
&=A(\WH{1},2\ldots j,-\WH{P}_L\,|\mathsf{Tr}_L\,\Vert\,\mathsf{H}_{L})=A_L\,,
\end{align}
where we have used the type-II recursive expansion~\eqref{eq:type2step1}. We note that $\W{A}_{L\ell}$ is independent of $\ell$, so that $\sum_{\ell}$ can pass through $\W{A}_{L\ell}$ and act only on $\W{A}_{R\ell}$:
\begin{align}
\W{A}_{bR}\equiv\sum_{\ell\in\mathsf{Tr}_{L}}\W{A}_{R\ell}&=\sum_{\substack{\pmb{h}_R| \mathsf{h}_R=\mathsf{H}_R\backslash h_1 \\ \pmb{\mathsf{Tr}}_{Rs}| \mathsf{Tr}_{Rp}=\mathsf{Tr}_R}}\Bigg[\,\underset{\{a_i,b_i\}\subset\pmb{t}_i}{\widetilde{\sum}}\,\Bigg]_{\pmb{t}_i\in\pmb{\mathsf{Tr}}_{Rs}} ~\sum_{\pmb\rho_R\in\pmb{h}_R\,\W{\shuffle}\,\pmb{K}(\pmb{\mathsf{Tr}}_{Rs},a,b)}~\sum_{\ell\in\mathsf{Tr}_{L}}~\sum_{b_{\ell}\in\pmb\ell}C^{\mu}_{h_1}(\pmb\rho_R)(k_{b_\ell})_{\mu}\nonumber\\
&\qquad\qquad\times A(\WH{P}_L,\{j+1\ldots r-1\}\shuffle\{\pmb\rho_R,h_1\},\WH{r}\,|\,\mathsf{Tr}_{Rp}\,\Vert\,\mathsf{h}_R)\nonumber\\
&=\sum_{\substack{\pmb{h}_R| \mathsf{h}_R=\mathsf{H}_R\backslash h_1 \\ \pmb{\mathsf{Tr}}_{Rs}| \mathsf{Tr}_{Rp}=\mathsf{Tr}_R}}\Bigg[\,\underset{\{a_i,b_i\}\subset\pmb{t}_i}{\widetilde{\sum}}\,\Bigg]_{\pmb{t}_i\in\pmb{\mathsf{Tr}}_{Rs}} ~\sum_{\pmb\rho_R\in\pmb{h}_R\,\W{\shuffle}\,\pmb{K}(\pmb{\mathsf{Tr}}_{Rs},a,b)}C^{\mu}_{h_1}(\pmb\rho_R)(K[\mathsf{Tr}_L])_{\mu}\nonumber\\
&\qquad\qquad\times A(\WH{P}_L,\{j+1\ldots r-1\}\shuffle\{\pmb\rho_R,h_1\},\WH{r}\,|\,\mathsf{Tr}_{Rp}\,\Vert\,\mathsf{h}_R)\,,
\label{eq:ARb}
\end{align}
where we have used the fact that $\sum_{\ell\in\mathsf{Tr}_L}\sum_{b_{\ell}\in\pmb\ell}k_{b_\ell}$ gives the total momentum of the set $\mathsf{Tr}_L$, denoted as $K[\mathsf{Tr}_L]$. Therefore, the contribution of this case is:
\begin{equation}
\text{Res}_b=A_L\times\frac{i}{P_L^2}\times\sum_{\ell\in\mathsf{Tr}_L}\W{A}_{R\ell}=A_L\times\frac{i}{P_L^2}\times \W{A}_{bR}\,.
\end{equation}
\paragraph{(c) The last transmuted particle in the left sub-amplitude is $h_i\in\mathsf{H}_L$}~\\
For the last case, as shown in the right panel of figure~\ref{fig:casetwo}, the summations factorize as:
\begin{gather}
\sum_{\pmb{h}|\mathsf{h}=\mathsf{H}\backslash h_1}=\sum_{\pmb{h}_L|\mathsf{h}_L=\mathsf{H}_L\backslash h_i}~\sum_{\pmb{h}_R|\mathsf{h}_R=\mathsf{H}_R\backslash h_1}\qquad\sum_{\pmb{\mathsf{Tr}}_s|\mathsf{Tr}_p=\mathsf{Tr}\backslash 1}=\sum_{\pmb{\mathsf{Tr}}_{Ls}|\mathsf{Tr}_{Lp}=\mathsf{Tr}_L}~\sum_{\pmb{\mathsf{Tr}}_{Rs}|\mathsf{Tr}_{Rp}=\mathsf{Tr}_R}\nonumber\\
\Bigg[\,\underset{\{a_i,b_i\}\subset\pmb{t}_i}{\widetilde{\sum}}\,\Bigg]_{\pmb{t}_i\in\pmb{\mathsf{Tr}}_s}=\Bigg[\,\underset{\{a_i,b_i\}\subset\pmb{t}_i}{\widetilde{\sum}}\,\Bigg]_{\pmb{t}_i\in\pmb{\mathsf{Tr}}_{Ls}}\Bigg[\,\underset{\{a_i,b_i\}\subset\pmb{t}_i}{\widetilde{\sum}}\,\Bigg]_{\pmb{t}_i\in\pmb{\mathsf{Tr}}_{Rs}}\,.
\end{gather}
However, the coefficients $C_{h_1}$ does not factorize, but gets separated into two terms:
\begin{align}
\label{eq:Ccase3}
C_{h_1}(\pmb\rho)=\epsilon_{h_1}\ldots F_{h_i}\ldots Y=C_{h_1}^{\mu}(\pmb\rho_R)(k_{h_i})_{\mu}C_{h_i}(\pmb\rho_L)+(\epsilon_{h_1}\ldots\epsilon_{h_i})\WH{C}_{h_i}(\pmb\rho_L)\,.
\end{align}
Correspondingly, the contribution to the residue should contain two terms:
\begin{equation}
\text{Res}_c=\sum_{h_i\in\mathsf{H}_L}\W{A}_{Lh_i}\times\frac{i}{P_L^2}\times\W{A}_{Rh_i}+\sum_{h_i\in\mathsf{H}_L}I_{h_i}\times\frac{i}{P_L^2}\times I'_{h_i}\,,
\end{equation}
where the second term of $\text{Res}_c$ is contributed by the second term of~\eqref{eq:Ccase3}. More specifically, we have:
\begin{align}
I_{h_i}&=\sum_{\substack{\pmb{h}_L| \mathsf{h}_L=\mathsf{H}_L\backslash h_i \\ \pmb{\mathsf{Tr}}_{Ls}| \mathsf{Tr}_{Lp}=\mathsf{Tr}_L}}\Bigg[\,\underset{\{a_i,b_i\}\subset\pmb{t}_i}{\widetilde{\sum}}\,\Bigg]_{\pmb{t}_i\in\pmb{\mathsf{Tr}}_{Ls}}~\sum_{\pmb\rho_L\in\pmb{h}_L\,\W{\shuffle}\,\pmb{K}(\pmb{\mathsf{Tr}}_{Ls},a,b)}\!\!\WH{C}_{h_i}(\pmb\rho_L)A(\WH{1},\{2\ldots j\}\shuffle\{\pmb\rho_L,h_i\},-\WH{P}_L\,|\,\mathsf{Tr}_{Lp}\,\Vert\,\mathsf{h}_L)
\end{align}
while $I'_{h_i}$ is another combination that contains $(\epsilon_{h_1}\ldots\epsilon_{h_i})$ as coefficients, whose exact form does not matter here. Due to the type-I generalized BCJ relation~\eqref{Eq:GeneralizedBCJType1}, we have actually $I_{h_i}=0$, such that the second term of $\text{Res}_c$ vanishes. Now for the first term of $\text{Res}_c$, we have:
\begin{align}
\W{A}_{Lh_i}&=\sum_{\substack{\pmb{h}_L| \mathsf{h}_L=\mathsf{H}_L\backslash h_i \\ \pmb{\mathsf{Tr}}_{Ls}| \mathsf{Tr}_{Lp}=\mathsf{Tr}_L}}\!\!\Bigg[\,\underset{\{a_i,b_i\}\subset\pmb{t}_i}{\widetilde{\sum}}\,\Bigg]_{\pmb{t}_i\in\pmb{\mathsf{Tr}}_{Ls}}\,\sum_{\pmb\rho_L\in\pmb{h}_L\,\W{\shuffle}\,\pmb{K}(\pmb{\mathsf{Tr}}_{Ls},a,b)}\!\!C_{h_i}(\pmb\rho_L)A(\WH{1},\{2\ldots j\}\shuffle\{\pmb\rho_L,h_i\},-\WH{P}_L\,|\,\mathsf{Tr}_{Lp}\,\Vert\,\mathsf{h}_L)\nonumber\\
&=A(\WH{1},2\ldots j,-\WH{P}_L\,|\mathsf{Tr}_L\,\Vert\,\mathsf{H}_{L})=A_L\,,
\end{align}
because of the type-I recursive expansion. Since it is independent of $h_i$, e can again pass the $\sum_{h_i}$ through and act it onto $\W{A}_{Rh_i}$:
\begin{align}
\W{A}_{Rc}\equiv\sum_{h_i\in\mathsf{H}_L}\W{A}_{Rh_i}&=\sum_{\substack{\pmb{h}_R| \mathsf{h}_R=\mathsf{H}_R\backslash h_1 \\ \pmb{\mathsf{Tr}}_{Rs}| \mathsf{Tr}_{Rp}=\mathsf{Tr}_R}}\Bigg[\,\underset{\{a_i,b_i\}\subset\pmb{t}_i}{\widetilde{\sum}}\,\Bigg]_{\pmb{t}_i\in\pmb{\mathsf{Tr}}_{Rs}} \,\sum_{\pmb\rho_R\in\pmb{h}_R\,\W{\shuffle}\,\pmb{K}(\pmb{\mathsf{Tr}}_{Rs},a,b)}\sum_{h_i\in\mathsf{H}_L}C^{\mu}_{h_1}(\pmb\rho_R)(k_{h_i})_{\mu}\nonumber\\
&\qquad\qquad\times A(\WH{P}_L,\{j+1\ldots r-1\}\shuffle\{\pmb\rho_R,h_1\},\WH{r}\,|\,\mathsf{Tr}_{Rp}\,\Vert\,\mathsf{h}_R)\nonumber\\
&=\sum_{\substack{\pmb{h}_R| \mathsf{h}_R=\mathsf{H}_R\backslash h_1 \\ \pmb{\mathsf{Tr}}_{Rs}| \mathsf{Tr}_{Rp}=\mathsf{Tr}_R}}\Bigg[\,\underset{\{a_i,b_i\}\subset\pmb{t}_i}{\widetilde{\sum}}\,\Bigg]_{\pmb{t}_i\in\pmb{\mathsf{Tr}}_{Rs}} ~\sum_{\pmb\rho_R\in\pmb{h}_R\,\W{\shuffle}\,\pmb{K}(\pmb{\mathsf{Tr}}_{Rs},a,b)}C^{\mu}_{h_1}(\pmb\rho_R)(K[\mathsf{H}_L])_{\mu}\nonumber\\
&\qquad\qquad\times A(\WH{P}_L,\{j+1\ldots r-1\}\shuffle\{\pmb\rho_R,h_1\},\WH{r}\,|\,\mathsf{Tr}_{Rp}\,\Vert\,\mathsf{h}_R)\,,
\label{eq:ARc}
\end{align}
where $K[\mathsf{H}_L]$ is the total momentum of the set $\mathsf{H}_L$. The contribution of this case can thus be written as:
\begin{equation}
\text{Res}_c=A_{L}\times\frac{i}{P_L^2}\times\W{A}_{Rc}\,.
\end{equation}

\subsubsection{Summary of the proof}
Now we summarize our proof. For the fiducial graviton $h_1\in\mathsf{H}_{L}$, we have shown that the residue matches in Eq.~\eqref{eq:resultcase1}. For the case $h_1\in\mathsf{H}_{R}$, our above calculation shows that
\begin{equation}
\text{Res}\left[\frac{{\tt Expansion}}{z}\right]=\text{Res}_a+\text{Res}_b+\text{Res}_c=A_L\times\frac{i}{P_L^2}\times(\W{A}_{Ra}+\W{A}_{Rb}+\W{A}_{Rc})\,.
\end{equation}
Using the expressions of $\W{A}_{Ra}$, $\W{A}_{Rb}$ and $\W{A}_{Rc}$ presented respectively in~\eqref{eq:ARa}, \eqref{eq:ARb} and~\eqref{eq:ARc}, we find that
\begin{align}
\W{A}_{Ra}+\W{A}_{Rb}+\W{A}_{Rc}&=\sum_{\substack{\pmb{h}_R| \mathsf{h}_R=\mathsf{H}_R\backslash h_1 \\ \pmb{\mathsf{Tr}}_{Rs}| \mathsf{Tr}_{Rp}=\mathsf{Tr}_R}}\!\!\Bigg[\,\underset{\{a_i,b_i\}\subset\pmb{t}_i}{\widetilde{\sum}}\,\Bigg]_{\pmb{t}_i\in\pmb{\mathsf{Tr}}_{Rs}} \,\sum_{\pmb\rho\in\pmb{h}_R\,\W{\shuffle}\,\pmb{K}(\pmb{\mathsf{Tr}}_{Rs},a,b)}\!\!C^{\mu}_{h_1}(\pmb\rho)\left(\W{Y}_{\rho_1}+K[\mathsf{Tr}_L]+K[\mathsf{H}_L]\right)_{\mu}\nonumber\\
&\qquad\qquad\times A(\WH{P}_L,\{j+1\ldots r-1\}\shuffle\{\pmb\rho,h_1\},\WH{r}\,|\,\mathsf{Tr}_{Rp}\,\Vert\,\mathsf{h}_R)\,.
\end{align}
If $\rho_1$ is right after $i\in\{j+1\ldots r-1\}$, we have
\begin{align}
(Y_{R})_{\rho_1}&\equiv\W{Y}_{\rho_1}+K[\mathsf{Tr}_L]+K[\mathsf{H}_L]=\WH{k}_1+k_2\ldots k_j+K[\mathsf{Tr}_L]+K[\mathsf{H}_L]+k_{j+1}+\ldots+k_i\nonumber\\
&=\WH{P}_L+k_{j+1}+\ldots +k_i\,.
\end{align}
Therefore, for $\pmb\rho\in\pmb{h}_R\,\W{\shuffle}\,\pmb{K}(\pmb{\mathsf{Tr}}_{Rs},a,b)$, we have $C^{\mu}_{h_1}(\pmb\rho)[(Y_R)_{\rho_1}]_{\mu}=C_{h_1}(\pmb\rho)$ and consequently
\begin{equation}
\W{A}_{Ra}+\W{A}_{Rb}+\W{A}_{Rc}=A(\WH{P}_L,j+1\ldots r\,|\,\mathsf{Tr}_R\,\Vert\,\mathsf{H}_R)=A_R\,.
\end{equation}
We find again that for $h_1\in\mathsf{H}_R$, the residues match:
\begin{equation}
\text{Res}\left[\frac{{\tt Expansion}}{z}\right]=\text{Res}\left[\frac{1}{z}\,A(\WH 1,2\ldots \WH r\,|\,\pmb{2}\,|\ldots|\,\pmb{m}\,\Vert\,\mathsf{H})\right]\,.
\end{equation}
We thus conclude that the residues match at all the physical poles under the shift~\eqref{Eq:MomentaDeformation}.

\subsection{The cancellation of spurious poles}

In the previous subsection, we have shown that all the physical poles at the right hand side of Eq.~\eqref{eq:rightexpansion} match with those at the left hand side. However, there is another class of factorization channel in~\eqref{eq:rightexpansion} when
a gluon trace in  $\{\pmb{h}\,\W{\shuffle}\,\pmb{K}(\pmb{\mathsf{Tr}}_s,a,b),h_1\}$ is separated into two parts. Since there is no such factorization channel on the left hand side of~\eqref{eq:type1step1} under the shift~\eqref{Eq:MomentaDeformation}, these poles should have zero residue, namely, they are spurious. Now we show that this is indeed true.


More specifically, we consider the case that the cut is in the middle a gluon trace $K^{\pmb{t_i}}_{a_ib_i}\in\pmb{K}$. More specifically, the cut is after $c_l\in\pmb{t}_i$ but before another gluon of $\pmb{t}_i$. The set $\pmb{K}$ and $\pmb{h}$ are then separated into:
\begin{equation}
\pmb{K}=\pmb{K}_{L}\cup \{(K_L)^{\pmb{t}_i}_{c_kc_l}\}\cup \{(K_R)^{\pmb{t}_i}_{c_{l+1}c_q}\}\cup\pmb{K}_{R}\,,\qquad\pmb{h}=\pmb{h}_{L}\cup\pmb{h}_{R}\,,
\end{equation}
where $({K}_{L})^{\pmb{t}_{i}}_{c_kc_l}=\{c_k,\{c_{k-1}\ldots c_1\}\shuffle\{c_{k+1}\ldots c_{l-1}\},c_l\}$ and $K_R$ is similarly defined. The coefficients factorize as:
\begin{equation}
C_{h_1}(\pmb{h}\,\W{\shuffle}\,\pmb{K})=(\epsilon_{h_1}\ldots k_{c_l})\WH{C}_{c_k}(\pmb{h}_{L}\,\W{\shuffle}\,\pmb{K}_{L})\,,
\end{equation}
while the residues have the form:
\begin{align}
&\propto\WH{C}_{c_k}(\pmb{h}_{L}\,\W{\shuffle}\,\pmb{K}_{L})\nonumber\\
&\quad \times A(1,\{2\ldots j\}\shuffle\{\pmb{h}_{L}\,\W{\shuffle}\,\pmb{K}_{L},c_k,\{c_{k-1}\ldots c_1\}\shuffle\{c_{k+1}\ldots c_{l-1}\},c_l\},-\WH{P}_{L}\,|\,\mathsf{Tr}_{Lp}\,\Vert\,{\mathsf{h}}_{L})\times\frac{i}{P_L^2}\times \W{A}_R\,.
\end{align}
It is important to notice that now $c_l$ is fixed and we need to sum over $c_k$. Once it is done, these contributions sum to zero due to the type-II generalized BCJ relation~\eqref{Eq:GeneralizedBCJType2}. Therefore, the spurious poles indeed cancel.

\section{Expansion to color ordered YM amplitudes}\label{sec:YMexpansion}

Having thoroughly studied the recursive expansion of general multitrace EYM amplitudes, we can iteratively expand every
EYM amplitude to color ordered YM amplitudes.  The procedure will be similar to the one given in~\cite{Fu:2017uzt}, thus we will
not give the detail of derivation and just present the final results. We will present it to two forms. The first one
is the expansion to ordered splitting. The second one is the expansion to KK basis. For the second one there
are two different expressions: one is to reconstruct the ordered splitting and the another one is to use a set of graphic rules based on increasing trees~\cite{Teng:2017tbo}.



\subsection{The expansion to ordered splitting}

First let us review the expansion of single trace EYM amplitudes to color ordered YM amplitudes in the ordered
splitting form. The expansion can be written as
\bea
A_{1,|\mathsf{H}|}(1,2\ldots r\,\Vert\,\mathsf{H})=\sum_{\pmb{\a}\in {\cal S}[\mathsf{H}]}
\Big[C_{\pmb{\a}}A_{1,0}
(1,\{2\ldots r-1\}\shuffle\pmb{\a}_1\shuffle\pmb{\a}_2\shuffle...\shuffle\pmb{\a}_k,r)\Big]\,.\Label{1trace-OS-1}
\eea
Now we explain the meaning of \eqref{1trace-OS-1}:
\begin{enumerate}[label=(\alph*)]
	
	\item First the ${\cal S}[\mathsf{H}]$ is the set of all \emph{ordered splittings} $\{\pmb{\a}_1,\pmb{\a}_2\ldots \pmb{\a}_k\}$ of $|\mathsf{H}|$ gravitons.
	To define the ordering splitting, we need to first prescribe an reference ordering, for example, $h_1\prec h_2\prec \cdots \prec h_m$. The choice of reference ordering can be arbitrary and we call this as the ``reference gauge'' denoted as $\pmb{\mathcal{R}}$. Given a reference gauge, we can divide $|\mathsf{H}|$ gravitons into $k$ non-empty {ordered} subsets $\{\pmb{\a}_1,\pmb{\a}_2\ldots \pmb{\a}_k\}$ with $1\leqslant k\leqslant |\mathsf{H}|$. These ordered subsets $\pmb{\a}_i$ should satisfy the following conditions: (1) Inside each subset, the last element must be the smallest one according to the reference gauge $\pmb{\mathcal{R}}$ (while other elements inside
each subset can be arbitrary ordering); (2) The last element of the subset $\pmb{\a}_i$ must be smaller than the last element of the subset $\pmb{\a}_j$ when $i<j$.
	
	To make thing clear, let us write down all ordered splittings of three gravitons with the reference gauge $h_1\prec h_2\prec h_3$:
	\bea k=1:&~~~& \pmb{\a}_1=\{h_2, h_3, h_1\}\,,\quad \{h_3, h_2, h_1\} \nn
	k=2: & ~~~ & \{\pmb{\a}_1, \pmb{\a}_2\}= \{ \{h_1\}, \{h_3, h_2\}\}\,,\quad\{ \{h_2, h_1\}, \{h_3\}\}\,,\quad\{ \{h_3, h_1\}, \{ h_2\}\} \nn
	k=3:& ~~~& \{\pmb{\a}_1, \pmb{\a}_2, \pmb{\a}_3\}= \{ \{h_1\}, \{h_2\}, \{h_3\}\}\,.~~~~\Label{3g-order-split} \eea
	For this example, the $\mathcal{S}[\{h_1,h_2,h_3\}]$ thus has six ordered splittings as explicitly given above.
	\item Having found the set ${\cal S}[\mathsf{H}]$, we need to sum over all ordered splittings. For a given
	ordered splitting $\{\pmb{\a}_1,\pmb{\a}_2\ldots \pmb{\a}_k\}$, the corresponding coefficient $C_{\pmb{\a}}$ is given by
	\bea C_{\pmb{\a}} &= &\prod_{t=1}^k c(\pmb{\a}_t)~~~~\Label{1trace-OS-2-1} \eea
	where for each subset $\pmb{\a}_t=\{h_{t,1}, h_{t,2},..., h_{t,p}\}$, the coefficient is
	\bea c(\a_t)=\left\{ \begin{array}{ll}\eps_{h_{t,1}}\cdot F_{h_{t,2}}\ldots F_{h_{t,p}}\cdot Z_{h_{t,p}},~~~~& p\geqslant 2\\
		\eps_{h_{t,1}}\cdot Z_{h_{t,1}}~~~~& p=1\end{array}\right.\,,\Label{1trace-OS-2-2}\eea
	where for each $\pmb{\rho}\in\{1,\{2\ldots r-1\}\shuffle\pmb{\a}_1\shuffle\pmb{\a}_2\shuffle...\shuffle\pmb{\a}_k,r\}$, $Z_{h}$ is the sum of momenta of those particles satisfying simultaneously the following two conditions: (1) They are at the left hand side of $h$ in $\pmb{\rho}$; (2) They are at the left hand side of $h$ in the following ordered list
	\bea \{1,2\ldots r-1,\pmb{\a}_1,\pmb{\a}_2\ldots\pmb{\a}_k,r\}\,.\Label{1trace-OS-2-3}\eea
	For example, for the splitting $\{ \{h_1\}, \{h_3, h_2\}\}$ given in  \eqref{3g-order-split}, $Z_{h_3}$ includes the momenta at the left $h_3$ in both $\pmb\rho$ and the ordered list $\{1,2\ldots r-1, h_1, h_3, h_2, r\}$.
	
\end{enumerate}
Given the expansion of single trace EYM amplitudes, we can immediately write down the
expansion of multitrace EYM amplitudes as
\bea A_{m,|\mathsf{H}}(1,2\ldots r\,|\,\pmb{2}\,|\ldots|\,\pmb{m}\,\Vert\,\mathsf{H})=\sum_{\pmb{\a}\in {\cal S}[\mathsf{H}]}
\Big[C_{\pmb{\a}}A_{1,0}
(1,\{2\ldots r-1\}\shuffle\pmb{\a}_1\shuffle\pmb{\a}_2\shuffle...\shuffle\pmb{\a}_k,r)\Big]\Label{Mtrace-OS-1}
\eea
by the trick ``turning a graviton into a trace of gluons'':
\begin{enumerate}[label=(\alph*)]
	
	\item Writing down the expansion \eqref{1trace-OS-1} of single trace EYM amplitudes with $(|\mathsf{H}|+m-1)$ gravitons;
	
	\item Replacing related gravitons in  \eqref{1trace-OS-1} to the corresponding traces of gluons according to
	the rule  \eqref{eq:gluonrule} if the graviton is not the last element of the ordered subset $\pmb{\a}_i$, or to the rule \eqref{eq:lasthrule} if the graviton is the last element of the ordered subset $\pmb{\a}_i$ (because it is the fiducial graviton in the expansion). The only difference is that now we need to use $Z_{h_i}$ or $Z_{a_i}$ instead of the $Y$ symbols.
	
\end{enumerate}
One thing we want to emphasize is that when applying the rule \eqref{eq:lasthrule}, the fiducial gluon in that trace of gluons must be same for all terms.\footnote{Following the iterative expansion, one may find that the fiducial gluon need not be the same for all terms. However, which group of terms have same fiducial gluon is not so easy to visualize. Thus to avoid the mistakes, the simplest way is to make them be the same for all terms. }
%

\subsection{Expansion to KK basis of color ordered YM amplitudes via ordered splitting}

Having understood the expansion to ordered splitting, we can write down the expansion to the KK basis of color ordered YM amplitudes for multi-trace EYM amplitude $A_{m,|\mathsf{H}|}(1,2\ldots r\,|\,\pmb{2}\,|\ldots|\,\pmb{m}\,\Vert\,\mathsf{H})$. Let us
denote the set ${\cal G}= \pmb{2}\cup\ldots\cup\pmb{m}\cup\mathsf{H}$, the KK basis expansion can be
written as
\bea A_{m,|\mathsf{H}|}(1,2\ldots r\,|\,\pmb{2}\,|\ldots|\,\pmb{m}\,\Vert\,\mathsf{H})=\sum_{\pmb\rho\in\rho({\cal G})}
\Big[C(\pmb\rho)A_{1,0}
(1,\{2\ldots r-1\}\shuffle\pmb\rho,r)\Big]\,.~~~\Label{Mtrace-KK-1}
\eea
where the $\sum_\rho$ is over all the permutations of the set ${\cal G}$, denoted as $\rho(\mathcal{G})$. We note that the length of $\pmb\rho$ is $$|\pmb\rho|=|\pmb{2}|+\ldots+|\pmb{m}|+|\mathsf{H}|=n-r\,.$$
The coefficient $C(\pmb{\rho})$ is given by:
\bea C(\pmb\rho)=\sum_{\pmb{\a}\in {\cal S}[\mathsf{H}]\supset \pmb\rho} C_{\pmb{\a}}~~~\Label{Mtrace-KK-2} \eea
where $C_{\pmb{\a}}$ is given in \eqref{1trace-OS-2-1}, and the summation $\sum_{\pmb{\a}\in {\cal S}[\mathsf{H}]\supset \pmb\rho}$ is over all the ordered splittings $\pmb\alpha$ that are consistent with $\pmb\rho$.
Namely, given $\pmb\a=\{\pmb{\a}_1,\pmb{\a}_2\ldots\pmb{\a}_k\}$, we must have
$\pmb\rho\in\pmb{\a}_1\shuffle\pmb{\a}_2\shuffle...\shuffle\pmb{\a}_k$. We now give an explicit algorithm that constructs all the ordered splittings $\pmb\a$ that are consistent with a given $\pmb\rho\in\rho(\mathcal{G})$:
%
\begin{enumerate}[label=(\arabic*)]
	
	\item At the first step, we need to determine a reference gauge $\pmb{\cal R}$ among the $(|\mathsf{H}|+m-1)$ objects
	$\{\pmb{2}\ldots\pmb{m},\mathsf{H}\}$. Furthermore, we need to determine the fiducial gluons for the $(m-1)$ traces $\{\pmb{2}\ldots\pmb{m}\}$, which will be denoted by $d_i$ for trace $\pmb{i}$.
	
	\item At the second step, we reconstruct the ordered list $\pmb{\W\a}$ of the $(|\mathsf{H}|+m-1)$ objects $\{\pmb{2}\ldots\pmb{m},\mathsf{H}\}$ from the ordered list $\pmb\rho\in\rho({\cal G})$. The method is the following:
	\begin{enumerate}
		
		\item Assuming $\pmb\rho=\{\rho_1, \rho_2\ldots\rho_{n-r}\}$, we start with the first element $\rho_1$. If it is graviton, we put $\rho_1$ as the first element in $\pmb{\W\a}$. If $\rho_1$ is a gluon of, for example, trace $\pmb{2}$,  we collect all the gluons of the trace $\pmb{2}$ according to ordering of $\pmb\rho$. In particular, we denote the first element we meet as $a_2=\rho_1$, and the last element as $b_2$. The obtained ordered list can be written as $\pmb{2}^{\pmb\rho}_{a_2b_2}\equiv\{a_2,\pmb{\kappa},b_2\}$.
		
		\item Now the key step is that we need to check if the ordering $\pmb{\kappa}$ of $\pmb{2}^{\pmb\rho}_{a_2b_2}$ belongs to $\mathsf{KK}[\pmb{2},a_2,b_2]$, namely, if $\pmb{2}^{\pmb{\rho}}_{a_2b_2}$ can be realized in a certain KK expansion in which $a_2$ and $b_2$ are anchored at the ends.
		If $\pmb{\kappa}\in\mathsf{KK}[\pmb{2},a_2,b_2]$, we then put $\pmb{2}^{\pmb\rho}_{a_2b_2}$ as the first element of $\pmb{\W{\a}}$.
		If $\pmb{\kappa}\notin\mathsf{KK}[\pmb{2},a_2,b_2]$, then the ordering $\pmb\rho$ is inconsistent with all the ordered splittings $\pmb{\a}$. Consequently, this coefficient $C(\pmb\rho)=0$. For example, if we have $\pmb{2}=\{r_1, r_2, r_3, r_4, r_5, r_6\}$ and $\pmb{2}^{\pmb\rho}_{r_2r_5}=\{r_2, r_4, r_6, r_1, r_3,r_5\}$, it is obvious that $\pmb{2}^{\pmb{\rho}}_{r_2r_5}$ does not belong to the KK expansion $\{r_2, \{r_3, r_4\}\shuffle\{r_1, r_6\},r_5\}$. Consequently, all those $\pmb{\rho}$ that contain the above $\pmb{2}^{\pmb{\rho}}_{r_2r_5}$ do not contribute.
		
		\item In summary, the first element of $\pmb{\W{\a}}$ can be:
		\begin{equation}
		\W{\a}_1=\left\{\begin{array}{ll}
		h_i & \text{if }\rho_1=h_i \\
		\pmb{i}^{\pmb\rho}_{a_ib_i} &\text{if }\rho_1=a_i\in\pmb{i}
		\end{array}\right.\,.
		\end{equation}
		Now we can remove everything contained in $\W{\a}_1$ from $\pmb\rho$, and study the first element of this reduced $\pmb{\rho}$ list. Iterating this procedure, we can construct the $(m+|\mathsf{H}|-1)$-objects ordered list $\pmb{\W{\a}}$.
		
	\end{enumerate}

	\item Having constructed the list $\pmb{\W\a}$, we carry on to construct all the consistent ordered splitting $\pmb{\a}$.
	The procedure is the following:
	\begin{enumerate}
		
		\item We start from the first element ${\cal R}_1$ of the reference gauge $\pmb{\cal R}$, and find its position in the list $\pmb{\W\a}$. Then any subset of $\{\W{\a}_1,\W{\a}_2\ldots\W{\a}_{\text{last}}=\mathcal{R}_1\}$ with $\mathcal{R}_1$ as the last element can be a candidate of $\pmb{\a}_1$, the first component of $\pmb\a$.
		
		\item Among all the candidates of $\pmb{\a}_1$ obtained in the last step, we need to delete all the inconsistent ones. This check includes two steps. The first step is to check if ${\cal R}_1$ represents a trace of gluons: the last element must be the prescribed fiducial gluon $d_i$. Otherwise, there would be no consistent ordered splitting $\pmb{\a}$ constructed from $\pmb{\W\a}$, and consequently $C(\pmb\rho)=0$. The second step is to check if the gluon traces represented by the elements in $\{\W \a_1, \W\a_2\ldots\W\a_{\text{last}}={\cal R}_1 \}$ are consistent with each other. For the two traces represented by $\W \a_i$ and $\W\a_j$ with $i<j$, the consistent condition is that all elements of the trace represented by $\W \a_i$ must be
		at the left hand side of all elements of the trace represented by $\W \a_j$ in the original list $\pmb\rho$.\footnote{Namely, only gluon traces that belong to different subsets of the ordered splitting can mix.}
		
		\item For each consistent subset $\pmb{\a}_1$, we remove it from  $\pmb{\W\a}$ and $\pmb{\mathcal{R}}$, resulting in
		a reduced list $\pmb{\W\a}'=\pmb{\W{\a}}\backslash\pmb{\a}_{1}$ and a reduced reference gauge $\pmb{\mathcal{R}}'=\pmb{\mathcal{R}}\backslash\pmb{\a}_1$. We can repeat the above process to obtain the second subset $\pmb{\a}_2$ of $\pmb{\a}$.
		
		\item By iterating the above procedures, we can eventually find all the consistent ordered splittings $\pmb{\a}$ from the list $\pmb{\W\a}$. We can then use Eq.~\eqref{Mtrace-KK-2} to evaluate $C(\pmb\rho)$.
		
	\end{enumerate}

\end{enumerate}
\subsection{Increasing trees and graphic rules for expansion coefficients}\label{sec:graphicrules}
Finally, we give a set of graph theoretical algorithms that evaluate the coefficients in Eq.~\eqref{Mtrace-KK-1} directly. These rules generalize the ones for single trace EYM~\cite{Teng:2017tbo}, and reveal some intricate structures in the DDM form of BCJ numerators. For simplicity, we fixed the reference gauge as:
\begin{equation}
\pmb{\mathcal{R}}=h_1\prec h_2\prec\ldots\prec h_{|\mathsf{H}|}\prec\pmb{2}\prec\ldots\prec\pmb{m}\,.
\end{equation}
This choice means that we first carry out the type-I recursive expansion until we arrive at an expansion in terms of pure gluon multitrace amplitudes. Then we carry out the type-II recursive expansion until we arrive at the pure YM expansion, which is our final result. Moreover, we always choose the last element of each gluon trace as our fiducial gluon during the type-II expansion:
\begin{equation}
\text{for gluon trace $\pmb{i}$:}\qquad d_i=\pmb{i}_{\text{last}}\,.
\end{equation}
This effectively exhausts the redundancy associated to the cyclicity of the gluon traces. Of course, there are many other gauge choices, but we find that our choice leads to the simplest rules. Before presenting the rules, we first study what a typical $\pmb\rho\in\rho(\mathcal{G})$ looks like. From our recursive expansions, one can easily tell that $\pmb{\rho}$ must be produced by shuffling gravitons and KK basis gluon traces:
\begin{equation}
\pmb{\rho}\in\pmb{2}^{\pmb\rho}_{a_2b_2}\shuffle\pmb{3}^{\pmb\rho}_{a_3b_3}\shuffle\ldots\shuffle\pmb{m}^{\pmb\rho}_{a_mb_m}\shuffle\{h_1\}\shuffle\{h_2\}\shuffle\ldots\shuffle\{h_{|\mathsf{H}|}\}\,,
\end{equation}
where, for example, $\pmb{2}^{\pmb\rho}_{a_2b_2}$ is a segment of $\pmb{\rho}$ whose elements are identical to that of the trace $\pmb{2}$. In the following, we represent it as:
\begin{equation}
\pmb{2}^{\pmb{\rho}}_{a_2b_2}\equiv
\adjustbox{raise=-0.5cm}{\begin{tikzpicture}
	\draw [line width=3pt,red] (0,0) -- (1,0);
	\filldraw (0,0) circle (2pt);
	\node at (0,-0.27) {$a_2$};
	\filldraw (1,0) circle (2pt);
	\node at (1,-0.27) {$b_2$};
	\end{tikzpicture}}
\in K^{\pmb{2}}_{a_2b_2}=\{a_2,\mathsf{KK}[\pmb{2},a_2,b_2],b_2\}\,,
\end{equation}
such that a typical $\pmb{\rho}$ can be graphically represented as:
\begin{equation}
\label{eq:typicalrho}
\pmb\rho=\adjustbox{raise=-0.5cm}{\begin{tikzpicture}
	\draw [line width=3pt,red] (0,0) -- (1,0) (1.5,0) -- (2,0) (4.25,0) -- (4.75,0);
	\filldraw (0,0) circle (2pt) (4.75,0) circle (2pt);
	\node at (0,-0.33) {$a_2$};
	\node at (4.75,-0.33) {$b_2$};
	\node at (1.25,0) {$\pmb\times$};
	\node at (1.25,-0.33) {$h_1$};
	\draw [line width=3pt,blue] (2.25,0) -- (3,0) (3.5,0) -- (4,0) (5,0) -- (5.5,0);
	\node at (2.25,-0.33) {$a_3$};
	\node at (5.5,-0.33) {$b_3$};
	\filldraw (2.25,0) circle (2pt) (5.5,0) circle (2pt);
	\node at (3.25,0) {$\pmb\times$};
	\node at (3.25,-0.33) {$h_2$};
	\end{tikzpicture}}\;.
\end{equation}
Any $\pmb\rho$ that is NOT of this form has vanishing $C(\pmb\rho)$. Now we give the graphic algorithm that computes the coefficient $C(\pmb\rho)$:
\paragraph{Step 1: Construct the ordering $\pmb\beta$. }Intuitively, a gluon trace resembles a single graviton according to the compactification understanding given in section~\ref{sec:compact}. Thus our first step is to extract another ordering $\pmb\beta$ from $\pmb\rho$. In $\pmb\beta$, there are exactly $|\mathsf{H}|+m-1$ objects, while the gluon traces are tentatively treated on the same footing as the gravitons, such that we can reuse here many features of the single trace construction~\cite{Teng:2017tbo}. The rule to obtain $\pmb\beta$ is the following:
\begin{align}
& \text{in $\pmb\rho$: }a_i\prec h_j\;\rightarrow\;\text{in $\pmb\beta$: }\pmb{i}^{\pmb{\rho}}_{a_ib_i}\prec h_{j}& &\text{in $\pmb\rho$: } a_i\prec a_j\;\rightarrow\;\text{in $\pmb\beta$: } \pmb{i}^{\pmb\rho}_{a_ib_i}\prec\pmb{j}^{\pmb\rho}_{a_jb_j}\,.
\end{align}
Namely, $\pmb\beta$ is just the ordering of the leading elements $a_i$ of each trace and the gravitons $h_i$ in $\pmb{\rho}$. For example, the $\pmb\rho$ of Eq.~\eqref{eq:typicalrho} gives:
\begin{equation}
\label{eq:betaorder}
\text{$\pmb\rho$ of Eq.~\eqref{eq:typicalrho}}\;\rightarrow\;\pmb\beta=\{\pmb{2}^{\pmb\rho}_{a_2b_2},h_1,\pmb{3}^{\pmb\rho}_{a_3b_3},h_2\}\,.
\end{equation}

\paragraph{Step 2: Construct the relevant increasing trees of $\pmb\beta$. } Next, we draw all the increasing trees with respect to the order:\footnote{In an increasing tree, if we draw a path from the root to a leaf, the vertices along the path must have $A_1\prec A_2\prec A_3\ldots$.}
\begin{equation}
\mathsf{g}\prec\beta_1\prec\beta_2\prec\ldots\prec\beta_{|\mathsf{H}|+m-1}\,,
\end{equation}
where $\mathsf{g}$ is the collection of gluons $\{1,2\ldots r-1\}$ in the first trace. In fact, what we have is a forest rooted in $\mathsf{g}$, but we can pack the structure at the roots into the $Y$ symbols such that our graphic rules become more compact and simple. Naively, there are $(|\mathsf{H}|+m-1)!$ increasing trees. If we were dealing with the single trace case, all the $\beta_i$'s would be gravitons, and all the trees would contribute. However, this is not true for our multitrace case, and our next job is to extract all the contributing trees.

Given a tree, similar to the single trace expansion given in~\cite{Teng:2017tbo}, we pick out the first element in the reference gauge $\pmb{\mathcal{R}}$, called $\mathcal{R}_1$, and draw a path $\mathcal{P}[1]=\{\phi_1=\mathcal{R}_1,\phi_2\ldots\phi_{\ell-1},\mathsf{g}\}$ to the root. We then delete all the elements in $\mathcal{P}[1]$ from $\pmb{\mathcal{R}}$, resulting in a reduce reference gauge $\pmb{\W{\mathcal{R}}}$, from which we can construct the second path $\mathcal{P}[2]=\{\W{\phi}_1=\W{\mathcal{R}}_1,\W{\phi}_2\ldots\W{\phi}_t,V_2\}$ from $\W{\mathcal{R}}_1$ towards the root. This path can either end on the root or a point on $\mathcal{P}[1]$. Therefore, the end point $V_2$ can either be the root $\mathsf{g}$ or a vertex in $\mathcal{P}[1]$:
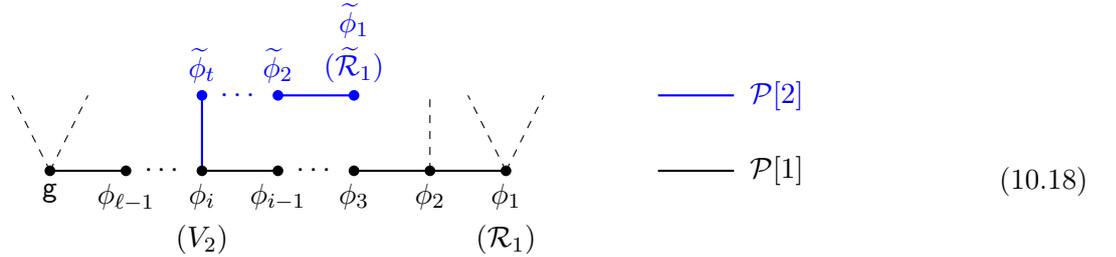
\begin{equation}
\label{eq:path2}
\adjustbox{raise=-1cm}{\begin{tikzpicture}
	\coordinate (g) at (0,0);
	\fill (g) circle (2pt) node[below=1pt]{$\mathsf{g}$};
	\coordinate (a) at (1,0);
	\fill (a) circle (2pt) node[below=1pt]{$\phi_{\ell-1}$};
	\draw [thick] (g) -- (a);
	\node at (1.5,0) {$\cdots$};
	\coordinate (b) at (2,0);
	\draw [thick,blue] (b) -- ++(0,1);
	\fill (b) circle (2pt) node[below=1pt,align=center]{$\phi_i$ \\ $(V_2)$};
	\coordinate (c) at (3,0);
	\fill (c) circle (2pt) node[below=1pt]{$\phi_{i-1}$};
	\draw [thick] (b) -- (c);
	\node at (3.5,0) {$\cdots$};
	\coordinate (d) at (4,0);
	\fill (d) circle (2pt) node[below=1pt]{$\phi_3$};
	\coordinate (e) at (5,0);
	\fill (e) circle (2pt) node[below=1pt]{$\phi_2$};
	\coordinate (f) at (6,0);
	\fill (f) circle (2pt) node[below=1pt,align=center]{$\phi_1$ \\ $(\mathcal{R}_1)$};
	\coordinate (h) at (2,1);
	\fill [blue] (h) circle (2pt) node[above=1pt]{$\W{\phi}_t$};
	\node at (2.5,1) [text=blue]{$\cdots$};
	\coordinate (i) at (3,1);
	\fill [blue] (i) circle (2pt) node[above=1pt]{$\W{\phi}_2$};
	\coordinate (j) at (4,1);
	\fill [blue] (j) circle (2pt) node[above=1pt,align=center]{$\W{\phi}_1$ \\ $(\W{\mathcal{R}}_1)$};
	\draw [thick,blue] (i) -- (j);
	\draw [thick] (d) -- (f);
	\draw [dashed] (e) -- ++(0,1);
	\draw [dashed] (g) -- ++(0.5,1) (g) -- ++(-0.5,1);
	\draw [dashed] (f) -- ++(0.5,1) (f) -- ++(-0.5,1);
	\draw [thick,black] (8,0) -- ++(1,0) node[right=2pt]{${\cal P}[1]$};
	\draw [thick,blue] (8,1) -- ++(1,0) node[right=2pt]{${\cal P}[2]$};
	\end{tikzpicture}}
\end{equation}
We can iterate this process until all elements in the tree are traversed. Now the critical observation is that only those trees that satisfy the following criteria are nonzero:
\begin{itemize}
	\item Given a path $\mathcal{P}[i]$, the starting point, say $\W{\phi}_1$ for $\mathcal{P}[2]$, is a gluon trace $\pmb{i}^{\pmb\rho}_{a_ib_i}$, then we must have $b_i=d_i$. Namely, the last element must be the prescribed fiducial gluon.
	\item Given a path $\mathcal{P}[i]$, all the elements, except for the end point (for example, $V_2$ of $\mathcal{P}[2]$), do not mix in the original $\pmb\rho$.
\end{itemize}
These criteria are very easy to understand. First, each path $\mathcal{P}[i]$ corresponds to the coefficients generated by a single use of our type-I or type-II recursive expansion. Then from the explicit expressions~\eqref{eq:type1step1} and \eqref{eq:type2step1}, one can observe that the elements contained in the coefficients do not mix. Moreover, if the first element of a path is a gluon trace, it means that this path is generated by using the type-II recursive expansion, such that the last element of that trace must be the fiducial gluon $d_i$. An immediate consequence is that if $b_i\neq d_i$, then the trace $\pmb{i}^{\pmb\rho}_{a_ib_i}$ cannot appear as a leaf. By the way, if a path starts with a gluon trace, then due to our specific choice of $\pmb{\mathcal{R}}$, all its vertices, except for the end point, must be gluon traces.

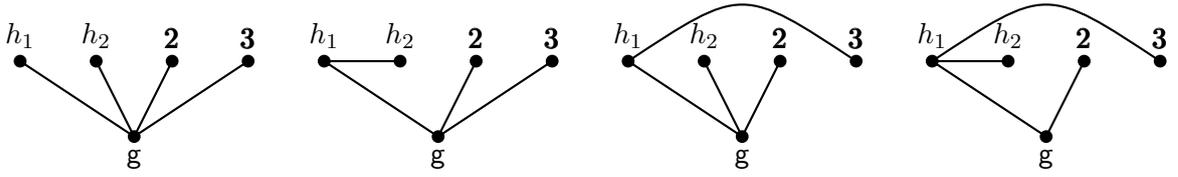
\begin{figure}[t]
	\centering
	\begin{tikzpicture}[starnode/.style={text=blue,font=\Huge}]
	\begin{scope}
	\coordinate (a) at (0,0);
	\filldraw [thick] (a) -- ++(-1.5,1) circle (2pt) node[above=1pt]{$h_1$} (a) -- ++(-0.5,1) circle(2pt) node[above=1pt]{$h_2$} (a) -- ++(0.5,1) circle (2pt) node[above=1pt]{$\pmb{2}$};
	\filldraw [thick] (a) circle (2pt) node[below=1pt]{$\mathsf{g}$} -- ++(1.5,1) circle (2pt) node[above=1pt]{$\pmb{3}$};
	\end{scope}
	\begin{scope}[xshift=4cm,yshift=0cm]
	\coordinate (a) at (0,0);
	\filldraw [thick] (a) -- ++(-1.5,1) circle (2pt) node[above=1pt]{$h_1$} -- ++(1,0) circle (2pt) node[above=1pt]{$h_2$} (a) -- ++(0.5,1) circle (2pt) node[above=1pt]{$\pmb{2}$};
	\filldraw [thick] (a) circle (2pt) node[below=1pt]{$\mathsf{g}$} -- ++(1.5,1) circle (2pt) node[above=1pt]{$\pmb{3}$};
	\end{scope}
	\begin{scope}[xshift=8cm]
	\coordinate (a) at (0,0);
	\filldraw [thick] (a) -- ++(-0.5,1) circle (2pt) node[above=1pt]{$h_2$} (a) -- ++(0.5,1) circle (2pt) node[above=1pt]{$\pmb{2}$};
	\filldraw [thick] (a) circle (2pt) node[below=1pt]{$\mathsf{g}$} -- ++(-1.5,1) circle (2pt) node[above=1pt]{$h_1$} (a) ++(1.5,1) circle (2pt) node[above=1pt]{$\pmb{3}$};
	\draw [thick] (-1.5,1) .. controls (0,2) and (0,2) .. (1.5,1);
	\end{scope}
	\begin{scope}[xshift=12cm,yshift=0cm]
	\coordinate (a) at (0,0);
	\filldraw [thick] (a) -- ++(0.5,1) circle (2pt) node[above=1pt]{$\pmb{2}$} ++ (-1,0) circle (2pt) node[above=1pt]{$h_2$} -- ++(-1,0);
	\filldraw [thick] (a) circle (2pt) node[below=1pt]{$\mathsf{g}$} -- ++(-1.5,1) circle (2pt) node[above=1pt]{$h_1$} (a) ++(1.5,1) circle (2pt) node[above=1pt]{$\pmb{3}$};
	\draw [thick] (-1.5,1) .. controls (0,2) and (0,2) .. (1.5,1);
	\end{scope}
	\end{tikzpicture}
	\caption{The four trees that contribute to $\pmb\rho$ of \eqref{eq:typicalrho} when $b_2=d_2$ and $b_3=d_3$.}
	\label{fig:exampletree}
\end{figure}

Before moving on, we go back to our example~\eqref{eq:typicalrho}. Among the $4!=24$ increasing trees of the $\pmb\beta$ given in~\eqref{eq:betaorder}, only the four given in figure~\ref{fig:exampletree} can possibly contribute. It is now clear that this $C(\pmb\rho)\neq 0$ only if $b_2=d_2$ and $b_3=d_3$.

\paragraph{Step 3: Evaluation. }Finally, the surviving spanning trees can be evaluated as follows:
\begin{itemize}
	\item For each path, if the starting vertex is a graviton $h_i$, then we replace it by $\epsilon_{h_i}$. If the starting vertex is a gluon trace $\pmb{i}^{\pmb\rho}_{c_id_i}$, then we replace it by $(-1)^{|\pmb{i}_{d_i,c_i}|}(-k_{c_i})$. Notice that we have restored the phase generated by the KK relations, which we have hidden all the way through in $\W{\sum}$.
	\item For the vertices in the middle of the path, we replace each graviton $h_i$ by $F_{h_i}$, and each gluon trace $\pmb{j}^{\pmb\rho}_{a_jb_j}$ by $(-1)^{|\pmb{j}_{b_j,a_j}|}(-k_{b_j}k_{a_j})$.
	\item For the end point, if it is the root $\mathsf{g}$, we replace it by $Y_{h_t}$ if the previous vertex is the graviton $h_t$, or $Y_{a_t}$ if the previous vertex is the gluon trace $\pmb{t}^{\pmb\rho}_{a_tb_t}$. If the end point is a graviton $h_i$, then we simply replace it by $k_{h_i}$.
	\item If the end point is a gluon trace $\pmb{j}^{\pmb\rho}_{a_jb_j}$, then we replace it by $K[\pmb{i}^{\pmb{\rho}}_{a_j\rightarrow a_i}]$ if the previous vertex is a gluon trace whose first element is $a_i$, or $K[\pmb{i}^{\pmb{\rho}}_{a_j\rightarrow h_i}]$ if the previous vertex is a graviton $h_i$. Here, we use $K[\pmb{j}^{\pmb{\rho}}_{a_j\rightarrow x}]$ to denote the sum of all the momenta of the trace $\pmb{j}^{\pmb\rho}_{a_jb_j}$ appearing before the element $x$ in $\pmb\rho$. For example, if $\pmb{i}^{\pmb\rho}_{a_ib_i}=\{3,4,5,6\}$ and $\pmb\rho=\{\ldots 3\ldots 4\ldots h\ldots 5\ldots 6\}$, then $K[\pmb{i}^{\pmb\rho}_{a_i\rightarrow h}]=k_3+k_4$.
\end{itemize}
According to these rules, the $\pmb\rho$ in Eq.~\eqref{eq:typicalrho} with $b_2=d_2$ and $b_3=d_3$ gives:
\begin{align}
C(\pmb\rho)&=(-1)^{|\pmb{2}_{b_2,a_1}|+|\pmb{3}_{b_3,a_3}|}(k_{a_2}\cdot Y_{a_2})(\epsilon_{h_1}\cdot Y_{h_1})\nonumber\\
&\quad\times\left[(\epsilon_{h_2}\cdot Y_{h_2})(k_{a_3}\cdot Y_{a_3})+(\epsilon_{h_2}\cdot k_{h_1})(k_{a_3}\cdot Y_{a_3})+(\epsilon_{h_2}\cdot Y_{h_2})(k_{a_3}\cdot k_{h_1})+(\epsilon_{h_2}\cdot k_{h_1})(k_{a_3}\cdot k_{h_1})\right]\,.
\end{align}
As a more concrete example, we consider the pure YM expansion of the amplitude $A_{3,0}(1,2\,|\,3,4,5\,|\,6,7)$, in which the coefficients $C(\pmb\rho)$ are identified directly with the  DDM form  of BCJ numerators:
\begin{equation}
A_{3,0}(1,2\,|\,3,4,5\,|\,6,7)=\sum_{\pmb\rho}n(1,\pmb\rho,2)A_{1,0}(1,\pmb\rho,2)\,.
\end{equation}
The reference gauge $\pmb{\mathcal{R}}$ we use is $\pmb{2}\prec\pmb{3}$. For the second and third trace, the fiducial gluons are chosen as $d_2=5$ and $d_3=7$. There are six numerators in which the two gluon traces $\{3,4,5\}$ and $\{6,7\}$ do not mix:
\begingroup
\allowdisplaybreaks
\begin{subequations}
	\begin{align}
	& n(1,\{6,7,3,4,5\},2)=(k_3\cdot k_7)(k_6\cdot k_1)+(k_3\cdot k_1)(k_6\cdot k_1)\\*
	& n(1,\{7,6,3,4,5\},2)=(k_3\cdot k_6)(k_7\cdot k_1)\\
	& n(1,\{6,7,4,3,5\},2)=-(k_4\cdot k_7)(k_6\cdot k_1)-(k_4\cdot k_1)(k_6\cdot k_1)\\*
	& n(1,\{7,6,4,3,5\},2)=-(k_4\cdot k_6)(k_7\cdot k_1)\\
	& n(1,\{3,4,5,6,7\},2)=(k_3\cdot k_1)(k_6\cdot k_1)+(k_3\cdot k_1)\left[k_6\cdot(k_3+k_4+k_5)\right]\\*
	& n(1,\{4,3,5,6,7\},2)=-(k_4\cdot k_1)(k_6\cdot k_1)-(k_4\cdot k_1)\left[k_6\cdot (k_3+k_4+k_5)\right]\,.
	\end{align}
\end{subequations}
\endgroup
The other sixteen numerators involve gluon trace mixing, and they can be grouped into:
\begingroup
\allowdisplaybreaks
\begin{subequations}
	\begin{align}
	& n(1,\{6,3,7,4,5\},2)=n(1,\{6,3,4,7,5\},2)=n(1,\{6,3,4,5,7\},2)=(k_3\cdot k_1)(k_6\cdot k_1)\\*
	& n(1,\{6,4,7,3,5\},2)=n(1,\{6,4,3,7,5\},2)=n(1,\{6,4,3,5,7\},2)=-(k_4\cdot k_1)(k_6\cdot k_1)\\
	& n(1,\{3,6,7,4,5\},2)=n(1,\{3,6,4,7,5\},2)=n(1,\{3,6,4,5,7\},2)=(k_3\cdot k_1)(k_6\cdot k_1)+(k_3\cdot k_1)(k_6\cdot k_3)\\*
	& n(1,\{4,6,7,3,5\},2)=n(1,\{4,6,3,7,5\},2)=n(1,\{4,6,3,5,7\},2)=-(k_4\cdot k_1)(k_6\cdot k_1)-(k_4\cdot k_1)(k_6\cdot k_4)\\
	& n(1,\{3,4,6,7,5\},2)=n(1,\{3,4,6,5,7\},2)=(k_3\cdot k_1)(k_6\cdot k_1)+(k_3\cdot k_1)\left[k_6\cdot(k_3+k_4)\right]\\*
	& n(1,\{4,3,6,7,5\},2)=n(1,\{4,3,6,5,7\},2)=-(k_4\cdot k_1)(k_6\cdot k_1)-(k_4\cdot k_1)\left[k_6\cdot (k_4+k_3)\right]\,.
	\end{align}
\end{subequations}
\endgroup
It is interesting to notice that many of them share the same expression.

\section{Conclusion and outlook}\label{sec:conclusion}
In this paper, we have presented two types of recursive expansions of generic EYM amplitudes, and given a lot of supporting evidences. Then we proved it using the BCFW recursive relation.
By repeatedly using these two recursive expansions, we can finally express any multitrace EYM amplitude by a linear combination of pure YM amplitudes in KK basis, the coefficients of which are just the BCJ numerators in KK basis. In addition, we have provided another recursive expansion for pure gluon EYM amplitudes using the integrand of the squeezing-form, and proved its equivalence with our type-II expansion. Together with the techniques presented in~\cite{Fu:2017uzt,Teng:2017tbo,Du:2017kpo}, we have obtained the full toolkit to write down the BCJ numerators for {pure YM, YM-scalar with $\phi^3+\phi^4$ interaction and NLSM.}

We have also studied two types of multitrace BCJ relations obtained from these recursive expansions. A full understanding of these BCJ relations provide us some important pictures on the interplay between gauge invariance and collinear behavior in the multitrace sector. Our result raises an interesting question. As pointed out by Arkani-Hamed, Rodina and Trnka~\cite{Arkani-Hamed:2016rak}, if we impose that, for each trivalent diagram used to calculate $n$-point spin one amplitudes, the kinematic numerator has mass dimension at most $n-2$, then gauge invariance uniquely fixes the amplitudes to be the YM ones. Interestingly, the numerators of $n$-point pure gluon EYM have mass dimension at least $n$, so that gauge invariance alone does not provide enough constraints~\cite{Arkani-Hamed:2016rak}. Then it will be interesting to see whether by imposing both the collinear behavior and gauge invariance, we can obtain enough constraints to uniquely pin down the pure gluon EYM amplitudes, or the most generic ones with gravitons.

Next, we would like to discuss the trick ``turning a graviton into a trace of gluons'' presented in section~\ref{sec:recursion}. It was known before that a graviton can be viewed as the ejection of two collinear gluons~\cite{Stieberger:2014cea,Stieberger:2015qja}. It will be interesting to see whether their results can be used to understand our recursive expansion, by a careful study of multi-collinear gluon amplitudes, and compare it with the amplitude in which these multi-collinear gluons are replaced by a single graviton.

Finally, as we mentioned in section~\ref{sec:compact}, we can picture the multitrace EYM integrand through a dimensional reduction of Eq.~\eqref{eq:PabDR}. As a possible future project, we would like to ask which theory, if any, corresponds to such integrand before imposing the dimension reduction conditions. One can start by first investigating the factorization behavior.


\acknowledgments
We would like to thank Song He, Rijun Huang, Henrik Johansson and Oliver Schlotterer for very useful discussions. In particular, we thank Oliver Schlotterer for valuable comments during the preparation of our manuscript. YD would like to acknowledge Jiangsu Ministry of Science and Technology under contract BK20170410, National Natural Science Foundation of China (NSFC) under Grant No. 111547310, as well as the 351 program of Wuhan University. BF is supported by Qiu-Shi Funding and the National Natural Science Foundation of China with Grant No.11575156, No.11135006 and No.11125523.

\appendix

\section{\boldmath World sheet \texorpdfstring{$SL(2,\mathbb{C})$}{SL(2,C)} transformation of \texorpdfstring{$\Pfp(\Pi)$}{Pf(Pi)}}\label{sec:PfPi}
In this appendix, we study the $SL(2,\mathbb{C})$ transformation of the EYM integrand
\begin{equation}
\label{eq:EYMintegrand}
\mathcal{I}_{\text{EYM}}(n)=\left[\,\prod_{i=1}^{m}\frac{1}{\langle\pmb{i}\rangle}\right]\Pfp(\Pi)\Pfp(\Psi)\,.
\end{equation}
Under the transformation \eqref{eq:Mobius}, it is straightforward to check that the entries of $X\in\{A,B,C\}$ transform as:
\begin{align}
\label{eq:ABCtrans}
X_{ab}\;\rightarrow\;(\gamma\sigma_a+\delta)(\gamma\sigma_b+\delta)X_{ab} \,.
\end{align}
Note that $C_{aa}\;\rightarrow(\gamma\sigma_a+\delta)^2C_{aa}$ holds only on the support of $d\Omega_{\text{CHY}}$. Then if we take the reduced Pfaffian of $\Psi$, the extra factors in \eqref{eq:ABCtrans} can be pulled out of the Pfaffian as:
\begin{align}
\Pfp(\Psi)\quad \rightarrow\quad&(\gamma\sigma_a+\delta)(\gamma\sigma_b+\delta)\frac{(-1)^{a+b}}{\sigma_{ab}}\left[\,\prod_{i=1}^{n}(\gamma\sigma_i+\delta)\right]\left[\prod_{i\neq a,b}^{n}(\gamma\sigma_i+\delta)\right]\Pf(\Psi_{ab}^{ab})\nonumber\\
&=\left[\,\prod_{i=1}^{n}(\gamma\sigma_i+\delta)^2\right]\Pfp(\Psi)\,.
\end{align}
The color trace part of $\mathcal{I}_{\text{EYM}}$ also has a simple transformation property:
\begin{equation}
\left[\,\prod_{i=1}^{m}\frac{1}{\langle\pmb{i}\rangle}\right]\;\rightarrow\;\left[\,\prod_{a\in\mathsf{G}}(\gamma\sigma_a+\delta)^2\right]\left[\,\prod_{i=1}^{m}\frac{1}{\langle\pmb{i}\rangle}\right]\,,
\end{equation}
since each Parke-Taylor factor contributes a factor $\prod_{a\in\pmb{i}}(\gamma\sigma_a+\delta)^2$, while the union of these traces gives the gluon set $\mathsf{G}$.

Then our main task now is to study the transformation of $\Pfp(\Pi)$. We choose $i=j=m$ in Eq.~\eqref{eq:PfpPi} when calculating the reduced Pfaffian, namely, we delete the two rows and columns that correspond to the $m$-th gluon trace in $\Pi$. After the transformation \eqref{eq:Mobius}, the $A_{\mathsf{H}}$, $B_{\mathsf{H}}$ and $C_{\mathsf{H}}$ part of $\Pi$ transform according to \eqref{eq:ABCtrans}, so that we can pull out the $(\gamma\sigma+\delta)$ factors out:
\begin{align}
\Pfp(\Pi)\quad\rightarrow\quad\left[\prod_{a\in\mathsf{H}}(\gamma\sigma_a+\delta)^2\right]\Pf\left(\begin{array}{cccc}
A_{\mathsf{H}} & -(\W{A}_{\mathsf{Tr,H}})^T & -(C_{\mathsf{H}})^T & -(\W{C}_{\mathsf{Tr,H}})^T \\
\W{A}_{\mathsf{Tr,H}} & \W{A}_{\mathsf{Tr}} & -(\W{C}_{\mathsf{H,Tr}})^T & -(\W{C}_{\mathsf{Tr}})^T \\
C_{\mathsf{H}} & \W{C}_{\mathsf{H,Tr}} & B_{\mathsf{H}} & -(\W{B}_{\mathsf{Tr,H}})^T \\
\W{C}_{\mathsf{Tr,H}} & \W{C}_{\mathsf{Tr}} & \W{B}_{\mathsf{Tr,H}} & \W{B}_{\mathsf{Tr}} \\
\end{array}\right)\,,
\end{align}
where the tilded blocks are
\begingroup
\allowdisplaybreaks
\begin{subequations}
    \begin{align}
    & (\W{A}_{\mathsf{Tr,H}})_{ib}=\sum_{c\in\pmb{i}}\frac{(\gamma\sigma_c+\delta)k_c\cdot k_b}{\sigma_{cb}} & & (\W{A}_{\mathsf{Tr}})_{ij}=\sum_{c\in\pmb{i}}\sum_{d\in\pmb{j}}\frac{(\gamma\sigma_c+\delta)(\gamma\sigma_d+\delta)k_c\cdot k_d}{\sigma_{cd}} \\
    & (\W{B}_{\mathsf{Tr,H}})_{ib}=\sum_{c\in\pmb{i}}\frac{(\alpha\sigma_c+\beta)k_c\cdot \epsilon_b}{\sigma_{cb}} & & (\W{B}_{\mathsf{Tr}})_{ij}=\sum_{c\in\pmb{i}}\sum_{d\in\pmb{j}}\frac{(\alpha\sigma_c+\beta)(\alpha\sigma_d+\beta)k_c\cdot k_d}{\sigma_{cd}} \\
    & (\W{C}_{\mathsf{Tr,H}})_{ib}=\sum_{c\in\pmb{i}}\frac{(\alpha\sigma_c+\beta)k_c\cdot k_b}{\sigma_{cb}} & & (\W{C}_{\mathsf{Tr}})_{ij}=\sum_{\substack{c\in\pmb{i}~d\in\pmb{j} \\ c\neq d}}\frac{(\alpha\sigma_c+\beta)(\gamma\sigma_d+\delta)k_c\cdot k_d}{\sigma_{cd}} \\*
    & (\W{C}_{\mathsf{H,Tr}})_{ai}=\sum_{c\in\pmb{i}}\frac{(\gamma\sigma_c+\delta)\epsilon_a\cdot k_c}{\sigma_{ac}}\,, & &
    \end{align}
\end{subequations}
\endgroup
where $1\leqslant i,j\leqslant m-1$ and $a,b\in\mathsf{h}$.\footnote{When $i=j$, it is understood that $(\W{A}_{\mathsf{Tr}})_{ii}=(\W{B}_{\mathsf{Tr}})_{ii}=0$.} Now we can multiply each row of the submatrix
\begin{equation*}
\left(\begin{array}{cccc}
\W{C}_{\mathsf{Tr,H}} & \W{C}_{\mathsf{Tr}} & \W{B}_{\mathsf{Tr,H}} & \W{B}_{\mathsf{Tr}}
\end{array}\right)
\end{equation*}
by $\gamma/\alpha$ and subtract them from the submatrix
\begin{equation*}
\left(\begin{array}{cccc}
\W{A}_{\mathsf{Tr,h}} & \W{A}_{\mathsf{Tr}} & -(\W{C}_{\mathsf{H,Tr}})^T & -(\W{C}_{\mathsf{Tr}})^T
\end{array}\right)\,.
\end{equation*}
After we perform the same manipulation to the corresponding columns, the Pfaffian is unchanged, while the affected matrix elements become:
\begingroup
\allowdisplaybreaks
\begin{subequations}
    \label{eq:submatrix1}
    \begin{align}
    (\W{A}_{\mathsf{Tr,H}})_{ib}& &\rightarrow & & &\sum_{c\in\pmb{i}}\left(\delta-\frac{\beta\gamma}{\alpha}\right)\frac{k_c\cdot k_b}{\sigma_{cb}}=\frac{1}{\alpha}(A_{\mathsf{Tr,H}})_{ib}\\
    (\W{C}_{\mathsf{H,Tr}})_{ai}& &\rightarrow & & &\sum_{c\in\pmb{i}}\left(\delta-\frac{\beta\gamma}{\alpha}\right)\frac{\epsilon_a\cdot k_c}{\sigma_{ac}}=\frac{1}{\alpha}(C_{\mathsf{H,Tr}})_{ai} \\
    \left[-(\W{A}_{\mathsf{Tr,H}})^T\right]_{ai}=-(\W{A}_{\mathsf{Tr,H}})_{ia} & &\rightarrow & & &\sum_{c\in\pmb{i}}\left(\delta-\frac{\beta\gamma}{\alpha}\right)\frac{k_c\cdot k_a}{\sigma_{ac}}=-\frac{1}{\alpha}(A_{\mathsf{Tr,H}})_{ia}=\frac{1}{\alpha}\left[-(A_{\mathsf{Tr,H}})^T\right]_{ai} \\
    \left[-(\W{C}_{\mathsf{H,Tr}})^T\right]_{ib}=-(\W{C}_{\mathsf{H,Tr}})_{bi}& &\rightarrow & & &\sum_{c\in\pmb{i}}\left(\delta-\frac{\beta\gamma}{\alpha}\right)\frac{\epsilon_b\cdot k_c}{\sigma_{cb}}=-\frac{1}{\alpha}(C_{\mathsf{H,Tr}})_{bi}=\frac{1}{\alpha}\left[-(C_{\mathsf{H,Tr}})^T\right]_{ib} \\
    (\W{C}_{\mathsf{Tr}})_{ij}& &\rightarrow & & & \sum_{\substack{c\in\pmb{i}~d\in\pmb{j} \\ c\neq d}}\left(\delta-\frac{\beta\gamma}{\alpha}\right)\frac{(\alpha\sigma_c+\beta)k_c\cdot k_d}{\sigma_{cd}}=\frac{1}{\alpha}\sum_{\substack{c\in\pmb{i}~d\in\pmb{j} \\ c\neq d}}\frac{(\alpha\sigma_c+\beta)k_c\cdot k_d}{\sigma_{cd}}\\
    \left[-(\W{C}_{\mathsf{Tr}})^T\right]_{ij}=-(\W{C}_{\mathsf{Tr}})_{ji}& &\rightarrow & & &\sum_{\substack{c\in\pmb{i}~d\in\pmb{j} \\ c\neq d}}\left(\delta-\frac{\beta\gamma}{\alpha}\right)\frac{(\alpha\sigma_d+\beta)k_c\cdot k_d}{\sigma_{cd}}=\frac{1}{\alpha}\sum_{\substack{c\in\pmb{i}~d\in\pmb{j} \\ c\neq d}}\frac{(\alpha\sigma_d+\beta)k_c\cdot k_d}{\sigma_{cd}}\,.
    \end{align}
    Finally, with a slightly more involved calculation, we get
    \begin{align}
    (\W{A}_{\mathsf{Tr}})_{ij}\quad\rightarrow\quad \sum_{c\in\pmb{i}}\sum_{d\in\pmb{j}}\left[\left(\delta-\frac{\beta\gamma}{\alpha}\right)(\gamma\sigma_d+\delta)-\frac{1}{\alpha}\left(\gamma\sigma_d+\frac{\beta\gamma}{\alpha}\right)\right]\frac{k_c\cdot k_d}{\sigma_{cd}}&=\frac{1}{\alpha^2}\sum_{c\in\pmb{i}}\sum_{d\in\pmb{j}}\frac{k_c\cdot k_d}{\sigma_{cd}}\nonumber\\
    &=\frac{1}{\alpha^2}(A_{\mathsf{Tr}})_{ij}
    \end{align}
\end{subequations}
\endgroup
for $i\neq j$. Next, we multiply each row of
\begin{equation*}
\left(\begin{array}{cccc}
\W{A}_{\mathsf{Tr,H}} & \W{A}_{\mathsf{Tr}} & -(\W{C}_{\mathsf{H,Tr}})^T & -(\W{C}_{\mathsf{Tr}})^T
\end{array}\right)
\end{equation*}
by $\alpha\beta$ and subtract them from
\begin{equation*}
\left(\begin{array}{cccc}
\W{C}_{\mathsf{Tr,H}} & \W{C}_{\mathsf{Tr}} & \W{B}_{\mathsf{Tr,H}} & \W{B}_{\mathsf{Tr}}
\end{array}\right)\,,
\end{equation*}
and perform the same transformation to the corresponding columns. Again, the Pfaffian is unchange, while the affected matrix elements become
\begingroup
\allowdisplaybreaks
\begin{subequations}
    \label{eq:submatrix2}
    \begin{align}
    (\W{C}_{\mathsf{Tr,H}})_{ib}& &\rightarrow & & &\sum_{c\in\pmb{i}}\left(\alpha\sigma_c+\beta-\beta\right)\frac{k_c\cdot k_b}{\sigma_{cb}}=\alpha (C_{\mathsf{Tr,H}})_{ib} \\
    (\W{B}_{\mathsf{Tr,H}})_{ib}& &\rightarrow & & &\sum_{c\in\pmb{i}}\left(\alpha\sigma_c+\beta-\beta\right)\frac{k_c\cdot \epsilon_b}{\sigma_{cb}}=\alpha(B_{\mathsf{Tr,H}})_{ib} \\
    \left[-(\W{C}_{\mathsf{Tr,H}})^T\right]_{ai}=-(\W{C}_{\mathsf{Tr,H}})_{ia} & &\rightarrow & & &\sum_{c\in\pmb{i}}\left(\alpha\sigma_c+\beta-\beta\right)\frac{k_c\cdot k_a}{\sigma_{ac}}=-\alpha(C_{\mathsf{Tr,H}})_{ai}=\alpha\left[-(C_{\mathsf{Tr,H}})^T\right]_{ai} \\
    \left[-(\W{B}_{\mathsf{Tr,H}})^T\right]_{ai}=-(\W{B}_{\mathsf{Tr,H}})_{ia} & &\rightarrow & & &\sum_{c\in\pmb{i}}\left(\alpha\sigma_c+\beta-\beta\right)\frac{k_c\cdot\epsilon_a}{\sigma_{ac}}=-\alpha(B_{\mathsf{Tr,H}})_{ia}=\alpha\left[-(B_{\mathsf{Tr,H}})^T\right]_{ai} \\
    (\W{C}_{\mathsf{Tr}})_{ij}& &\rightarrow & & &\frac{1}{\alpha}\sum_{\substack{c\in\pmb{i}~d\in\pmb{j} \\ c\neq d}}\left(\alpha\sigma_c+\beta-\beta\right)\frac{k_c\cdot k_d}{\sigma_{cd}}=(C_{\mathsf{Tr}})_{ij} \\
    \left[-(\W{C}_{\mathsf{Tr}})^T\right]_{ij}=-(\W{C}_{\mathsf{Tr}})_{ji} & &\rightarrow & & &\frac{1}{\alpha}\sum_{\substack{c\in\pmb{i}~d\in\pmb{j} \\ c\neq d}}\left(\alpha\sigma_d+\beta-\beta\right)\frac{k_c\cdot k_d}{\sigma_{cd}}=-(C_{\mathsf{Tr}})_{ji} \\
    (\W{B}_{\mathsf{Tr}})_{ij}& &\rightarrow & & &\sum_{c\in\pmb{i}~d\in\pmb{j}}\left[(\alpha\sigma_c+\beta)(\alpha\sigma_d+\beta)-\beta(\alpha\sigma_d+\beta)-\alpha\beta\sigma_c\right]\frac{k_c\cdot k_d}{\sigma_{cd}}\nonumber\\*
    & & & & &=\alpha^2(B_{\mathsf{Tr}})_{ij} \qquad (i\neq j)\,.
    \end{align}
\end{subequations}
\endgroup
Finally, if we pull out the $\alpha$ factors from all the matrix elements in Eq.~\eqref{eq:submatrix1} and \eqref{eq:submatrix2}, they get cancelled outside the Pfaffian, such that we get:
\begin{equation}
\Pf\left(\begin{array}{cccc}
A_{\mathsf{H}} & -(\W{A}_{\mathsf{Tr,H}})^T & -(C_{\mathsf{H}})^T & -(\W{C}_{\mathsf{Tr,H}})^T \\
\W{A}_{\mathsf{Tr,H}} & \W{A}_{\mathsf{Tr}} & -(\W{C}_{\mathsf{H,Tr}})^T & -(\W{C}_{\mathsf{Tr}})^T \\
C_{\mathsf{H}} & \W{C}_{\mathsf{H,Tr}} & B_{\mathsf{H}} & -(\W{B}_{\mathsf{Tr,H}})^T \\
\W{C}_{\mathsf{Tr,H}} & \W{C}_{\mathsf{Tr}} & \W{B}_{\mathsf{Tr,H}} & \W{B}_{\mathsf{Tr}} \\
\end{array}\right)=\Pfp(\Pi)\,.
\end{equation}
This leads to the desired $SL(2,\mathbb{C})$ transformation property of $\Pfp(\Pi)$:
\begin{equation}
\Pfp(\Pi)\;\rightarrow\;\left[\prod_{a\in\mathsf{H}}(\gamma\sigma_a+\delta)^2\right]\Pfp(\Pi)\,.
\end{equation}
It is now obvious that the EYM integrand \eqref{eq:EYMintegrand} indeed transforms according to \eqref{eq:Mobius}:
\begin{equation}
\mathcal{I}_{L}(n)\;\rightarrow\;\mathcal{I}_{L}(n)\prod_{i=1}^{n}(\gamma\sigma_i+\delta)^2\,.
\end{equation}

\section{The equivalence of different recursive expansions}\label{sec:equiv}


The main purpose of this appendix is to show the equivalence between the squeezing-form recursive expansion~\eqref{eq:puregluonexp} and the type-II recursive expansion~\eqref{eq:type2step1} for pure gluon multitrace EYM amplitudes. In the first subsection, we prove a very useful identity, and then apply it to show the equivalence at the double trace level. We then give a brief discussion on how the color ordering information is encoded in our expansions. In the second subsection, we give a general proof of the equivalence.

\subsection{The pure double trace case}

We want to show the equivalence between the double trace expression~\eqref{eq:dtAn} and~\eqref{Double-result} in this subsection. Since we will focus on this simple case, we will use more straightforward notations. The expression~\eqref{eq:dtAn}, obtained by using the cross-ratio identity, can be rephrased
as
\begin{align} &\quad\frac{s_{12\ldots n}}{\Spaa{1,2\ldots n}\Spaa{r_1,r_2\ldots r_m}} \nn
& = \sum_{i=2}^n \sum_{j=1}^{m-1}\frac{
s_{i r_j}(-1)^{n-i}(-1)^{j-1}}{\Spaa{1,\{2\ldots i-1\}\shuffle\{i+1\ldots n\}^T,i,r_j,
\{r_{j+1}\ldots r_{m-1}\}\shuffle\{r_1\ldots r_{j-1}\}^T,r_m}}\,.\Label{Merge-2-1}
\end{align}
We need to show that Eq.~\eqref{Merge-2-1} actually equals:
\begin{equation}
\label{eq:type2double}
\sum_{j=1}^{m-1}\frac{2k_j\cdot Y_j\,(-1)^{j-1}}{\Spaa{1,\{2\ldots n-1\}\shuffle\{r_j,\{r_{j+1}\ldots r_{m-1}\}\shuffle\{r_1\ldots r_{j-1}\}^{T},r_m\},n}}\,.
\end{equation}
The proof can be carried out by using the following algebraic identity:
%
\bea  \frac{1}{ \Spaa{1,
\{\a_1\ldots\a_a\}\shuffle\{n,\gamma_1\ldots \gamma_b\}, A,
\pmb\b}}=\frac{(-1)^{b+|\pmb\b|+1}}{ \Spaa{1, \{\a_1\ldots\a_a\}\shuffle\pmb{\b}^T
,A, \gamma_b\ldots
\gamma_2,\gamma_1,n}}\,,\Label{gen-simplify}\eea
where $A$ is an arbitrary element, and $\pmb\b$ is an arbitrary ordered set whose length is $|\pmb\b|$.
The proof of Eq.~\eqref{gen-simplify} is straightforward. First we put $n$ to the end by using the KK relation:
\bea T& \equiv & \frac{1}{\Spaa{1,
\{\a_1\ldots\a_a\}\shuffle\{n,\gamma_1\ldots\gamma_b\}, A, \pmb\b}}\nn
& = & \sum_{i=0}^a \frac{1}{ \Spaa{1, \{\a_1\ldots\a_i\}, n,
\{\a_{i+1}\ldots\a_a\}\shuffle\{\gamma_1\ldots\gamma_b\}, A, \pmb\b}}\nn
& = & \sum_{i=0}^a \frac{(-1)^{a-i+b+|\pmb\b|+1}}{\Spaa{1,
\{\a_1\ldots\a_i\}\shuffle\{ \pmb\b^T, A,\{\gamma_b\ldots\gamma_1\}
 \shuffle\{\a_a\ldots\a_{i+1}\}\}, n}}\,.\eea
In the above shuffle product, only $\a_i$, $\a_{i+1}$ and $\gamma_1$ can be the last element. We can thus expand the sum over shuffle accordingly into following three terms:
\bea T &= &\sum_{i=1}^a  \frac{(-1)^{a-i+b+|\pmb\b|+1}}{ \Spaa{1,
\{\a_1\ldots\a_{i-1}\}\shuffle\{ \pmb\b^T, A,\{\gamma_b\ldots\gamma_1\}
 \shuffle\{\a_a,...,\a_{i+1}\}\},a_i, n}}  \nn
 & & +\sum_{i=0}^{a-1} \frac{(-1)^{a-i+b+|\pmb\b|+1}}{ \Spaa{1, \{\a_1\ldots\a_i\}\shuffle\{ \pmb\b^T, A,\{\gamma_b\ldots\gamma_1\}
 \shuffle\{\a_a\ldots a_{i+2}\}\}, a_{i+1}, n}} \nn
 & & +\sum_{i=0}^a \frac{(-1)^{a-i+b+|\pmb\b|+1}}{ \Spaa{1, \{\a_1\ldots\a_i\}\shuffle\{ \pmb\b^T, A,\{\gamma_b\ldots\gamma_2\}
 \shuffle\{\a_a\ldots\a_{i+1}\}\}, \gamma_1,n}}\,. \eea
After shifting the dummy variable $i$, we easily see that the first two terms cancel each other, while only the third term is left. By this manipulation, we have moved the $\gamma_1$ to the right of the shuffle product. Iterating the above procedure to move out the entire $\{\gamma_1\ldots \gamma_b\}$ list, eventually we arrive at
\bea  T &= & \sum_{i=0}^a \frac{(-1)^{a-i+b+|\pmb\b|+1}}{ \Spaa{1,
\{\a_1\ldots\a_i\}\shuffle\{ \pmb\b^T, A,\a_a\ldots\a_{i+1}\},\gamma_b\ldots
\gamma_2,\gamma_1,n}}\nn
& = & \sum_{i=1}^a \frac{(-1)^{a-i+b+|\pmb\b|+1}}{ \Spaa{1,
\{\a_1\ldots\a_{i-1}\}\shuffle\{ \pmb\b^T,
A,\a_a\ldots\a_{i+1}\},\a_i,\gamma_b\ldots \gamma_2,\gamma_1,n}}\nn
& & + \sum_{i=0}^{a-1} \frac{(-1)^{a-i+b+|\pmb\b|+1}}{ \Spaa{1,
\{\a_1\ldots\a_i\}\shuffle\{ \b^T,
A,\a_a\ldots\a_{i+2}\},\a_{i+1},\gamma_b\ldots
\gamma_2,\gamma_1,n}}\nn
& & +  \frac{(-1)^{b+|\pmb\b|+1}}{ \Spaa{1, \{\a_1\ldots\a_a\}\shuffle\{ \pmb\b^T
\},A, \gamma_b\ldots \gamma_2,\gamma_1,n}}\,.\eea
The last term comes from the special case of $i=a$ in the first line, while the last element of the shuffle can be $\a_a$ or $A$. The summation of the first two terms in our result is again zero, so that we have proved
\eqref{gen-simplify}.

Now we apply Eq.~\eqref{gen-simplify} to each term in the summation of Eq.~\eqref{Merge-2-1}. The result is:
\bea & & \frac{s_{12\ldots n}}{ \Spaa{1,2\ldots n}\Spaa{r_1,r_2\ldots r_m}} \nn
& = & \sum_{i=2}^n \sum_{j=1}^{m-1} \frac{
s_{i r_j}(-1)^{m+j+1}}{ \Spaa{1,\{2\ldots i-1\}\shuffle\{r_m,
\{r_1\ldots r_{j-1}\}\shuffle\{r_{j+1}\ldots r_{m-1}\}^T,r_j \} , i,
i+1\ldots n-1,n}} \nn
& = & \sum_{j=1}^{m-1} \frac{ 2 k_{r_j}\cdot \W Y_{r_j}\, (-1)^{m+j+1}}{ \Spaa{1,\{2\ldots n-1\}\shuffle\{r_m,
\{r_1\ldots r_{j-1}\}\shuffle\{r_{j+1}\ldots r_{m-1}\}^T,r_j \},
n}}\,, \Label{Merge-2-simplify-1}\eea
where $\W Y_{r_j}$ is the sum of momentum at the right hand side of the leg $r_j$ in the shuffle sum.
To go from the second line to third line, we need to carry out the summation over $i$, and some algebraic
manipulations are used. Finally, we perform an ordering reverse to the last line of Eq.~\eqref{Merge-2-simplify-1}, and at the same time, reverse $\Spaa{1,2\ldots n}$ on the left hand side. After that, we relabel the elements of $\{1,2\ldots n\}$ according to $i\rightarrow n-i+1$. These manipulations give
\begin{align}
\frac{s_{12\ldots n}}{\Spaa{1,2\ldots n}\Spaa{r_1,r_2\ldots r_m}}=\sum_{j=1}^{m-1}\frac{2k_{r_j}\cdot Y_{r_j}\,(-1)^{j-1}}{\Spaa{1,\{2\ldots n-1\}\shuffle\{r_j,\{r_{j+1}\ldots r_{m-1}\}\shuffle\{r_1\ldots r_{j-1}\}^{T},r_m\},n}}\,,\Label{Merge-2-simplify-2}
\end{align}
which is nothing but Eq.~\eqref{eq:type2double}. We thus proved the equivalence between Eq.~\eqref{Double-result} and Eq.~\eqref{eq:dtAn}.
%
%
%
%
%
In summary, we have shown the equivalence of three expressions: \eqref{Merge-2-1}, \eqref{Merge-2-simplify-1}
and \eqref{Merge-2-simplify-2}. It is worth to notice the different location of the fixed $r_m$ in~\eqref{Merge-2-simplify-1}
and~\eqref{Merge-2-simplify-2}.
When we use the cross-ratio identities at the first time, there are two fixed choices. This is the reason why $1$ and $r_m$ have been fixed in \eqref{Merge-2-1}, \eqref{Merge-2-simplify-1} and \eqref{Merge-2-simplify-2}. This gives an brief explanation for the difference between type-I and II recursive expansions. For type-I, we take $1$ and $h_1$ as our fixed elements in the first use of the cross-ratio identity, while in type-II, it is $1$ and $r_m$. This also explains why for the second trace $\pmb{2}$, only $d_2$ is fixed while $c_2$ should be summed over in type-II expansion.


Now we give some discussion on the color ordering. Although both gluon traces are
color ordered, neither of which is manifest at the right hand side of~\eqref{Merge-2-1}. In~\eqref{Merge-2-simplify-1}
and~\eqref{Merge-2-simplify-2} the color ordering is kept for the first trace, but not manifest for the
second. Thus we want to understand how the color ordering information is coded in these expressions. This can be done by considering the divergent behavior under the collinear limit: divergence appears for nearby pairs, but not for non-nearby pairs.

We first consider a nearby pair, for example, $(r_{m-1}, r_m)$. When $k_{r_m}\to t q$ and $k_{r_{m-1}}\to (1-t) q$, the term with $j=m-1$ does not contribute at the right hand side of \eqref{Merge-2-simplify-2}, such that we are left with:
\bea  & & \sum_{j=1}^{m-1} \frac{2 k_{r_j}\cdot  Y_{r_j}\,(-1)^{j-1}}{\Spaa{1,\{2\ldots n-1\}\shuffle\{r_j,
\{r_1\ldots r_{j-1}\}^T\shuffle\{r_{j+1}\ldots r_{m-1}\},r_m \},
n}}\nn
& \to & \sum_{j=1}^{m-2} \frac{2 k_{r_j}\cdot  Y_{r_j}\,(-1)^{j-1}}{\Spaa{1,\{2\ldots n-1\}\shuffle\{r_j,
\{r_1\ldots r_{j-1}\}^T\shuffle\{r_{j+1}\ldots r_{m-2}\},r_{m-1},r_m \},
n}}\nn
& \to & \mathcal{S}_{r_{m-1},r_m;q}\sum_{j=1}^{m-2} \frac{2 k_{r_j}\cdot  Y_{r_j}\,(-1)^{j-1}}{ \Spaa{1,\{2\ldots n-1\}\shuffle\{r_j,
\{r_1\ldots r_{j-1}\}^T\shuffle\{r_{j+1}\ldots r_{m-2}\},q\}, n}}\eea
which is indeed the expected collinear behavior.

For a non-nearby pair, for example, $(r_1, r_3)$. At
the right handed side of Eq.~\eqref{Merge-2-simplify-2}, only the $j=2$ term could give singular contribution:
\bea & & \frac{2 k_{r_j}\cdot  Y_{r_j}\,(-1)^{j-1}}{ \Spaa{1,\{2\ldots n-1\}\shuffle\{r_2,
\{r_1\}\shuffle\{r_{3}\ldots r_{m-1}\},r_m \}, n}}\nn
& \to & \frac{2 k_{r_j}\cdot  Y_{r_j}\,(-1)^{j-1}}{
\Spaa{1,\{2,...,n-1\}\shuffle\{r_2, r_1,r_{3},...,r_{m-1},r_m
\}, n}}+ \frac{2 k_{r_j}\cdot  Y_{r_j}\,(-1)^{j-1}}{
\Spaa{1,\{2\ldots n-1\}\shuffle\{r_2, r_3,r_{1}\ldots r_{m-1}\},r_m
\}, n}}\nn
& \to & \left(\mathcal{S}_{r_1,r_3;q}+\mathcal{S}_{r_3,r_1;q}\right)\frac{2 k_{r_j}\cdot  Y_{r_j}\,(-1)^{j-1}}{
    \Spaa{1,\{2\ldots n-1\}\shuffle\{r_2, q\ldots r_{m-1}\},r_m
        \}, n}}\,.\eea
Since the universal split function satisfies $\mathcal{S}_{r_1,r_3;q}=-\mathcal{S}_{r_3,r_1;q}$, this collinear limit is regular. Although the above calculations are simple, they have shown the reason why the KK structure
appears in our recursive expansion: \emph{it codes the color ordering information!}

\subsection{ General Proof}

Proof given in the previous subsection can easily be generalized to generic pure gluon multitrace EYM amplitudes.
Now we prove the equivalence between the general formula~\eqref{eq:type2step1} of type-II recursive expansion and the squeezing-form expansion presented in Eq.~\eqref{eq:puregluonexp}, repeated here as:
\begin{align}
&\quad A_{m,0}(1,\dots,r\,|\,\pmb{2}\,|\ldots|\,\pmb{m})=A_{m,0}(r,1,\dots,r-1\,|\,\pmb{2}\,|\ldots|\,\pmb{m})\nn
&=\Sl_{\pmb{\mathsf{Tr}}_s|\mathsf{Tr}_p=\mathsf{Tr}\backslash\{1,2\}}\underset{\{l,\underline{r}\}\subset\pmb{1}}{\widetilde{\sum}}\Bigg[\,\underset{\{a_i,b_i\}\subset\pmb{t}_i}{\widetilde{\sum}}
\,\Bigg]_{i=1}^s~\underset{\{c_2,\underline{d_2}\}\subset\pmb{2}}{\widetilde{\sum}}(-k_{c_2})\cdot(- k_{b_s}k_{a_s})\ldots(-k_{b_1}k_{a_1})\cdot k_{l}\nn
&\qquad\qquad\times A_{p+1,0}(r,\mathsf{KK}[\pmb{1},r,l],l,\pmb{K}(\pmb{\mathsf{Tr}}_s,a,b),c_2,\mathsf{KK}[\pmb{2},c_2,d_2],d_2|\mathsf{Tr}_p)\,,\Label{Tei-gen}
\end{align}
where we set $r$ as the first element in the first trace, instead of $1$\footnote{It is worth to compare our result with that of the previous subsection, where $1$ has been chosen as the first element when using the cross-ratio identity. With this choice, we arrive at Eq.~\eqref{Merge-2-simplify-1} first and then \eqref{Merge-2-simplify-2}. In this subsection, we choose $r$ as the first element, which will directly lead to Eq.~\eqref{Merge-2-simplify-2}.}. In this formula, both $r$ and $d_2$ are fixed. More explicitly, we have
\bea
\mathsf{KK}[\pmb{1},r,l]=\{r-1,\dots,l+2,l+1\}\shuffle\{1,2,\dots,l-1\}\qquad 1\leqslant l<r\,.
\eea
If we use $\pmb\b$ to represent the list $\{\pmb{K}(\pmb{\mathsf{Tr}}_s,a,b),c_2,\mathsf{KK}[\pmb{2},c_2,d_2],d_2\}$, the first trace at the last line of Eq.~\eqref{Tei-gen} becomes $\{ r,\{r-1\ldots l+2,l+1\}\shuffle\{1,2\ldots l-1\},l, \pmb\b\}$. We can now use the relation~\eqref{gen-simplify} to get:
\bea  \frac{1}{ \Spaa{r,\{r-1\ldots l+2,l+1\}\shuffle\{1,2\ldots l-1\},l, \pmb\b}}
 & = & \frac{ (-1)^{l-1+|\pmb\b|} }{ \Spaa{r,\{r-1\ldots l+2,l+1\}\shuffle\{\b^T\}, l, l-1\ldots 2,1 }}\nn
 & = & \frac{ (-1)^{r+l-1}}{ \Spaa{1,2\ldots l-1,l, \{l+1\ldots r-1\}\shuffle\{\b\}, r}}\,.\Label{gen-simplify-1}\eea
Putting it back to~\eqref{Tei-gen}, we get:
\begin{align} &\quad A_{m,0}(1\ldots r\,|\,\pmb{2}\,|\dots|\,\pmb{m})\nn
&= \Sl_{\pmb{\mathsf{Tr}}_s|\mathsf{Tr}_p=\mathsf{Tr}\backslash\{1,2\}}\Bigg[\,\underset{\{a_i,b_i\}\subset\pmb{t}_i}{\widetilde{\sum}}
\,\Bigg]_{i=1}^s~\underset{\{c_2,\underline{d_2}\}\subset\pmb{2}}{\widetilde{\sum}}~\sum_{l=1}^{r-1} (-1)^{r-(l+1)} (-1)^{r+l-1}(-k_{c_2})\cdot (-k_{b_s}k_{a_s})\ldots(-k_{b_1}\cdot k_{a_1})_{\mu}\nn
&\qquad\quad \times(k_{l})^{\mu}A(1,2\ldots l-1,l, \{l+1\ldots r-1\}\shuffle\{\pmb{K}(\pmb{\mathsf{Tr}}_s,a,b),c_2,\mathsf{KK}[\pmb{2},c_2,d_2],d_2\}, r,|\mathsf{Tr}_p) \nn
& = \Sl_{\pmb{\mathsf{Tr}}_s|\mathsf{Tr}_p=\mathsf{Tr}\backslash\{1,2\}}\Bigg[\,\underset{\{a_i,b_i\}\subset\pmb{t}_i}{\widetilde{\sum}}
\,\Bigg]_{i=1}^s~\underset{\{c_2,\underline{d_2}\}\subset\pmb{2}}{\widetilde{\sum}}(-k_{c_2})\cdot (-k_{b_s}k_{a_s})\ldots (-k_{b_1}\cdot k_{a_1})\cdot Y_{a_1}\nn
&\qquad\quad\times A_{p+1,0}(1, \{2,...,r-1\}\shuffle\{\pmb{K}(\pmb{\mathsf{Tr}}_s,a,b),c_2,\mathsf{KK}[\pmb{2},c_2,d_2],d_2\}, r,|\mathsf{Tr}_p)\,.
\end{align}
Thus we have shown the equivalence between the squeezing-form expansion~\eqref{eq:puregluonexp} and type-II recursive expansion~\eqref{eq:type2step1}.

\section{Fun with Pfaffians}

\subsection{General matrices}\label{sec:Pfaffian1}
In this appendix, we develop a systematic way to expand a Pfaffian in terms of its blockwise principal minors: suppose we have a generic $(2m-2)$ dimensional antisymmetric matrix
\begin{equation}
  M=\begin{pmatrix}
  U & \widetilde{V} \\
  V & W
  \end{pmatrix}\qquad\widetilde{V}=-V^T\,,
\end{equation}
we want to find out the expansion coefficients $\mathcal{C}$ of
\begin{equation}
\label{eq:Mexp}
  \Pf[M]=\sum_{\substack{\mathsf{a}|\mathsf{b}=\{1,2\ldots m-1\} \\ \mathsf{b}\neq \varnothing}}\mathcal{C}[\mathsf{b}]\Pf[M(\mathsf{a})]\,,
\end{equation}
where $\Pf[M(\mathsf{a})]$ is a blockwise principal minor of $\Pf[M]$ defined in section~\ref{sec:genexp}. We use the notation:
\begin{equation}
  \Pf\begin{bmatrix}
  i_1 & i_2 & \cdots \\
  j_1 & j_2 & \cdots
  \end{bmatrix}
\end{equation}
to denote a generic minor of $\Pf[M]$ with the rows (columns) $\{i_1,i_2\ldots\}$ deleted in the upper (left) half of $M$, and the rows (columns) $\{j_1,j_2\ldots\}$ deleted in the lower (right) half. We get a blockwise principal minor if the two sets coincide: $\{i_1,i_2\ldots\}=\{j_1,j_2\ldots\}$. In general, the $\mathcal{C}$'s in Eq.~\eqref{eq:Mexp} are not unique, since we can recursively expand, for example, $\Pf[M(\mathsf{a})]$ in terms of the smaller minors. In this case, $\Pf[M(\mathsf{a})]$ is eliminated, and we must have the corresponding $\mathcal{C}[\mathsf{b}]=0$. Such an observation indicates that to make the expansion unique, we need to require that $\mathcal{C}\neq 0$ for all $\mathsf{a}$.\footnote{By writing down this statement, we need to forbid the manipulations that make no combinitorical sense. For example, we should not take a fraction of $\Pf[M(\mathsf{a})]$ and perform the previously mentioned recursive expansion.}

As outlined in section~\ref{sec:genexp}, we first expand $\Pf[M]$ along the row of $V_{11}$, which gives:
\begin{equation}
  \Pf[M]=(-1)^{m+1}V_{11}\Pf[M(2\ldots m-1)]+(-1)^{m}\sum_{i=2}^{m-1}(-1)^{i}V_{1i}\Pf\begin{bmatrix}
  i \\ 1
  \end{bmatrix}+\sum_{i=2}^{m-1}(-1)^{i}W_{1i}\Pf\begin{bmatrix}
  & \\ 1 & i
  \end{bmatrix}\,.
\end{equation}
From this equation, we can simply read out the coefficient $\mathcal{C}[1]$:
\begin{equation}
  \mathcal{C}_m[1]=(-1)^{m+1}\mathscr{C}[1]=(-1)^{m+1}V_{11}\,.
\end{equation}
For future convenience, we also define:
\begin{align}
  \mathcal{C}_{1}^{\,\;i}=(-1)^{m+i}\mathscr{C}_{1}^{\,\,i}=(-1)^{m+i}V_{1i}& &\mathcal{C}_{1i}=(-1)^{i}\mathscr{C}_{1i}=(-1)^{i}W_{1i}\,.
\end{align}
In particular, we have $\mathcal{C}_{m}[1]=\mathcal{C}_{1}^{\,\;1}$. Using these new symbols, we can rewrite the above expansion as:
\begin{equation}
  \Pf[M]=\mathcal{C}_m[1]\Pf[M(2\ldots m-1)]+\sum_{i=2}^{m-2}\left(\mathcal{C}_{1}^{\,\;i}\Pf\begin{bmatrix}
  i \\ 1
  \end{bmatrix}+\mathcal{C}_{1i}\Pf\begin{bmatrix}
  & \\ 1 & i
  \end{bmatrix}\right)\,.
\end{equation}
Next, we expand $\Pf\left[\begin{array}{c} i \\ 1 \end{array}\right]$ along the row $i$ in the lower half and $\Pf\left[\begin{array}{cc} & \\ 1 & i \end{array}\right]$ along the row $i$ in the upper half, such that a new set of blockwise principal minors appear:
\begin{align}
  \Pf[M]&=\mathcal{C}_m[1]\Pf[M(2\ldots m-1)]+\sum_{i=2}^{m-1}\mathcal{C}_m[1i]\Pf[M(2\ldots\slashed{i}\ldots m-1)]\nonumber\\
  &\quad+\sum_{i=2}^{m-1}\sum_{\substack{ j=2 \\ j\neq i}}^{m-1}\left(\mathcal{C}_{1i}^{\,\;ij}\Pf\begin{bmatrix} i & j \\ 1 & i \end{bmatrix}+\mathcal{C}_{1ij}^{\,\;i}\Pf\begin{bmatrix} i & & \\ 1 & i & j \end{bmatrix}\right)\,,
\end{align}
where the new coefficients are
\begin{align}
  &\mathcal{C}_{1i}^{\,\;ij}=(-1)^{j+\theta(i-j)}\mathscr{C}_{1i}^{\,\,ij}& &\mathscr{C}_{1i}^{\,\,ij}=\mathscr{C}_{1}^{\,\,i}V_{ij}-\mathscr{C}_{1i}U_{ij}=V_{1i}V_{ij}-W_{1i}U_{ij} \nonumber\\
  &\mathcal{C}_{1ij}^{\,\;i}=(-1)^{j+\theta(i-j)}(-1)^{m+1}\mathscr{C}_{1ij}^{\,\,i}& &\mathscr{C}_{1ij}^{\,\,i}=\mathscr{C}_{1}^{\,\,i}W_{ij}-\mathscr{C}_{1i}\widetilde{V}_{ij}=V_{1i}W_{ij}-W_{1i}\widetilde{V}_{ij}\,.
\end{align}
The step function $\theta(i-j)$ makes the sign alternation correctly continues across $i$. In particular, we have
\begin{align}
  \mathcal{C}_m[1i]=\mathscr{C}[1i]=\mathcal{C}_{1i}^{\,\;i1}=\mathscr{C}_{1i}^{\,\,i1}=V_{1i}V_{i1}-W_{1i}U_{i1}\,.
\end{align}
The calculation gets more involved if we carry out this expansion one step further, since we need to carefully handle the orders in the upper and lower indices of $\mathcal{C}$:
\begin{align}
  \Pf[M]&=\mathcal{C}_m[1]\Pf[M(2\ldots m-1)]+\sum_{i=2}^{m-1}\mathcal{C}_m[1i]\Pf[M(2\ldots\slashed{i}\ldots m-1)]\nonumber\\
  &\quad+\sum_{\{i,j\}\subset\{2\ldots m-1\}}\mathcal{C}_m[1ij]\Pf[M(2\ldots\slashed{i}\ldots\slashed{j}\ldots m-1)]\nonumber\\
  &\quad+\sum_{\{i,j\}\subset\{2\ldots m-1\}}\sum_{k\notin\{i,j\}}\left(\mathcal{C}_{1ij}^{\,\;ijk}\Pf\begin{bmatrix}
  i & j & k \\ 1 & i & j
  \end{bmatrix}+\mathcal{C}_{1ijk}^{\,\;ij}\Pf\begin{bmatrix}
  i & j & & \\ 1 & i & j & k
  \end{bmatrix}\right)\,.
\end{align}
All the new coefficients in the last two lines should not depend on the order of $\{i,j\}$, since the associated minors do not. Indeed, we have
\begin{align}
  \mathcal{C}^{\,\;ijk}_{1ij}&=(-1)^{m+k+\theta(j-k)+\theta(k-i)}\left(\mathscr{C}^{\,\,ij}_{1i}V_{jk}-\mathscr{C}^{\,\,i}_{1ij}U_{jk}\right)+(-1)^{m+k+\theta(i-k)+\theta(k-j)}\left(\mathscr{C}^{\,\,ji}_{1j}V_{ik}-\mathscr{C}^{\,\,j}_{1ji}U_{ik}\right)\nonumber\\
  \mathcal{C}^{\,\;ij}_{1ijk}&=(-1)^{k+\theta(j-k)+\theta(k-i)}\left(\mathscr{C}^{\,\,ij}_{1i}W_{jk}-\mathscr{C}^{\,\,i}_{1ij}\widetilde{V}_{jk}\right)+(-1)^{k+\theta(i-k)+\theta(k-j)}\left(\mathscr{C}^{\,\,ji}_{1j}W_{ik}-\mathscr{C}^{\,\,j}_{1ji}\widetilde{V}_{ik}\right)\,.
\end{align}
The origin of this complexity is that we can reach at the same minor by deleting the rows in different orders. By observing that the phase factor $(-1)^{k+\theta(j-k)+\theta(k-i)}$ is invariant under the exchange of $i$ and $j$, we can simplify the above two relations as:
\begin{align}
  &\mathcal{C}^{\,\;ijk}_{1ij}=(-1)^{k+1+\theta(j-k)+\theta(i-k)}(-1)^{m}\mathscr{C}^{\,\,ijk}_{1ij}& &\mathscr{C}_{1ij}^{\,\,ijk}=\mathscr{C}_{1i}^{\,\,ij}V_{jk}-\mathscr{C}^{\,\,i}_{1ij}U_{jk}+\mathscr{C}^{\,\,ji}_{1j}V_{ik}-\mathscr{C}^{\,\,j}_{1ji}U_{ik}\nonumber\\
  &\mathcal{C}^{\,\;ij}_{1ijk}=(-1)^{k+1+\theta(j-k)+\theta(i-k)}\mathscr{C}^{\,\,ij}_{1ijk}& &\mathscr{C}_{1ijk}^{\,\,ij}=\mathscr{C}^{\,\,ij}_{1i}W_{jk}-\mathscr{C}^{\,\,i}_{1ij}\widetilde{V}_{jk}+\mathscr{C}^{\,\,ji}_{1j}W_{ik}-\mathscr{C}^{\,\,j}_{1ji}\widetilde{V}_{ik}\,.
\end{align}
Finally, we have
\begin{align}
  \mathcal{C}_{m}[1ij]=(-1)^{m}\mathscr{C}[1ij]=\mathcal{C}^{\,\;ij1}_{1ij}=(-1)^{m}\mathscr{C}_{1ij}^{\,\,ij1}\,.
\end{align}
This process can be repeatedly carried out until we have deleted all rows and columns in original matrix $M$, which leads to the following two recursive relations
\begin{align}
\label{eq:recursiveC}
  &\mathscr{C}_{1t_2\ldots t_s}^{\,\,t_2\ldots t_sk}=\left(\mathscr{C}_{1t_2\ldots t_{s-1}}^{\,\,t_2\ldots t_{s-1}t_s}V_{t_sk}-\mathscr{C}_{1t_2\ldots t_{s-1}t_s}^{\,\,t_2\ldots t_{s-1}}U_{t_sk}\right)+\sum_{\ell=2}^{s-1}\left(t_s\leftrightarrow t_\ell\right)\nonumber\\
  &\mathscr{C}_{1t_2\ldots t_sk}^{\,\,t_2\ldots t_s}=\left(\mathscr{C}_{1t_2\ldots t_{s-1}}^{\,\,t_2\ldots t_{s-1}t_s}W_{t_sk}-\mathscr{C}_{1t_2\ldots t_{s-1}t_s}^{\,\,t_2\ldots t_{s-1}}\widetilde{V}_{t_sk}\right)+\sum_{\ell=2}^{s-1}\left(t_s\leftrightarrow t_\ell\right)\,.
\end{align}
By construction, the result does not depend on the order of $\{i_2\ldots i_s\}$. Consequently, if we choose $\mathsf{b}=\{t_1,t_2\ldots t_s\}$, we have
\begin{equation}
\label{eq:Coe}
  \mathcal{C}_m[t_1t_2\ldots t_s]=-(-1)^{\frac{s(s-1)}{2}}(-1)^{sm}\mathscr{C}[t_1t_2\ldots t_s]\qquad\qquad\mathscr{C}[t_1t_2\ldots t_s]\equiv\mathscr{C}_{t_1t_2\ldots t_s}^{~\;t_2\ldots t_st_1}\,.
\end{equation}
This gives the first line of Eq.~\eqref{eq:C1s}. To obtain the expression of $\mathscr{C}$, we need to use the information in the CHY integrand.

\subsection{Pure gluon EYM integrands}\label{sec:Pfaffian2}
With the preparation of the previous subsection, we can now calculate $\mathcal{C}_m$ of Eq.~\eqref{eq:Pfexpansion} simply by identifying
\begin{align}
  M=\Pi_{m}(\pmb{1,2\ldots m-1})& &U=A_{\mathsf{tr}}& & V=C_{\mathsf{tr}}& &W=B_{\mathsf{tr}}\,.
\end{align}
Actually, the first few cases have already been calculated in Eq.~\eqref{eq:tt2} and \eqref{eq:Cpqrp}. Our first job is to prove inductively that
\begin{subequations}
\label{eq:Cassumption}
\begin{align}
  \mathscr{C}_{1t_2\ldots t_s}^{\,\,t_2\ldots t_sk}&=\sum_{a\in\pmb{1}}\sigma_a\Bigg[\sum_{\{i_\ell,j_\ell\}\subset\pmb{t}_{\ell}}\sigma_{i_\ell j_\ell}\Bigg]_{\ell=2}^{s}\sum_{b\in\pmb{k}}\left[\mathcal{F}(a|\{j_2i_2\}\ldots\{j_si_s\}|b)+\text{permutations of }(2\ldots s)\right]\\
  \mathscr{C}_{1t_2\ldots t_sk}^{\,\,t_2\ldots t_s}&=\sum_{a\in\pmb{1}}\sigma_a\Bigg[\sum_{\{i_\ell,j_\ell\}\subset\pmb{t}_\ell}\sigma_{i_\ell j_\ell}\Bigg]_{\ell=2}^{s}\sum_{b\in\pmb{k}}\sigma_b\left[\mathcal{F}(a|\{j_2i_2\}\ldots\{j_si_s\}|b)+\text{permutations of }(2\ldots s)\right]\,,
\end{align}
\end{subequations}
where the function $\mathcal{F}$ is given by
\begin{equation}
\label{eq:calF}
  \mathcal{F}(a|\{j_2i_2\}\ldots \{j_si_s\}|b)=\frac{k_a\cdot k_{j_2}}{\sigma_{aj_2}}\left(\prod_{c=2}^{s-1}\frac{k_{i_c}\cdot k_{j_{c+1}}}{\sigma_{i_cj_{c+1}}}\right)\frac{k_{i_s}\cdot k_b}{\sigma_{i_sb}}\,.
\end{equation}
With the above inductive assumption, we come to calculate
\begin{subequations}
\label{eq:Cs+1}
\begin{align}
  \mathscr{C}_{1t_2\ldots t_st_{s+1}}^{\,\,t_2\ldots t_st_{s+1}k}&=\left(\mathscr{C}_{1t_2\ldots t_s}^{\,\,t_2\ldots t_st_{s+1}}(C_{\mathsf{tr}})_{t_{s+1}k}-\mathscr{C}_{1t_2\ldots t_st_{s+1}}^{\,\,t_2\ldots t_s}(A_{\mathsf{tr}})_{t_{s+1}k}\right)+\sum_{\ell=2}^{s}(t_{s+1}\leftrightarrow t_{\ell})\\
  \mathscr{C}_{1t_2\ldots t_st_{s+1}k}^{\,\,t_2\ldots t_st_{s+1}}&=\left(\mathscr{C}_{1t_2\ldots t_s}^{\,\,t_2\ldots t_st_{s+1}}(B_{\mathsf{tr}})_{t_{s+1}k}+\mathscr{C}_{1t_2\ldots t_st_{s+1}}^{\,\,t_2\ldots t_s}(C_{\mathsf{tr}})_{kt_{s+1}}\right)+\sum_{\ell=2}^{s}(t_{s+1}\leftrightarrow t_{\ell})
\end{align}
\end{subequations}
according to the recursive definition \eqref{eq:recursiveC}. All the $\mathscr{C}$'s on the right hand side are at level $s$, which contain the sum over all permutations of the repeated indices according to the inductive assumption \eqref{eq:Cassumption}. The prescription $\sum(t_{s+1}\leftrightarrow t_{\ell})$ then guarantees that the $\mathscr{C}$'s on the left hand side, which are at level $s+1$, contain all the permutations of $\{t_2\ldots t_st_{s+1}\}$. More explicitly, we write:
\begin{subequations}
\begin{align}
  &\mathscr{C}_{1t_2\ldots t_s}^{\,\,t_2\ldots t_st_{s+1}}=\sum_{a\in\pmb{1}}\sigma_a\Bigg[\sum_{\{i_\ell,j_\ell\}\subset\pmb{t}_\ell}\sigma_{i_\ell j_\ell}\Bigg]_{\ell=2}^{s}\sum_{j_{s+1}\in\pmb{t}_{s+1}}\left[\mathcal{F}(a|\{j_2i_2\}\ldots \{j_si_s\}|j_{s+1})+\text{permutations of }(2\ldots s)\right]\nonumber\\
  &\mathscr{C}_{1t_2\ldots t_st_{s+1}}^{\,\,t_2\ldots t_s}=\sum_{a\in\pmb{1}}\sigma_a\Bigg[\sum_{\{i_\ell,j_\ell\}\subset\pmb{t}_\ell}\sigma_{i_\ell j_\ell}\Bigg]_{\ell=2}^{s}\sum_{j_{s+1}\in\pmb{t}_{s+1}}\!\sigma_{j_{s+1}}\left[\mathcal{F}(a|\{j_2i_2\}\ldots\{j_si_s\}|j_{s+1})+\text{permutations of }(2\ldots s)\right]\nonumber\\
  &(C_{\mathsf{tr}})_{t_{s+1}k}=\sum_{i_{s+1}\in\pmb{t}_{s+1}}\sum_{b\in\pmb{k}}\frac{\sigma_{i_{s+1}}k_{i_{s+1}}\cdot k_b}{\sigma_{i_{s+1}b}}\qquad\qquad\quad(A_{\mathsf{tr}})_{t_{s+1}k}=\sum_{i_{s+1}\in\pmb{t}_{s+1}}\sum_{b\in\pmb{k}}\frac{k_{i_{s+1}}\cdot k_b}{\sigma_{i_{s+1}b}}\nonumber\\
  &(B_{\mathsf{tr}})_{t_{s+1}k}=\sum_{i_{s+1}\in\pmb{t}_{s+1}}\sum_{b\in\pmb{k}}\frac{\sigma_{i_{s+1}}\sigma_bk_{i_{s+1}}\cdot k_b}{\sigma_{i_{s+1}b}}\qquad\qquad(C_{\mathsf{tr}})_{kt_{s+1}}=-\sum_{i_{s+1}\in\pmb{t}_{s+1}}\sum_{b\in\pmb{k}}\frac{\sigma_{b}k_{i_{s+1}}\cdot k_b}{\sigma_{i_{s+1}b}}\,,
\end{align}
\end{subequations}
Because of the simple identity:
\begin{align}
  \sigma_{i_{s+1}j_{s+1}}\mathcal{F}(a|\{j_2i_2\}\ldots\{j_si_s\},\{j_{s+1}i_{s+1}\}|b)&=\mathcal{F}(a|\{j_2i_2\}\ldots\{j_si_s\}|j_{s+1})\frac{\sigma_{i_{s+1}}k_{i_{s+1}}\cdot k_b}{\sigma_{i_{s+1}b}}\nonumber\\
  &\quad-\mathcal{F}(a|\{j_2i_2\}\ldots\{j_si_s\}|j_{s+1})\frac{\sigma_{j_{s+1}}k_{i_{s+1}}\cdot k_b}{\sigma_{i_{s+1}}b}\,,
\end{align}
Eq.~\eqref{eq:Cs+1} reduces to the form of Eq.~\eqref{eq:Cassumption} at level $s+1$ and thus our inductive proof completes. More importantly, according to the definition \eqref{eq:Coe} of $\mathscr{C}[1t_2\ldots t_s]$, Eq.~\eqref{eq:Cassumption} gives us
\begin{equation}
\label{eq:Ctemp1}
  \mathscr{C}[1t_2\ldots t_s]=\sum_{a,b\in\pmb{1}}\sigma_a\Bigg[\sum_{\{i_\ell,j_\ell\}\subset\pmb{t}_\ell}\sigma_{i_\ell j_\ell}\Bigg]_{\ell=2}^{s}\left[\mathcal{F}(a|\{j_2i_2\}\ldots\{j_si_s\}|b)+\text{permutations of }(2\ldots s)\right]\,.
\end{equation}
Under the exchange of dummy indices $a\leftrightarrow b$ and $i_\ell\leftrightarrow j_\ell$, we only need to make the following replacements in the above equation:
\begin{align}
\label{eq:Creplace}
&\sigma_a\rightarrow\sigma_b & &\Bigg[\sum_{\{i_\ell,j_\ell\}\subset\pmb{t}_\ell}\sigma_{i_\ell j_\ell}\Bigg]_{\ell=1}^{s}\rightarrow (-1)^{s-1}\Bigg[\sum_{\{i_\ell,j_\ell\}\subset\pmb{t}_\ell}\sigma_{i_\ell j_\ell}\Bigg]_{\ell=1}^{s}\nonumber\\
& & &\mathcal{F}(a|\{j_2i_2\}\ldots\{j_si_s\}|b)\rightarrow (-1)^{s}\mathcal{F}(a|\{j_si_s\}\ldots\{j_2i_2\}|b)\,.
\end{align}
If we average over \eqref{eq:Ctemp1} and the one after the replacement \eqref{eq:Creplace}, we get our final expression
\begin{equation}
\label{eq:C12s}
  \mathscr{C}[t_1t_2\ldots t_s]=\frac{1}{2}\Bigg[\sum_{\{i_\ell,j_\ell\}\subset\pmb{t}_\ell}\sigma_{i_\ell j_\ell}\Bigg]_{\ell=1}^{s}\left[\prod_{c=1}^{s}\frac{k_{i_c}\cdot k_{j_{c+1}}}{\sigma_{i_cj_{c+1}}}+\text{noncyclic permutations of }(12\ldots s)\right]\,,
\end{equation}
where we have defined $1\equiv t_1$ and $\{j_{s+1},i_{s+1}\}\equiv \{j_1,i_1\}$ to fully manifest the cyclicity. At the point, we have derived the second line of Eq.~\eqref{eq:C1s}, and completed the derivation of Eq.~\eqref{eq:Pfexpansion}. Before moving on, we remark that the above calculation is still at the general algebraic level, which holds without any assumption on the kinematics.

\section{Cross-ratio identity and Pfaffian expansion}\label{sec:CrossRatio}
In this appendix, we show how to derive Eq.~\eqref{eq:finalexpansion} from Eq.~\eqref{eq:Pfexpansion} using the cross-ratio identity. We first prove a very useful identity as a lemma in the first subsection, and use it to give the general inductive proof in the second subsection.
\subsection{An interesting identity}\label{sec:identity}
The purpose of this subsection is to show that the Pfaffian expansion coefficients derived in the previous two appendices can also be obtained by repeatedly using of the cross-ratio identity.
We start with $(C_{\mathsf{tr}})_{t_1t_1}=(k_{\pmb{t_1}})^2/2$, which can be rewritten as
\begin{align}
  &(C_{\mathsf{tr}})_{t_1t_1}+\sum_{\ell=2}^{s}\mathscr{R}(t_1|t_\ell)+\sum_{t_\ell\notin\{t_1\ldots t_s\}}\mathscr{R}(t_1|t_\ell)=0\nonumber\\
  &\mathscr{R}(t_1|t_\ell)=\sum_{i\in\pmb{t}_1}\sum_{j\in\pmb{t}_\ell}(k_i\cdot k_j)\frac{\sigma_{nj}\sigma_{ia}}{\sigma_{ij}\sigma_{na}}\quad(a\in\pmb{t}_1~~ n\notin\pmb{t}_1)\,,
\end{align}
according to the cross-ratio identity. More generally, we can recursively define the symmetrized $\mathscr{R}$ symbol as follows:
\begin{align}
\label{eq:Rrecursion}
  &\quad\mathscr{R}(t_1\{t_2\ldots t_{s-1}\}t_s)\left[-\frac{(k_{\pmb{t}_s})^2}{2}\right]+\sum_{\ell=2}^{s-1}(t_s\leftrightarrow t_{\ell})\nonumber\\
  &=\sum_{t_\ell\notin\{t_1\ldots t_s\}}\mathscr{R}(t_1\{t_2\ldots t_{s-1}t_s\}t_\ell)+\mathscr{R}[t_1t_2\ldots t_s]+\sum_{\ell=2}^{s}\,\sum_{\substack{\mathsf{a}|\mathsf{b}=\{t_2\ldots t_{s}\}\backslash\{t_\ell\} \\ \mathsf{b}\neq\varnothing}}\mathscr{R}(t_1\,\mathsf{a}\,t_\ell)\mathscr{R}[t_\ell\,\mathsf{b}]\,.
\end{align}
We have already used the $s=2$ and $s=3$ relation in Eq.~\eqref{eq:R12C22} and \eqref{eq:R123C}. Now we claim that the solution to this recursive relation is
\begin{align}
\label{eq:R12s}
  \mathscr{R}(t_1\{t_2\ldots t_{s-1}\}t_s)&=\sum_{j_s\in\pmb{t}_s}\mathscr{R}(t_1\{t_2\ldots t_{s-1}\}t_s)_{j_s}\nonumber\\
  \mathscr{R}(t_1\{t_2\ldots t_{s-1}\}t_s)_{j_s}&=\mathscr{R}(t_1|t_2\ldots t_{s-1}|t_s)_{j_s}+\text{permutations of }(2\ldots s-1)\nonumber\\
  \mathscr{R}(t_1|t_2\ldots t_{s-1}|t_s)_{j_s}&=\sum_{i_1\in\pmb{t}_1}\frac{\sigma_{i_1a}\sigma_{nj_s}}{\sigma_{na}}\Bigg[\sum_{\{i_\ell,j_\ell\}\subset\pmb{t}_\ell}\sigma_{i_\ell j_\ell}\Bigg]_{\ell=2}^{s-1}\mathcal{F}(i_1|\{j_2i_2\}\ldots \{j_{s-1}i_{s-1}\}|j_s)\,,
\end{align}
where the function $\mathcal{F}$ has been defined in Eq.~\eqref{eq:calF}. Some explicit results for small $s$ can be found in Eq.~\eqref{eq:R12}, \eqref{eq:R124} and \eqref{eq:R1234}. More importantly, if we close the trace cycle by setting $t_s=t_1$, then we are allowed to exchange the dummy indices $i_\ell$ and $j_\ell$. In the above equation, the consequence is just the replacement:
\begin{equation}
  \frac{\sigma_{i_1a}\sigma_{nj_1}}{\sigma_{na}}\;\rightarrow\;-\frac{\sigma_{j_1a}\sigma_{ni_1}}{\sigma_{na}}
\end{equation}
Averaging over the two equivalent expressions, we get:
\begin{equation}
  \frac{1}{2}\left(\frac{\sigma_{i_1a}\sigma_{nj_1}}{\sigma_{na}}-\frac{\sigma_{j_1a}\sigma_{ni_1}}{\sigma_{na}}\right)=\frac{1}{2}\sigma_{i_1j_1}\,.
\end{equation}
Now we reach a very interesting result:
\begin{equation}
  \mathscr{R}[t_1t_2\ldots t_s]=\frac{1}{2}\Bigg[\sum_{\{i_\ell,j_\ell\}\subset\pmb{t}_\ell}\sigma_{i_\ell j_\ell}\Bigg]_{\ell=1}^{s}\left[\prod_{c=1}^{s}\frac{k_{i_c}\cdot k_{j_{c+1}}}{\sigma_{i_cj_{c+1}}}+\text{noncyclic permutations of }(12\ldots s)\right]\,.
\end{equation}
Looking back to Eq.~\eqref{eq:C12s}, we find that:
\begin{equation}
  \mathscr{R}[t_1t_2\ldots t_s]=\mathscr{C}[t_1t_2\ldots t_s]\,.
\end{equation}
In the next appendix, we are going to use this identity to simplify the Pfaffian expansion \eqref{eq:Pfexpansion}.

Before moving on, we provide an inductive proof to Eq.~\eqref{eq:R12s}. As the starting point, the $s=2$ and $s=3$ case have already been explicitly calculated in section~\ref{sec:multi_example}. Now we assume that Eq.~\eqref{eq:R12s} holds at level $s$, and plug it into Eq.~\eqref{eq:Rrecursion}. On the left hand side, we have
\begin{align}
\label{eq:Rstep1}
  &\quad\mathscr{R}(t_1\{t_2\ldots t_{s-1}\}t_s)\left[-\frac{(k_{\pmb{t}_s})^2}{2}\right]+\sum_{\ell=2}^{s-1}(t_s\leftrightarrow t_\ell)\nonumber\\
  &=\sum_{j_s\in\pmb{t}_s}\mathscr{R}(t_1\{t_2\ldots t_{s-1}\}t_s)_{j_s}\sum_{\ell=1}^{s-1}\sum_{i_s\in\pmb{t}_s}\sum_{j_\ell\in\pmb{t}_\ell}(k_{i_s}\cdot k_{j_\ell})\frac{\sigma_{nj_\ell}\sigma_{i_sj_s}}{\sigma_{i_sj_\ell}\sigma_{nj_s}}\nonumber\\
  &\quad+\sum_{j_s\in\pmb{t}_s}\mathscr{R}(t_1\{t_2\ldots t_{s-1}\}t_s)_{j_s}\sum_{t_\ell\notin\{t_1\ldots t_s\}}\sum_{i_s\in\pmb{t}_s}\sum_{j_\ell\in\pmb{t}_\ell}(k_{i_s}\cdot k_{j_\ell})\frac{\sigma_{nj_\ell}\sigma_{i_sj_s}}{\sigma_{i_sj_\ell}\sigma_{nj_s}}+\sum_{\ell=2}^{s-1}(t_s\leftrightarrow t_\ell)\,.
\end{align}
The prescription $\sum_{\ell=2}^{s-1}(t_s\leftrightarrow t_\ell)$ keeps both sides invariant under the permutations of $\{t_1\ldots t_s\}$. In the first line of Eq.~\eqref{eq:Rstep1}, the $\ell=1$ terms gives:
\begin{align}
  &\quad\sum_{j_1\in\pmb{t}_1}\sum_{i_s,j_s\in\pmb{t}_s}\mathscr{R}(t_1|t_2\ldots t_{s-1}|t_s)(k_{i_s}\cdot k_{j_1})\frac{\sigma_{nj_1}\sigma_{i_sj_s}}{\sigma_{i_sj_1}\sigma_{nj_s}}\nonumber\\
  &=\sum_{i_1,j_1\in\pmb{t_1}}\frac{\sigma_{i_1a}\sigma_{nj_1}}{\sigma_{na}}\Bigg[\sum_{\{i_q,j_q\}\subset\pmb{t}_q}\sigma_{i_qj_q}\Bigg]_{q=2}^{s}\mathcal{F}(i_1|\{j_2i_2\}\ldots\{j_si_s\}|j_1)\xrightarrow{\text{symmetrization}}\mathscr{R}[t_1t_2\ldots t_s]\,.
\end{align}
Next, for the other $\ell$'s, we have
\begin{align}
  &\ell=2\ldots s-1: & &\quad\sum_{i_s,j_s\in\pmb{t}_s}\sum_{j'_\ell\in\pmb{t}_\ell}\mathscr{R}(t_1|t_2\ldots t_\ell\ldots t_{s-1}|t_s)_{j_s}(k_{i_s}\cdot k_{j'_\ell})\frac{\sigma_{nj'_\ell}\sigma_{i_sj_s}}{\sigma_{i_sj'_\ell}\sigma_{nj_s}}\nonumber\\
  & & &=\sum_{i_1\in\pmb{t}_1}\sum_{j_\ell\in\pmb{t}_\ell}\frac{\sigma_{i_1a}\sigma_{nj_\ell}}{\sigma_{na}}\Bigg[\sum_{\{i_q,j_q\}\subset\pmb{t_q}}\sigma_{i_qj_q}\Bigg]_{q=2}^{\ell-1}\mathcal{F}(i_1|\{j_2i_2\}\ldots\{j_{\ell-1}i_{\ell-1}\}|j_\ell)\nonumber\\
  & & &\quad\times\sum_{i_\ell,j'_\ell\in\pmb{t}_\ell}\frac{\sigma_{nj'_\ell}\sigma_{i_\ell j_\ell}}{\sigma_{nj_\ell}}\Bigg[\sum_{\{i_q,j_q\}\subset\pmb{t}_q}\sigma_{i_qj_q}\Bigg]_{q=\ell+1}^{s}\mathcal{F}(i_\ell|\{j_{\ell+1}i_{\ell+1}\}\ldots\{j_si_s\}|j'_\ell)\nonumber\\
  & & &\xrightarrow{\text{symmetrization}}\sum_{\mathsf{a}|\mathsf{b}=\{t_2\ldots\slashed{t}_\ell\ldots t_{s}\}}\mathscr{R}(t_1\,\mathsf{a}\,t_\ell)\mathscr{R}[t_\ell\,\mathsf{b}]\,.
\end{align}
Now Eq.~\eqref{eq:Rrecursion} tells that our proof will complete if
\begin{equation}
  \sum_{j_s\in\pmb{t}_s}\mathscr{R}(t_1\{t_2\ldots t_{s-1}\}t_s)_{j_s}\sum_{i_s\in\pmb{t}_s}\sum_{j_\ell\in\pmb{t}_\ell}(k_{i_s}\cdot k_{j_\ell})\frac{\sigma_{nj_\ell}\sigma_{i_sj_s}}{\sigma_{i_sj_\ell}\sigma_{nj_s}}+\sum_{\ell=2}^{s-1}(t_s\leftrightarrow t_\ell)
\end{equation}
with $t_\ell\notin\{t_1\ldots t_s\}$ gives the correct form of $\mathscr{R}(t_1\{t_2\ldots t_s\}t_\ell)$. Indeed, we have
\begin{align}
  \sum_{i_s,j_s\in\pmb{t}_s}\mathscr{R}(t_1|t_2\ldots t_{s-1}|t_s)_{j_s}\sum_{j_\ell\in\pmb{t}_\ell}(k_{i_s}\cdot k_{j_\ell})\frac{\sigma_{nj_\ell}\sigma_{i_sj_s}}{\sigma_{i_sj_\ell}\sigma_{nj_s}}&=\sum_{i_1\in\pmb{t}_1}\sum_{j_\ell\in\pmb{t}_\ell}\frac{\sigma_{i_1a}\sigma_{nj_\ell}}{\sigma_{na}}\Bigg[\sum_{i_q,j_q\in\pmb{t}_q}\sigma_{i_qj_q}\Bigg]_{q=2}^{s}\mathcal{F}(t_1|t_2\ldots t_s|t_\ell)\nonumber\\
  &\xrightarrow{\text{symmetrization}}\mathscr{R}(t_1\{t_2\ldots t_s\}t_\ell)\,.
\end{align}
Therefore, we find that $\mathscr{R}(t_1\{t_2\ldots t_s\}t_\ell)$ has the form~\eqref{eq:R12s}, and the inductive proof completes.

\subsection{Pfaffian expansion}\label{sec:pfexp}
The purpose of this subsection is to prove Eq.~\eqref{eq:finalexpansion} with the preparation of the last subsection, starting from Eq.~\eqref{eq:Pfexpansion}.
In all the $\mathcal{C}_m$'s, the diagonal $C_{\mathsf{tr}}$ only appears at the level $m-1$, which has been shown separately in Eq.~\eqref{eq:Pfexpansion}. The cross-ratio identity gives:
\begin{equation}
\label{eq:CRstep1}
(-1)^{m+1}(C_{\mathsf{tr}})_{11}\Pf[\Pi_{m-1}(\mathsf{A}_{\cancel{1m}})]=(-1)^{m}\mathscr{R}(1|m)\Pf[\Pi_{m-1}(\mathsf{A}_{\cancel{1m}})]+(-1)^{m}\sum_{i=2}^{m-1}\mathscr{R}(1|i)\Pf[\Pi_{m-1}(\mathsf{A}_{\cancel{1m}})]\,.
\end{equation}
The term $\mathscr{R}(1|m)$ correctly connects $\langle\pmb{1}\rangle^{-1}$ and $\langle\pmb{m}\rangle^{-1}$, and it should remain in our final result. We thus define the following quantities:
\begin{align}
\label{eq:ABC1}
&\mathbb{A}_1\equiv(-1)^{m}\mathscr{R}(1|m)\Pf[\Pi_{m-1}(\mathsf{A}_{\cancel{1m}})]& &\mathbb{B}_1\equiv (-1)^{m}\sum_{i=2}^{m-1}\mathscr{R}(1|i)\Pf[\Pi_{m-1}(\mathsf{A}_{\cancel{1m}})]\nonumber\\
& & &\mathbb{C}_1\equiv\sum_{\substack{\mathsf{a}|\mathsf{b}=\mathsf{A}_{\cancel{1m}} \\ |\mathsf{a}|\leqslant m-3}}\mathcal{C}_m[1\,\mathsf{b}]\Pf[\Pi_{|\mathsf{a}|+1}(\mathsf{a})]\,,
\end{align}
such that $\Pf[\Pi_{m}(\mathsf{A}_{\cancel{m}})]=\mathbb{A}_1+\mathbb{B}_1+\mathbb{C}_1$. Next, we recursively expand $\Pf[\Pi_{m-1}(\mathsf{A}_{\cancel{1m}})]$ in $\mathbb{B}_1$ using Eq.~\eqref{eq:Pfexpansion}:
\begin{subequations}
    \begin{align}
    \mathbb{B}_1&=(-1)^{m}\sum_{i=2}^{m-1}\mathscr{R}(1|i)\Pf[\Pi_{m-1}(\mathsf{A}_{\cancel{1m}})]\nonumber\\
    &=(-1)\sum_{i=2}^{m-1}\mathscr{R}(1|i)[-(C_{\mathsf{tr}})_{ii}]\Pf[\Pi_{m-2}(\mathsf{A}_{\cancel{1im}})]+(-1)^{m}\sum_{i=2}^{m-1}\mathscr{R}(1|i)\sum_{\substack{\mathsf{a}|\mathsf{b}=\mathsf{A}_{\cancel{1im}} \\ \mathsf{b}\neq\varnothing}}
    \mathcal{C}_{m-1}[i\,\mathsf{b}]\Pf[\Pi_{|\mathsf{a}|+1}(\mathsf{a})]\nonumber\\
    \label{eq:levelm-2a}
    &=-\sum_{i=2}^{m-1}\mathscr{R}(1|i|m)\Pf[\Pi_{m-2}(\mathsf{A}_{\cancel{1im}})]-\sum_{i=2}^{m-1}\mathscr{R}[1i]\Pf[\Pi_{m-2}(\mathsf{A}_{\cancel{1im}})]\\
    \label{eq:levelm-2b}
    &\quad
    -\sum_{\{i,j\}\subset\mathsf{A}_{\cancel{1m}}}\mathscr{R}(1|i|j)\Pf[\Pi_{m-2}(\mathsf{A}_{\cancel{1im}})]+(-1)^{m}\sum_{i=2}^{m-1}\mathscr{R}(1|i)\sum_{\substack{\mathsf{a}|\mathsf{b}=\mathsf{A}_{\cancel{1im}} \\ \mathsf{b}\neq\varnothing}}\mathcal{C}_{m-1}[i\,\mathsf{b}]\Pf[\Pi_{|\mathsf{a}|+1}(\mathsf{a})]\,.
    \end{align}
\end{subequations}
To go from the second equality to the third, we have used Eq.~\eqref{eq:Rrecursion} to recursively generate the $\mathscr{R}$ symbols. In Eq.~\eqref{eq:levelm-2a}, the first term again connects the Parke-Taylor factors $\langle\pmb{1}\rangle^{-1}$, $\langle\pmb{i}\rangle^{-1}$ and $\langle\pmb{m}\rangle^{-1}$, and it should remain in the final result. We thus define:
\begin{align}
\label{eq:Am-3}
\mathbb{A}_2&\equiv\mathbb{A}_1-\sum_{i=2}^{m-1}\mathscr{R}(1|i|m)\Pf[\Pi_{m-2}(\mathsf{A}_{\cancel{1im}})]\nonumber\\
&=(-1)^{m}\mathscr{R}(1|m)\Pf[\Pi_{m-1}(\mathsf{A}_{\cancel{1m}})]-\sum_{i=2}^{m-1}\mathscr{R}(1|i|m)\Pf[\Pi_{m-2}(\mathsf{A}_{\cancel{1im}})]\,.
\end{align}
Because of the identity $\mathscr{R}[1i]=\mathscr{C}[1i]$, the second term of Eq.~\eqref{eq:levelm-2a} exactly cancels the term with $\mathsf{b}=\{i\}$ in $\mathbb{C}_1$ defined in Eq.~\eqref{eq:ABC1}: $\mathcal{C}_{m}[1i]=\mathscr{C}[1i]$.
All the terms in Eq.~\eqref{eq:levelm-2b} are emerged from the use of cross-ratio identity, and we group them into:
\begin{equation}
\label{eq:Bm-3}
\mathbb{B}_2\equiv-\sum_{\{i,j\}\subset\mathsf{A}_{\cancel{1m}}}\mathscr{R}(1|i|j)\Pf[\Pi_{m-2}(\mathsf{A}_{\cancel{1im}})]+(-1)^{m}\sum_{i=2}^{m-1}\mathscr{R}(1|i)\sum_{\substack{\mathsf{a}|\mathsf{b}=\mathsf{A}_{\cancel{1im}} \\ |\mathsf{a}|\leqslant m-4}}\mathcal{C}_{m-1}[i\,\mathsf{b}]\Pf[\Pi_{|\mathsf{a}|+1}(\mathsf{a})]\,.
\end{equation}
We then group the left-over terms in $\mathbb{C}_1$ into:
\begin{equation}
\label{eq:Cm-3}
\mathbb{C}_2\equiv\sum_{\substack{\mathsf{a}|\mathsf{b}=\mathsf{A}_{\cancel{1m}} \\ |\mathsf{a}|\leqslant m-4}}\mathcal{C}_m[1\,\mathsf{b}]\Pf[\Pi_{|\mathsf{a}|+1}(\mathsf{a})]\,.
\end{equation}
Now we have again $\Pf[\Pi_{m}(\mathsf{A}_{\cancel{m}})]=\mathbb{A}_2+\mathbb{B}_2+\mathbb{C}_2$. We can observe that in $\mathbb{A}_2$, all the Pfaffians have dimension $\geqslant 2(m-3)$ while in $\mathbb{B}_2$, all the Pfaffians have dimension $\leqslant 2(m-3)$, and in $\mathbb{C}_2$, $\leqslant 2(m-4)$.

To make this recursive pattern more clear, we perform the calculation for one more step. In $\mathbb{B}_2$, we again expand the first term using Eq.~\eqref{eq:Pfexpansion}:
\begin{align}
\label{eq:B2first}
-\sum_{\{i,j\}\subset\mathsf{A}_{\cancel{1m}}}\mathscr{R}(1|i|j)\Pf[\Pi_{m-2}(\mathsf{A}_{\cancel{1im}})]&=(-1)^{m-1}\sum_{\{i,j\}\subset\mathsf{A}_{\cancel{1m}}}\mathscr{R}(1|i|j)[-(C_{\mathsf{tr}})_{jj}]\Pf[\Pi_{m-3}(\mathsf{A}_{\cancel{1ijm}})]\nonumber\\
&\quad-\sum_{\{i,j\}\subset\mathsf{A}_{\cancel{1m}}}\,
\sum_{\substack{\mathsf{a}|\mathsf{b}=\mathsf{A}_{\cancel{1ijm}} \\ |\mathsf{a}|\leqslant m-5}}\mathscr{R}(1|i|j)\mathcal{C}_{m-2}[j\,\mathsf{b}]\Pf[\Pi_{|\mathsf{a}|+1}(\mathsf{a})]\,.
\end{align}
Now using  a special case of Eq.~\eqref{eq:Rrecursion}, we write the first term of Eq.~\eqref{eq:B2first} as:
\begin{align}
\label{eq:m-3expansion}
\sum_{\{i,j\}\subset\mathsf{A}_{\cancel{1m}}}\mathscr{R}(1|i|j)[-(C_{\mathsf{tr}})_{jj}]\Pf[\Pi_{m-3}(\mathsf{A}_{\cancel{1ijm}})]&=\sum_{\{i,j\}\subset\mathsf{A}_{\cancel{1m}}}\Big(\mathscr{R}(1\{ij\}m)+\mathscr{R}[1ij]\Big)\Pf[\Pi_{m-3}(\mathsf{A}_{\cancel{1ijm}})]\nonumber\\
&\quad+\sum_{\{i,j\}\subset\mathsf{A}_{\cancel{1m}}}\,\sum_{q\in\mathsf{A}_{\cancel{1ijm}}}\mathscr{R}(1\{ij\}q)\Pf[\Pi_{m-3}(\mathsf{A}_{\cancel{1ijm}})]\nonumber\\
&\quad+\sum_{\{i,j\}\subset\mathsf{A}_{\cancel{1m}}}\left[\mathscr{R}(1|i)+\mathscr{R}(1|j)\right]\mathscr{R}[ij]\Pf[\Pi_{m-3}(\mathsf{A}_{\cancel{1ijm}})]\,.
\end{align}
In the first line of this equation, the first term goes to the final result, such that we define
\begin{align}
\label{eq:Am-4}
\mathbb{A}_3&=\mathbb{A}_2-(-1)^{m}\sum_{\{i,j\}\subset\mathsf{A}_{\cancel{1m}}}\mathscr{R}(1\{ij\}m)\Pf[\Pi_{m-3}(\mathsf{A}_{\cancel{1ijm}})]\nonumber\\
&=(-1)^{m}\mathscr{R}(1|m)\Pf[\Pi_{m-1}(\mathsf{A}_{\cancel{1m}})]-\sum_{i=2}^{m-1}\mathscr{R}(1|i|m)\Pf[\Pi_{m-2}(\mathsf{A}_{\cancel{1im}})]\nonumber\\
&\quad-(-1)^{m}\sum_{\{i,j\}\subset\mathsf{A}_{\cancel{1m}}}\mathscr{R}(1\{ij\}m)\Pf[\Pi_{m-3}(\mathsf{A}_{\cancel{1ijm}})]\,.
\end{align}
while the second term cancels the $|\mathsf{a}|=m-4$ terms in Eq.~\eqref{eq:Cm-3}, because of the identity $\mathscr{C}[1ij]=\mathscr{R}[1ij]$ and $\mathcal{C}_{m}[1ij]=(-1)^{m}\mathscr{C}[1ij]$. We can thus define $\mathbb{C}_3$ as
\begin{equation}
\label{eq:Cm-4}
\mathbb{C}_3=\sum_{\substack{\mathsf{a}|\mathsf{b}=\mathsf{A}_{\cancel{1m}} \\ |\mathsf{a}|\leqslant m-5}}\mathcal{C}_m[1\,\mathsf{b}]\Pf[\Pi_{|\mathsf{a}|+1}(\mathsf{a})]\,.
\end{equation}
Next, because of the relation $\mathcal{C}_{m-1}[ij]=\mathcal{C}_{m}[ij]=\mathscr{R}[ij]$, the last line of Eq.~\eqref{eq:m-3expansion} cancels the $|\mathsf{a}|=m-4$ term in $\mathbb{B}_2$ as given in Eq.~\eqref{eq:Bm-3}. The second line of Eq.~\eqref{eq:B2first} and~\eqref{eq:m-3expansion}, together with the remaining terms in $\mathbb{B}_2$, gives us: (with a slight rearrangement)
\begin{align}
\mathbb{B}_3&=(-1)^{m}
\sum_{\{i,j\}\subset\mathsf{A}_{\cancel{1m}}}\,\sum_{\substack{\mathsf{a}|\mathsf{b}=\mathsf{A}_{\cancel{1ijm}} \\ |\mathsf{a}|\leqslant m-5}}\mathscr{R}(1|i)\mathcal{C}_{m-1}[ij\,\mathsf{b}]\Pf[\Pi_{|\mathsf{a}|+1}(\mathsf{a})]\nonumber\\
&\quad -\!\sum_{\{i,j\}\subset\mathsf{A}_{\cancel{1m}}}\,
\sum_{\substack{\mathsf{a}|\mathsf{b}=\mathsf{A}_{\cancel{1ijm}} \\ |\mathsf{a}|\leqslant m-5}}\mathscr{R}(1|i|j)\mathcal{C}_{m-2}[j\,\mathsf{b}]\Pf[\Pi_{|\mathsf{a}|+1}(\mathsf{a})]-(-1)^{m}\!\!\sum_{\{i,j,q\}\subset\mathsf{A}_{\cancel{1m}}}\mathscr{R}(1\{ij\}q)\Pf[\Pi_{m-3}(\mathsf{A}_{\cancel{1ijm}})]\,.
\end{align}
At this point, we have constructed $\mathbb{A}_3$, $\mathbb{B}_3$ and $\mathbb{C}_3$ such that $\Pf[\Pi_{m}(\mathsf{A}_{\cancel{m}})]=\mathbb{A}_3+\mathbb{B}_3+\mathbb{C}_3$. Meanwhile, all the Pfaffians contained in $\mathbb{A}_3$ have dimension $\geqslant 2(m-4)$ while in $\mathbb{B}_3$, all the Pfaffians have dimension $\leqslant 2(m-4)$, and in $\mathbb{C}_3$, $\leqslant 2(m-5)$. Now a recursive pattern emerges: we expand the Pfaffian of the highest dimension in $\mathbb{B}$ at each step, which will produce terms that appear in the final results, which are collected in $\mathbb{A}$, as well as certain cancellations between $\mathbb{C}$ and within $\mathbb{B}$. After each step, we reduce the highest Pfaffian dimension of $\mathbb{B}$ and $\mathbb{C}$ by two, while reducing the lowest Pfaffian dimension of $\mathbb{A}$ by two. This observation leads to the following inductive proof of Eq.~\eqref{eq:finalexpansion}.

Our inductive scheme is set up as follows. We define $\mathbb{A}$, $\mathbb{B}$ and $\mathbb{C}$ for a generic $r$ in the range $1\leqslant r\leqslant m-2$:
\begin{align}
\label{eq:ABCr}
&\mathbb{A}_r=\sum_{\substack{\mathsf{a}|\mathsf{b}=\mathsf{A}_{\cancel{1m}} \\ |\mathsf{a}|\geqslant m-r-1}}\mathcal{R}(1\,\mathsf{b}\,m)\Pf[\Pi_{|\mathsf{a}|+1}(\mathsf{a})]\qquad\qquad\mathbb{C}_{r}=\sum_{\substack{\mathsf{a}|\mathsf{b}=\mathsf{A}_{\cancel{1m}} \\ |\mathsf{a}|\leqslant m-r-2}}\mathcal{C}_{m}[1\,\mathsf{b}]\Pf[\Pi_{|\mathsf{a}|+1}(\mathsf{a})]\nonumber\\
&\mathbb{B}_{r}=\sum_{q\in\mathsf{A}_{\cancel{1m}}}\sum_{\substack{\mathsf{c}\subset\mathsf{A}_{\cancel{1qm}} \\ |\mathsf{c}|=r-1}}\!\mathcal{R}(1\,\mathsf{c}\,q)\Pf[\Pi_{m-r}(\mathsf{A}_{\cancel{1\mathsf{c}m}})]+\sum_{q\in\mathsf{A}_{\cancel{1m}}}\sum_{\substack{\mathsf{a}|\mathsf{b}|\mathsf{c}=\mathsf{A}_{\cancel{1qm}} \\ |\mathsf{c}|\leqslant r-2\,,\,|\mathsf{a}|\leqslant m-r-2}}\!\!\!\!\mathcal{R}(1\,\mathsf{c}\,q)\mathcal{C}_{m-|\mathsf{c}|-1}[q\,\mathsf{b}]\Pf[\Pi_{|\mathsf{a}|+1}(\mathsf{a})]\,,
\end{align}
where $\mathcal{R}$ and $\mathcal{C}_m$ are nothing but $\mathscr{R}$ and $\mathscr{C}$ dressed with a phase factor, which are given respectively in Eq.~\eqref{eq:R1s} and~\eqref{eq:C1s}. We want to prove that
\begin{equation}
\label{eq:levelr}
\Pf[\Pi_{m}(\mathsf{A}_{\cancel{m}})]=\mathbb{A}_r+\mathbb{B}_r+\mathbb{C}_r\,.
\end{equation}
The case with $r=1$, $2$ and $3$ have been explicitly verified before. Next, we assume that Eq.~\eqref{eq:levelr} holds at level $r$, and try to show that it also holds at level $r+1$. Our proof starts with the expansion of the Pfaffian along the row of $(C_{\mathsf{tr}})_{qq}$ in the first term of $\mathbb{B}_r$ (since we always have $q\in\mathsf{A}_{\cancel{1\mathsf{c}m}}$):
\begin{align}
\label{eq:Bexpansion1}
\sum_{q\in\mathsf{A}_{\cancel{1m}}}\,\sum_{\substack{\mathsf{c}\subset\mathsf{A}_{\cancel{1qm}} \\ |\mathsf{c}|=r-1}}\!\!\mathcal{R}(1\,\mathsf{c}\,q)\Pf[\Pi_{m-r}(\mathsf{A}_{\cancel{1\mathsf{c}m}})]&=\sum_{q\in\mathsf{A}_{\cancel{1m}}}\,\sum_{\substack{\mathsf{c}\subset\mathsf{A}_{\cancel{1qm}} \\ |\mathsf{c}|=r-1}}\!\!\mathcal{R}(1\,\mathsf{c}\,q)[-(C_{\mathsf{tr}})_{qq}]\Pf[\Pi_{m-r-1}(\mathsf{A}_{\cancel{1\mathsf{c}qm}})]\nonumber\\
&\quad +\sum_{q\in\mathsf{A}_{\cancel{1m}}}\sum_{\substack{\mathsf{a}|\mathsf{b}|\mathsf{c}=\mathsf{A}_{\cancel{1qm}} \\ |\mathsf{c}|= r-1\,,\,|\mathsf{a}|\leqslant m-r-3}}\!\!\!\!\mathcal{R}(1\,\mathsf{c}\,q)\mathcal{C}_{m-r}[q\,\mathsf{b}]\Pf[\Pi_{|\mathsf{a}|+1}(\mathsf{a})]\,.
\end{align}
We then apply the recursive relation~\eqref{eq:Rrecursion} onto the first term of Eq.~\eqref{eq:Bexpansion1}, which becomes:
\begin{subequations}
\label{eq:Bexp}
\begin{align}
\label{eq:Bexpa}
\text{first term of \eqref{eq:Bexpansion1}}&=\sum_{\substack{\mathsf{a}|\mathsf{b}=\mathsf{A}_{\cancel{1m}} \\ |\mathsf{a}|=m-r-2}}\left[\mathcal{R}(1\,\mathsf{b}\,m)\Pf[\Pi_{|\mathsf{a}|+1}(\mathsf{a})]+(-1)^{\frac{(r+1)(2m+r)}{2}}\mathscr{R}[1\,\mathsf{b}]\Pf[\Pi_{|\mathsf{a}|+1}(\mathsf{a})]\right]\\
\label{eq:Bexpb}
&\quad+\sum_{q\in\mathsf{A}_{\cancel{1m}}}\sum_{\substack{\mathsf{c}\subset\mathsf{A}_{\cancel{1qm}} \\ |\mathsf{c}|=r}}\mathcal{R}(1\,\mathsf{c}\,q)\Pf[\Pi_{m-r-1}(\mathsf{A}_{\cancel{1\mathsf{c}m}})]\\
\label{eq:Bexpc}
&\quad +\sum_{q\in\mathsf{A}_{\cancel{1m}}}\sum_{\substack{\mathsf{a}|\mathsf{b}|\mathsf{c}=\mathsf{A}_{\cancel{1qm}} \\ |\mathsf{c}|\leqslant r-2\,,\,|\mathsf{a}|= m-r-2}}\!\!\!\!(-1)^{\frac{(r+1)(2m+r)}{2}}\mathscr{R}(1\,\mathsf{c}\,q)\mathscr{R}[q\,\mathsf{b}]\Pf[\Pi_{|\mathsf{a}|+1}(\mathsf{a})]\,.
\end{align}
\end{subequations}
Now we distribute the above terms to the right places: we add the first term of Eq.~\eqref{eq:Bexpa} to $\mathbb{A}_{r}$, the result of which just equals to $\mathbb{A}_{r+1}$:
\begin{equation}
\mathbb{A}_r+\text{first term of \eqref{eq:Bexpa}}=\sum_{\substack{\mathsf{a}|\mathsf{b}=\mathsf{A}_{\cancel{1m}} \\ |\mathsf{a}|\geqslant m-r-2}}\mathcal{R}(1\,\mathsf{b}\,m)\Pf[\Pi_{|\mathsf{a}|+1}(\mathsf{a})]=\mathbb{A}_{r+1}\,.
\end{equation}
In the second term of Eq.~\eqref{eq:Bexpa}, we have $|\mathsf{b}|=m-2-|\mathsf{a}|=r$, such that
\begin{equation*}
(-1)^{\frac{(r+1)(r+2m)}{2}}\mathscr{R}[1\,\mathsf{b}]=\mathcal{R}[1\,\mathsf{b}]=-\mathcal{C}_{m}[1\,\mathsf{b}]\,.
\end{equation*}
When we add it to $\mathbb{C}_r$, it cancels exactly the $|\mathsf{a}|=m-r-2$ term in the sum, and gives:
\begin{equation}
\mathbb{C}_r+\text{second term of \eqref{eq:Bexpa}}=\sum_{\substack{\mathsf{a}|\mathsf{b}=\mathsf{A}_{\cancel{1m}} \\ |\mathsf{a}|\leqslant m-r-3}}\mathcal{C}_{m}[1\,\mathsf{b}]\Pf[\Pi_{|\mathsf{a}|+1}(\mathsf{a})]=\mathbb{C}_{r+1}\,.
\end{equation}
Next, comparing with Eq.~\eqref{eq:ABCr}, we find that Eq.~\eqref{eq:Bexpb} is just the first term of $\mathbb{B}_{r+1}$:
\begin{equation}
{\eqref{eq:Bexpb}}=\text{first term of $\mathbb{B}_{r+1}$}\,.
\end{equation}
Finally, in \eqref{eq:Bexpc}, we have $|\mathsf{b}|+|\mathsf{c}|=m-3-|\mathsf{a}|=r-1$ in the sum. Remarkably, we have exactly the correct phase factor to achieve:
\begin{equation}
(-1)^{\frac{(r+1)(r+2m)}{2}}\mathscr{R}(1\,\mathsf{c}\,q)\mathscr{R}[q\,\mathsf{b}]=(-1)^{\frac{(r+1)(r+2m)}{2}}\mathscr{R}(1\,\mathsf{c}\,q)\mathscr{C}[q\,\mathsf{b}]=-\mathcal{R}(1\,\mathsf{c}\,q)\mathcal{C}_{m-|\mathsf{c}|-1}[q\,\mathsf{b}]
\end{equation}
for any $|\mathsf{b}|+|\mathsf{c}|=r-1$. Therefore, Eq.~\eqref{eq:Bexpc} cancels exactly the $|\mathsf{a}|=m-r-2$ terms in the second term of $\mathbb{B}_r$. We can then easily show that:
\begin{align}
\text{second term of $\mathbb{B}_r$}\;+&\;\text{second term of \eqref{eq:Bexpansion1}}+\eqref{eq:Bexpc}\nonumber\\
&=\sum_{q\in\mathsf{A}_{\cancel{1m}}}\sum_{\substack{\mathsf{a}|\mathsf{b}|\mathsf{c}=\mathsf{A}_{\cancel{1qm}} \\ |\mathsf{c}|\leqslant r-1\,,\,|\mathsf{a}|\leqslant m-r-3}}\!\!\!\!\mathcal{R}(1\,\mathsf{c}\,q)\mathcal{C}_{m-|\mathsf{c}|-1}[q\,\mathsf{b}]\Pf[\Pi_{|\mathsf{a}|+1}(\mathsf{a})]\nonumber\\
&=\;\text{second term of $\mathbb{B}_{r+1}$}\,.
\end{align}
Now we have processed all the terms in Eq.~\eqref{eq:Bexpansion1} and \eqref{eq:Bexp}. Since all the terms can be arranged exactly into $\mathbb{A}_{r+1}$, $\mathbb{B}_{r+1}$ and $\mathbb{C}_{r+1}$, we have thus proved that:
\begin{equation}
\mathbb{A}_{r+1}+\mathbb{B}_{r+1}+\mathbb{C}_{r+1}=\mathbb{A}_r+\mathbb{B}_r+\mathbb{C}_r=\Pf[\Pi_{m}(\mathsf{A}_{\cancel{m}})]\,.
\end{equation}
Our inductive proof of Eq.~\eqref{eq:levelr} thus completes. To obtain our final goal Eq.~\eqref{eq:finalexpansion}, we simply take $r=m-1$. Since nothing is contained in $\mathbb{B}_{m-1}$ and $\mathbb{C}_{m-1}$, we have $\mathbb{B}_{m-1}=\mathbb{C}_{m-1}=0$ and:
\begin{equation}
\Pf[\Pi_{m}(\mathsf{A}_{\cancel{m}})]=\mathbb{A}_{m-1}=\sum_{\mathsf{a}|\mathsf{b}=\mathsf{A}_{\cancel{1m}}}\mathcal{R}(1\,\mathsf{b}\,m)\Pf[\Pi_{a+1}(\mathsf{a})]\,,
\end{equation}
which is nothing but Eq.~\eqref{eq:finalexpansion}.

\section{Basis by  BCJ relations}\label{sec:BCJblock}

In the discussion of the type-II BCJ relation, we have defined the building block~\eqref{BCJ-base}. Now using the generalized fundamental BCJ relation~\eqref{Gen-BCJ}, we can find the basis for these building blocks. Using the basis, we can
show different recursive expansions are equivalent to each other, and give the proof for both type-I and type-II BCJ relation.

To find the basis, we need to write down all the generalized fundamental BCJ relations for $k$ gravitons $L=\{a_1,...,a_k\}$. To do so, we separate the set $L$ into two ordered subsets $L_1=\{a_{i_1},...,a_{i_t}\}$ and $L_2=\{a_{i_{t+1}},...,a_{i_k}\}$ with $1\leqslant t\leqslant k$. With a given splitting, we have the corresponding BCJ relation
\bea 0=\sum_{\shuffle}(\sum_{s=1}^t k_{a_{i_s}}\cdot X_{a_{i_s}}) A\left(1,\{2\dots,r-1\}\shuffle \{a_{i_{t+1}},...,a_{i_k}\} \shuffle \{a_{i_1},...,a_{i_t}\},r\right)~~~
\Label{Gen-BCJ-2} \eea
If we consider  all possible ordered splitting of $L$ to $L_1 $ and $L_2$, we get all possible BCJ relations, then from them we can find the basis of building blocks. It is easy to see that the total number of relations is $k\times k!$ where $k!$ is the number of permutations of $L$ list, while the other $k$ is the number of splittings, with $L_1$ nonempty. Similarly, it is easy to see that the total number of building blocks is $k\times k!$. Although the number of equations is the same as the number of variables, since these equations are not independent to each other, we do have nonzero solutions to the basis.


To demonstrate our strategy, let us present several examples.

\subsection{The case of \texorpdfstring{$k=2$}{k=2}}

For this case, we have following BCJ relations. When $L_1=\{a_1\}$, we get:\footnote{We have omitted the common subscript $\{1,2\ldots r\}$ of $\mathcal{B}$.}
\bea 0 &= & \sum_{\shuffle}( k_{a_{1}}\cdot X_{a_{1}}) A\left(1,\{2\dots,r-1\}\shuffle \{a_2\} \shuffle \{a_{1}\},r\right) \nn
& = & \sum_{\shuffle}( k_{a_{1}}\cdot X_{a_{1}}) A\left(1,\{2\dots,r-1\}\shuffle \{a_1,a_2\},r\right)+\sum_{\shuffle}( k_{a_{1}}\cdot X_{a_{1}}) A\left(1,\{2\dots,r-1\}\shuffle \{a_2,a_1\} ,r\right)\nn
& = & {\cal B}[\{a_1, a_2\};1]+{\cal B}[\{a_2, a_1\};2]\,. \eea
When $L_1=\{a_2\}$, we get:
\bea 0 &= & \sum_{\shuffle}( k_{a_{2}}\cdot X_{a_{2}}) A\left(1,\{2\dots,r-1\}\shuffle \{a_1\} \shuffle \{a_{2}\},r\right)= {\cal B}[\{a_1, a_2\};2]+{\cal B}[\{a_2, a_1\};1]\,. \eea
When $L_1=\{a_1, a_2\}$, the BCJ relation is
\bea 0 ={\cal B}[\{a_1, a_2\};1]+{\cal B}[\{a_1, a_2\};2]\,. \eea
When $L_1=\{a_2, a_1\}$, the BCJ relation is
\bea 0 ={\cal B}[\{a_2, a_1\};1]+{\cal B}[\{a_2, a_1\};2]\,. \eea
From these four relations, we can see that among these four building blocks, three of them can solve by one as follows:
\bea {\cal B}[\{a_1, a_2\};2]={\cal B}[\{a_2, a_1\};2]=-{\cal B}[\{a_1, a_2\};1],~~~~~{\cal B}[\{a_2, a_1\};1]={\cal B}[\{a_1, a_2\};1]\,.\Label{B-block-n=2}\eea
Using this result, one can easily see that two different representations \eqref{Merge-2-simplify-1}
and \eqref{Merge-2-simplify-2} are equivalent for the special case of pure gluon double trace.

\subsection{The case \texorpdfstring{$k=3$}{k=3}}

For this case, there are $3\times 3!=18$ equations, which are given by
\bea L_1/L_2 & = & \{a_1\}/\{a_2, a_3\};~~~~{\rm plus~~permutation}\nn
 L_1/L_2 & = & \{a_1,a_2\}/\{a_3\};~~~~{\rm plus~~permutation}\nn
 L_1/L_2 & = & \{a_1,a_2, a_3\}/\varnothing;~~~~{\rm plus~~permutation.}\eea
From these equations, we can solve $11$ building blocks by the other $7$. One of the choice is
\begingroup
\allowdisplaybreaks
\bea
{\cal B}[\{a_1,a_2,a_3\};3]& = &
   -{\cal B}[\{a_1,a_2,a_3\};1]-{\cal B}[\{a_1,a_2,a_3\};2],\nn
   {\cal B}[\{a_1,a_3,a_2\};2]& = &
   {\cal B}[\{a_1,a_2,a_3\};2]-{\cal B}[\{a_1,a_3,a_2\};1]+{\cal B}[\{
   a_2,a_1,a_3\};1],\nn
   {\cal B}[\{a_1,a_3,a_2\};3] & = &
   -{\cal B}[\{a_1,a_2,a_3\};2]-{\cal B}[\{a_2,a_1,a_3\};1],\nn
   {\cal B}[\{a_2,a_1,a_3\};3]& = &
   -{\cal B}[\{a_2,a_1,a_3\};1]-{\cal B}[\{a_2,a_1,a_3\};2],\nn
   {\cal B}[\{a_2,a_3,a_1\};2]& = &
   {\cal B}[\{a_1,a_2,a_3\};1]+{\cal B}[\{a_2,a_1,a_3\};2]-{\cal B}[\{
   a_2,a_3,a_1\};1],\nn
   {\cal B}[\{a_2,a_3,a_1\};3] & = &
   -{\cal B}[\{a_1,a_2,a_3\};1]-{\cal B}[\{a_2,a_1,a_3\};2],\nn
   {\cal B}[\{a_3,a_1,a_2\};1] & = &
   {\cal B}[\{a_1,a_2,a_3\};1]+{\cal B}[\{a_1,a_3,a_2\};1]-{\cal B}[\{
   a_2,a_1,a_3\};1],\nn
   {\cal B}[\{a_3,a_1,a_2\};3] & = &
   -{\cal B}[\{a_1,a_2,a_3\};1]-{\cal B}[\{a_1,a_3,a_2\};1]+{\cal B}[\{a_2,a_1,a_3\};1]-{\cal B}[\{a_3,a_1,a_2\};2],\nn
   {\cal B}[\{a_3,a_2,a_1\};1] & = &
   -{\cal B}[\{a_1,a_2,a_3\};1]+{\cal B}[\{a_2,a_1,a_3\};1]+{\cal B}[\{a_2,a_3,a_1\};1],\nn
   {\cal B}[\{a_3,a_2,a_1\};2]& = &
   {\cal B}[\{a_1,a_2,a_3\};1]+{\cal B}[\{a_1,a_3,a_2\};1]-{\cal B}[\{
   a_2,a_1,a_3\};1]\nonumber\\*
   &&-{\cal B}[\{a_2,a_3,a_1\};1]+{\cal B}[\{a_3,a_1,a_2\};2]\,,\nn
   {\cal B}[\{a_3,a_2,a_1\};3]& = &
   -{\cal B}[\{a_1,a_3,a_2\};1]-{\cal B}[\{a_3,a_1,a_2\};2]\,.~~~~\Label{T3-solve}
   \eea
\endgroup
Now we use above result to discuss the pure double trace case given in \eqref{Merge-2-simplify-1}
and \eqref{Merge-2-simplify-2}. Assuming the second cycle is given by $\Spaa{a_1 a_2 a_3}$, by
the known cyclic and order reversing symmetries, we can conjecture
\bea A(1,2,...,r\,\Vert\,a_1, a_2, a_3) & = &  \a_1 A_1+\a_2 A_2+\a_3 A_3-\a_1 B_1-\a_2 B_2-\a_3 B_3~~~~\Label{sym-T3}\eea
where $\a_1,\a_2, \a_3$ are unknown coefficients and
\bea A_1 & = & {\cal B}[\{a_1, a_2, a_3\};1]+ {\cal B}[\{a_3, a_1, a_2\};1]+{\cal B}[\{a_2, a_3, a_1\};1] \nn
A_2 & = & {\cal B}[\{a_1, a_2, a_3\};2]+ {\cal B}[\{a_3, a_1, a_2\};2]+{\cal B}[\{a_2, a_3, a_1\};2] \nn
A_3 & = & {\cal B}[\{a_1, a_2, a_3\};3]+ {\cal B}[\{a_3, a_1, a_2\};3]+{\cal B}[\{a_2, a_3, a_1\};3] \nn
B_1 & = & {\cal B}[\{a_3, a_2, a_1\};1]+ {\cal B}[\{a_1, a_3, a_2\};1]+{\cal B}[\{a_2, a_1, a_3\};1] \nn
B_2 & = & {\cal B}[\{a_3, a_2, a_1\};2]+ {\cal B}[\{a_1, a_3, a_2\};2]+{\cal B}[\{a_2, a_1, a_3\};2] \nn
B_3 & = & {\cal B}[\{a_3, a_2, a_1\};3]+ {\cal B}[\{a_1, a_3, a_2\};3]+{\cal B}[\{a_2, a_1, a_3\};3]~~~~\Label{sym-T3-AB} \eea
When we use \eqref{T3-solve} to \eqref{sym-T3}, we find that
\bea \text{the right hand side of \eqref{sym-T3}}= 3 (\a_1-\a_3) ( {\cal B}[\{a_1,a_2,a_3\};1] -{\cal B}[\{a_2,a_1,a_3\};1])\,.\Label{T3-form-final}\eea
If we impose $3(\a_1-\a_3)=-1$, it is nothing but the
result given in~\eqref{Merge-2-simplify-2}. The arbitrary of $\a_2, \a_3$ coefficients
 shows again there are many different ways to represent the same
result. Using the method presented in this appendix, one can study the equivalence between different representation, some of which may have interesting properties.

\section{Unifying relations}\label{sec:unify}

In the paper \cite{Cheung:2017ems}, the so-called unifying relations are
established between amplitudes of different theories. According to
one of such relations, the tree level multitrace EYM amplitudes
$A_{m,|\mathsf{H}|}(\pmb{1}\,|\,\pmb{2}\,|\dots|\,m\,\Vert\,\mathsf{H})$ can be derived
by acting $m$ trace operators $\mathcal{T}[\pmb{i}]$ onto the Einstein gravity amplitude
$A_{0,|\pmb{1}|+\dots+|\pmb{m}|+|\mathsf{H}|}$:
\bea
A_{m,|\mathsf{H}|}(\pmb{1}\,|\,\pmb{2}\,|\dots|\,\pmb{m}\,\Vert\,\mathsf{H})
=\prod_{i=1}^{m}\mathcal{T}[\pmb{i}]A_{0,|\pmb{1}|+\dots+|\pmb{m}|+|\mathsf{H}|}.
\eea
Assuming that the trace $\pmb{i}$ is given by the ordered set
$\pmb{i}=\{\alpha_1,\alpha_2\ldots\alpha_l\}$, the trace
operator $\mathcal{T}[\pmb{i}]$ that constructs the trace $\pmb{i}$ is defined as:
\bea \mathcal{T}[\pmb{i}]\equiv
\mathcal{T}[\{\alpha_1,\alpha_2\ldots\alpha_l\}]=\mathcal{T}[\alpha_1\alpha_l]
\prod\limits_{i=2}^{l-1}\mathcal{T}[\alpha_{i-1}\alpha_i\alpha_{i+1}]\,,\Label{T-i-Ope}
\eea
where
\bea
\mathcal{T}[ab]\equiv\frac{\partial}{\partial(\epsilon_a\cdot\epsilon_b)}\qquad\qquad\mathcal{T}[abc]\equiv
\frac{\partial}{ \partial(k_a\cdot\epsilon_b)}-\frac{\partial}{
	\partial(k_c\cdot\epsilon_b)}\,.\Label{T-i-Ope-1} \eea
It is worth to emphasize that the ordering of action is from the
left to the right in Eq.~\eqref{T-i-Ope}, namely, we should act
$\mathcal{T}[\alpha_1\alpha_l]$ first and then
$\mathcal{T}[\alpha_{i-1}\alpha_i\alpha_{i+1}]$ from $i=2$ to
$i=l-1$. One can understand this unifying relation by directly acting the $\mathcal{T}$'s onto the reduced Pfaffian $\Pfp(\Psi)$, and compare the result with the two forms of the EYM integrands presented in section~\ref{sec:integrand}.
Alternatively, one can act the $\mathcal{T}$'s onto the recursive expansion of Einstein gravity amplitude~\cite{Fu:2017uzt,Du:2017kpo}, and see how the result groups into the recursive expansion of EYM presented in this paper.
The idea is the following: both recursive expansions can very conveniently give us polynomial BCJ numerators $n^{\text{YM}}$ and $n^{\text{EYM}}$. Meanwhile, $n^{\text{EYM}}$ can also be obtained by acting the $\mathcal{T}$'s onto the YM ones:
\begin{equation}
\left(\prod\mathcal{T}\right)\cdot n^{\text{YM}}=n^{\text{EYM}}\,.
\end{equation}
Our calculation gives an idea on the simplicity of $n^{\text{EYM}}$ obtained by the recursive expansion presented in this paper and the direct $\mathcal{T}$ action on $n^{\text{YM}}$.

Our study shows that a proper expansion of gravity amplitudes, as done in~\cite{Fu:2017uzt,Du:2017kpo}, is crucial to obtain a well-organized result from the unifying relation. The reason is that some of the $\epsilon\cdot\epsilon$ and $\epsilon\cdot k$ factors, where the $\epsilon$'s are ``half polarization'' of gravitons, have been pulled out into the
coefficients, so that the action of the $\mathcal{T}$ operators are significantly simplified. As a concrete example, we show how to obtain the type-I recursive expansion of $A_{2,1}(1,2,3\,|\,4,5,6,7\,\Vert\,8)$ from the gravity amplitude $A_{0,1}(1,2\ldots 8)$. We will give some details on how the outcome of the $\mathcal{T}$-action leads to the patterns like $\pmb{h}\,\W{\shuffle}\,\pmb{K}$ in our recursive expansion.


Let us start with the construction of the first trace
$\pmb{1}=\{1,2,3\}$. To make the calculation as simple as possible,
we must use the proper expansion of graviton amplitude, which is
given by
\begin{align}
\label{eq:A8exp}
&A_{0,8}(1,2,3,4,5,6,7,8)=(\epsilon_1\cdot\epsilon_3)A_{1,6}(1,3\,\Vert\,2,4,5,6,7,8)\nn
&\quad+\sum_{l=1}^6(-1)^l\sum_{\{i_1\dots i_l\}\subset\{2,4,5,6,7,8\}}(\epsilon_1\cdot
F_{\sigma_1}\cdot F_{\sigma_2}\dots\cdot F_{\sigma_l} \cdot
\epsilon_3)A_{1,6-l}(1,\sigma,3\,\Vert\,\{2,4,5,6,7,8\}\backslash
\{i_1\dots i_l\})\,,
\end{align}
where the summation is over all the ordered subset $\{i_1\ldots i_l\}$ of $\{2,4,5,6,7,8\}$. 
The advantage of such a choice is as follows. When applying
the operator $\mathcal{T}[\{1,2,3\}]$ to Eq.~\eqref{eq:A8exp}, only the
first term gives nonzero contribution after acting
$\mathcal{T}[13]=\frac{\partial}{\partial(\epsilon_1\cdot\epsilon_3)}$:
\bea
\mathcal{T}[123]\Big(\mathcal{T}[13]A_{0,8}(1,2,3,4,5,6,7,8)\Big)=\left[\frac{\partial}{
	\partial(k_1\cdot\epsilon_2)}-\frac{\partial}{
	\partial(k_3\cdot\epsilon_2)}\right]A_{1,6}(1,3\,\Vert\,2,4,5,6,7,8)\,.
\eea
To continue, we expand $A_{1,6}(1,3\Vert\,2,4,5,6,7,8)$ by
the single trace recursive expansion~\eqref{eq:singletrace} with $2$
as the fiducial graviton:
\bea A_{1,6}(1,3\,\Vert\,2,4,5,6,7,8)=(\epsilon_2\cdot
k_1)A_{1,5}(1,2,3\,\Vert\,4,5,6,7,8)+\dots\,, \eea
where the ellipsis denotes other terms in the expansion which do not
contain $\epsilon_2\cdot k_1$ (noting that all the coefficients do not
contain $\epsilon_2\cdot k_3$). Again this very careful choice makes
the action of $\mathcal{T}[123]$ very simple. The result of acting $\mathcal{T}[\{1,2,3\}]$ is just the single trace EYM amplitude with five gravitons:
%
\bea
\label{eq:A15_1}
\mathcal{T}[\{1,2,3\}]A_{0,8}(1,2,3,4,5,6,7,8)=A_{1,5}(1,2,3\,\Vert\,4,5,6,7,8)\,.
\eea
We remark that although our example is very
simple, it does show a general pattern on how to reduce the gravity
amplitude to the single trace EYM amplitude using the trace operator $\mathcal{T}$: a proper choice of the single trace recursive expansion given in~\cite{Fu:2017uzt,Teng:2017tbo} is crucial.

Now we construct the second gluon trace $\pmb{2}=\{4,5,6,7\}$ by
acting $\mathcal{T}[\{4,5,6,7\}]$ onto Eq.~\eqref{eq:A15_1}. For this case, the proper choice
to expand it is to take $8$ as the fiducial graviton. Thus we have
\bea &&A_{1,5}(1,2,3\,\Vert\,4,5,6,7,8)=T_1+T_2+T_3+T_4+T_5\,,\eea
where
\begingroup
\allowdisplaybreaks
\begin{subequations}
\bea
T_1&=&(\epsilon_8\cdot Y_8)A_{1,4}(1,\{2\}\shuffle\{8\},3\,\Vert\,4,5,6,7)\\
T_2&=&\sum_{i\in \{4,5,6,7\}}(\epsilon_8\cdot F_i\cdot Y_i)A_{1,3}(1,\{2\}\shuffle\{i,8\},3\,\Vert\,\{4,5,6,7\}\backslash \{i\})\\
T_3&=&\sum_{\substack{i,j\in\{4,5,6,7\}\\i\neq j}}(\epsilon_8\cdot F_i\cdot F_j\cdot Y_j)A_{1,2}(1,\{2\}\shuffle\{j,i,8\},3\,\Vert\,\{4,5,6,7\}\backslash \{i,j\})\\
T_4&=&\sum_{\substack{i,j,k\in\{4,5,6,7\}\\i\neq j\neq k}}(\epsilon_8\cdot F_i\cdot F_j\cdot F_k\cdot Y_k)A_{1,1}(1,\{2\}\shuffle\{k,j,i,8\},3\,\Vert\,\{4,5,6,7\}\backslash \{i,j,k\})\\
T_5&=&\sum_{\substack{i,j,k,l\in\{4,5,6,7\}\\i\neq j\neq k\neq
		l}}(\epsilon_8\cdot F_i\cdot F_j\cdot F_k\cdot F_l\cdot
Y_l)A_{1,0}(1,\{2\}\shuffle\{l,k,j,i,8\},3)\,. \eea
\end{subequations}
\endgroup
We note that each term contains a sum over all the permutations. For example, $T_5$ sums over all the $4!=24$ permutations of $i$, $j$, $k$ and $l$. Now we act the operator $\mathcal{T}[\{4,5,6,7\}]=\mathcal{T}[47]\mathcal{T}[456]\mathcal{T}[567]$ to above terms one by one:
\begin{itemize}
	
	\item First, $T_2$ has to vanish when acted
	by $\mathcal{T}[\{4,5,6,7\}]$, because in $T_2$, $\epsilon_7$ can only be
	contracted with either $\epsilon_8$ or $Y_i$. Neither case survives
	under the action of $\mathcal{T}[47]$.
	
	\item In $T_3$, only the contribution
	with $\{i,j\}=\{4,7\}$ and $\{7,4\}$ survive under the action of $\mathcal{T}[4,5,6,7]$ and gives rise
	\begin{equation}
	\mathcal{T}[\{4,5,6,7\}]\,T_3=\mathcal{T}[456]\mathcal{T}[567]\mathcal{T}[47]\Big[(\epsilon_8\cdot
	F_7\cdot F_4\cdot
	Y_4)A_{1,2}(1,\{2\}\shuffle\{4,7,8\},3\,\Vert\,5,6)+(4\leftrightarrow7)\Big]\,,
	\end{equation}
	where it is easy to show that:
	\begin{align}
	\mathcal{T}[47](\epsilon_8\cdot F_7\cdot F_4\cdot Y_4)=(\epsilon_8\cdot k_7)(-k_4\cdot Y_4)\,,& &\mathcal{T}[47](\epsilon_8\cdot F_4\cdot F_7\cdot Y_7)=(\epsilon_8\cdot k_4)(-k_7\cdot Y_7)\,.
	\end{align}
	The  expression
	$\mathcal{T}[456]\mathcal{T}[567]A_{1,2}(1,\{2\}\shuffle\{4,7,8\},3\,\Vert\,5,6)$
	should be evaluated by further using the single-trace recursive expansion. The
	calculation is be similar and we do not give the
	details. One can find that the final result of this calculation is:
	\begin{align}
	\mathcal{T}[\{4,5,6,7\}]T_3&=(\epsilon_8\cdot k_7)(-k_4\cdot
	Y_4)A_{1,0}(1,\{2\}\shuffle\{4,5,6,7,8\},3)\nonumber\\
	&\quad +(\epsilon_8\cdot k_4)(-k_7\cdot
	Y_7)A_{1,0}(1,\{2\}\shuffle\{7,6,5,4,8\},3)\,.\Label{eq:A14-21}
	\end{align}	
	%
	%
	%
	%

	\item In $T_4$, only those terms satisfying both the following
	two conditions survive under the action of $\mathcal{T}[4,5,6,7]$: (1) Both $4$ and $7$
	belong to the ordered set $\{i,j,k\}$ (2) The ordered set $\{i,j,k\}$ is a segment of
	either $\pmb{2}=\{4,5,6,7\}$ or its reflection $\pmb{2}^T=\{7,6,5,4\}$.
	The $\{i,j,k\}$ satisfying these conditions are
	\bea \{7,4,5\},~~\{6,7,4\},~~\{5,4,7\},~~\{4,7,6\}, \eea
	where the first two are segments of $\pmb{2}$ while the
	last two are segments of $\pmb{2}^T$. Doing similar
	calculation we find
	\begin{subequations}
	\label{eq:A15}
	\bea &&\mathcal{T}[\{4,5,6,7\}]\Bigl[(\epsilon_8\cdot F_7\cdot
	F_4\cdot F_5\cdot
	Y_5)A_{1,1}(1,\{2\}\shuffle\{5,4,7,8\},3\,\Vert\,6)\Bigr]\nn
	&=&-(\epsilon_8\cdot k_7)(-k_5\cdot Y_5)
	A_{1,0}(1,\{2\}\shuffle\{5,\{4\}\shuffle\{6\},7,8\},3)\,,\Label{eq:A15-31}
	\eea
	\bea &&\mathcal{T}[\{4,5,6,7\}]\Bigl[(\epsilon_8\cdot F_6\cdot
	F_7\cdot F_4\cdot
	Y_4)A_{1,1}(1,\{2\}\shuffle\{4,7,6,8\},3\,\Vert\,5)\Bigr]\nn
	&=&-(\epsilon_8\cdot k_6)(-k_4\cdot
	Y_4)A_{1,0}(1,\{2\}\shuffle\{4,\{7\}\shuffle\{5\},6,8\},3)\,,\Label{eq:A15-32}
	\eea
	\bea &&\mathcal{T}[\{4,5,6,7\}]\Bigl[(\epsilon_8\cdot F_5\cdot
	F_4\cdot F_7\cdot
	Y_7)A_{1,1}(1,\{2\}\shuffle\{7,4,5,8\},3\,\Vert\,6)\Bigr]\nn
	&=&-(\epsilon_8\cdot k_5)(-k_7\cdot
	Y_7)A_{1,0}(1,\{2\}\shuffle\{7,\{4\}\shuffle\{6\},5,8\},3)\,,\Label{eq:A15-33}
	\eea
	\bea &&\mathcal{T}[\{4,5,6,7\}]\Bigl[(\epsilon_8\cdot F_4\cdot
	F_7\cdot F_6\cdot
	Y_6)A_{1,1}(1,\{2\}\shuffle\{6,7,4,8\},3\,\Vert\,5)\Bigr]\nn
	&=&-(\epsilon_8\cdot k_4)(-k_6\cdot
	Y_6)A_{1,0}(1,\{2\}\shuffle\{6,\{7\}\shuffle\{5\},4,8\},3)\,.\Label{eq:A15-34}
	\eea
		\end{subequations}
	%


	\item In $T_5$, similar to the previous case, only the terms satisfying
	(1) Both $4$ and $7$ belong to $\{i,j,k,l\}$ (2) $\{i,j,k,l\}$ is a segment of either the trace
	$\pmb{2}=\{4,5,6,7\}$ or its reflection $\pmb{2}^T=\{7,6,5,4\}$ survive under the action
	of $\mathcal{T}[\{4,5,6,7\}]$. There are six different choices:
	\bea
	\{7,4,5,6\},~~\{6,7,4,5\},~~\{5,6,7,4\},~~\{4,7,6,5\},~~\{5,4,7,6\},~~\{6,5,4,7\}.
	\eea
	The first three are cyclic equivalent with $\pmb{2}$, while
	the last three are cyclic equivalent with $\pmb{2}^T$. When
	acted by $\mathcal{T}[\{4,5,6,7\}]$, we obtain
	\begin{align}
	&\quad\mathcal{T}[\{4,5,6,7\}]T_5\nn
	&=(\epsilon_8\cdot k_7)(-k_6\cdot
	Y_6)A_{1,0}(1,\{2\}\shuffle\{6,5,4,7,8\},3)+(\epsilon_8\cdot
	k_6)(-k_5\cdot
	Y_5)A_{1,0}(1,\{2\}\shuffle\{5,4,7,6,8\},3)\nn
	&+(\epsilon_8\cdot k_5)(-k_4\cdot
	Y_4)A_{1,0}(1,\{2\}\shuffle\{4,7,6,5,8\},3)+(\epsilon_8\cdot
	k_4)(-k_5\cdot
	Y_5)A_{1,0}(1,\{2\}\shuffle\{5,6,7,4,8\},3)\nn
	&+(\epsilon_8\cdot k_5)(-k_6\cdot
	Y_6)A_{1,0}(1,\{2\}\shuffle\{6,7,4,5,8\},3)+(\epsilon_8\cdot
	k_6)(-k_7\cdot
	Y_7)A_{1,0}(1,\{2\}\shuffle\{7,4,5,6,8\},3)\,.\Label{eq:A15-41}
	\end{align}

	\item Now we are left with $\mathcal{T}[4,5,6,7]\,T_1$. To carry
	out, we use the recursive expansion of single trace EYM
	amplitudes as above and do similar calculations. There will be
	many terms, but one can check that when summing all them
	together, we get nothing but
	\bea \mathcal{T}[4,5,6,7]\,T_1=(\epsilon_8\cdot
	Y_8)A_{2,0}(1,\{2\}\shuffle\{8\},3\,|\,4,5,6,7)\,.\Label{eq:A15-1}
	\eea
	%
	

\end{itemize}
When putting everything together and compare with our type-I
recursive expansion
\begin{align}
A_{2,1}(1,2,3\,|\,4,5,6,7\,\Vert 8)&=(\epsilon_8\cdot F_8)A_{2,0}(1,\{2\}\shuffle\{8\},3\,|\,4,5,6,7)\nn
&\quad +\underset{\{a,b\}\subset\pmb{2}}{\widetilde{\sum}}(\epsilon_8\cdot
k_{b})(-k_{a_2}\cdot
Y_{a_2})A_{1,0}(1,\{2\}\shuffle\{a_2,\mathsf{KK}[\pmb{2},a_2,b_2],b_2,8\},3).\Label{Eq:A21}
\end{align}
we find the exact matching:
\begin{itemize}
	\item Eq.~\eqref{eq:A15-1} gives the first line of the
	expansion~\eqref{Eq:A21}.
	\item The two terms in Eq.~\eqref{eq:A14-21} give the $(a_2,b_2)=(4,7)$ and
	$(7,4)$ parts of the second line of Eq.~\eqref{Eq:A21}.
	\item The four terms in Eq.~\eqref{eq:A15} give the $(a_2,b_2)=(5,7)$, $(7,5)$, $(4,6)$ and $(6,4)$ parts of the second line of Eq.~\eqref{Eq:A21}.
	\item The six terms in Eq.~\eqref{eq:A15-41} give the $(a_2,b_2)=(6,7)$, $(5,6)$, $(4,5)$, $(5,4)$,
	$(6,5)$ and $(7,6)$ parts of the second line of Eq.~\eqref{Eq:A21}.
\end{itemize}

\bibliographystyle{JHEP}
\bibliography{Refs}

\end{document}